\documentclass{article}
\usepackage{amsfonts,lscape,inputenc}
\usepackage{amsmath}
\usepackage{amssymb}
\usepackage{graphicx}
\usepackage[onehalfspacing]{setspace}
\usepackage[top=1.0in, bottom=1.0in, left=1in, right=1in]{geometry}

\providecommand{\U}[1]{\protect\rule{.1in}{.1in}}

\newenvironment{proof}[1][Proof]{\noindent\textbf{#1.} }{\ \rule{0.5em}{0.5em}}
\begin{document}

\title{{\huge Double robust inference for continuous updating GMM}}
\author{Frank Kleibergen\thanks{Corresponding author. Email: f.r.kleibergen@uva.nl. Phone: (+31)205254397. Amsterdam School of Economics, University of Amsterdam, Roetersstraat 11, 1018 WB Amsterdam, The Netherlands. }\ \ \ \ \ \ \ \ \ Zhaoguo Zhan\thanks{Email: zzhan@kennesaw.edu. Department of Economics, Finance, and Quantitative Analysis, Coles College of Business, Kennesaw State University, GA 30144, USA.}}
\date{\today}
\maketitle

\begin{abstract} 
We propose the double robust Lagrange multiplier (DRLM)\ statistic for
testing hypotheses specified on the pseudo-true value of the structural
parameters in the generalized method of moments.  The pseudo-true value is defined as the minimizer of the population continuous updating objective function  and equals the true value of the structural parameter in the absence of misspecification.\nocite{hhy96} The (bounding) $\chi ^{2}$ limiting distribution of the DRLM\
statistic is robust to both misspecification and weak identification of the
structural parameters, hence its name. To emphasize its
importance for applied work, we use the DRLM test to analyze the return on
education, which is often perceived to be weakly identified, using data from
Card (1995) where misspecification occurs in case of treatment heterogeneity; and to analyze the risk premia associated with risk factors
proposed in Adrian et al. (2014) and He et al. (2017), where both misspecification and weak
identification need to be addressed.
\end{abstract}



\setstretch{1.63}


\section{Introduction}

Little more than twenty years ago, inference procedures for analyzing
possibly weakly identified structural parameters using the generalized
method of moments (GMM)\ of Hansen (1982)\nocite{han82} were mostly lacking.
Since then huge progress has been made to develop such procedures, see e.g.
Staiger and Stock (1997),\nocite{stst97} Dufour (1997),\nocite{duf97} Stock
and Wright (2000),\nocite{sw00} Kleibergen (2002, 2005, 2009),\nocite{kf00b}
Moreira (2003),\nocite{mor01} Andrews and Cheng (2012), \nocite{andchen12}
Andrews and Mikusheva (2016a, b), and Han and McCloskey (2019).\nocite{am16}\nocite{am16b}\nocite{HanMcClos19} At present,
we therefore have a variety of so-called weak identification robust
inference methods. Given the prevalence of weak identification in applied
work, a lot of emphasis has also been put in raising awareness amongst
practitioners, see e.g. Kleibergen and
Mavroeidis (2009),\nocite{kmav09} Beaulieu et al. (2013),\nocite{bdk13} Mavroeidis et al. (2014),\nocite{mpms14}
Andrews et al. (2019),\nocite{ass19} and Kleibergen and Zhan\ (2020).\nocite%
{kz19}

The weak identification robust inference procedures in GMM lead to inference
that is centered around the continuous updating estimator (CUE) of Hansen et
al. (1996).\nocite{hhy96} GMM requests the moment condition to hold at a
(unknown) true value of the parameter which is then also the minimizer of
the population continuous updating objective function. The inference
resulting from weak identification robust inference procedures concerning
hypotheses specified on the true value of the structural parameters remains
reliable under varying degrees of identification. When there is no value of
the structural parameters where the GMM\ moment conditions exactly hold, the
structural model is rendered misspecified and we refer to the minimizer of
the (population continuous updating) GMM objective function as the
pseudo-true value. The pseudo-true value depends on the (population)
objective function at hand and different objective functions lead to
distinct pseudo-true values. We use the minimizer of the population
continuous updating objective function as the pseudo-true value because of
its invariance properties and since weak identification robust tests lead to
inference that is centered around it. In case of misspecification, these
inference procedures for testing hypotheses specified on the pseudo-true
value become size distorted for just small amounts of misspecification. This
would not sound as much of a problem if it was possible to efficiently
detect such misspecification. This is, however, not so since
misspecification tests, like the Sargan-Hansen test (Sargan (1958) and
Hansen (1982)), \nocite{sar58}\nocite{han82} are virtually powerless in
settings of joint misspecification and weak identification; see Gospodinov
et al. (2017).\nocite{gkr17} Weak identification robust inference procedures
thus came about to overcome the general critique of non-robustness of
traditional inference procedures to varying identification strengths, see
e.g. Staiger and Stock (1997) and Dufour (1997),\nocite{duf97}\nocite{stst97}
but are similarly non-robust to misspecification.

Arguably, the first to emphasize the importance of misspecification in the
presence of weak (or no) identification were Kan and Zhang (1999).\nocite%
{kz99} With the surge in applied work on structural estimation, awareness of
misspecification has grown further, see Hall and Inoue (2003).\nocite%
{hall2003large} In asset pricing models, for example, it is now generally
accepted that misspecification, alongside weak identification, is an
important empirical issue, see e.g. Kan et al. (2013) and Kleibergen and
Zhan (2020). Kan et al. (2013) therefore developed misspecification robust $t
$-statistics for the Fama-MacBeth (FM) (1973)\nocite{fm73} two-pass
estimator, $i.e.$ the typical estimator employed to estimate risk premia in
linear asset pricing models. Similarly, Hansen and Lee (2021)\nocite{hl21}
construct the limiting distribution of an iterated GMM\ estimator in
misspecified GMM, which can be used to conduct Wald tests on the pseudo-true
value of the structural parameters; furthermore, Evdokimov and Koles\'{a}r (2018)\nocite%
{evkol18} and Lee (2018)\nocite{lee18} analyze testing the treatment effect
resulting from multiple instruments whose local average treatment
effects might differ, leading to misspecification of the moment equation of
the underlying linear instrumental variables (IV) regression model. These
misspecification robust tests on the pseudo-true
value are, however, not robust to weak
identification, so identical to the weak identification robust inference
procedures, they cannot deal with the empirically relevant setting of both
misspecification and weak identification for which Hansen and Lee (2021)
state \textquotedblleft \ldots\ this extension would be desirable but
considerably more challenging.\textquotedblright\ 

In this paper, we therefore extend the
weak identification robust score or Lagrange multiplier (KLM) test from
Kleibergen (2002, 2005, 2009) to a double robust Lagrange multiplier (DRLM)
test. This DRLM\ test is size correct and robust to both misspecification
and weak identification, hence its name.  The DRLM\ statistic is a quadratic
form of the score function, which equals zero at all stationary points of the
CUE sample objective function. This is also the case for the KLM\ statistic
and explains the power problems of the KLM\ test, see e.g. Andrews et al.
(2006).\nocite{andms05} To overcome the power problems of the KLM test, the
KLM statistic can be combined in a conditional or unconditional manner with
the Anderson-Rubin (AR) (1949) statistic, see e.g. Andrews (2016). \nocite%
{and15}\nocite{AR49} Andrews et al. (2006) show that the conditional
likelihood ratio test of Moreira (2003) provides the optimal manner of
combining these statistics for the homoskedastic linear IV regression model with one included endogenous variable. We use the
maximal invariant to show that in case of misspecification, it is not
obvious how to improve the power of the DRLM test by such combination
arguments, since the statistics with which the DRLM statistic is to be
combined to improve power have non-central limiting distributions with
parameters that cannot be consistently estimated under misspecification. We
therefore improve the power of the DRLM\ test by exploiting the
specification of the derivative of the DRLM\ statistic with respect to the
structural parameters.

The rest of the paper is organized as follows. In the second section, we
present continuous updating GMM with misspecification, and discuss how and when a
structural interpretation can be obtained from the pseudo-true value. We introduce a measure of the identification strength which has to (considerably) exceed the minimal value of the population continuous updating objective function for the pseudo-true value to be structurally interpretable. In the
third section, we introduce the DRLM test and prove that it is size correct. For ease of exposition,  we also illustrate the latter using a simulation experiment. The
fourth section conducts a power study of the DRLM\ test and other weak
identification robust tests. It shows that weak identification robust tests
on the pseudo-true value of the structural parameters are size distorted for
just small amounts of misspecification while the DRLM\ test is not. It also
proposes the power improvement rule and shows that the resulting test
procedure has generally good power. The fifth section  conducts a simulation
experiment using nonlinear GMM with an asset pricing Euler moment equation
that results from a constant relative rate of risk aversion (CRRA) utility
function. The sixth section applies the DRLM\ test to risk premia using
asset pricing data from Adrian et al. (2014)\nocite{aem14} and He et al.
(2017)\nocite{hkm17}, and to analyze the return on education using data from
Card (1995) for which local average treatment effects that differ over the
instruments can lead to misspecification, see Imbens and Angrist (1994).\nocite{imang94}
Especially for the risk premium parameters, we show that usage of other
inference procedures understates the uncertainty of the risk measures
because of the misspecification and weak identification present. The seventh
section concludes. Technical details and additional material are relegated to the Online
Appendix.

\section{GMM with potential misspecification}

We analyze the $m\times 1$ parameter vector $\theta =(\theta _{1}\ldots
\theta _{m})^{\prime }$ whose parameter region is the $\mathbb{R}^{m}.$ The $%
k_{f}\times 1$ dimensional function $f(.,.)$ is a continuously
differentiable function of the parameter vector $\theta $ and a Borel
measurable function of a data vector $X_{t}$ which is observed for
time/individual $t.$ Since we focus on misspecification, the model is overidentified, i.e. there are more moment equations than structural parameters so 
$k_{f}>m.$ The population moment function of $f(\theta ,X_{t})$ equals $\mu
_{f}(\theta ):$%
\begin{equation}
E_{X}(f(\theta ,X_{t}))=\mu _{f}(\theta ),  \label{euler}
\end{equation}%
with $\mu _{f}(\theta )$ a $k_{f}$-dimensional continuously differentiable
function. Unlike regular GMM, see Hansen (1982)\nocite{han82}, we do not
request that there is a specific value of $\theta ,$ say $\theta _{0},$ at
which $\mu _{f}(\theta _{0})=0.$ Our analysis thus differs from a recent one
proposed by Cheng et al. (2021),\nocite{cdl21} who construct a model
selection procedure for evaluating  potentially misspecified models
with possibly weakly identified structural parameters, which explicitly uses
a set of base moments contained in all considered models that are guaranteed
to hold. We analyze $\theta $ using the continuous updating setting of
Hansen et al. (1996).\nocite{hhy96} We use it because of its invariance
properties and since it leads to inference using identification robust
statistics in standard GMM, see e.g. Stock and Wright (2000)\nocite{sw00}
and Kleibergen (2005). The accompanying population continuous updating
objective function is:%
\begin{equation}
Q_{p}(\theta )=\mu _{f}(\theta )^{\prime }V_{ff}(\theta )^{-1}\mu
_{f}(\theta ),  \label{popobj}
\end{equation}%
with $V_{ff}(\theta )$ the covariance matrix of the sample moment $f_{T}(\theta ,X)=\frac{1}{T}\sum_{t=1}^{T}f_{t}(\theta ),$ $%
f_{t}(\theta )=f(\theta ,X_{t})$:\footnote{%
Throughout the paper, we use recentered covariance matrices while the
continuous updating estimator is identical under a recentered or uncentered
version of the covariance matrix estimator; see Theorem 1 of Hansen and Lee (2021).%
\nocite{hl21} }
\begin{equation}
\begin{array}{rl}
V_{ff}(\theta )= & \lim_{T\rightarrow \infty }E\left[ T\left( f_{T}(\theta
,X)-\mu _{f}(\theta )\right) \left( f_{T}(\theta ,X)-\mu _{f}(\theta
)\right) ^{\prime }\right] ,%
\end{array}
\label{covmat}
\end{equation}%
so $f_{T}(\theta ,X)$ is the sample analog of $\mu _{f}(\theta )$ for a data
set of $T$ observations: $X_{t},$ $t=1,\ldots ,T.$

We define the pseudo-true value of $\theta ,$ $\theta ^{\ast },$ as the
minimizer of the population objective function:%
\begin{equation}
\theta ^{\ast }=\arg \min_{\theta \in \mathbb{R}^{m}}\text{ }Q_{p}(\theta ).
\label{pseudo}
\end{equation}%
Lateron we discuss if this $\theta ^{\ast }$ is our object of interest, which
depends amongst others on whether a measure of the amount of misspecification is less than a
measure of the strength of identification. The minimizer of the population
objective function satisfies the first order condition (FOC) stated in
Theorem 1.

\paragraph{Theorem 1:}
The FOC (divided by two) for a stationary point $\theta ^{s}$ of the
population objective function reads:%
\begin{equation}
\begin{array}{c}
\frac{1}{2}\frac{\partial }{\partial \theta ^{\prime }}Q_{p}(\theta ^{s})=0\
\ \ \Leftrightarrow \ \ \ \mu _{f}(\theta ^{s})^{\prime }V_{ff}(\theta
^{s})^{-1}D(\theta ^{s})=0,%
\end{array}
\label{foc cue}
\end{equation}%
with%
\begin{equation}
\begin{array}{rl}
D(\theta )= & J(\theta )-\left[ V_{\theta _{1}f}(\theta )V_{ff}(\theta
)^{-1}\mu _{f}(\theta )\ldots V_{\theta _{m}f}(\theta )V_{ff}(\theta
)^{-1}\mu _{f}(\theta )\right]%
\end{array}
\label{dcue}
\end{equation}%
and $J(\theta )=\frac{\partial }{\partial \theta ^{\prime }}\mu _{f}(\theta
),$%
\begin{equation}
\begin{array}{rlll}
V_{\theta _{i}f}(\theta )= & \lim_{T\rightarrow \infty }E\left[ T(\frac{%
\partial }{\partial \theta _{i}}(f_{T}(\theta ,X)-\mu _{f}(\theta )))\left(
f_{T}(\theta ,X)-\mu _{f}(\theta )\right) ^{\prime }\right] , &  & 
i=1,\ldots ,m.%
\end{array}
\label{covd}
\end{equation}

\begin{proof}
See the Online Appendix and Kleibergen (2005)\nocite{kf00a}.\smallskip
\end{proof}

Theorem 1 shows that if there is a unique value of $\theta ,$ $\theta _{0},$
for which $\mu _{f}(\theta _{0})=0$, then also $\theta ^{\ast }=\theta _{0}$
and $D(\theta _{0})=J(\theta _{0}).$ The misspecification thus implies that
the recentered Jacobian $D(\theta ^{\ast })$ differs from the population
Jacobian that results from the moment equations, $J(\theta ^{\ast }),$ in
other instances.

\subsection{Running example 1: Linear asset pricing model}

The linear asset pricing model shows the extent to which the mean of an $%
(N+1) $-dimensional vector of asset returns $\mathcal{R}_{t}$ is spanned by
the betas of $m$ risk factors contained in the $m$-dimensional vector $%
F_{t}. $ It is reflected by the moment function:%
\begin{equation}
\mu_{f}(\lambda_{0},\lambda_{F})=E(\mathcal{R}_{t})-\iota_{N+1}\lambda _{0}-%
\mathcal{B\lambda}_{F},  \label{mean with zero beta}
\end{equation}
with $\iota_{N+1}$ an $(N+1)$-dimensional vector of ones, $\mathcal{B}$ an $%
(N+1)\times m$ dimensional matrix:%
\begin{equation}
\mathcal{B=}\text{cov(}\mathcal{R}_{t},F_{t})\text{var(}F_{t})^{-1},
\label{betaspec}
\end{equation}
and $\lambda_{0}$ is the zero-beta return, $\lambda_{F}$ is the $m$%
-dimensional vector of risk premia.

The asset pricing moment equation in (\ref{mean with zero beta}) can be more
compactly written by removing the zero-beta return which we accomplish by
taking the asset returns in deviation of the $(N+1)$-th asset return:%
\footnote{
Our results are invariant with
respect to the asset return which is subtracted; see Kleibergen and Zhan
(2020).\nocite{kz19}} 
\begin{equation}
\begin{array}{ccccc}
R_{t}=\left( 
\begin{array}{c}
\mathcal{R}_{1t} \\ 
\vdots \\ 
\mathcal{R}_{Nt}%
\end{array}
\right) -\iota_{N}\mathcal{R}_{(N+1)t}, &  &  &  & \beta=\left( 
\begin{array}{c}
\mathcal{B}_{1} \\ 
\vdots \\ 
\mathcal{B}_{N}%
\end{array}
\right) -\iota_{N}\mathcal{B}_{N+1},%
\end{array}
\label{returns}
\end{equation}
for $\mathcal{R}_{t}=(\mathcal{R}_{1t}\ldots\mathcal{R}_{(N+1)t})^{\prime},$ 
$\mathcal{B}=(\mathcal{B}_{1}^{\prime}\ldots\mathcal{B}_{N+1}^{\prime
})^{\prime}.$ The removal of the zero-beta return leads to the moment function:%
\begin{equation}
\begin{array}{c}
\mu_{f}(\lambda_{F})=\mu_{R}-\beta\lambda_{F},%
\end{array}
\label{meandef}
\end{equation}
with $\mu_{R}=E(R_{t})$ and $\beta=cov(R_{t},F_{t})var(F_{t})^{-1}$.

The mean asset returns are not necessarily fully spanned by the $\beta $'s. We therefore analyze the pseudo-true value of the risk premia $\lambda
_{F}^{\ast }$ which is the minimizer of the population continuous updating
objective function:%
\begin{equation}
Q_{p}(\lambda _{F})=(\mu _{R}-\beta \lambda _{F})^{\prime }\left[ \text{Var}%
\left( \sqrt{T}\left( \bar{R}-\hat{\beta}\lambda _{F}\right) \right) \right]
^{-1}(\mu _{R}-\beta \lambda _{F}),  \label{objective}
\end{equation}%
since $f_{T}(\lambda _{F},X)=\bar{R}-\hat{\beta}\lambda _{F},$ with $\bar{R}=%
\frac{1}{T}\sum_{t=1}^{T}R_{t}$ and $\hat{\beta}=\frac{1}{T}\sum_{t=1}^{T}%
\bar{R}_{t}\bar{F}_{t}^{\prime }\left( \frac{1}{T}\sum_{j=1}^{T}\bar{F}_{j}%
\bar{F}_{j}^{\prime }\right) ^{-1},$ $\bar{R}_{t}=R_{t}-\bar{R},$ $\bar{F}%
_{t}=F_{t}-\bar{F},$ $\bar{F}=\frac{1}{T}\sum_{t=1}^{T}F_{t}.$   The
population continuous updating objective function results from a generalized
reduced rank problem, see also Kleibergen (2007):\nocite{kf04}%
\begin{equation}
\begin{array}{c}
Q_{p}(\lambda _{F})=\min_{D\in \mathbb{R}^{N\times m}}Q_{p}(\lambda _{F},D)%
\end{array}
\label{rank 1}
\end{equation}%
with $D(\lambda _{F})=\arg \min_{D\in \mathbb{R}^{N\times m}}Q_{p}(\lambda
_{F},D)$ and%
\begin{equation}
\begin{array}{cl}
Q_{p}(\lambda _{F},D)= & \left[ \text{vec}\left( \left( \mu _{R}\text{ }%
\vdots \text{ }\beta \right) +D\left( \lambda _{F}\text{ }\vdots \text{ }%
I_{m}\right) \right) \right] ^{\prime }\left[ \text{Var}\left( \sqrt{T}%
\left( \bar{R}^{\prime }\text{ }\vdots \text{ vec(}\hat{\beta})^{\prime
}\right) ^{\prime }\right) \right] ^{-1} \\ 
& \left[ \text{vec}\left( \left( \mu _{R}\text{ }\vdots \text{ }\beta
\right) +D\left( \lambda _{F}\text{ }\vdots \text{ }I_{m}\right) \right) %
\right] .%
\end{array}
\label{rank 2}\textbf{}
\end{equation}%

The minimal value of  (\ref%
{rank 2}) over $(\lambda _{F},$ $D)$  is invariant  to the reduced rank specification
implied by $D(\lambda _{F}$ $\vdots $ $I_{m}).$ When using another reduced
rank specification, say, $A(I_{m}$ $\vdots $ $\phi ),$ with $A$ an $N\times m
$ matrix and $\phi $ an $m$-dimensional vector, it leads to an identical value
of the optimized objective function over $(\phi ,$ $A)$. Hence, restrictions
imposed on this specification, like, for example, $\phi _{1}=0,$ with $\phi
_{1}$ the top element of $\phi ,$ which imposes a reduced rank value on just 
$\beta ,$ lead to a larger (or equal) value of the minimized objective
function. This restricted specification is thus such that the objective
function reflects the identification strength of $\lambda _{F}$ as reflected
by the distance of $\beta $ from a reduced rank value. If the minimal value of the
objective function in (\ref{rank 1}) coincides with the one resulting from 
this restricted specification, some or even all elements of the resulting pseudo-true value $\lambda _{F}^{\ast }$ will be very large or even infinite since they now result from a reduced rank value of $\beta$, and do not reflect risk premia. For the pseudo-true
value $\lambda _{F}^{\ast }$ to reflect risk premia, so to have a structural
interpretation and be of interest, the strength of identification has
 to exceed  a measure of the amount of
misspecification. We can therefore use  the minimal value of the population continuous updating objective function resulting from (\ref{rank 1}) as a measure of the amount of
misspecification, and  compare  it  with a measure of the identification strength ($IS$), whose sample analog corresponds with a statistic testing the rank of $\beta$, see e.g. Cragg and Donald (1997), Kleibergen and Paap (2006), and Robin and
Smith (2000):\footnote{In the homoskedastic linear IV regression model with one included endogenous variable, the counterpart for the identification measure $``IS"$ in (\ref{rank 3}) equals the number of instruments times the population analog of the first stage $F$-statistic.}\nocite{cradon97}\nocite{kpaap02}\nocite{robsmit00}
\begin{equation}%
\begin{array}
[c]{rl}%
IS= & \min_{\xi\in\mathbb{R}^{(m-1)}}Q_{r}(\xi)\\
Q_{r}(\xi)= & \binom{1}{-\xi}^{\prime}\beta^{\prime}\left[  \left(  \binom{1}{-\xi
}\otimes I_{N}\right)  ^{\prime}\text{Var}\left(  \sqrt{T}\text{vec(}\hat{\beta})\right)
\left(  \binom{1}{-\xi}\otimes I_{N}\right)  \right]  ^{-1}\beta\binom{1}%
{-\xi}\\
= & \min_{G\in\mathbb{R}^{N\times(m-1)}}Q_{r}(\xi,G)\\
Q_{r}(\xi,G)= & \left[  \text{vec}\left(  \beta+G\left(  \xi\text{ }%
\vdots\text{ }I_{m-1}\right)  \right)  \right]  ^{\prime}\left[
\text{Var}\left(  \sqrt{T}\text{vec(}\hat{\beta})\right)  \right]
^{-1}\left[  \text{vec}\left(  \beta+G\left(  \xi\text{ }\vdots\text{ }%
I_{m-1}\right)  \right)  \right].
\end{array}
\label{rank 3}%
\end{equation}
The Online Appendix provides a proof that the $IS$ identification strength measure in (\ref{rank 3}) equals the minimal value of the restricted objective function alluded to previously, where we used the reduced rank specification $A(I_{m}$ $\vdots $ $\phi ),$ with $A$ an $N\times m
$ matrix and $\phi $ an $m$-dimensional vector with its top element restricted to zero. The identification strength measure is thus always larger than or equal to the minimal value of the population continuous updating objective function. When the minimal value of the population continuous updating objective function is then just slightly smaller than $IS$ in (\ref{rank 3}), we have to be cautious with interpreting the pseudo-true value as risk premia which is then also reflected by their very large values. We next further illustrate this for a simplified setting of the linear asset pricing model.


When $\mu_{F}=E(F_{t})=0$ and $\hat{\beta}$ results from the regression of $%
\bar{R}_{t}$ on $\bar{F}_{t}$ in which the error term is assumed to be
i.i.d. with $N\times N$ dimensional covariance matrix $\Omega,$ Lemma 1 in
the Online Appendix shows that $\bar{R}$ and $\hat{\beta}$ are independently
normally distributed in large samples, see also Shanken (1992) and
Kleibergen (2009). \nocite{sh92}\nocite{kf09} The population continuous
updating objective function (\ref{objective}) then simplifies to:%
\begin{equation}
\begin{array}{c}
Q_{p}(\lambda_{F})=\frac{1}{1+\lambda_{F}^{\prime}Q_{\bar{F}\bar{F}%
}^{-1}\lambda_{F}}(\mu_{R}-\beta\lambda_{F})^{\prime}\Omega^{-1}(\mu_{R}-%
\beta\lambda_{F}),%
\end{array}
\label{ob iid}
\end{equation}
with $Q_{\bar{F}\bar{F}}=$var($F_{t}),$ so its minimal value equals the
smallest root of the characteristic polynomial:%
\begin{equation}
\begin{array}{cc}
\left\vert \tau\left( 
\begin{array}{cc}
1 & 0 \\ 
0 & Q_{\bar{F}\bar{F}}^{-1}%
\end{array}
\right) -\left( \mu_{R}\text{ }\vdots\text{ }\beta\right)
^{\prime}\Omega^{-1}\left( \mu_{R}\text{ }\vdots\text{ }\beta\right)
\right\vert & =0.%
\end{array}
\label{popchar}
\end{equation}

\paragraph{Proposition 1.}

Using a value of $\lambda_{F},$ $\lambda_{F}^{s},$ that satisfies the FOC in
Theorem 1, the smallest root of the characteristic polynomial in (\ref%
{popchar})\ equals either 
\begin{equation}
\begin{array}{c}
\frac{1}{1+\lambda_{F}^{s\prime}Q_{\bar{F}\bar{F}}^{-1}\lambda_{F}^{s}}%
\left( \mu_{R}-\beta\lambda_{F}^{s}\right) ^{\prime}\Omega^{-1}\left(
\mu_{R}-\beta\lambda_{F}^{s}\right)%
\end{array}
\label{root1}
\end{equation}
or the smallest root of the characteristic polynomial:%
\begin{equation}
\begin{array}{lc}
\left\vert \tau(Q_{\bar{F}\bar{F}}+\lambda_{F}^s\lambda_{F}^{s%
\prime})^{-1}-D(\lambda_{F}^{s})^{\prime}\Omega^{-1}D\left(
\lambda_{F}^{s}\right) \right\vert & =0,%
\end{array}
\label{root2}
\end{equation}
with $D(\lambda_{F})=-\beta-\left( \mu_{R}-\beta\lambda_{F}\right)
\lambda_{F}^{\prime}Q_{\bar{F}\bar{F}}^{-1}(1+\lambda_{F}^{\prime}Q_{\bar {F}%
\bar{F}}^{-1}\lambda_{F})^{-1}=-(\beta Q_{\bar{F}\bar{F}}+\mu_{R}%
\lambda_{F}^{\prime})(Q_{\bar{F}\bar{F}}+\lambda_{F}\lambda_{F}^{%
\prime})^{-1}.$\smallskip

\begin{proof}
The rewriting of (\ref{popchar}) to obtain the above is conducted in the
Online Appendix.\smallskip
\end{proof}

Without misspecification, there is a value of $\lambda _{F}^{s}$ for which (%
\ref{root1}) is equal to zero, so it is the smallest root of the
characteristic polynomial. Proposition 1 therefore shows that in models with
misspecification, the minimizer of the population objective function is not
necessarily our object of interest or put differently, has a structural interpretation. For example, when $m=1,$ $\beta =0$ and $\mu _{R}\neq 0,$
the roots of the characteristic polynomial in (\ref{popchar}) equal zero,
attained when $\lambda _{F}\rightarrow \pm \infty $; and $\mu _{R}^{\prime
}\Omega ^{-1}\mu _{R},$ attained at $\lambda _{F}=0$. The smallest root
then corresponds with the $IS$ identification strength measure in (\ref{rank 3}), so the resulting pseudo-true value $\lambda _{F}\rightarrow \pm \infty $ cannot be interpreted as a risk premium. The pseudo-true
value is only of interest when it has a structural interpretation, so it
represents risk premia, which occurs when the $IS$ identification strength measure (\ref{rank 3}) strictly exceeds the minimal value of the population objective function. This condition clearly fails when $\beta =0$ so $IS=0$, but $\mu _{R}\neq 0$. This setting is used in
Kan and Zhang (1999)\nocite{kz99} to point at the misbehavior of traditional
inference methods; see also Gospodinov et al. (2017).\nocite{gkr17}

\subsection{Running example 2: Linear IV regression model}

For the linear IV regression model:
\begin{equation}
\begin{array}{rl}
y= & X\theta +\varepsilon,  \\ 
X= & Z\Pi +V,%
\end{array}
\label{liniv}
\end{equation}%
with $\theta $ and $\Pi $ $m\times 1$ and $k\times m$ matrices containing
unknown parameters, $y=(y_{1}\ldots y_{T})^{\prime }$ and $X=(X_{1}\ldots
X_{T})^{\prime }$ $T\times 1$ and $T\times m$ dimensional matrices
containing the endogenous variables, $Z=(Z_{1}\ldots Z_{T})^{\prime }$ a $%
T\times k$ matrix containing the instrumental variables, $\varepsilon
=(\varepsilon _{1}\ldots \varepsilon _{T})^{\prime }$ and $V=(V_{1}\ldots
V_{T})^{\prime }$ are $T\times 1$ and $T\times m$ matrices of errors. The
population moment function is:%
\begin{equation}
\begin{array}{c}
\mu _{f}(\theta )=\sigma _{Zy}-\Sigma _{ZX}\theta ,%
\end{array}
\label{mean iv}
\end{equation}%
with $\sigma _{Zy}=E((Z_{t}-\mu _{Z})(y_{t}-\mu _{y})),$ $\Sigma
_{ZX}=E((Z_{t}-\mu _{Z})(X_{t}-\mu _{X})^{\prime })=Q_{\bar{Z}\bar{Z}}\Pi ,$ 
$Q_{\bar{Z}\bar{Z}}=E((Z_{t}-\mu _{Z})(Z_{t}-\mu _{Z})^{\prime }),$ $\mu
_{y}=E(y_{t}),$ $\mu _{X}=E(X_{t}),$ $\mu _{Z}=E(Z_{t}).$ When $%
u_{t}=\varepsilon _{t}+V_{t}^{\prime }\theta $ and $V_{t}$ are i.i.d.
distributed with mean zero and covariance matrix $\Omega=\left(
\genfrac{}{}{0pt}{}{\omega_{uu}}{\omega_{Vu}}%
\genfrac{}{}{0pt}{}{\omega_{uV}}{\Omega_{VV}}%
\right)  $, the
population continuous updating objective function of the linear IV
regression model is:%
\begin{equation}
\begin{array}{c}
Q_{p}(\theta )=\frac{1}{\omega _{uu}-2\omega _{uV}\theta +\theta ^{\prime
}\Omega _{VV}\theta }(\sigma _{Zy}-\Sigma _{ZX}\theta )^{\prime }Q_{\bar{Z}%
\bar{Z}}^{-1}(\sigma _{Zy}-\Sigma _{ZX}\theta ).%
\end{array}
\label{ob iv}
\end{equation}%
Along the same lines as for the linear asset pricing model, the minimal
value of this population continuous updating objective function equals the
smallest root of a characteristic polynomial:%

\begin{equation}
\begin{array}{cc}
\left\vert \tau \Omega -\left( \sigma _{Zy}\text{ }\vdots \text{ }\Sigma
_{ZX}\right) ^{\prime }Q_{\bar{Z}\bar{Z}}^{-1}\left( \sigma _{Zy}\text{ }%
\vdots \text{ }\Sigma _{ZX}\right) \right\vert  & =0.%
\end{array}%
\end{equation}%
If there is no value of $\theta $ for which $\mu _{f}(\theta )=0,$ identical
to the characteristic polynomial of the linear asset pricing model, the
smallest root of the characteristic polynomial is only associated with
misspecification when the amount of misspecification is less than the
identification strength, so the $IS$ identification strength measure (\ref{rank 3}) adapted to the linear IV regression model exceeds the minimal value of the population objective function.\footnote{This adaptation is just the population analog of a rank statistic testing for a reduced rank value of $\Pi$.}

Misspecified linear IV regression models are of interest in several
settings, for example, when analyzing treatment effects. In case of multiple
discrete instruments and heterogeneous treatment effects, the local average
treatment effects of Imbens and Angrist (1994) differ over the instruments, so the linear IV
regression model using all these instruments is misspecified. The
pseudo-true value is then a function of these local average treatment
effects. We lateron provide an empirical illustration of this using data
from Card (1995) in Section 6. Koles\'{a}r et al. (2015)\nocite{kcfgi15}
provide another example of how a misspecified linear IV regression model can
render a structural interpretation. Similarly, Kan et al. (2013)\nocite%
{krs13}\ give a structural interpretation to the misspecified linear factor
model as minimizing the pricing errors. In the Online Appendix, we provide
further discussions on how a structural interpretation can be given to these
models in case of misspecification. It is also important to realize that the
identification of the structural parameters is often rather weak in applied
settings in which case misspecification tests have very little power, see
Gospodinov et al. (2017).\nocite{gkr17} The identification robust tests
needed because of weak identification then become size distorted for testing
the pseudo-true value in the presence of misspecification, so it is
important to have tests which remain size correct for these empirically
relevant settings.

\section{Double robust score test}

The sample analog of the population continuous updating objective function
is the sample objective function for the continuous updating estimator (CUE)
of Hansen et al. (1996):\nocite{hhy96}%

\begin{equation}
\hat{Q}_{s}(\theta)=f_{T}(\theta,X)^{\prime}\hat{V}_{ff}(\theta)^{-1}f_{T}(%
\theta,X),  \label{cue sample}
\end{equation}
with $\hat{V}_{ff}(\theta)$ a consistent estimator of $V_{ff}(\theta)$, $%
\hat{V}_{ff}(\theta)\underset{p}{\rightarrow}V_{ff}(\theta)$, so the CUE, $%
\hat{\theta},$ is:%
\begin{equation}
\hat{\theta}=\arg\min_{\theta\in\mathbb{R}^{m}}\hat{Q}_{s}(\theta).
\end{equation}

To construct the large sample behavior of test statistics centered around
the CUE, we make Assumption 1 as in Kleibergen (2005) except that it
concerns the large sample behavior of the sample moments and their
derivative at the pseudo-true value $\theta^{\ast} $ instead of the true
value.

\paragraph{Assumption 1.}

\textit{For a value of }$\theta $\textit{\ equal to the minimizer of the
continuous updating population objective function,} $\theta ^{\ast },$%
\textit{\ the }$k_{f}\times 1$\textit{\ dimensional derivative of }$%
f_{t}(\theta )$\textit{\ with respect to }$\theta _{i},$\textit{\ } 
\begin{equation}
\begin{array}{c}
q_{it}(\theta )=\frac{\partial f_{t}(\theta )}{\partial \theta _{i}}%
:k_{f}\times 1,\qquad \qquad i=1,\ldots ,m,%
\end{array}
\label{ptspec}
\end{equation}%
\textit{is such that the joint limiting behavior of the sums of the series }$%
\bar{f}_{t}(\theta )=f_{t}(\theta )-E(f_{t}(\theta ))$ \textit{and }$\bar{q}%
_{t}(\theta )=(\bar{q}_{1t}(\theta )^{\prime }\ldots \bar{q}_{mt}(\theta
)^{\prime })^{\prime },$ \textit{with} $\bar{q}_{it}(\theta )=q_{it}(\theta
)-E(q_{it}(\theta )),$ \textit{accords with the central limit theorem:}%
\begin{equation}
\begin{array}{rl}
\frac{1}{\sqrt{T}}\sum_{t=1}^{T}\left( 
\begin{array}{c}
\bar{f}_{t}(\theta ) \\ 
\bar{q}_{t}(\theta )%
\end{array}%
\right) & \underset{d}{\rightarrow }\left( 
\begin{array}{c}
\psi _{f}(\theta ) \\ 
\psi _{\theta }(\theta )%
\end{array}%
\right) \sim N(0,V(\theta )),%
\end{array}
\label{central}
\end{equation}%
\noindent \textit{where }$\psi _{f}:k_{f}\times 1,$\textit{\ }$\psi _{\theta
}:k_{\theta }\times 1,$\textit{\ }$k_{\theta }=mk_{f},$ \textit{and }$%
V(\theta )$ \textit{is a positive semi-definite symmetric }$(k_{f}+k_{\theta
})\times (k_{f}+k_{\theta })$ \textit{matrix,} 
\begin{equation}
\begin{array}{c}
V(\theta )=\left( 
\begin{array}{cc}
V_{ff}(\theta ) & V_{f\theta }(\theta ) \\ 
V_{\theta f}(\theta ) & V_{\theta \theta }(\theta )%
\end{array}%
\right) ,%
\end{array}
\label{ospec}
\end{equation}%
\textit{with }$V_{\theta f}(\theta )=V_{f\theta }(\theta )^{\prime
}=(V_{\theta _{1}f}(\theta )^{\prime }\ldots V_{\theta _{m}f}(\theta
)^{\prime })^{\prime },$\textit{\ }$V_{\theta \theta }(\theta )=(V_{\theta
_{i}\theta _{j}}(\theta )):i,j=1,\ldots ,m;$ \textit{and }$V_{ff}(\theta ),$%
\textit{\ }$V_{\theta _{i}f}(\theta ),$ $V_{\theta _{i}\theta _{j}}(\theta )$
\textit{are }$k_{f}\times k_{f}$ \textit{dimensional matrices for} $%
i,j=1,\ldots ,m,$ \textit{and }%
\begin{equation}
\begin{array}{rl}
V(\theta )= & \lim_{T\rightarrow \infty }\text{var}\left( \sqrt{T}\left( 
\begin{array}{c}
f_{T}(\theta ,X) \\ 
vec(q_{T}(\theta ,X))%
\end{array}%
\right) \right) ,%
\end{array}
\label{ds1}
\end{equation}%
\textit{with }$q_{T}(\theta ,X)=\frac{\partial f_{T}(\theta ,X)}{\partial
\theta ^{\prime }}|_{\theta }=\frac{1}{T}\sum_{t=1}^{T}(q_{1t}(\theta
)\ldots q_{mt}(\theta ))$.

Assumption 1 requests a joint central limit theorem to hold for the sample
moments and their derivative with respect to $\theta $. It is satisfied
under mild conditions which are listed in Kleibergen (2005), like, for
example, finite $r$-th moments for $r>2,$ mixing conditions for the sample
moments in case of time-series data. Allowing for a positive semi-definite
covariance matrix $V(\theta )$ is important for applications, like, for
example, dynamic linear panel data models. We next also use Assumption 2
from Kleibergen (2005) which concerns the convergence of the covariance
matrix estimator $\hat{V}(\theta ).$

\paragraph{Assumption 2.}

\textit{The convergence behavior of the covariance matrix estimator }$\hat {V%
}(\theta)$ \textit{towards }$V(\theta)$ \textit{is such that}%
\begin{equation}
\begin{array}{c}
\hat{V}(\theta)\underset{p}{\rightarrow}V\left( \theta\right) \text{ }and%
\text{ }\frac{\partial\text{vec(}\hat{V}_{ff}(\theta))}{\partial
\theta^{\prime}}\underset{p}{\rightarrow}\frac{\partial\text{vec(}%
V_{ff}(\theta))}{\partial\theta^{\prime}}.%
\end{array}
\label{ass2}
\end{equation}

The CUE satisfies the FOC for a minimum of the CUE sample objective function.

\paragraph{Theorem 2:}

The FOC (divided by two) for a stationary point $\hat{\theta}^{s}$ of the
CUE sample objective function reads:%
\begin{equation}
\frac{1}{2}\frac{\partial}{\partial\theta^{\prime}}\hat{Q}_{s}(\hat{\theta }%
^{s}) =0 \ \ \ \Leftrightarrow \ \ \ f_{T}(\hat{\theta}^{s},X)^{\prime}\hat{V%
}_{ff}(\hat{\theta}^{s})^{-1}\hat {D}(\hat{\theta}^{s}) =0,  \label{cue foc}
\end{equation}
with%
\begin{equation}
\begin{array}{rl}
\hat{D}(\theta)= & q_{T}(\theta,X)-\left[ \hat{V}_{\theta_{1}f}(\theta )\hat{%
V}_{ff}(\theta)^{-1}f_{T}(\theta,X)\ldots\hat{V}_{\theta_{m}f}(\theta)\hat{V}%
_{ff}(\theta)^{-1}f_{T}(\theta,X)\right]%
\end{array}%
\end{equation}
and 
\begin{equation}
\begin{array}{c}
\hat{V}(\theta)=\left( 
\begin{array}{cc}
\hat{V}_{ff}(\theta) & \hat{V}_{f\theta}(\theta) \\ 
\hat{V}_{\theta f}(\theta) & \hat{V}_{\theta\theta}(\theta)%
\end{array}
\right) ,%
\end{array}%
\end{equation}
with $\hat{V}_{\theta f}(\theta)=\hat{V}_{f\theta}(\theta)^{\prime}=(\hat {V}%
_{\theta_{1}f}(\theta)^{\prime}\ldots\hat{V}_{\theta_{m}f}(\theta)^{\prime
})^{\prime},$\textit{\ }$\hat{V}_{\theta\theta}(\theta)=(\hat{V}_{\theta
_{i}\theta_{j}}(\theta)):i,j=1,\ldots,m;$ and $\hat{V}_{ff}(\theta ),$%
\textit{\ }$\hat{V}_{\theta_{i}f}(\theta),$ $\hat{V}_{\theta_{i}\theta_{j}}(%
\theta)$ are $k_{f}\times k_{f}$ dimensional matrices for $i,j=1,\ldots ,m.$%
\smallskip

\begin{proof}
It follows along the lines of the proof of Theorem 1; see also Kleibergen
(2005)\nocite{kf00a}.\smallskip
\end{proof}

Theorem 2 shows that the FOC of the sample CUE objective function can in an
identical manner be factorized as the FOC of the population continuous
updating objective function provided in Theorem 1. Theorem 3 further shows that the two components in
which the FOC of the sample objective function factorizes are independently
distributed in large samples.

\paragraph{Theorem 3:}

When Assumptions 1 and 2 hold and for $\theta^{\ast}$ the pseudo-true value
minimizing the population continuous updating objective function:\textit{\ } 
\begin{equation}
\begin{array}{rl}
\sqrt{T}\left( f_{T}(\theta^{\ast},X)-\mu_{f}(\theta^{\ast})\right) & 
\underset{d}{\rightarrow}\psi_{f}(\theta^{\ast}), \\ 
\sqrt{T}\text{vec}\left( \hat{D}(\theta^{\ast})-D(\theta^{\ast})\right) & 
\underset{d}{\rightarrow}\psi_{\theta.f}(\theta^{\ast}),%
\end{array}
\label{lem1}
\end{equation}
where $\psi_{\theta.f}(\theta^{\ast})=\psi_{\theta}(\theta^{\ast})-V_{\theta
f}(\theta^{\ast})V_{ff}(\theta^{\ast})^{-1}\psi_{f}(\theta^{\ast})$ and 

\begin{equation}
\begin{array}{c}
\psi_{f}(\theta^{\ast})\sim N(0,V_{ff}(\theta^{\ast})), \\ 
\psi_{\theta.f}(\theta^{\ast})\sim N(0,V_{\theta\theta.f}(\theta^{\ast})),%
\end{array}
\label{pisthetf}
\end{equation}
with\textit{\ }$V_{\theta\theta.f}(\theta)=V_{\theta\theta}(\theta)-V_{%
\theta f}(\theta)V_{ff}(\theta)^{-1}V_{f\theta}(\theta),$\textit{\ }and%
\textit{\ }$\psi_{\theta.f}(\theta^{\ast})$\textit{\ }is independent of $%
\psi_{f}(\theta^{\ast}).$\textit{\medskip}

\begin{proof}
See the Online Appendix and Lemma 1 in Kleibergen (2005).\medskip
\end{proof}

In standard GMM using the CUE objective function, the sample moment $%
f_{T}(\theta,X)$ is centered at zero at the true value, so we can use
different identification robust statistics, like the score,
GMM-Anderson-Rubin and extensions of the conditional likelihood ratio
statistic of Moreira (2003);\nocite{mor01} see Stock and Wright (2000),%
\nocite{sw00} Kleibergen (2005)\nocite{kf00a}, Andrews (2016)\nocite{and15}
and Andrews and Mikusheva\ (2016a, b).\nocite{am16} In our misspecified GMM\
setting the sample moment is not centered at zero, so we can not use any of
these statistics. We therefore propose a misspecification robust score
statistic, which uses that the expected value of the limit of the derivative
of the sample objective function:%
\begin{equation}
\begin{array}{rl}
s(\theta)=\frac{1}{2}\frac{\partial}{\partial\theta^{\prime}}\hat{Q}%
_{s}(\theta)= & f_{T}(\theta,X)^{\prime}\hat{V}_{ff}(\theta)^{-1}\hat {D}%
(\theta),%
\end{array}
\label{sample score}
\end{equation}
is equal to zero at the pseudo-true value $\theta^{\ast}$, as shown in Theorem 4 below.

\paragraph{Theorem 4:}

When Assumptions 1 and 2 hold, $\theta ^{\ast }$ is the minimizer of the
population continuous updating objective function, and%
\begin{equation}
\begin{array}{rl}
\bar{\mu}_{f}(\theta ^{\ast })= & \lim_{T\rightarrow \infty }E\left[ \sqrt{T}%
f_{T}(\theta ^{\ast },X)\right] \\ 
\bar{D}(\theta ^{\ast })= & \lim_{T\rightarrow \infty }E\left[ \sqrt{T}%
q_{T}(\theta ^{\ast },X)\right] -\left[ V_{\theta _{1}f}(\theta ^{\ast })V_{ff}(\theta ^{\ast })^{-1}\bar{%
\mu}_{f}(\theta ^{\ast })\ldots V_{\theta _{m}f}(\theta ^{\ast
})V_{ff}(\theta ^{\ast })^{-1}\bar{\mu}_{f}(\theta ^{\ast })\right]%
\end{array}
\label{local meander}
\end{equation}%
with $\bar{\mu}_{f}(\theta ^{\ast })$ and $\bar{D}(\theta ^{\ast })$ finite
valued $k_{f}$ and $k_{f}\times m$ dimensional continuously differentiable
functions of $\theta ^{\ast },$ so $\bar{\mu}_{f}(\theta ^{\ast })^{\prime
}V_{ff}(\theta ^{\ast })^{-1}\bar{D}(\theta ^{\ast })\equiv 0,$ the limit
behavior of $s(\theta ^{\ast })$ is characterized by:%
\begin{equation}
\begin{array}{rl}
Ts(\theta ^{\ast })\underset{d}{\rightarrow } & \bar{\mu}_{f}(\theta ^{\ast
})^{\prime }V_{ff}(\theta ^{\ast })^{-1}\Psi _{\theta .f}(\theta ^{\ast
})+\psi _{f}(\theta ^{\ast })^{\prime }V_{ff}(\theta ^{\ast })^{-1}\bar{D}%
(\theta ^{\ast })+\psi _{f}(\theta ^{\ast })^{\prime }V_{ff}(\theta ^{\ast
})^{-1}\Psi _{\theta .f}(\theta ^{\ast }),%
\end{array}
\label{lim score}
\end{equation}%
with vec($\Psi _{\theta .f}(\theta ^{\ast }))=\psi _{\theta .f}(\theta
^{\ast }),$ so the expected value of the limit of the derivative of the
sample CUE objective function is equal to zero at the pseudo-true value $%
\theta ^{\ast }:$%
\begin{equation}
\begin{array}{r}
\lim_{T\rightarrow \infty }E\left[ T\times s(\theta ^{\ast })\right] =0.%
\end{array}
\label{expectation score}
\end{equation}

\begin{proof}
See the Online Appendix. \smallskip
\end{proof}

The limit behavior of the score in (\ref{lim score}) equals the
sum of three distinct elements. Since all normal random variables involved
in the limit expression are independently distributed according to Theorem 3, the mean of the limit
behavior of the score is equal to zero. Theorem 4 uses local to zero
sequences for $\mu _{f}(\theta )$ and $D(\theta )$ which are orthogonal at
the pseudo-true value $\theta ^{\ast }.$ This is without loss of generality. We just use them to save on notation, since it avoids that certain
bounded random variables get multiplied by diverging objects which would
imply that the expectation becomes ill defined. This treatment is analogous to the weak instrument asymptotics (see, e.g. Staiger and Stock (1997)) that lead to weak identification robust tests.

\subsection{DRLM statistic}

If the limit expression of the score in (\ref{lim score}) would just
consist of the first two elements, it would be straightforward to construct
the weight matrix for a score statistic, since these two components are
independently distributed. The weight matrix would then consist of the sum
of the covariance matrices of each of these two components, and the limiting
distribution of the score statistic would be $\chi ^{2}(m).$ Since $\bar{\mu}%
_{f}(\theta ^{\ast })$ and $\bar{D}(\theta ^{\ast })$ are not consistently
estimable and the third component present in the limit expression (\ref{lim
score}) is not independent of both the first and second component, we cannot
use this weight matrix for the score statistic. We provide this argument
since it provides the insight into how we do obtain an appropriate weight matrix for constructing our test statistic, as we show next.

We note that the sum of the second and third component of the limit
expression in (\ref{lim score}) equals the limit of the score used in the
KLM statistic from Kleibergen (2005). The limit behavior of the KLM
statistic can be expressed as:%
\begin{equation}
\begin{array}{lc}
\psi _{f}(\theta ^{\ast })^{\prime }V_{ff}(\theta ^{\ast })^{-1}\left[ \bar{D%
}(\theta ^{\ast })+\Psi _{\theta .f}(\theta ^{\ast })\right]  &  \\ 
\left( \left[ \bar{D}(\theta ^{\ast })+\Psi _{\theta .f}(\theta ^{\ast })%
\right] ^{\prime }V_{ff}(\theta ^{\ast })^{-1}\left[ \bar{D}(\theta ^{\ast
})+\Psi _{\theta .f}(\theta ^{\ast })\right] \right) ^{-1} &  \\ 
\left[ \bar{D}(\theta ^{\ast })+\Psi _{\theta .f}(\theta ^{\ast })\right]
^{\prime }V_{ff}(\theta ^{\ast })^{-1}\psi _{f}(\theta ^{\ast }) & \sim \chi
^{2}(m).%
\end{array}
\label{score lim1}
\end{equation}%
In an identical manner, we can add the first and third component of the
limit expression in (\ref{lim score}) to obtain:%
\begin{eqnarray}
&&\left[ \bar{\mu}_{f}(\theta ^{\ast })+\psi _{f}(\theta ^{\ast })\right]
^{\prime }V_{ff}(\theta ^{\ast })^{-1}\Psi _{\theta .f}(\theta ^{\ast })  \notag
\\ 
 & &\left( \left( I_{m}\otimes V_{ff}(\theta ^{\ast })^{-1}\left[ \bar{\mu}%
_{f}(\theta ^{\ast })+\psi _{f}(\theta ^{\ast })\right] \right) ^{\prime
}V_{\theta \theta .f}(\theta ^{\ast })\left( I_{m}\otimes V_{ff}(\theta
^{\ast })^{-1}\left[ \bar{\mu}_{f}(\theta ^{\ast })+\psi _{f}(\theta ^{\ast
})\right] \right) \right) ^{-1} \label{score lim2}\\ 
&& \Psi _{\theta .f}(\theta ^{\ast })^{\prime }V_{ff}(\theta ^{\ast })^{-1}
\left[ \bar{\mu}_{f}(\theta ^{\ast })+\psi _{f}(\theta ^{\ast })\right]  \ \ \ \ \ \ \ \ \ \ \ \ \ \ \ \ \ \ \ \ \ \ \ \ \ \ \ \ \ \ \ \ \ \ \  
\sim \chi ^{2}(m). \notag%
\end{eqnarray}

The weight function 

\begin{equation}
\begin{array}{c}
\left[ \bar{D}(\theta ^{\ast })+\Psi _{\theta .f}(\theta ^{\ast })\right]
^{\prime }V_{ff}(\theta ^{\ast })^{-1}\left[ \bar{D}(\theta ^{\ast })+\Psi
_{\theta .f}(\theta ^{\ast })\right] 
\end{array}
\label{weight1}
\end{equation}%
involved in the limit behavior of the KLM statistic in (\ref{score lim1})
takes account of the dependence between the second and third component of
the limit expression of the score  in (\ref{lim score}). Similarly,
the weight function 
\begin{equation}
\begin{array}{c}
\left( I_{m}\otimes V_{ff}(\theta ^{\ast })^{-1}\left[ \bar{\mu}_{f}(\theta
^{\ast })+\psi _{f}(\theta ^{\ast })\right] \right) ^{\prime }V_{\theta
\theta .f}(\theta ^{\ast })\left( I_{m}\otimes V_{ff}(\theta ^{\ast })^{-1}%
\left[ \bar{\mu}_{f}(\theta ^{\ast })+\psi _{f}(\theta ^{\ast })\right]
\right) 
\end{array}
\label{weight2}
\end{equation}%
in (\ref{score lim2}) takes account of the dependence between the first and
third component of the limit expression of the score in (\ref{lim score}).
Identical to the case where we just have the first two components present in
the limit expression of the score, we sum the weight functions in (\ref%
{weight1}) and (\ref{weight2}) for our score statistic. Hence, the limit
expression of our score statistic presented below in\ Definition 1 is:%
\begin{equation}
\begin{array}{l}
\left\{ \left[ \bar{\mu}_{f}(\theta ^{\ast })+\psi _{f}(\theta ^{\ast })%
\right] ^{\prime }V_{ff}(\theta ^{\ast })^{-1}\Psi _{\theta .f}(\theta
^{\ast })+\psi _{f}(\theta ^{\ast })^{\prime }V_{ff}(\theta ^{\ast
})^{-1}\bar{D}%
(\theta ^{\ast })\right\}  \\ 
\left\{ \left[ \bar{D}(\theta ^{\ast })+\Psi _{\theta .f}(\theta ^{\ast })%
\right] ^{\prime }V_{ff}(\theta ^{\ast })^{-1}\left[ \bar{D}(\theta ^{\ast
})+\Psi _{\theta .f}(\theta ^{\ast })\right] \right.  \textit{+}  \\ 
\left. \left( I_{m}\otimes V_{ff}(\theta ^{\ast })^{-1}\left[ \bar{\mu}%
_{f}(\theta ^{\ast })+\psi _{f}(\theta ^{\ast })\right] \right) ^{\prime
}V_{\theta \theta .f}(\theta ^{\ast })\left( I_{m}\otimes V_{ff}(\theta
^{\ast })^{-1}\left[ \bar{\mu}_{f}(\theta ^{\ast })+\psi _{f}(\theta ^{\ast
})\right] \right) \right\} ^{-1} \\ 
\left\{ \Psi _{\theta .f}(\theta ^{\ast })^{\prime }V_{ff}(\theta ^{\ast
})^{-1}\left[ \bar{\mu}_{f}(\theta ^{\ast })+\psi _{f}(\theta ^{\ast })%
\right] +\bar{D}%
(\theta ^{\ast })^{\prime }V_{ff}(\theta ^{\ast
})^{-1}\psi _{f}(\theta ^{\ast })\right\} .%
\end{array}
\label{score lim3}
\end{equation}%

The double robust score or Lagrange multiplier statistic then results by
plugging in estimators that lead to the appropriate limit behavior of the
score statistic in (\ref{score lim3}): $\sqrt{T}\hat{\mu}_{f}(\theta ^{\ast })=%
\sqrt{T}f_{T}(\theta ^{\ast },X)\underset{d}{\rightarrow }\bar{\mu}%
_{f}(\theta ^{\ast })+\psi _{f}(\theta ^{\ast }),$ $\sqrt{T}\hat{D}(\theta
^{\ast })\underset{d}{\rightarrow }\bar{D}(\theta ^{\ast })+\Psi _{\theta
.f}(\theta ^{\ast }).$

\paragraph{Definition 1.}

The double robust score or Lagrange multiplier (DRLM) statistic for testing H%
$_{0}:\theta=\theta^{\ast},$ with $\theta^{\ast}$ the pseudo-true value, is:%
\begin{eqnarray}
DRLM(\theta^{\ast})&=& T^{2}\times f_{T}(\theta^{\ast},X)^{\prime}\hat{V}_{ff}(\theta^{\ast})^{-1}%
\hat{D}(\theta^{\ast}) \notag\\ 
&&\left[ T\times\left( I_{m}\otimes\hat{V}_{ff}(\theta^{\ast})^{-1}f_{T}(%
\theta^{\ast},X)\right) ^{\prime}\hat{V}_{\theta\theta.f}(\theta^{\ast
})\left( I_{m}\otimes\hat{V}_{ff}(\theta^{\ast})^{-1}f_{T}(\theta^{\ast
},X)\right) +\right. \notag\\ 
&&\left. T\times\hat{D}(\theta^{\ast})^{\prime}\hat{V}_{ff}(\theta^{\ast})^{-1}%
\hat{D}(\theta^{\ast})\right] ^{-1}\hat{D}(\theta^{\ast})^{\prime}\hat{V}%
_{ff}(\theta^{\ast})^{-1}f_{T}(\theta^{\ast},X).%
\label{drlm}
\end{eqnarray}


The component resulting from the weight matrix (\ref{weight1}) in
the overall weight matrix makes the quadratic form of the second and third
component of the limit expression of the score  in (\ref{lim score}%
) with it $\chi ^{2}(m)$ distributed. Similarly, the weight matrix (\ref%
{weight2}) does so for the quadratic form of the first and third component
of the limit expression of the score  in (\ref{lim score}). Since the first and second component are independently distributed, the third component is therefore \textquotedblleft
double\textquotedblright\ counted in the overall weight matrix. This makes the limit behavior in (\ref{score lim3}%
) bounded by a $\chi ^{2}(m)$ distributed random variable. This bound is
sharp when the third component of the limit behavior of the score is
negligible, which occurs for large values of $\bar{\mu}_{f}(\theta ^{\ast })$
and/or $\bar{D}(\theta ^{\ast }).$ When $\bar{\mu}_{f}(\theta ^{\ast })$ and 
$\bar{D}(\theta ^{\ast })$ are both equal to zero, the limit behavior reduces to:%
\begin{equation}
\begin{array}{l}
\psi _{f}(\theta ^{\ast })^{\prime }V_{ff}(\theta ^{\ast })^{-1}\Psi
_{\theta .f}(\theta ^{\ast })\Big[ \Psi _{\theta .f}(\theta ^{\ast
})^{\prime }V_{ff}(\theta ^{\ast })^{-1}\Psi _{\theta .f}(\theta ^{\ast
}) \\ 
+ \left. \left( I_{m}\otimes V_{ff}(\theta ^{\ast })^{-1}\psi _{f}(\theta
^{\ast })\right) ^{\prime }V_{\theta \theta .f}(\theta ^{\ast })\left(
I_{m}\otimes V_{ff}(\theta ^{\ast })^{-1}\psi _{f}(\theta ^{\ast })\right)
\Big] ^{-1}\Psi _{\theta .f}(\theta ^{\ast })^{\prime }V_{ff}(\theta
^{\ast })^{-1}\psi _{f}(\theta ^{\ast }), \right.%
\end{array}
\label{score lim4}
\end{equation}%
which is obviously bounded by a $\chi ^{2}(m)$ distributed
random variable. For intermediate values of $\bar{\mu}_{f}(\theta ^{\ast })$
and $\bar{D}(\theta ^{\ast }),$ the $\chi ^{2}(m)$ bound remains, which is
further articulated in the proof of Theorem 5.

\paragraph{Theorem 5:}

When Assumptions 1 and 2 hold and given the specifications in (\ref{local
meander}), the limit behavior of DRLM($\theta ^{\ast })$ under H$_{0}:\theta
=\theta ^{\ast },$ with $\theta ^{\ast }$ the minimizer of the population
continuous updating objective function, is bounded according to:%
\begin{equation}
\begin{array}{c}
\lim_{T\rightarrow \infty }\Pr \left[ DRLM(\theta ^{\ast })>cv_{\chi
^{2}(m)}(\alpha )\right] \leq \alpha ,%
\end{array}
\label{bound lim drlm}
\end{equation}%
with $cv_{\chi ^{2}(m)}(\alpha )$ the $(1-\alpha )\times 100\%$ critical
value for the $\chi ^{2}(m)$ distribution.

\begin{proof}
See the Online Appendix, which also provides an extension to Assumptions 1
and 2 by stating the parameter space of the distributions which render the
DRLM test size correct; see also Andrews and Guggenberger (2017).\nocite%
{ag17}\smallskip
\end{proof}

\subsection{DRLM test for the linear asset pricing model}
For further exposition, we use the DRLM statistic to test the risk premia in the linear asset
pricing model with i.i.d. errors.

\paragraph{Running example 1: Linear asset pricing model}

For a DRLM\ test of the risk premia, we need the specification of the
different components of the DRLM\ statistic for the linear asset pricing
model with i.i.d. errors:%
\begin{equation}
\begin{array}{rl}
f_{T}(\lambda _{F},X)= & \bar{R}-\hat{\beta}\lambda _{F} \\ 
\hat{D}(\lambda _{F})= & -\hat{\beta}-(\bar{R}-\hat{\beta}\lambda
_{F})(1+\lambda _{F}^{\prime }\hat{Q}_{\bar{F}\bar{F}}^{-1}\lambda
_{F})^{-1}\lambda _{F}^{\prime }\hat{Q}_{\bar{F}\bar{F}}^{-1} \\ 
= & -\frac{1}{T}\sum_{t=1}^{T}R_{t}(\bar{F}_{t}+\lambda _{F})^{\prime }\left[
\frac{1}{T}\sum_{t=1}^{T}(\bar{F}_{t}+\lambda _{F})(\bar{F}_{t}+\lambda
_{F})^{\prime }\right] ^{-1} \\ 
\hat{V}_{ff}(\lambda _{F})= & (1+\lambda _{F}^{\prime }Q_{\bar{F}\bar{F}%
}^{-1}\lambda _{F})\hat{\Omega} \\ 
\hat{V}_{\theta \theta .f}(\lambda _{F})= & (\hat{Q}_{\bar{F}\bar{F}%
}+\lambda _{F}\lambda _{F}^{\prime })^{-1}\otimes \hat{\Omega},%
\end{array}
\label{lin mod def}
\end{equation}%
so the specification of the DRLM statistic reads:%
\begin{equation}
\begin{array}{rl}
DRLM(\lambda _{F}^{\ast })= & T(1+\lambda _{F}^{\ast \prime }\hat{Q}_{\bar{F}%
\bar{F}}^{-1}\lambda _{F}^{\ast })^{-1}(\bar{R}-\hat{\beta}\lambda
_{F}^{\ast })^{\prime }\hat{\Omega}^{-1}\hat{D}(\lambda _{F}^{\ast }) \\ 
& \left[ (1+\lambda _{F}^{\ast \prime }\hat{Q}_{\bar{F}\bar{F}}^{-1}\lambda
_{F}^{\ast })^{-1}(\bar{R}-\hat{\beta}\lambda _{F}^{\ast })^{\prime }\hat{%
\Omega}^{-1}(\bar{R}-\hat{\beta}\lambda _{F}^{\ast })(Q_{\bar{F}\bar{F}%
}+\lambda _{F}^{\ast }\lambda _{F}^{\ast \prime })^{-1}+\right.  \\ 
& \left. \hat{D}(\lambda _{F}^{\ast })^{\prime }\hat{\Omega}^{-1}\hat{D}%
(\lambda _{F}^{\ast })\right] ^{-1}\hat{D}(\lambda _{F}^{\ast })^{\prime }%
\hat{\Omega}^{-1}(\bar{R}-\hat{\beta}\lambda _{F}^{\ast }) \\ 
= & \hat{\mu}(\lambda _{F})^{\ast \prime }\hat{D}(\lambda _{F}^{\ast
})^{\ast }\left[ \hat{\mu}(\lambda _{F})^{\ast \prime }\hat{\mu}(\lambda
_{F})^{\ast }I_{m}+\hat{D}(\lambda _{F}^{\ast })^{\ast \prime }\hat{D}%
(\lambda _{F}^{\ast })^{\ast }\right] ^{-1}\hat{D}(\lambda _{F}^{\ast
})^{\ast \prime }\hat{\mu}(\lambda _{F})^{\ast },%
\end{array}
\label{lmr iid}
\end{equation}%
with $\hat{\mu}(\lambda _{F})^{\ast }=\sqrt{T}\hat{\Omega}^{-\frac{1}{2}}(%
\bar{R}-\hat{\beta}\lambda _{F})(1+\lambda _{F}^{\prime }\hat{Q}_{\bar{F}%
\bar{F}}^{-1}\lambda _{F})^{-\frac{1}{2}}=\sqrt{T}\hat{V}_{ff}(\lambda
_{F})^{-\frac{1}{2}}f_{T}(\lambda _{F},X),$ and $\hat{D}(\lambda _{F})^{\ast
}=\sqrt{T}\hat{\Omega}^{-\frac{1}{2}}\hat{D}(\lambda _{F})$  \ \ $(\hat{Q}_{\bar{F}%
\bar{F}}+\lambda _{F}\lambda _{F}^{\prime })^{\frac{1}{2}}.$

\paragraph{Corollary 1.}

When Assumptions 1 and 2 hold and under i.i.d. errors, the limit behavior of
the DRLM statistic under H$_{0}:\lambda _{F}=\lambda _{F}^{\ast }$ is
characterized by:%
\begin{equation}
\begin{array}{rl}
DRLM(\lambda _{F}^{\ast })\underset{d}{\rightarrow } & \left[ \psi
_{f}^{\prime }(\bar{D}+\Psi _{\theta .f})+\bar{\mu}^{\prime }\Psi _{\theta
.f}\right] \Big[ (\bar{\mu}+\psi _{f})^{\prime }(\bar{\mu}+\psi
_{f})I_{m}+  \\ 
& \left. \left( \bar{D}+\Psi _{\theta .f}\right) ^{\prime }\left( \bar{D}%
+\Psi _{\theta .f}\right) \Big] ^{-1}\left[ (\bar{D}+\Psi _{\theta
.f})^{\prime }\psi _{f}+\Psi _{\theta .f}^{\prime }\bar{\mu}\right] \right. \\ 
\preceq  & \chi ^{2}(m),%
\end{array}
\label{limit lmr iid}
\end{equation}%
with $\bar{\mu}=\Omega ^{-\frac{1}{2}}\bar{\mu}(\lambda _{F}^{\ast
})(1+\lambda _{F}^{\ast \prime }Q_{\bar{F}\bar{F}}^{-1}\lambda _{F}^{\ast
})^{-\frac{1}{2}},$ $\bar{D}=\Omega ^{-\frac{1}{2}}\bar{D}(\lambda
_{F}^{\ast })(Q_{\bar{F}\bar{F}}+\lambda _{F}^{\ast }\lambda _{F}^{\ast
\prime })^{\frac{1}{2}},$ $\bar{\mu}^{\prime }\bar{D}\equiv 0,$ $\psi _{f}$
and $\Psi _{\theta .f}$ $N\times 1$ and $N\times m$ dimensional random
matrices that consist of independent standard normal random variables,  and \textquotedblleft $\preceq $\textquotedblright\ indicates
stochastically dominated.\footnote{For a continuous non-negative scalar random
variable $u$: $u\preceq \chi ^{2}(m)$ implies that $\Pr \left[ u>cv_{\chi ^{2}(m)}(\alpha )%
\right] \leq \alpha $ for $\alpha \in (0,1].$} \smallskip 

The limit behavior of the DRLM statistic in Corollary 1 shows that it under H%
$_{0}$ only depends on two parameters, the \textquotedblleft lengths$"$ of $\bar{\mu}$ and $\bar{D%
}$, which reflect the amount of misspecification and the strength of identification respectively, and is dominated by a $\chi ^{2}(m)$ distribution. 

\subsection{Size of the DRLM test}

Next, we illustrate the size of the DRLM test for the linear asset pricing model discussed above. In particular, Figure 1 shows the
rejection frequencies of 5\% significance DRLM tests with a 95\% $\chi
^{2}(1)$ critical value as a function of the lengths of $\bar{\mu}$ and $%
\bar{D}$ for a single factor setting, so $m=1,$ and $N=25.$ The latter number
corresponds with the twenty-five Fama-French size and book-to-market sorted
portfolios, which are the default in the asset pricing literature; see Fama
and French (1993).\nocite{ff93} Also  when $m=1$, $\bar{D}$ reduces to a vector with the same dimension as $\bar{\mu}$, so their lengths result from the inner products of the elements in each vector.

\begin{equation*}
\begin{array}{c}
\text{Figure 1: Rejection frequency of 5\% significance DRLM tests of H}%
_{0}:\lambda _{F}=\lambda _{F}^{\ast }\text{ using } \\ 
\text{a }95\%\text{ }\chi ^{2}(1)\text{ critical value as a function of the
lengths of }\bar{\mu}\text{ and }\bar{D},\text{ }m=1,\text{ }N=25. \\ 
\\ 
\raisebox{-0pt}{\includegraphics[
height=2.1139in,
width=2.9386in
]%
{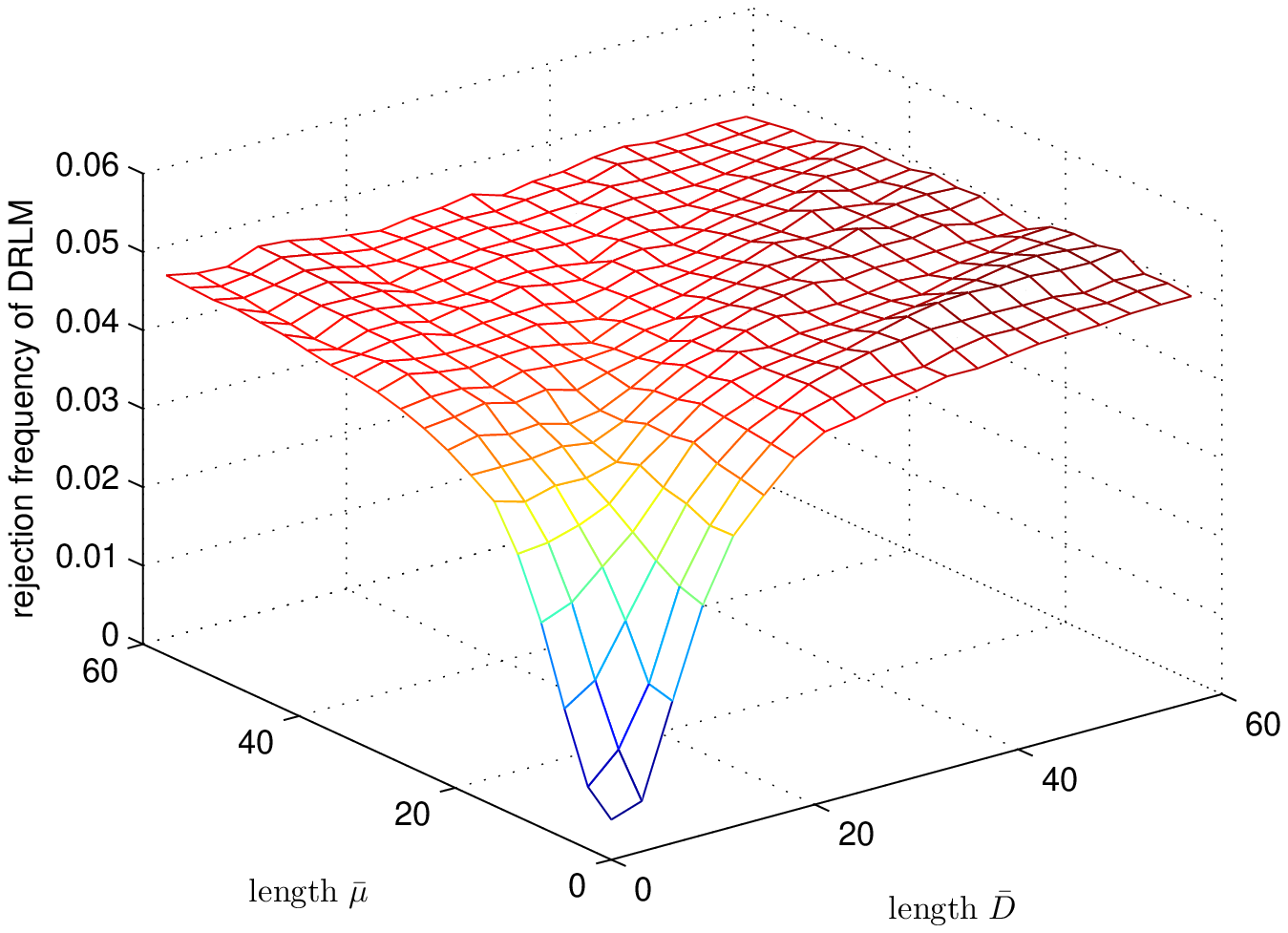}%
}
\raisebox{-0pt}{\includegraphics[
height=2.1139in,
width=2.9386in
]%
{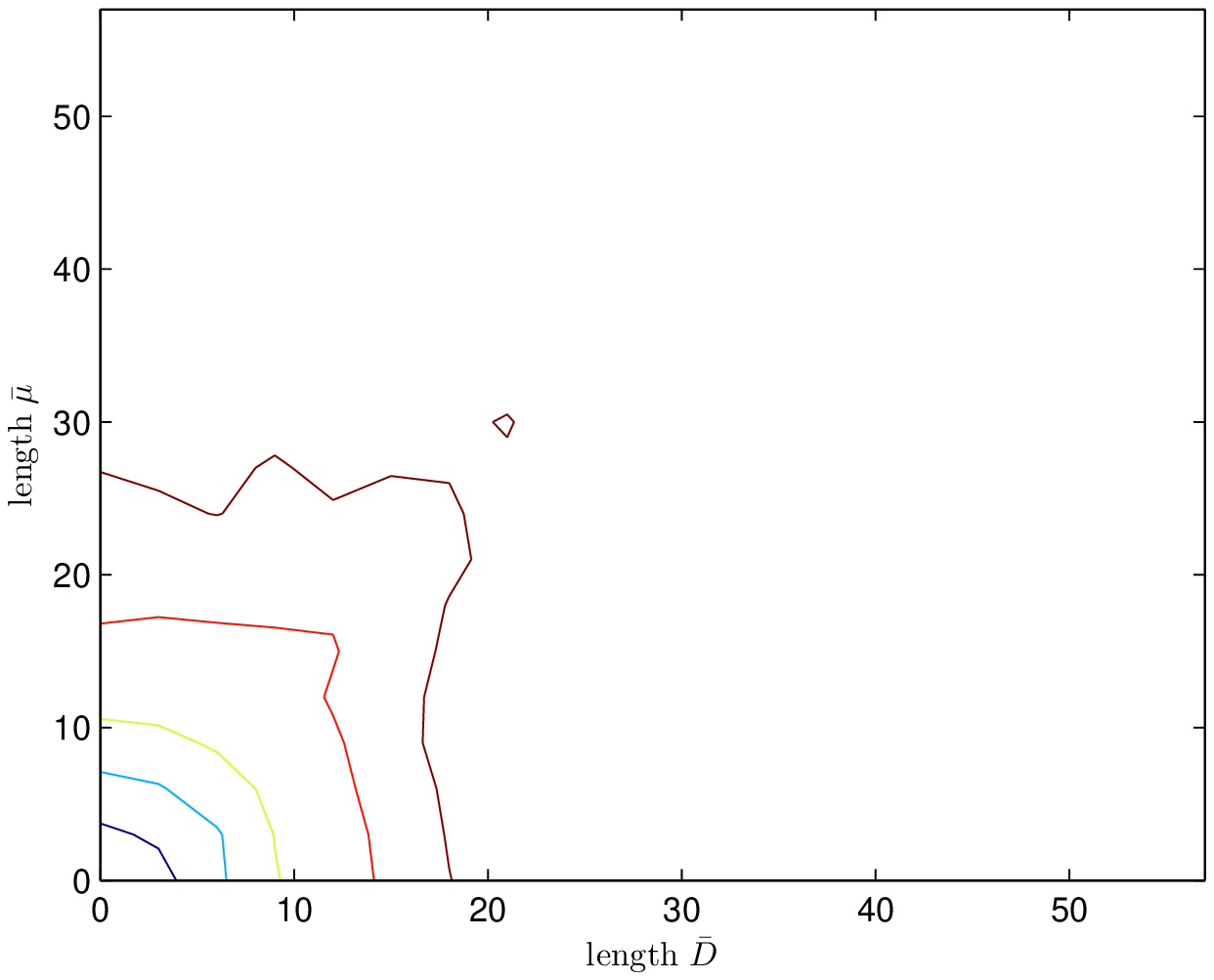}%
}
\end{array}%
\end{equation*}

\begin{equation*}
\begin{array}{c}
\text{Figure 2: Rejection frequency of 5\% significance KLM tests of H}%
_{0}:\lambda_{F}=\lambda_{F}^{\ast}\text{ using } \\ 
\text{a }95\%\text{ }\chi^{2}(1)\text{ critical value as a function of the
lengths of }\bar{\mu}\text{ and }\bar{D},\text{ }m=1,\text{ }N=25. \\ 
\\ 
\raisebox{-0pt}{\includegraphics[
height=2.1139in,
width=2.9386in
]%
{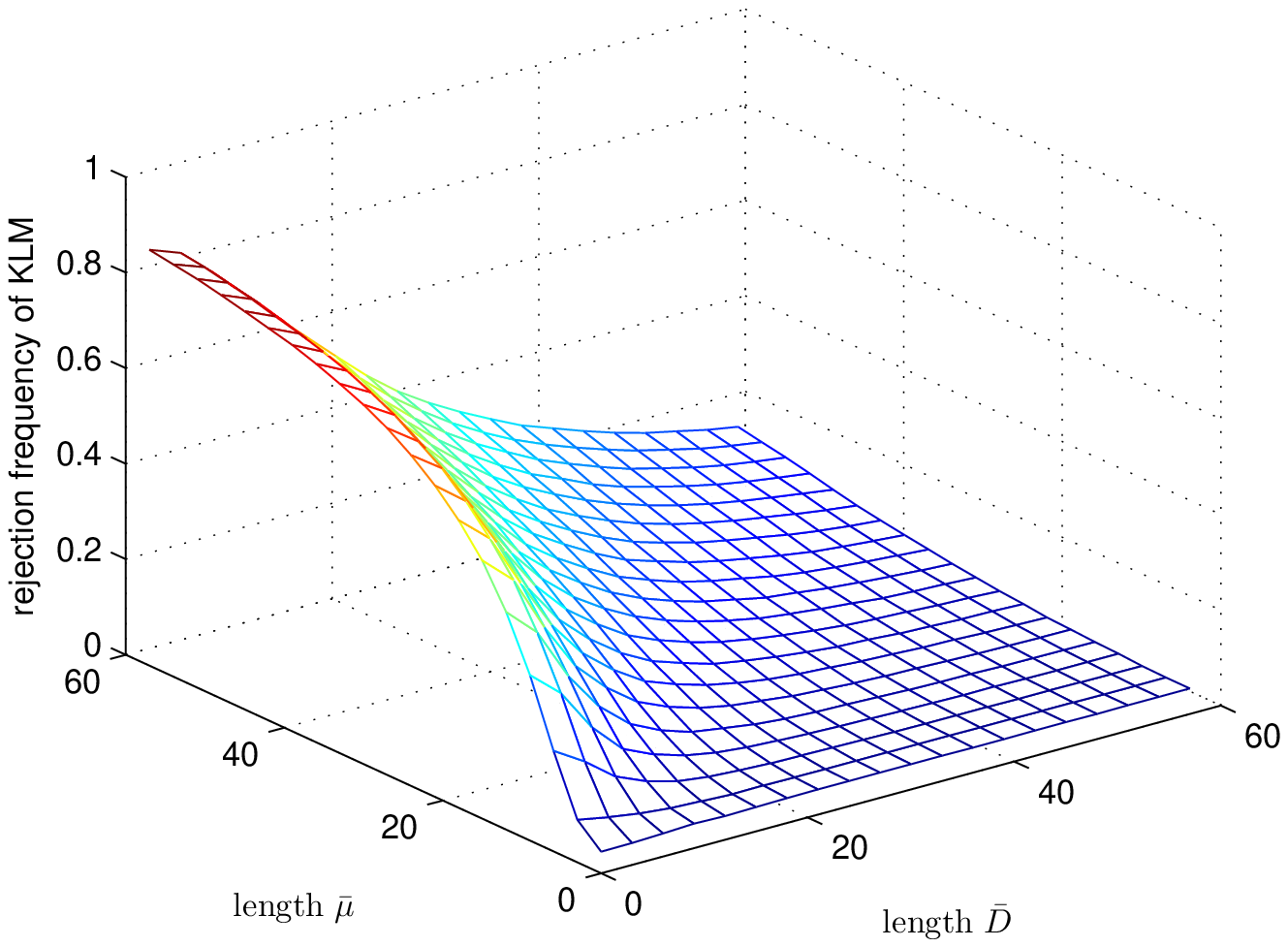}%
}
\raisebox{-0pt}{\includegraphics[
height=2.1139in,
width=2.9386in
]%
{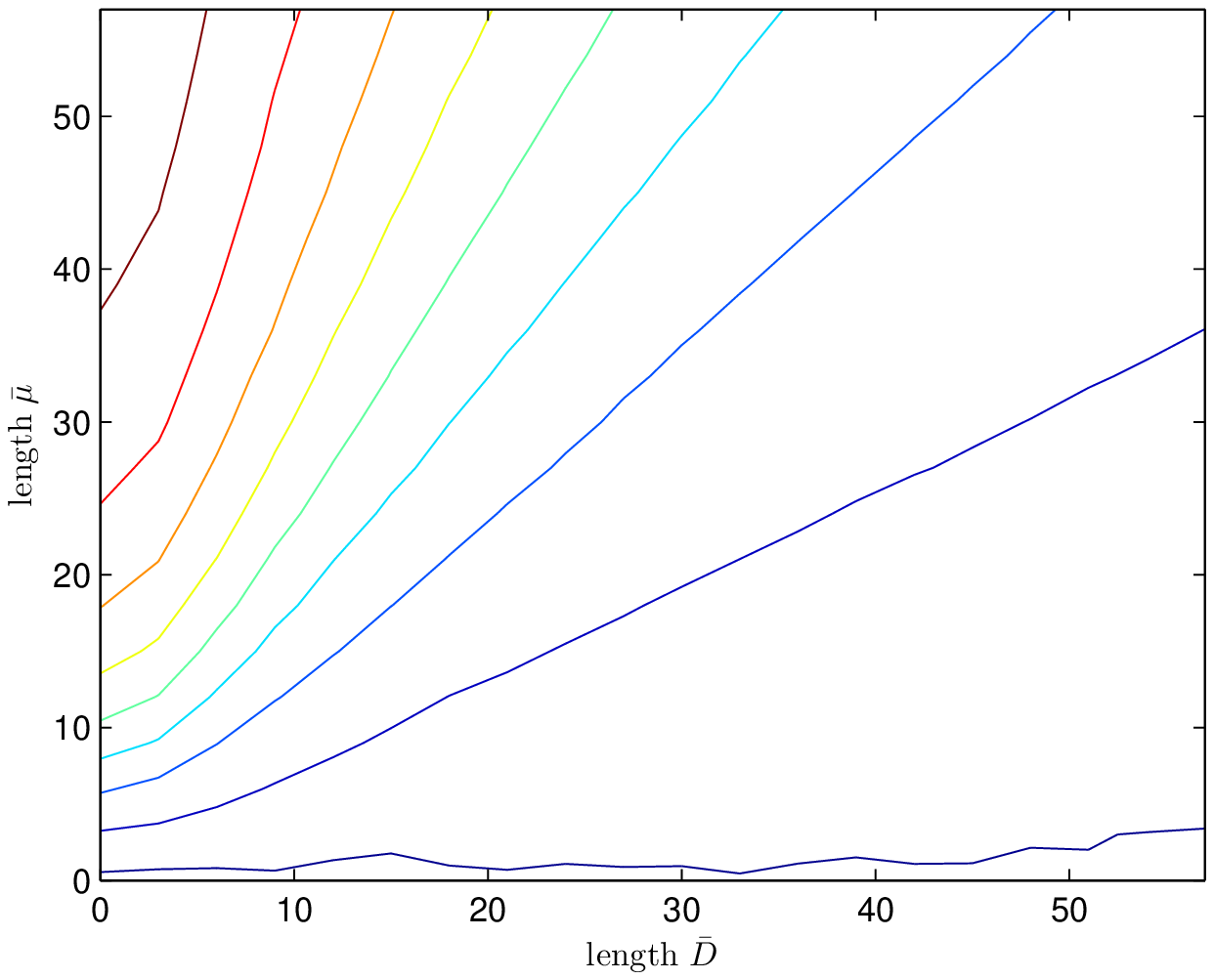}%
}
\end{array}%
\end{equation*}

Figure 1 shows that the DRLM test is size correct, since its rejection
frequency does not exceed 5\% for any length of $\bar{\mu}$ and $\bar{D}.$
For comparison, Figure 2 presents the rejection frequencies of the KLM test,
see Kleibergen (2005), as a function of the lengths of $\bar{\mu}$ and $\bar{%
D}.$ It shows that the KLM test is only size correct when there is no
misspecification so $\bar{\mu}=0$, and can be severely size distorted for
small values of the length of $\bar{\mu},$ especially when paired with small
values of the length of $\bar{D}.$

Figure 1 also shows that the DRLM test is conservative when the lengths of
both $\bar{\mu}$ and $\bar{D}$ are small. This is comparable to the subset 
Anderson-Rubin test for the homoskedastic linear IV regression model 
which Guggenberger et al. (2012) \nocite{gkmc12} show 
to be conservative in case of weak identification when using standard ${\chi}^2$ 
critical values. In Guggenberger et al. (2019) \nocite{gkm17} a data-dependent 
conditional critical value function is therefore proposed, which makes the subset 
Anderson-Rubin test near optimal. To reduce the conservativeness of the DRLM test, we  follow Guggenberger et al. (2019) and calibrate a feasible conditional critical value function based on the maximum of $\hat{\mu}(\lambda
_{F})^{\ast \prime }\hat{\mu}(\lambda _{F})^{\ast }$ and $\hat{D}(\lambda
_{F})^{\ast \prime }\hat{D}(\lambda _{F})^{\ast }$. Specifically, when the maximum of
these is less than two-hundred and fifty, we computed a 95\% conditional
critical value function based on $\max (\hat{\mu}(\lambda _{F})^{\ast \prime
}\hat{\mu}(\lambda _{F})^{\ast },$ $\hat{D}(\lambda _{F})^{\ast \prime }\hat{%
D}(\lambda _{F})^{\ast }).\footnote{%
The conditional critical value function we calibrated for Figure 3 is $%
f(r)=2.4+(\lfloor r\rfloor ^{0.35})\times (3.84-2.4)/(250^{0.35})$ for $%
r\leq 250$ and $f(r)=3.84$ for $r>250,$ with $r$ the conditioning variable
and $\lfloor .\rfloor $ the entier function.}$ Using the conditional
critical value, the contour lines in Figure 3
show that the conservativeness of a 5\% significance DRLM\ test has been
reduced substantially from an area where the maximal length of $\bar{\mu}$
and $\bar{D}$ is less than twenty to an area where their sum is less than
ten. 
\begin{equation*}
\begin{array}{c}
\text{Figure 3: Rejection frequency of 5\% significance DRLM tests of H}%
_{0}:\lambda _{F}=\lambda _{F}^{\ast }\text{ using } \\ 
\text{a conditional 95\% critical value as a function of the lengths of }%
\bar{\mu}\text{ and }\bar{D},\text{ }m=1,\text{ }N=25. \\ 
\\ 
\raisebox{-0pt}{\includegraphics[
height=2.2139in,
width=2.9386in
]%
{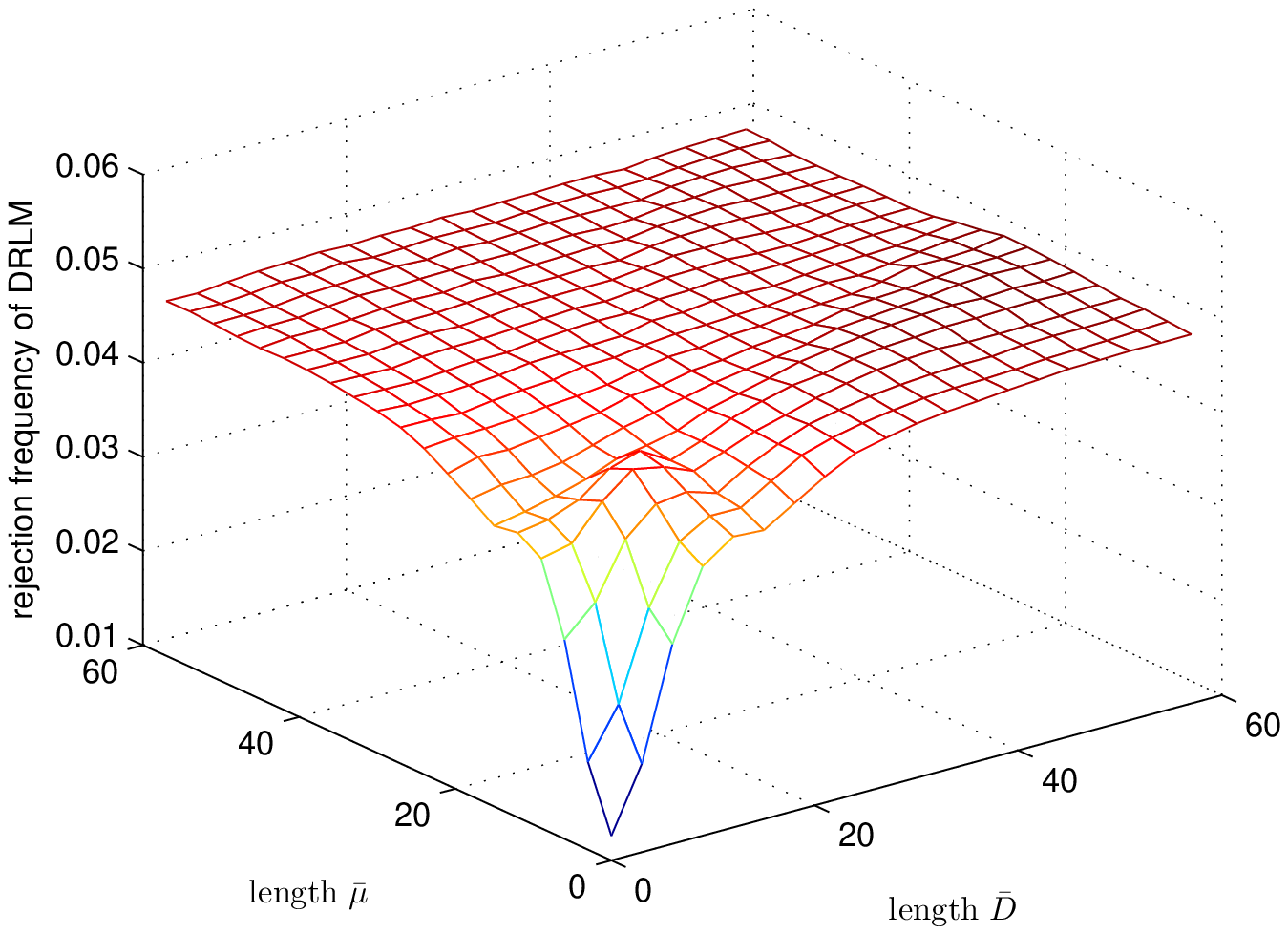}%
}
\raisebox{-0pt}{\includegraphics[
height=2.2139in,
width=2.9386in
]%
{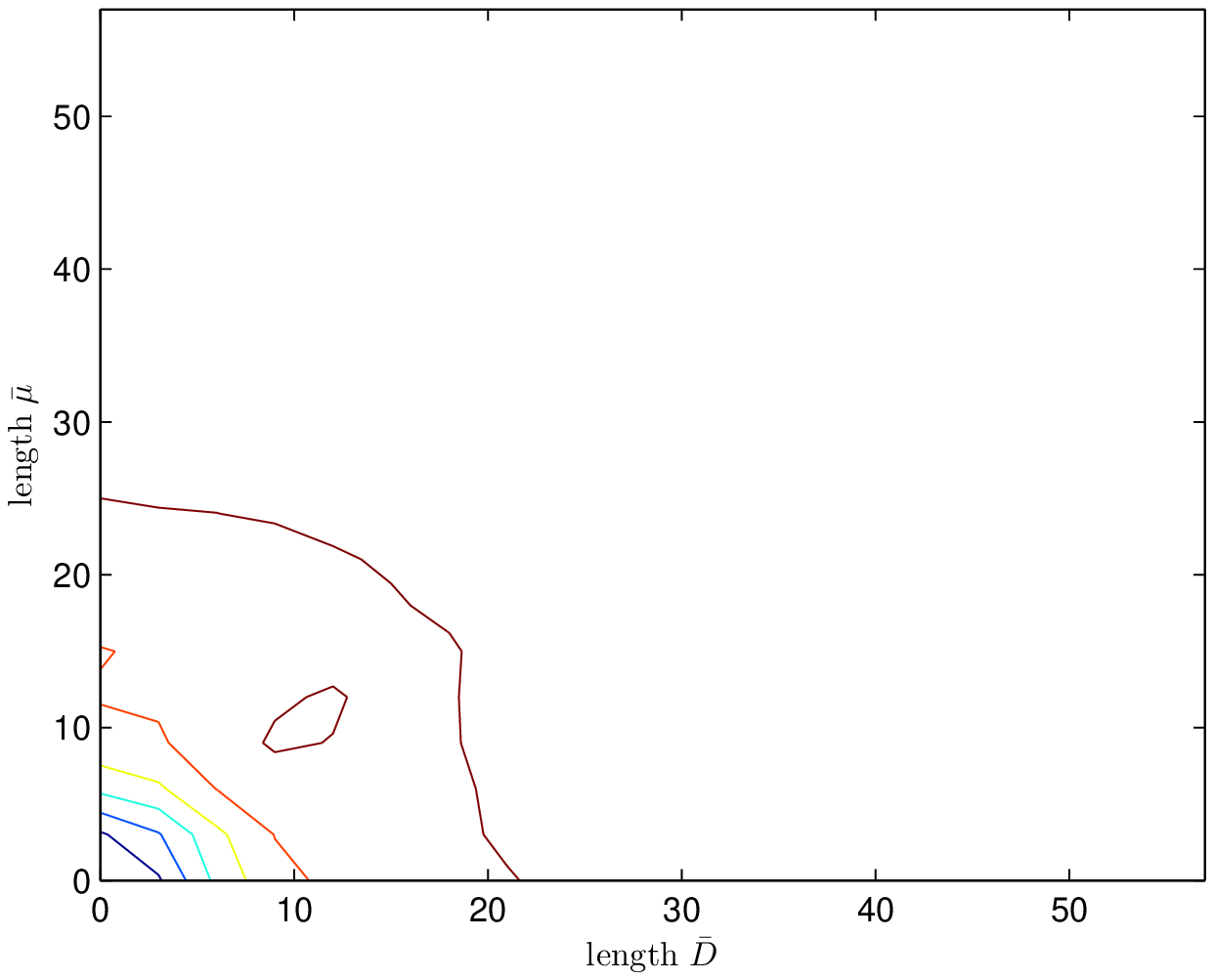}%
}

\end{array}%
\end{equation*}

\section{Power}

The score is equal to zero at all stationary points of the CUE sample
objective function, so the same holds for tests based on a quadratic form of
it, like, for example, the DRLM\ and KLM tests, as well. This leads to the
somewhat oddly behaved power of the KLM test in regular GMM. Tests with
better power properties therefore exist in GMM that, implicitly or
explicitly, combine the KLM test with an asymptotically independent $J$-test
in either a conditional or unconditional manner, see Moreira (2003),\nocite%
{mor01} Kleibergen (2005),\nocite{kf00a} Andrews et al. (2006),\nocite%
{andms05} Andrews (2016),\nocite{and15} and\ Andrews and Mikusheva (2016a, b).%
\nocite{am16} In our misspecified GMM setting, this is, however, not
possible since the limiting distribution of the $J$-statistic is a
non-central $\chi ^{2}$ distribution with an unknown non-centrality
parameter. Hence, we can not combine this limiting distribution with that of
the DRLM statistic to obtain the (conditional) critical values for a
combination test.

\subsection{Power improvement}

To improve the power of the DRLM test, we
can further reject hypothesized values of $\theta $  which are close to a stationary
point of the CUE sample objective function other than the CUE. This would be
similar to the, conditional or unconditional, identification robust
combination tests in regular GMM, which use that while the KLM\ test does not
reject at such values of $\theta ,$ $J$ and/or GMM Anderson-Rubin (AR)
tests, see Anderson and Rubin (1949)\nocite{AR49} and Stock and Wright
(2000),\nocite{sw00} likely do. For hypothesized values of $\theta $ close
to the CUE, these combination tests put most weight on the KLM test but
shift the weight towards the $J$ and GMM-AR tests when $\theta $ is close to
other stationary points, see Andrews (2016)\nocite{and15} and Kleibergen
(2007).\nocite{kf04} Since the limiting distributions of the $J$ and GMM-AR
statistics depend on unknown nuisance parameters in our misspecified GMM\
setting, it is not clear how we can use these statistics to improve power.
To improve the power of the DRLM\ test, we
can further reject values of $\theta $  when in between the hypothesized value and
the CUE there are significant values of the DRLM statistic. 

We next lay out
the  steps needed to turn the above idea into a size correct test for stylized linear GMM settings.

\paragraph{Theorem 6:}

\textbf{a. }For a given data set of realized values and a linear moment
equation, the sum of $f_{T}(\theta,X)^{\prime}\hat{V}_{ff}(%
\theta)^{-1}f_{T}(\theta,X)$ and vec($\hat{D}(\theta))^{\prime}\hat{V}%
_{\theta\theta .f}(\theta)^{-1}$vec($\hat{D}(\theta))$ does not vary over $%
\theta.\smallskip$

\noindent\textbf{b. } When $m=1$ and $f_{T}(\theta,X)$ is linear in $\theta,$
the derivative of DRLM($\theta)$ with respect to $\theta$ reads:%
\begin{equation}
\begin{array}{l}
\frac{1}{2}\frac{\partial}{\partial\theta}DRLM(\theta)=T\left( \frac {%
f_{T}(\theta,X)^{\prime}\hat{V}_{ff}(\theta)^{-1}\hat{D}(\theta)}{\left[
f_{T}(\theta,X)^{\prime}\hat{V}_{ff}(\theta)^{-1}\hat{V}_{\theta\theta
.f}(\theta)\hat{V}_{ff}(\theta)^{-1}f_{T}(\theta,X)+\hat{D}(\theta)^{\prime }%
\hat{V}_{ff}(\theta)^{-1}\hat{D}(\theta)\right] }\right) \times \\ 
\begin{array}{cl}
& \left\{ \hat{D}(\theta)^{\prime}\hat{V}_{ff}(\theta)^{-1}\hat{D}%
(\theta)-2f_{T}(\theta,X)^{\prime}\hat{V}_{ff}(\theta)^{-1}\hat{V}_{\theta
f}(\theta)\hat{V}_{ff}(\theta)^{-1}D_{T}(\theta,X)-\right. \\ 
& f_{T}(\theta,X)^{\prime}\hat{V}_{ff}(\theta)^{-1}\hat{V}_{\theta\theta
.f}(\theta)\hat{V}_{ff}(\theta)^{-1}f_{T}(\theta,X)+2f_{T}(\theta,X)^{\prime
}\hat{V}_{ff}(\theta)^{-1}\hat{D}(\theta)\times \\ 
& \left. \frac{f_{T}(\theta,X)^{\prime}\hat{V}_{ff}(\theta)^{-1}\hat {V}%
_{\theta f}(\theta)\hat{V}_{ff}(\theta)^{-1}\hat{V}_{\theta\theta.f}(\theta)%
\hat{V}_{ff}(\theta)^{-1}f_{T}(\theta,X)+\hat{D}(\theta)^{\prime}\hat{V}%
_{ff}(\theta)^{-1}\hat{V}_{\theta f}(\theta)\hat{V}_{ff}(\theta )^{-1}\hat{D}%
(\theta)}{f_{T}(\theta,X)^{\prime}\hat{V}_{ff}(\theta)^{-1}\hat{V}%
_{\theta\theta.f}(\theta)\hat{V}_{ff}(\theta)^{-1}f_{T}(\theta ,X)+\hat{D}%
(\theta)^{\prime}\hat{V}_{ff}(\theta)^{-1}\hat{D}(\theta)}\right\} .%
\end{array}%
\end{array}
\label{derdrlm}
\end{equation}

\noindent\textbf{c. }When the data is i.i.d., $m=1,$ and $f_{T}(\theta,X)$
is linear in $\theta:$ $\hat{V}(\theta)$ has a Kronecker product structure
so we can specify $\hat{V}_{ff}(\theta)=\hat{v}_{ff}(\theta)\hat{V},$ $\hat {%
V}_{\theta f}(\theta)=\hat{v}_{\theta f}(\theta)\hat{V}$ and $\hat{V}%
_{\theta\theta.f}(\theta)=\hat{v}_{\theta\theta.f}(\theta)\hat{V},$ with $%
\hat{v}_{ff}(\theta),$ $\hat{v}_{\theta f}(\theta),$ $\hat{v}_{\theta
\theta.f}(\theta)$ scalar functions of $\theta$ and $\hat{V}$ a $k_{f}\times
k_{f}$ dimensional covariance matrix estimator, and the derivative of DRLM($%
\theta)$ reduces to:

\begin{equation*}
\begin{array}{l}
\frac{1}{2}\frac{\partial }{\partial \theta }DRLM(\theta )=\left( \frac{%
\left( \hat{V}_{ff}(\theta )^{-\frac{1}{2}}f_{T}(\theta ,X)\right) ^{\prime
}\left( \hat{V}_{\theta \theta .f}(\theta )^{-\frac{1}{2}}\hat{D}(\theta
)\right) }{f_{T}(\theta ,X)^{\prime }\hat{V}_{ff}(\theta )^{-1}f_{T}(\theta
,X)+\hat{D}(\theta )^{\prime }\hat{V}_{\theta \theta .f}(\theta )^{-1}\hat{D}%
(\theta )}\right) \times  \\ 
\begin{array}{cl}
& \left( T\times \hat{D}(\theta )^{\prime }\hat{V}_{\theta \theta .f}(\theta
)^{-1}\hat{D}(\theta )-T\times f_{T}(\theta ,X)^{\prime }\hat{V}_{ff}(\theta
)^{-1}f_{T}(\theta ,X)\right) \left( \frac{\hat{v}_{\theta \theta .f}(\theta
)}{\hat{v}_{ff}(\theta )}\right) ^{\frac{1}{2}}.%
\end{array}%
\end{array}%
\end{equation*}

\begin{proof}
See the Online Appendix. $\smallskip$
\end{proof}

\paragraph{Running example 1: Linear asset pricing model}

Theorem 6c shows that for the one factor linear asset pricing model with
i.i.d. errors, the derivative of the DRLM\ statistic is proportional to the
difference between the GMM-AR statistic, $T\times f_{T}(\theta,X)^{\prime}%
\hat{V}_{ff}(\theta)^{-1}f_{T}(\theta,X),$ and an independently distributed
statistic reflecting the strength of identification, $T\times\hat{D}%
(\theta)^{\prime}\hat{V}_{\theta\theta.f}(\theta)^{-1}\hat{D}(\theta).$
Theorem 6a further shows that, for a given data set of realized values, the
sum of these two statistics does not depend on $\theta.$ Given a realized
data set, the DRLM\ statistic considered as a function of $\theta$ thus
attains its maximum when both statistics are identical so they equal half
their sum.

\paragraph{Corollary 2.}

For a given data set of realized values for the one factor linear asset
pricing model with i.i.d. errors, the maximal value of the DRLM statistic as
a function of $\lambda _{F}$ is attained at the value of $\lambda _{F}$
where the GMM-AR statistic, $T\times f_{T}(\lambda _{F},X)^{\prime }\hat{V}%
_{ff}(\lambda _{F})^{-1}f_{T}(\lambda _{F},X),$ equals half the sum of $%
T\times f_{T}(\lambda _{F},X)^{\prime }\hat{V}_{ff}(\lambda
_{F})^{-1}f_{T}(\lambda _{F},X)$ and $T\times \hat{D}(\lambda _{F})^{\prime }%
\hat{V}_{\theta \theta .f}(\lambda _{F})^{-1}\hat{D}(\lambda _{F}).$
\smallskip \smallskip 

Using Corollary 2 and the sample equivalent of the characteristic polynomial
in (\ref{popchar}), we can solve for the value of $\lambda_{F}$ that
maximizes the DRLM statistic for a given data set of realized values. We do
so by not equating the characteristic polynomial to zero but to half the sum
of $T\times f_{T}(\lambda_{F},X)^{\prime}\hat{V}_{ff}(%
\lambda_{F})^{-1}f_{T}(\lambda _{F},X)$ and $T\times\hat{D}%
(\lambda_{F})^{\prime}\hat{V}_{\theta\theta .f}(\lambda_{F})^{-1}\hat{D}%
(\lambda_{F}),$ which, as stated in Theorem 6a, is constant over $%
\lambda_{F}.$ We can then straightforwardly solve for the value of $%
\lambda_{F}$ that maximizes the DRLM statistic in a data set of realized
values. We use this maximizer to improve the power of the DRLM\ test, as follows.

The power of a $100\times\alpha\%$ significance DRLM\ test of H$%
_{0}:\lambda_{F}=\lambda_{F}^{1}$ can be improved by rejecting H$_{0}$
alongside for significant values of DRLM($\lambda_{F}^{1})$ also when both:

\begin{enumerate}
\item The maximal value of the DRLM\ statistic for the analyzed data set is
significant at the $100\times\alpha\%$ level.

\item The DRLM statistic evaluated at $\lambda _{F}^{1}$ is insignificant at
the $100\times \alpha \%$ level but $\lambda _{F}^{1}$ lies inside the
closed interval indicated by the significant maximizers of the DRLM\
statistic that does contain the CUE.
\end{enumerate}

The above algorithm rejects H$_{0}$ alongside for significant values of DRLM(%
$\lambda _{F}^{1})$ also when there is a significant value of the DRLM\
statistic on the line between $\lambda _{F}^{1}$ and the CUE. To show that
the above algorithm leads to a size correct test, we compute its rejection
frequency when testing H$_{0}:\lambda _{F}=0$ using the setup from Figures
1-3. While the generic specification of the DRLM\ test is for a stationary
point of the population continuous updating objective function, the above
algorithm explicitly tests for the minimizer. When computing the size of the
test at the hypothesized value, of, say, zero, we therefore have to
ascertain that it is the minimizer of the population objective function. For
the setup in Figures 1-3, which uses the limit expression of the DRLM\
statistic in (\ref{limit lmr iid}), the population minimizer is at zero if
the amount of misspecification is less than the strength of identification
so the length of $\bar{\mu}$ is less than that of $\bar{D}$, since the $IS$ identification measure (\ref{rank 3}) equals the quadratic form of $\bar{D}$. When the length
of $\bar{\mu}$ exceeds that of $\bar{D},$ the minimizer of the population
objective function is at $\pm \infty$ as discussed in Section 2. In standard GMM, there is no
misspecification so the amount of misspecification is then always less than or
equal to the identification strength, i.e.  the hypothesized value automatically
corresponds with the minimizer of the population objective function. 
\begin{equation*}
\begin{array}{c}
\text{Figure 4: Rejection frequency of 5\% significance tests of H}%
_{0}:\lambda _{F}=0\text{ using power improved DRLM} \\ 
\text{and a calibrated conditional 95\% critical value as a function of the
lengths of }\bar{\mu}\text{ and }\bar{D},\text{ }m=1,\text{ }N=25. \\ 
\\ 
\raisebox{-0pt}{\includegraphics[
height=2.2139in,
width=2.9386in
]%
{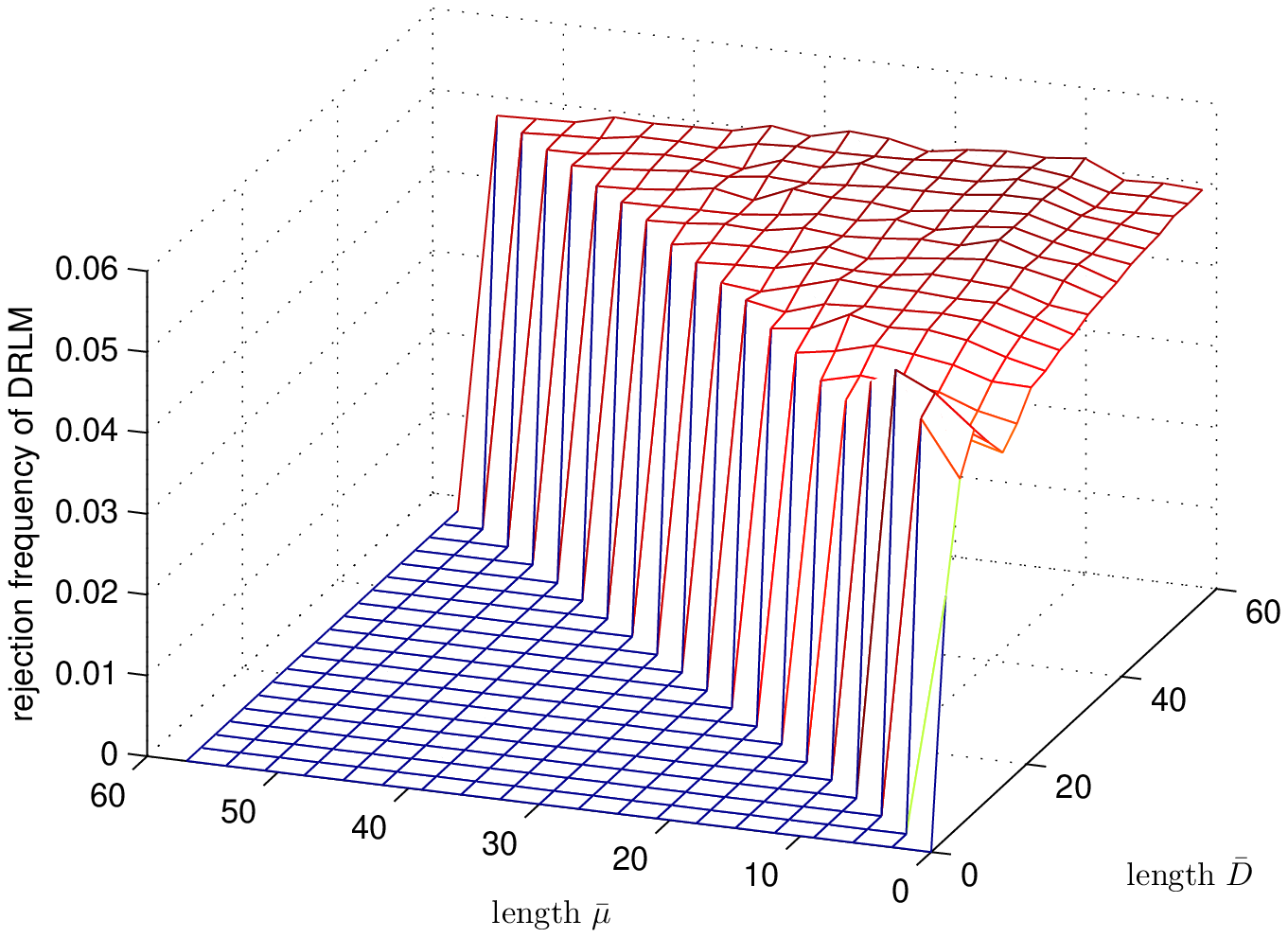}%
}
\raisebox{-0pt}{\includegraphics[
height=2.2139in,
width=2.9386in
]%
{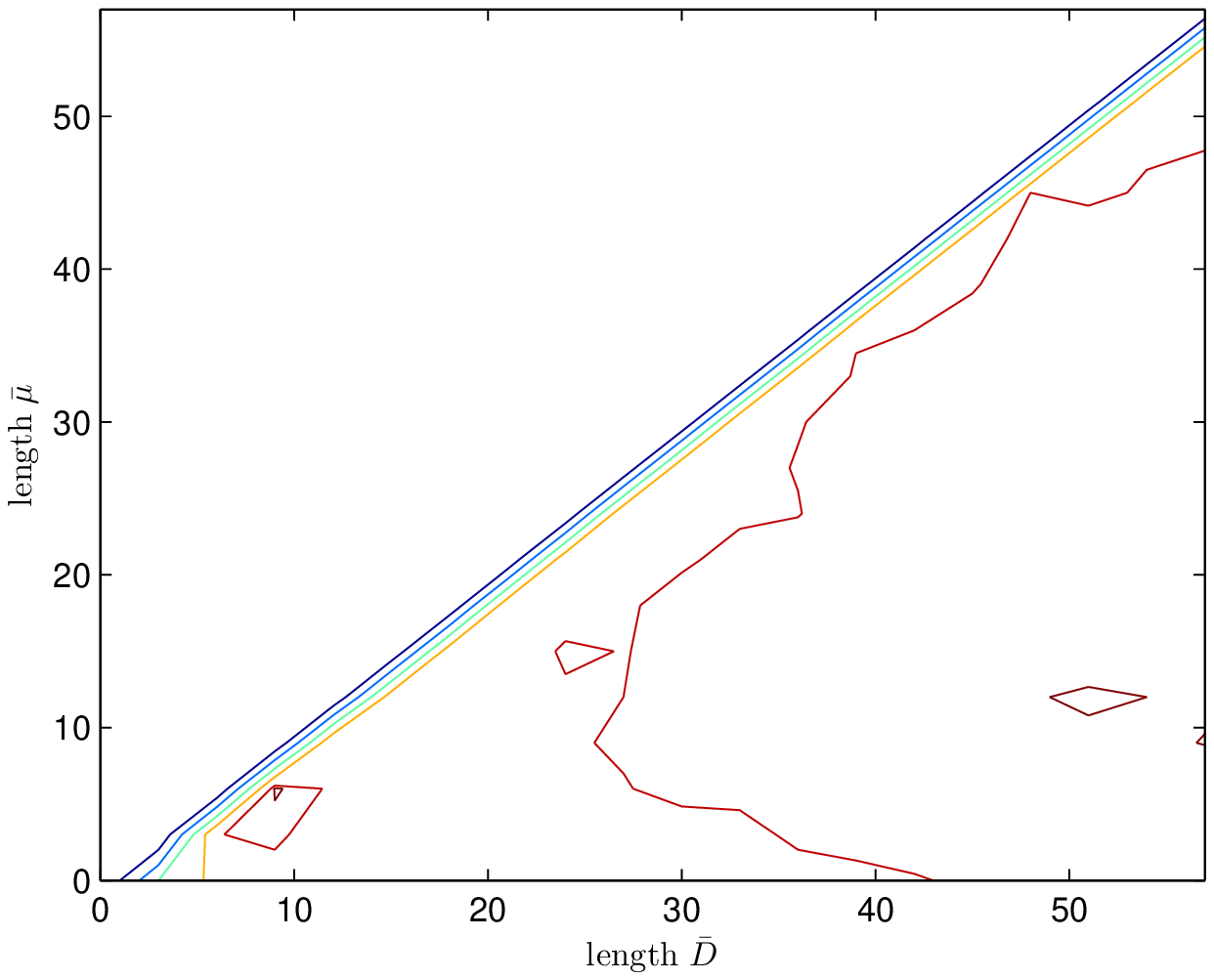}%
}
\end{array}%
\end{equation*}

Figure 4 shows the rejection frequency of the power improved DRLM\ test when
the minimizer of the population continuous updating objective function
equals the hypothesized value which is zero. Figure 4 does therefore not
show the rejection frequency for values where the length of $\bar{\mu}$
exceeds that of $\bar{D}$, since the hypothesized value does then not
correspond with the minimizer of the population objective function which is
at $\pm \infty .$ The rejection frequencies in Figure 4 are computed using
the calibrated conditional critical values explained previously. Figure 4
shows that the power improvement does not affect the size of the DRLM\ test
when the hypothesized value equals the minimizer of the population continuous updating
objective function.

\subsection{Power analysis}

We use the one factor linear asset pricing model to compare the power and
size of different identification robust test procedures with that of the
DRLM test. For the power analysis, the minimizer of the population
continuous updating objective function is the pseudo-true value $%
\lambda_{F}^{\ast}$ while we test for a zero value under the null
hypothesis. We then map out the power curve by changing the pseudo-true
value and keeping the hypothesized value, zero, fixed. Theorem 7 states the
limiting distributions of the different components of the DRLM\ statistic
for testing the hypothesis of interest used for the power analysis.

\paragraph{Theorem 7:}

For testing H$_{0}:\lambda _{F}=\lambda _{F}^{1}=0,$ the limit behaviors of
the components of the DRLM\ statistic in the one factor linear asset pricing
model with i.i.d. errors, $m=1$ and $Q_{\bar{F}\bar{F}}=1,$ while the
pseudo-true value equals $\lambda _{F}^{\ast },$ are characterized by:%
\begin{equation}
\begin{array}{rll}
\sqrt{T}\hat{\Omega}^{-\frac{1}{2}}\bar{R} & \underset{d}{\rightarrow } & 
\bar{\mu}(1+(\lambda _{F}^{\ast })^{2})^{-\frac{1}{2}}-\bar{D}(1+(\lambda
_{F}^{\ast })^{2})^{-\frac{1}{2}}\lambda _{F}^{\ast }+\psi _{f}^{\ast
}(\lambda _{F}^{1}=0), \\ 
\sqrt{T}\hat{\Omega}^{-\frac{1}{2}}\hat{D}(\lambda _{F}^{1}=0) & \underset{d}%
{\rightarrow } & \bar{D}(1+(\lambda _{F}^{\ast })^{2})^{-\frac{1}{2}}+\bar{%
\mu}(1+(\lambda _{F}^{\ast })^{2})^{-\frac{1}{2}}\lambda _{F}^{\ast }+\psi
_{\theta .f}^{\ast }(\lambda _{F}^{1}=0),%
\end{array}
\label{power spec}
\end{equation}%
with $\psi _{f}^{\ast }(\lambda _{F}^{1}=0)$, $\psi _{\theta .f}^{\ast
}(\lambda _{F}^{1}=0)$ independent standard normal $N$ dimensional random
vectors, $\mu ^{\ast }=\lim_{T\rightarrow \infty }\sqrt{T}\mu _{f}(\lambda
_{F}^{\ast }),$ $\mu _{f}(\lambda _{F}^{\ast })=\mu _{R}-\beta \lambda
_{F}^{\ast },$ $D^{\ast }=\lim_{T\rightarrow \infty }\sqrt{T}D(\lambda
_{F}^{\ast }),$ $D(\lambda _{F}^{\ast })=-\beta -\mu _{f}(\lambda _{F}^{\ast
})\lambda _{F}^{\ast \prime }(Q_{\bar{F}\bar{F}}+\lambda _{F}^{\ast }\lambda
_{F}^{\ast \prime })^{-1},$ $\bar{\mu}=\Omega ^{-\frac{1}{2}}\mu ^{\ast
}(1+\lambda _{F}^{\ast \prime }Q_{\bar{F}\bar{F}}^{-1}\lambda _{F}^{\ast
})^{-\frac{1}{2}},\mathbf{\ }\bar{D}=\Omega ^{-\frac{1}{2}}D^{\ast }(Q_{\bar{%
F}\bar{F}}+\lambda _{F}^{\ast }\lambda _{F}^{\ast \prime })^{\frac{1}{2}},$
so $\bar{\mu}^{\prime }\bar{D}\equiv 0.$ \smallskip\ 

\begin{proof}
See the Online Appendix.\smallskip
\end{proof}

The specification in Theorem 7 is such that, since $\bar{\mu}^{\prime }\bar{D%
}\equiv 0,$ $\lambda _{F}^{\ast }$ is the minimizer of the population
continuous updating objective function when the length of $\bar{D},$ whose quadratic form equals the $IS$ identification strength measure (\ref{rank 3}), is larger than or equal to the
length of $\bar{\mu},$ which reflects misspecification. The product of the
limit behavior of both components in (\ref{power spec}): 
\begin{equation}
\begin{array}{l}
T\bar{R}^{\prime }\hat{\Omega}^{-1}\hat{D}(\lambda _{F}^{1}=0)\underset{d}{%
\rightarrow }(1+(\lambda _{F}^{\ast })^{2})^{-1}\lambda _{F}^{\ast }\left( 
\bar{\mu}^{\prime }\bar{\mu}-\bar{D}^{\prime }\bar{D}\right) + \\ 
\quad (1+(\lambda _{F}^{\ast })^{2})^{-\frac{1}{2}}\left[ \psi _{f}^{\ast
}(\lambda _{F}^{1}=0)^{\prime }\left( \bar{D}+\bar{\mu}\lambda _{F}^{\ast
}\right) +\psi _{\theta .f}^{\ast }(\lambda _{F}^{1}=0)^{\prime }\left( \bar{%
\mu}-\bar{D}\lambda _{F}^{\ast }\right) \right] ,%
\end{array}
\label{cross theo 7}
\end{equation}%
further shows that identification is problematic when the lengths of $\bar{%
\mu}$ and $\bar{D}$ are equal so the amount of misspecification equals the
identification strength.

We next analyze the power of identification robust tests and the DRLM test
for two settings of misspecification: no misspecification, and weak
misspecification. The power analysis for a mildly misspecified setting is discussed in the Online Appendix.

\subsubsection{No misspecification}

We first compare the power of the DRLM test with existing identification
robust tests when no misspecification is present, so all of these tests are
size correct. Figures  5-7 show the different
power curves. Figure 5 shows the power curves of the KLM\ test of Kleibergen
(2002, 2005, 2009)\nocite{kf00b}\nocite{kf00a}\nocite{kf09} and the DRLM\
test for various identification strengths and no misspecification. The power
of the KLM\ test is known to be non-monotonic which is in line with Panel
5.1. Panel 5.2 shows that power curves of the DRLM test are non-monotonic
as well. 

\begin{equation*}
\begin{array}{c}
\text{Figure 5: Power of 5\% significance KLM and DRLM tests of} \\ 
\text{ H}_{0}:\lambda_{F}=0\text{ with no misspecification, }N=25,\text{ }Q_{%
\bar{F}\bar{F}}=1 \\ 
\begin{array}{cc}
\raisebox{-0pt}{\includegraphics[
height=2.0139in,
width=2.9386in
]%
{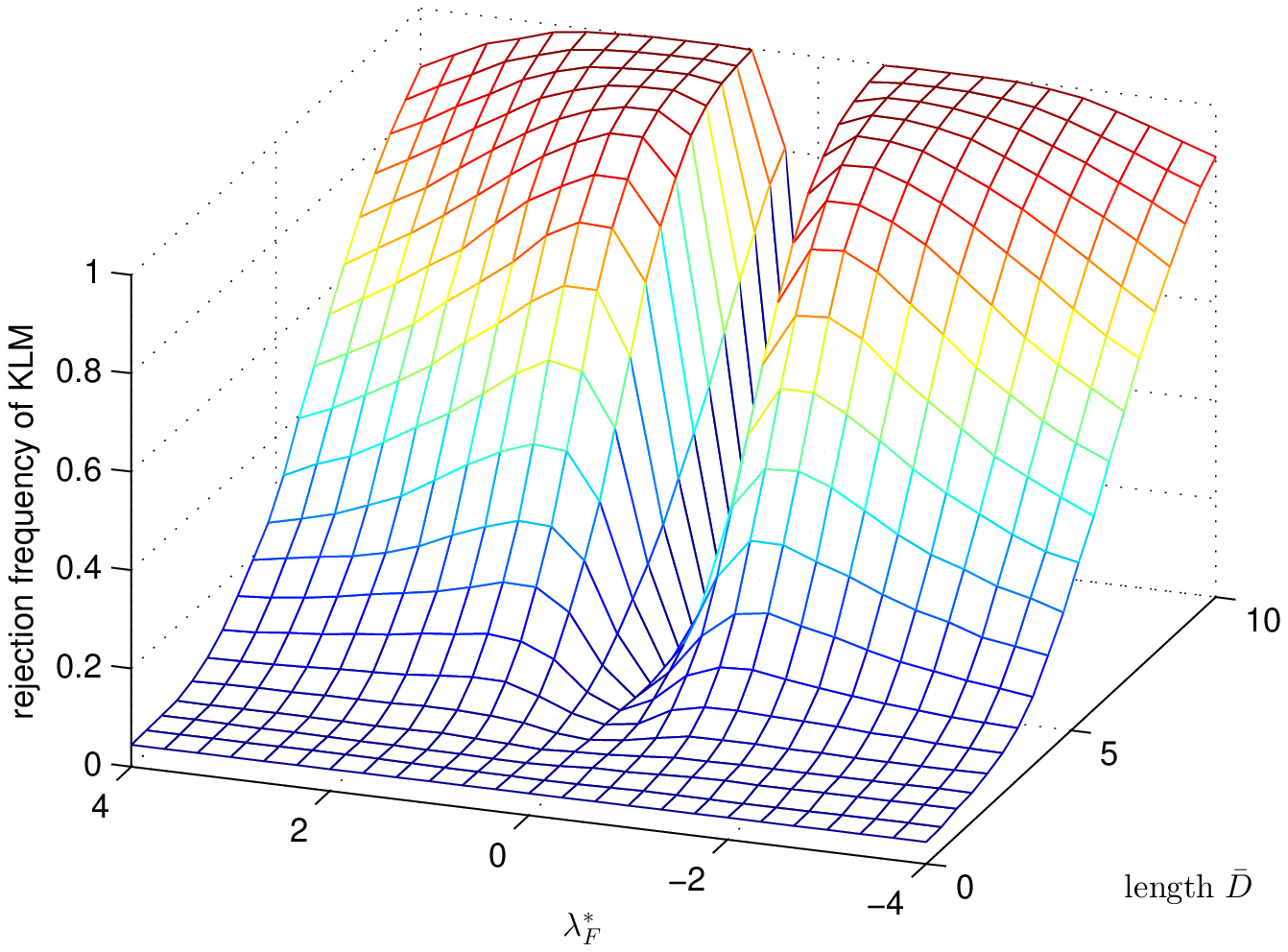}%
}
&
\raisebox{-0pt}{\includegraphics[
height=2.0139in,
width=2.9386in
]%
{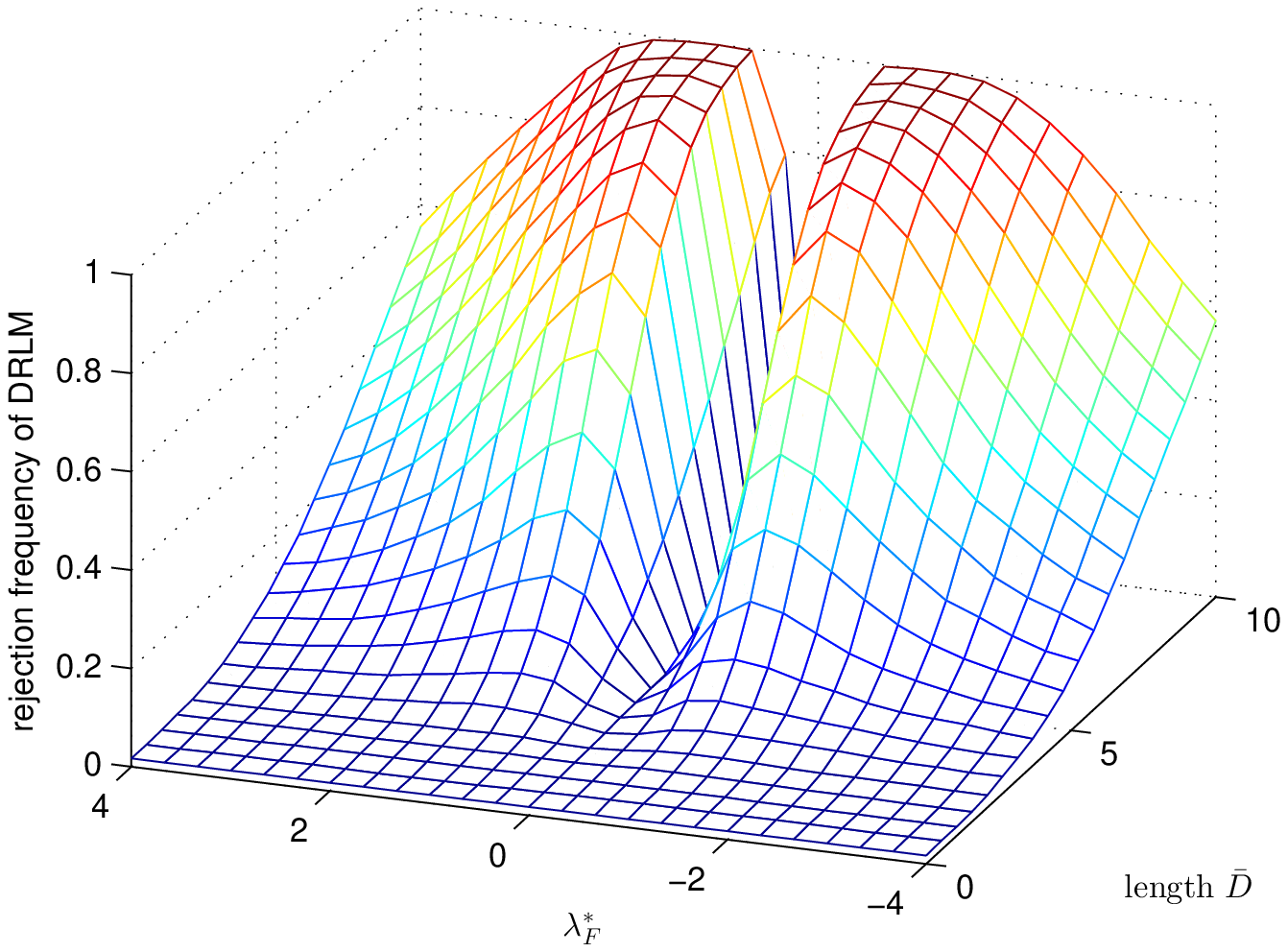}%
}
\\
\text{Panel 5.1: KLM} & \text{Panel 5.2: DRLM}%
\end{array}%
\end{array}%
\end{equation*}

\begin{equation*}
\begin{array}{c}
\text{Figure 6: Power of 5\% significance LR and size and power improved } \\ 
\text{DRLM tests of H}_{0}:\lambda_{F}=0\text{ with no misspecification, }%
N=25,\text{ }Q_{\bar{F}\bar{F}}=1 \\ 
\begin{array}{cc}
\raisebox{-0pt}{\includegraphics[
height=2.0139in,
width=2.9386in
]%
{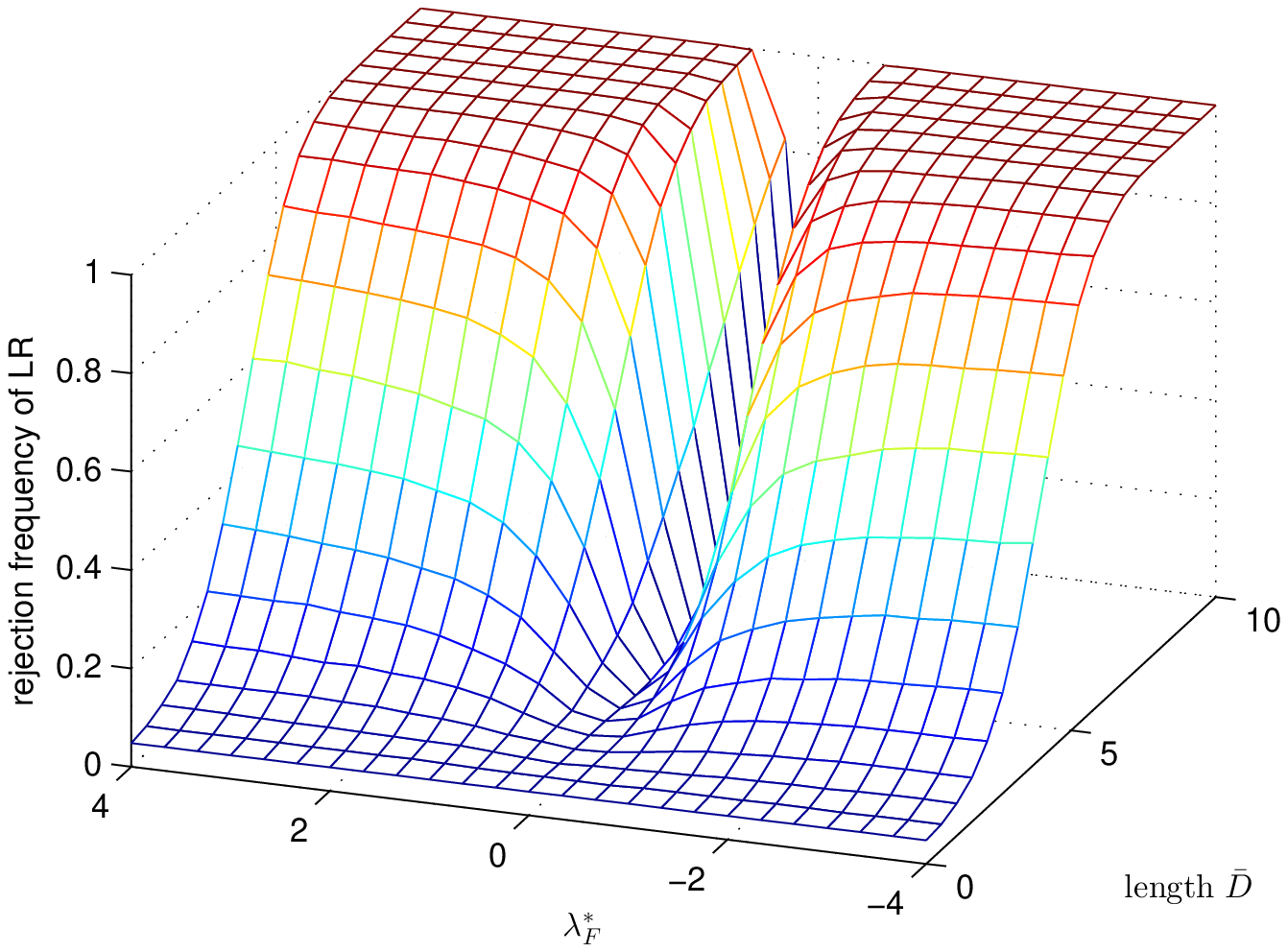}%
}
&
\raisebox{-0pt}{\includegraphics[
height=2.0139in,
width=2.9386in
]%
{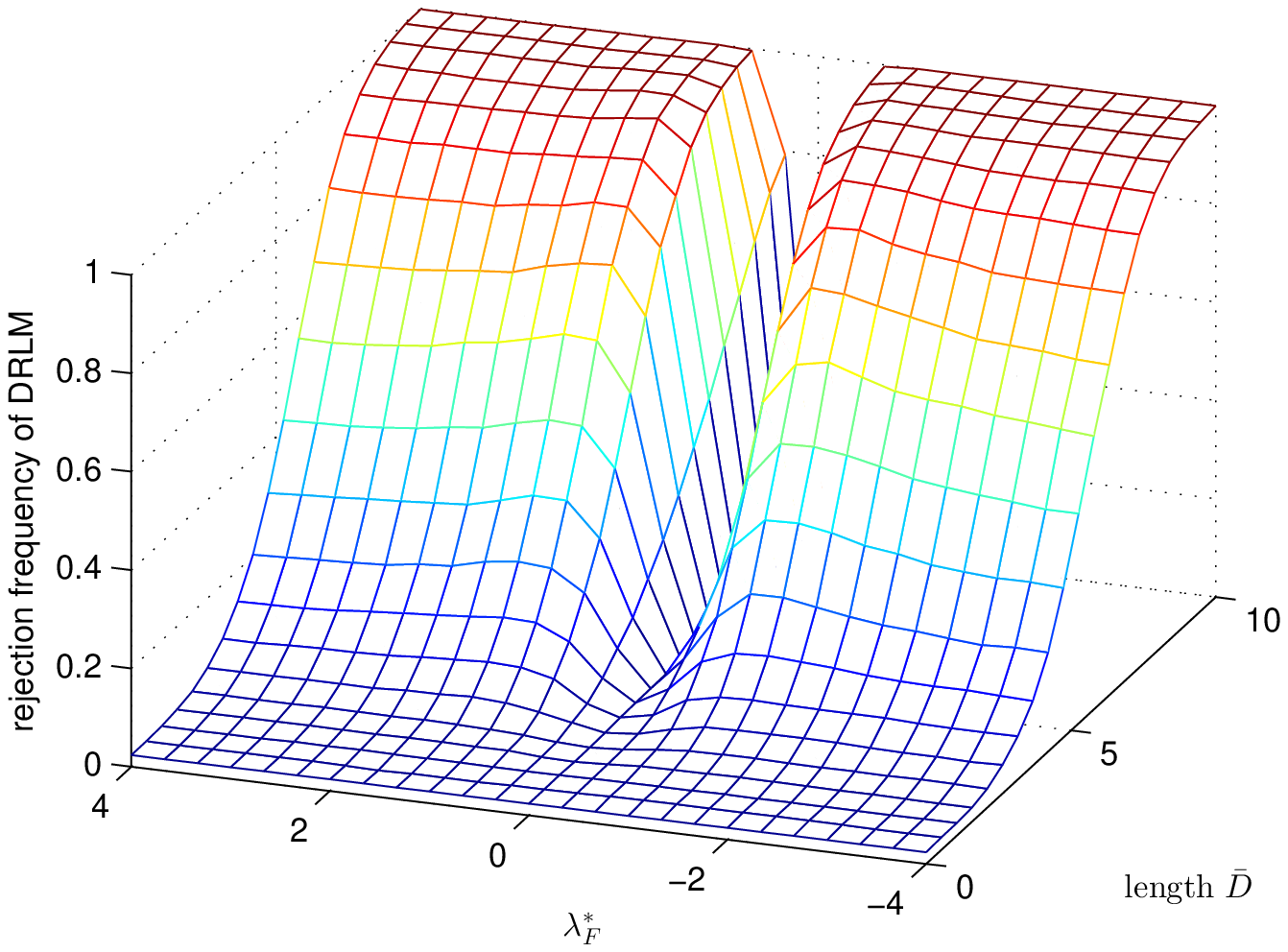}%
}
\\

\text{Panel 6.1: LR} & \text{Panel 6.2: DRLM with size and } \\ 
& \text{power improvements}%
\end{array}%
\end{array}%
\end{equation*}

Panel 6.2 in Figure 6 shows that the size and power improved DRLM test,
which uses the size and power improvement procedures discussed previously,
has a nearly monotonic power curve. Panel 6.1 in Figure 6 shows power curves
of the conditional likelihood ratio (LR) test of Moreira (2003) which is
known to be optimal for this setting, see Andrews et al. (2006).\nocite%
{andms05} Figure 7 shows power curves of the factor Anderson-Rubin (AR)
test, see Anderson and Rubin (1949) and Kleibergen (2009).\nocite{AR49}%
\nocite{kf09}  Overall, Figures 5-7 show that without misspecification, DRLM is comparable to several existing identification robust tests.
\begin{equation*}
\begin{array}{c}
\text{Figure 7: Power of 5\% significance  AR tests of H}_{0}:\lambda_{F}=0 \text{ with no misspecification, }N=25,\text{ }%
Q_{\bar{F}\bar{F}}=1 \\ 
\begin{array}{c}
\raisebox{-0pt}{\includegraphics[
height=2.0139in,
width=2.9386in
]%
{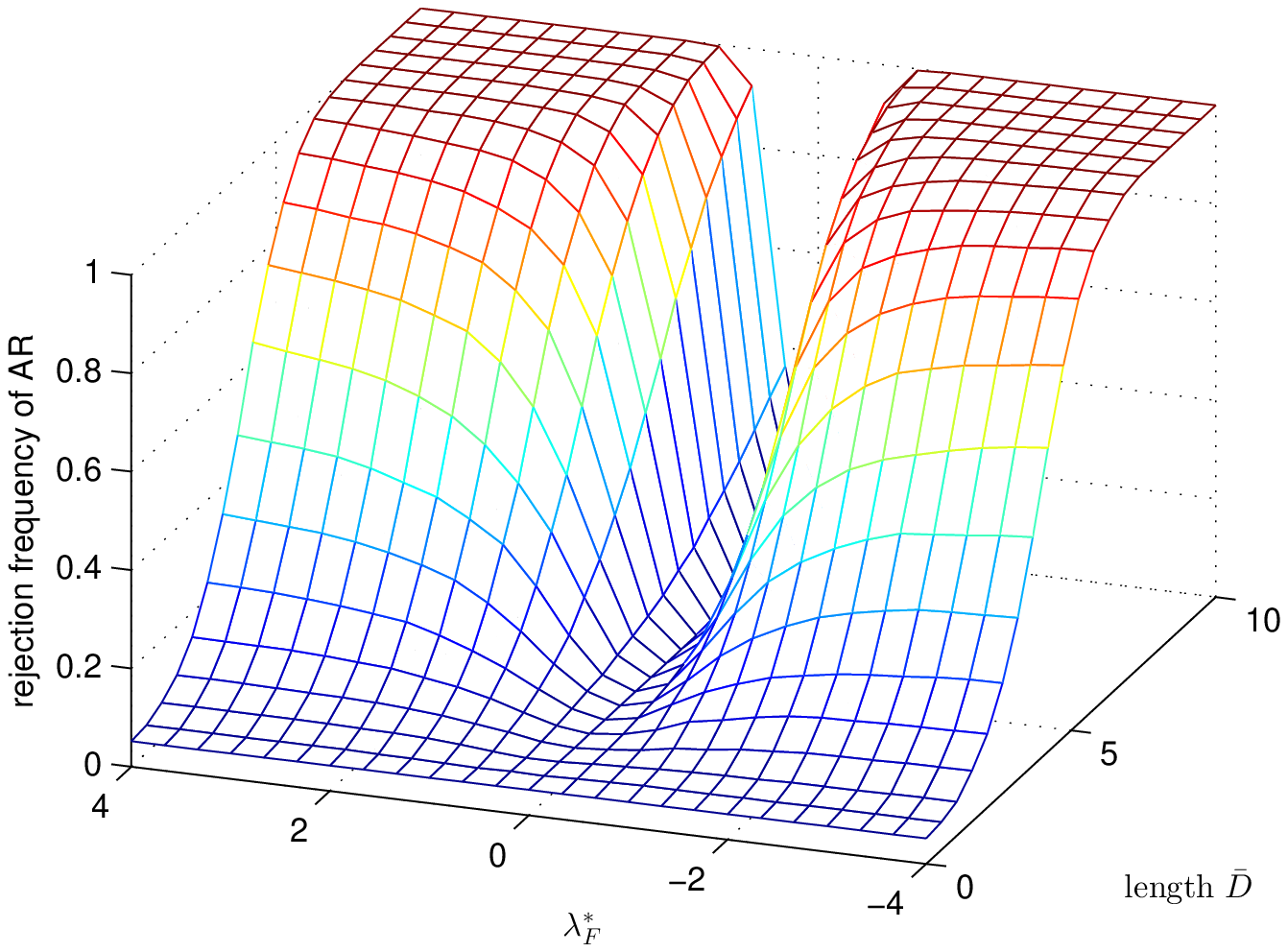}%
}
\end{array}%
\end{array}%
\end{equation*}

\subsubsection{Weak misspecification}

We next compare the power of the different test procedures in a setting of
weak misspecification where $\bar{\mu}^{\prime }\bar{\mu}=4.4$. Figure 8 therefore shows power curves of the KLM\ and DRLM tests
for various identification strengths, while  Figure 9
shows power curves of the LR and size and power corrected DRLM\ test. Figure
10 shows power curves of the factor AR test. The power curves of the different test
procedures are comparable to the ones in the previous Figures 5-7 except that we
observe size distortion of the identification robust factor AR, KLM and LR tests in Figures 8-10.
Except for the factor AR test, these size distortions become less when the
identification strength increases. For the conditional LR test, the rejection
frequency at zero decreases from 15\% to 9\% when the identification
strength increases. It equals 13\% when the amount of misspecification
equals the identification strength. For the KLM test, it decreases from 7\%\
to 5\%. For the factor AR test, the rejection frequency at zero equals 15\% for all
settings of the identification strength, since no estimator of the
identification strength is involved in the factor AR test. For the DRLM and size
and power improved DRLM tests, we observe no size distortion.

What is striking is that, for small values of the identification strength,
the power of the identification robust factor AR and LR tests decreases when $%
\lambda _{F}^{\ast }$ moves away from zero. This results since when the
amount of misspecification exceeds the identification strength, the
population continuous updating objective function is maximized at zero
instead of minimized. The population continuous updating objective function
is then minimized when $\lambda _{F}$ equals $\pm \infty .$ When the
strength of identification equals zero, so the length of $\bar{D}=0,$ the
moment equation (\ref{meandef}) is, however, still not satisfied at these
values of $\lambda _{F}$ so the LR, KLM and factor AR tests remain size distorted
even at these values. Moving away from zero at these settings of the
identification strength, however, in general reduces the sample continuous
updating objective function, which then leads to a lower rejection frequency
of these tests. For values of the identification strength exceeding the
amount of misspecification, the population continuous updating objective
function is minimized at zero, so we then no longer observe a reduction of
the rejection frequency when $\lambda _{F}^{\ast }$ moves away from zero.%
\begin{equation*}
\begin{array}{c}
\text{Figure 8: Power of 5\% significance KLM and DRLM tests of} \\ 
\text{ H}_{0}:\lambda _{F}=0\text{ with misspecification, }\bar{\mu}^{\prime
}\bar{\mu}=4.4,\text{ }N=25,\text{ }Q_{\bar{F}\bar{F}}=1 \\ 
\begin{array}{cc}
\raisebox{-0pt}{\includegraphics[
height=2.0139in,
width=2.9386in
]%
{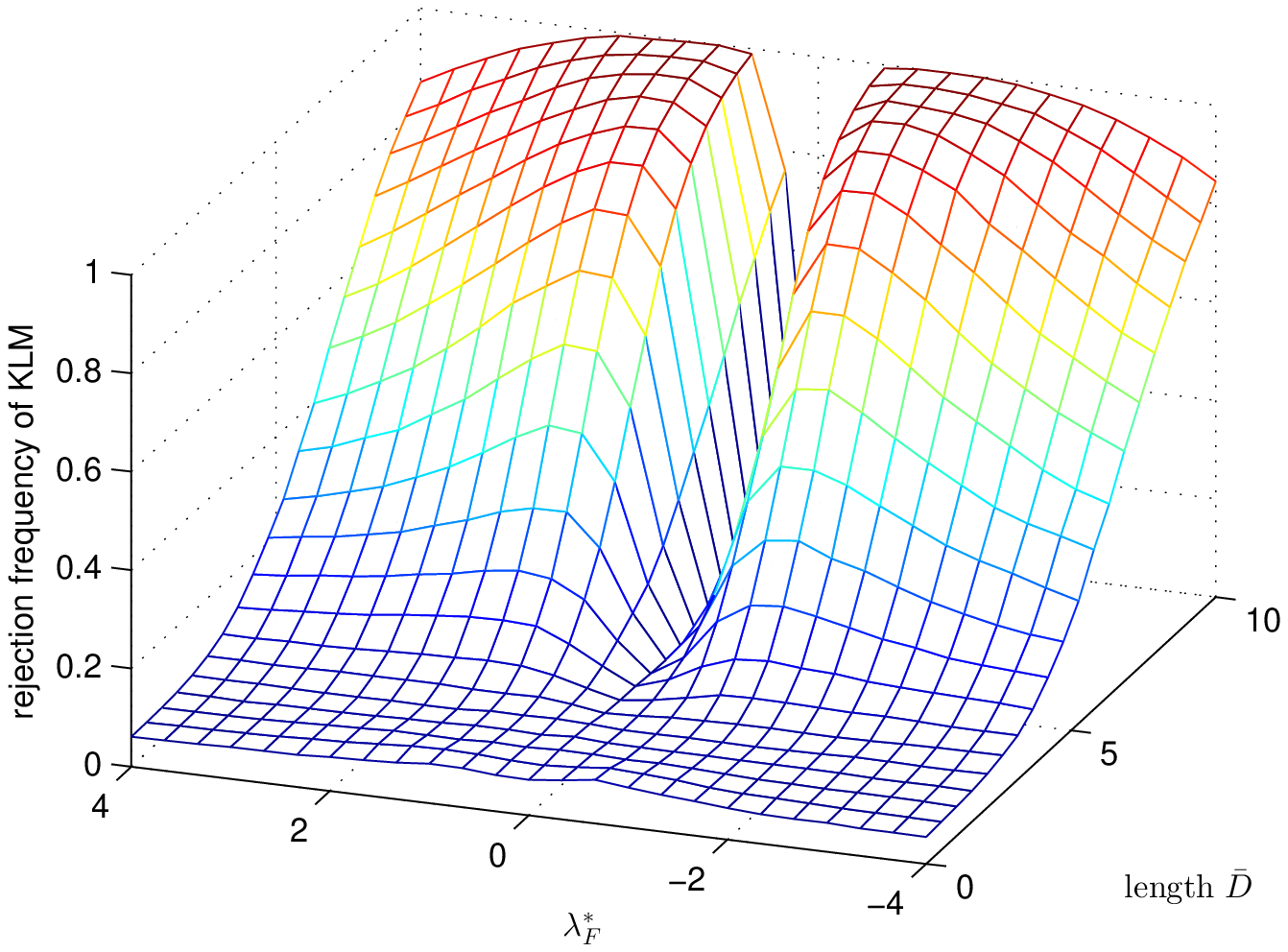}%
}
&
\raisebox{-0pt}{\includegraphics[
height=2.0139in,
width=2.9386in
]%
{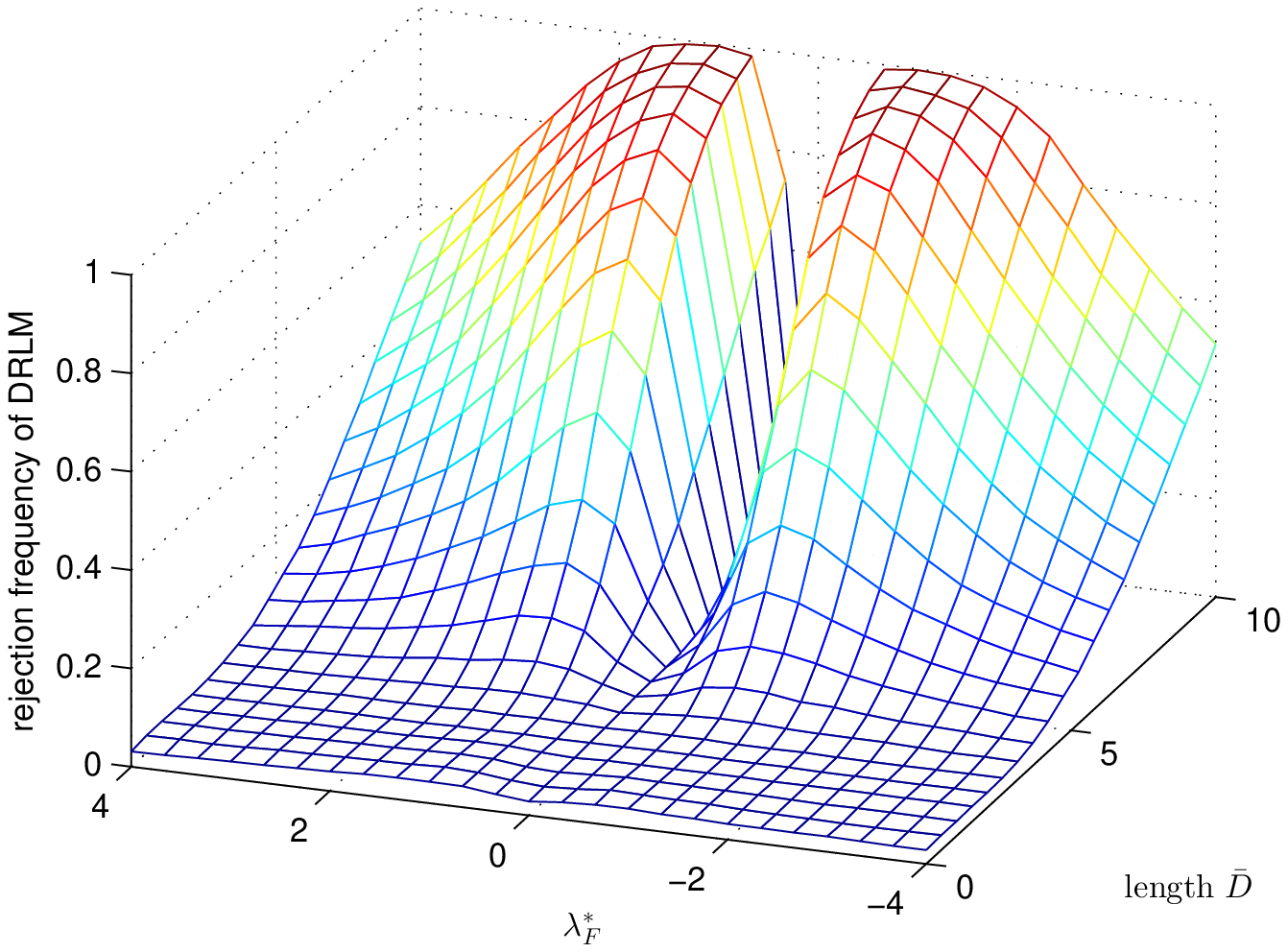}%
}
\\
\text{Panel 8.1: KLM} & \text{Panel 8.2: DRLM}%
\end{array}%
\end{array}%
\end{equation*}

\begin{equation*}
\begin{array}{c}
\text{Figure 9: Power of 5\% significance LR and size and power improved } \\ 
\text{DRLM tests of H}_{0}:\lambda_{F}=0\text{ with misspecification, }\bar{%
\mu}^{\prime}\bar{\mu}=4.4,\text{ }N=25,\text{ }Q_{\bar{F}\bar{F}}=1 \\ 
\begin{array}{cc}
\raisebox{-0pt}{\includegraphics[
height=2.0139in,
width=2.9386in
]%
{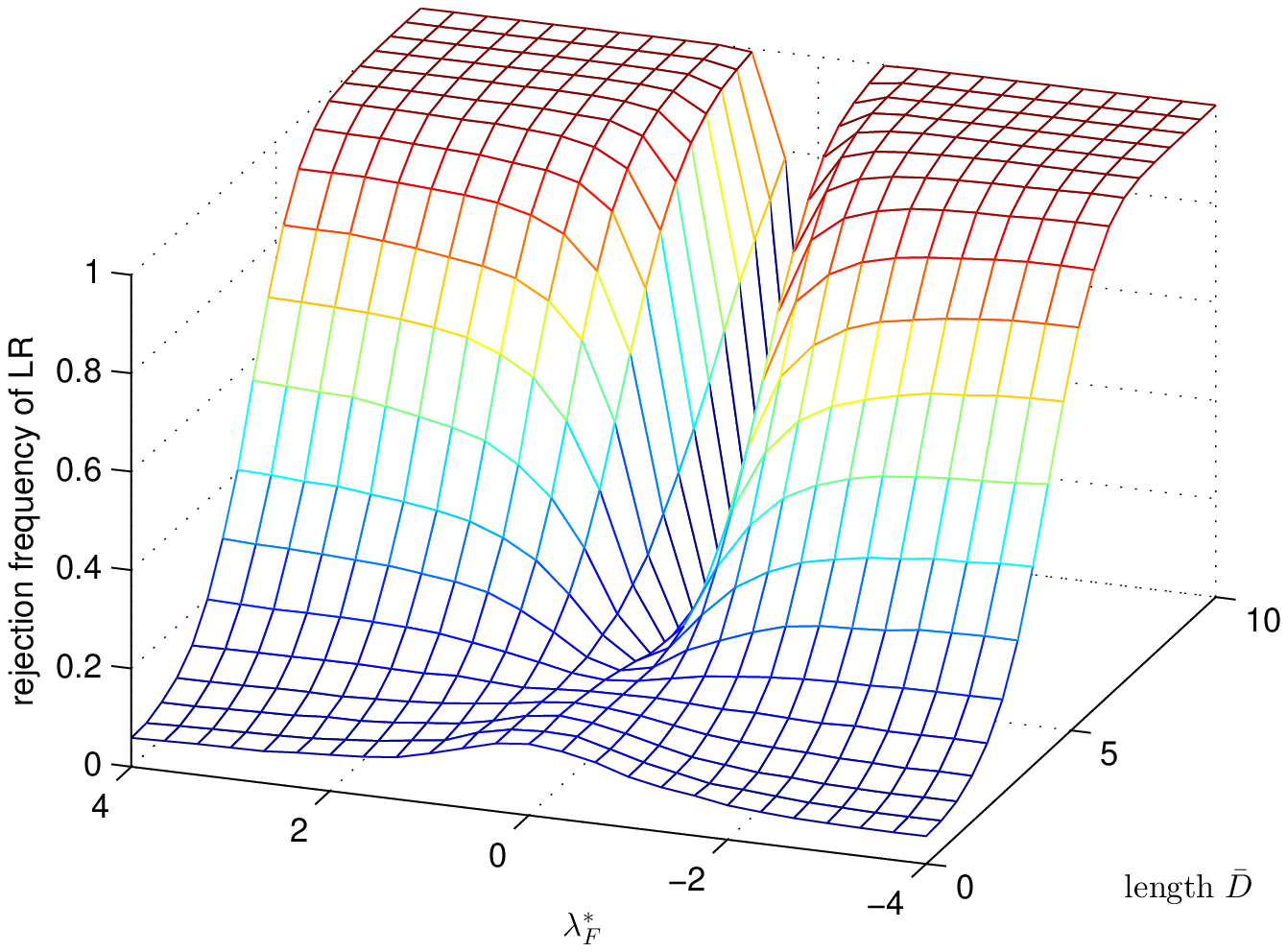}%
}
&
\raisebox{-0pt}{\includegraphics[
height=2.0139in,
width=2.9386in
]%
{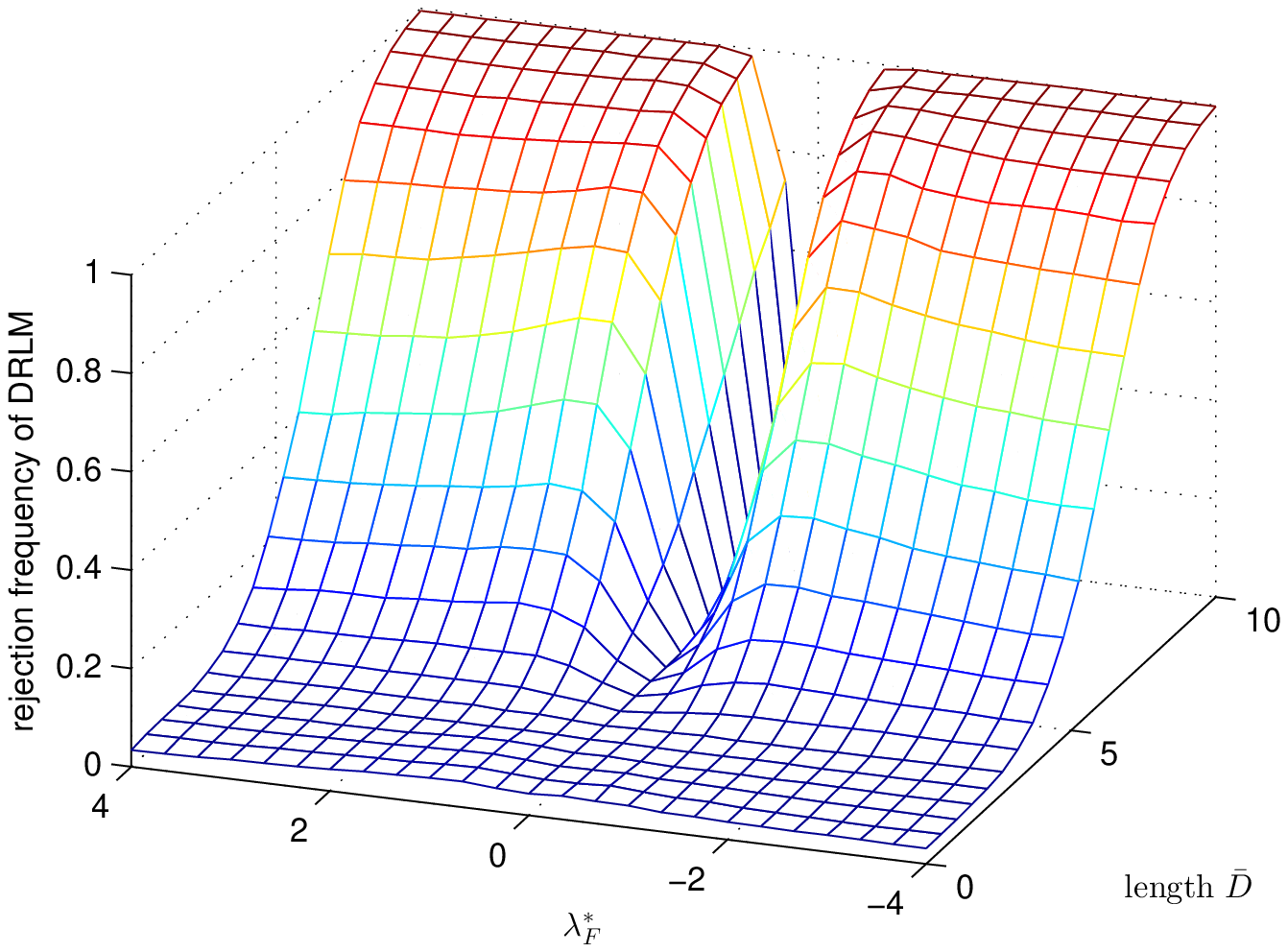}%
}\\ 
\text{Panel 9.1: LR} & \text{Panel 9.2: DRLM with size and } \\ 
& \text{power improvements}%
\end{array}%
\end{array}%
\end{equation*}

\begin{equation*}
\begin{array}{c}
\text{Figure 10: Power of 5\% significance AR tests of H}_{0}:\lambda_{F}=0 \ \text{with misspecification,  } \\ \bar{\mu}^{\prime}\bar{\mu}=4.4,\text{ }N=25,%
\text{ }Q_{\bar{F}\bar{F}}=1 \\ 
\begin{array}{c}
\raisebox{-0pt}{\includegraphics[
height=2.0139in,
width=2.9386in
]%
{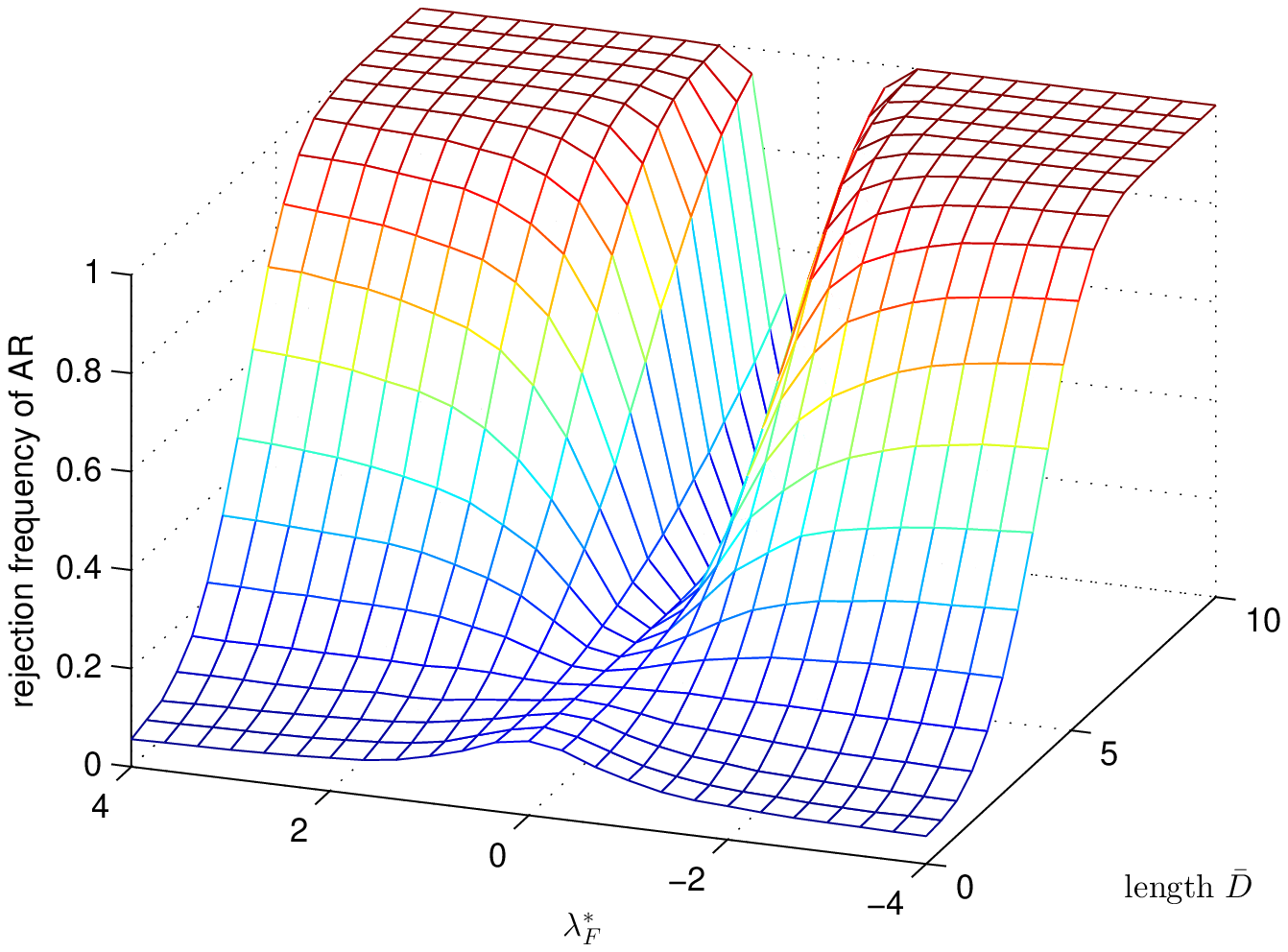}%
}

\end{array}%
\end{array}%
\end{equation*}
\begin{equation*}
\begin{array}{c}
\text{Figure 11: Distribution function of $J$-statistic for misspecification
when H}_{0}:\lambda _{F}=0\text{ holds, } \\ 
\text{solid line: }\bar{D}^{\prime }\bar{D}=0,\text{ dash-dot: }\bar{D}%
^{\prime }\bar{D}=4.4=\text{strength of misspecification, dashed: }\bar{D}%
^{\prime }\bar{D}=100. \\ 
\raisebox{-0pt}{\includegraphics[
height=2.0139in,
width=2.9386in
]%
{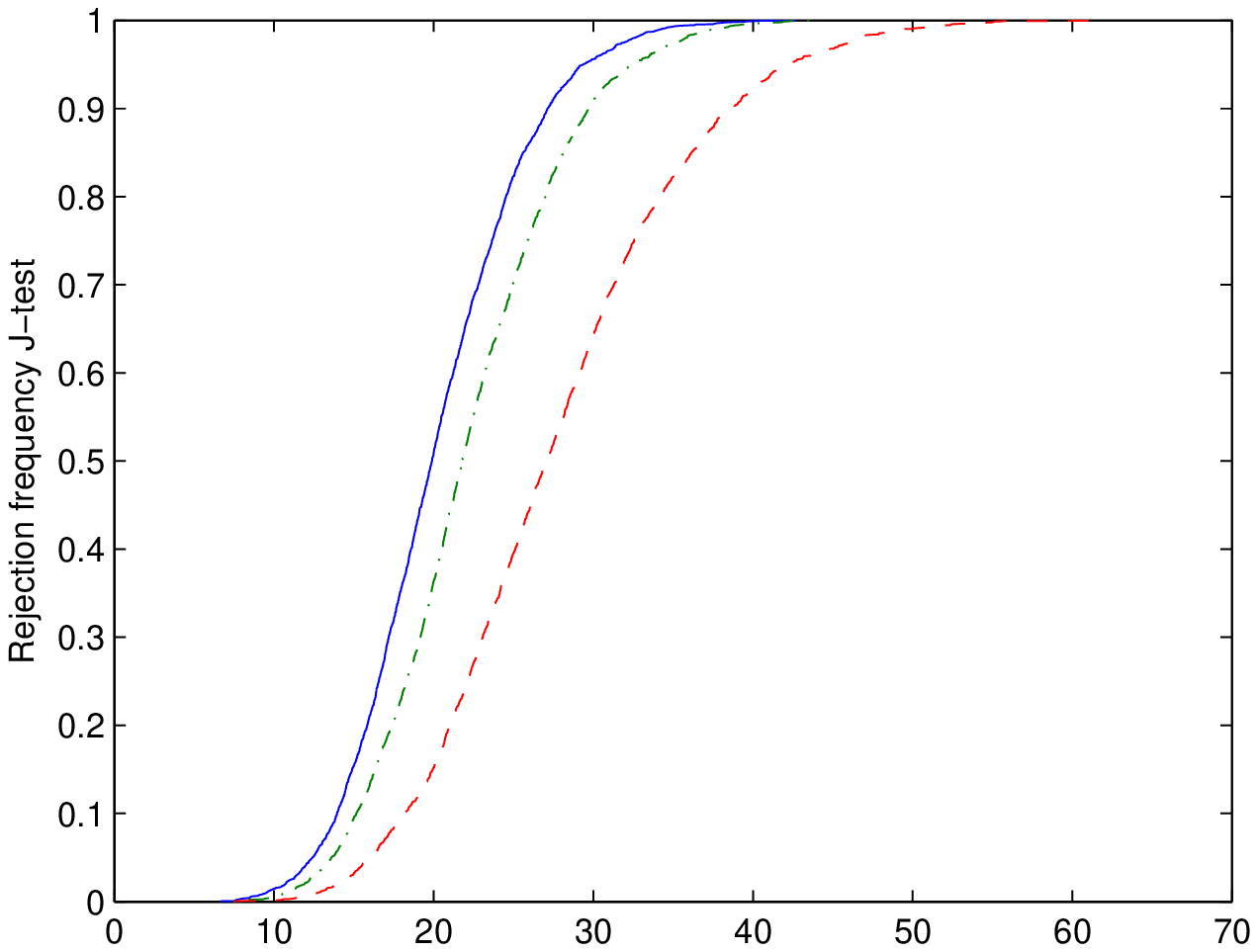}%
}
\\
\end{array}%
\end{equation*}

To show the difficulty of detecting the weak misspecification used in Figures 8-10, Figure 11 presents the simulated distribution function of
the misspecification $J$-statistic, which equals the minimal value of the factor AR
statistic for the simulated data, when the null hypothesis holds, so for
values of $\lambda _{F}^{\ast }$ equal to zero. In particular, Figure 11 shows the
distribution function of the  $J$-statistic for three
different values of the identification strength $\bar{D}^{\prime }\bar{D}:$
0, 4.4 and 100. In Guggenberger et al. (2012),\nocite{gkmc12} it is shown
that the distribution function of the $J$-statistic is a non-increasing
function of the identification strength. Recognizing that the 95\% critical
value of the $\chi ^{2}(24)$ distribution$,$ since $N-1=24,$ equals 36.42,
Figure 11 shows that we never reject no misspecification at the 5\%
significance level when $\bar{D}^{\prime }\bar{D}$ equals 0 or 4.4, and we
only do so in 15\% of the cases when $\bar{D}^{\prime }\bar{D}$ equals 100.
Thus, Figure 11 illustrates the difficulty of detecting weak misspecification. In the Online Appendix, we also discuss a setting of mild misspecification with $\bar{\mu}^{\prime
}\bar{\mu}=10$ where it is also very hard for the $J$-test to detect misspecification, and the size distortions of the weak identification robust tests become even more pronounced while the DRLM test remains size correct.

\subsection{More power improvements?}

We further analyze the power of invariant tests for which we use that they
are a function of the maximal invariant. We therefore construct the maximal
invariant for a stylized setting of the linear asset pricing model with
independent normal errors, a known value of the covariance matrix and a
fixed number of observations, see also Andrews et al. (2006) which uses an
identical setting for the linear IV regression model. In order to do so, we
first conduct a singular value decomposition of $\Omega ^{-\frac{1}{2}%
}\left( 
\begin{array}{cc}
\ddot{\mu}_{R} & \ddot{\beta}%
\end{array}%
\right) \left( 
\begin{array}{cc}
1 & 0 \\ 
0 & Q_{\bar{F}\bar{F}}^{\frac{1}{2}}%
\end{array}%
\right) ,$ with $\ddot{\mu}_{R}=\sqrt{T}\mu _{R},$ $\ddot{\beta}=\sqrt{T}%
\beta ,$ which is invariant to transformations and whose least squares
estimator has an identity covariance matrix.

\paragraph{Theorem 8:}

A singular value decomposition of $\Omega ^{-\frac{1}{2}}\left( 
\begin{array}{cc}
\ddot{\mu}_{R} & \ddot{\beta}%
\end{array}%
\right) \left( 
\begin{array}{cc}
1 & 0 \\ 
0 & Q_{\bar{F}\bar{F}}^{\frac{1}{2}}%
\end{array}%
\right) $ results in:%
\begin{equation}
\begin{array}{l}
\Omega ^{-\frac{1}{2}}\left( 
\begin{array}{cc}
\ddot{\mu}_{R} & \ddot{\beta}%
\end{array}%
\right) \left( 
\begin{array}{cc}
1 & 0 \\ 
0 & Q_{\bar{F}\bar{F}}^{\frac{1}{2}}%
\end{array}%
\right) =\mathcal{USV}^{\prime }= \\ 
-\Omega ^{-\frac{1}{2}}D(\lambda _{F}^{\ast })\left( 
\begin{array}{cc}
\lambda _{F}^{\ast } & I_{m}%
\end{array}%
\right) \left( 
\begin{array}{cc}
1 & 0 \\ 
0 & Q_{\bar{F}\bar{F}}^{\frac{1}{2}}%
\end{array}%
\right) +\Omega ^{\frac{1}{2}}D(\lambda _{F}^{\ast })_{\perp }\delta \left( 
\begin{array}{cc}
\lambda _{F}^{\ast } & I_{m}%
\end{array}%
\right) _{\perp }\left( 
\begin{array}{cc}
1 & 0 \\ 
0 & Q_{\bar{F}\bar{F}}^{-\frac{1}{2}}%
\end{array}%
\right) ,%
\end{array}
\label{svdlinfac}
\end{equation}%
with $\mathcal{U}$ an $N\times N$ dimensional orthonormal matrix, $\mathcal{V}
$ an $(m+1)\times (m+1)$ dimensional orthonormal matrix, and $S$ an $N\times
(m+1)$ dimensional diagonal matrix with the singular values in decreasing
order on the main diagonal:%
\begin{equation}
\mathcal{U}=\left( 
\begin{array}{ll}
\mathcal{U}_{11} & \mathcal{U}_{12} \\ 
\mathcal{U}_{21} & \mathcal{U}_{22}%
\end{array}%
\right) ,\ S=\left( 
\begin{array}{ll}
\mathcal{S}_{1} & 0 \\ 
0 & \mathcal{S}_{2}%
\end{array}%
\right) \text{ and }V=\left( 
\begin{array}{ll}
\mathcal{V}_{11} & \mathcal{V}_{12} \\ 
\mathcal{V}_{21} & \mathcal{V}_{22}%
\end{array}%
\right) ,  \label{svd elements}
\end{equation}%
where $\mathcal{U}_{11},$ $\mathcal{S}_{1},$ $\mathcal{V}_{21}$ are $m\times
m$ dimensional matrices$;$ $\mathcal{S}_{2}$ is an $(N-m)\times 1$
dimensional matrix, $\mathcal{V}_{11}^{\prime },$ $\mathcal{V}_{22}$ are $%
m\times 1$ dimensional vectors, $\mathcal{U}_{12},$ $\mathcal{U}_{21},$ and $%
\mathcal{U}_{22}$ are $m\times (N-m),$ $(N-m)\times m$ and $(N-m)\times (N-m)
$ dimensional matrices and $\mathcal{V}_{12}$ is a scalar. The $N\times (N-m)
$ dimensional matrix $D(\lambda _{F}^{\ast })_{\perp }$ is the orthogonal
complement of $D(\lambda _{F}^{\ast }),$ $D(\lambda _{F}^{\ast })_{\perp
}^{\prime }D(\lambda _{F}^{\ast })\equiv 0,$ $D(\lambda _{F}^{\ast })_{\perp
}^{\prime }\Omega D(\lambda _{F}^{\ast })_{\perp }\equiv I_{N-m};$ and $%
\left( 
\begin{array}{cc}
\lambda _{F}^{\ast } & I_{m}%
\end{array}%
\right) _{\perp }$ is the $1\times (m+1)$ dimensional orthogonal complement
of $\left( 
\begin{array}{cc}
\lambda _{F}^{\ast } & I_{m}%
\end{array}%
\right) ,$ $\left( 
\begin{array}{cc}
\lambda _{F}^{\ast } & I_{m}%
\end{array}%
\right) \left( 
\begin{array}{cc}
\lambda _{F}^{\ast } & I_{m}%
\end{array}%
\right) _{\perp }^{\prime }\equiv 0$ and $\left( 
\begin{array}{cc}
\lambda _{F}^{\ast } & I_{m}%
\end{array}%
\right) _{\perp }\left( 
\begin{array}{cc}
1 & 0 \\ 
0 & Q_{\bar{F}\bar{F}}^{-1}%
\end{array}%
\right) \left( 
\begin{array}{cc}
\lambda _{F}^{\ast } & I_{m}%
\end{array}%
\right) _{\perp }^{\prime}\equiv 1,$ so $\left( 
\begin{array}{cc}
\lambda _{F}^{\ast } & I_{m}%
\end{array}%
\right) _{\perp }=\left( 
\begin{array}{cc}
1 & -\lambda _{F}^{\ast \prime }%
\end{array}%
\right) \allowbreak \left( 1+\lambda _{F}^{\ast \prime }Q_{\bar{F}\bar{F}%
}^{-1}\lambda _{F}^{\ast }\right) ^{-\frac{1}{2}}:$ 

\begin{equation}
D(\lambda _{F}^{\ast })=-\Omega ^{\frac{1}{2}}\mathcal{U}_{1}S_{1}\mathcal{V}%
_{21}^{\prime }Q_{\bar{F}\bar{F}}^{-\frac{1}{2}},\ \lambda _{F}^{\ast }=Q_{%
\bar{F}\bar{F}}^{\frac{1}{2}}\mathcal{V}_{21}^{\prime -1}\mathcal{V}%
_{11}^{\prime },\ \delta =(\mathcal{U}_{22}\mathcal{U}_{22}^{\prime })^{-%
\frac{1}{2}}\mathcal{U}_{22}S_{2}\mathcal{V}_{12}^{\prime }(\mathcal{V}_{12}%
\mathcal{V}_{12}^{\prime })^{-\frac{1}{2}}.  \label{svd parameters}
\end{equation}

\begin{proof}
See the Online Appendix and also Kleibergen and Paap (2006).\nocite{kpaap02}%
\smallskip 
\end{proof}

The squared singular values are the roots of the characteristic polynomial
in (\ref{popchar}), so $\lambda _{F}^{\ast }$ in Theorem 8 is the pseudo-true
value of the risk premia. The population moment $\mu _{f}(\lambda _{F})$
results from post-multiplying $\Omega ^{-\frac{1}{2}}\left( 
\begin{array}{cc}
\mu _{R} & \beta 
\end{array}%
\right) \left( 
\begin{array}{cc}
1 & 0 \\ 
0 & Q_{\bar{F}\bar{F}}^{\frac{1}{2}}%
\end{array}%
\right) $ by $\left( 
\begin{array}{c}
1 \\ 
-Q_{\bar{F}\bar{F}}^{-\frac{1}{2}}\lambda _{F}%
\end{array}%
\right) ,$ which is spanned by $\left( 
\begin{array}{cc}
1 & 0 \\ 
0 & Q_{\bar{F}\bar{F}}^{-\frac{1}{2}}%
\end{array}%
\right) \left( 
\begin{array}{cc}
\lambda _{F}^{\ast } & I_{m}%
\end{array}%
\right) _{\perp }^{\prime },$ and pre-multiplying by $\Omega ^{\frac{1}{2}}.$
The derivative of the population continuous updating objective function at $%
\lambda _{F}$ then results as:%
\begin{equation}
\begin{array}{l}
\sqrt{T}\mu _{f}(\lambda _{F})^{\prime }\Omega ^{-1}D\left( \lambda _{F}\right)
=-\left( 
\begin{array}{c}
1 \\ 
-\lambda _{F}%
\end{array}%
\right) ^{\prime }\left( 
\begin{array}{cc}
\lambda _{F}^{\ast } & I_{m}%
\end{array}%
\right) ^{\prime }D(\lambda _{F}^{\ast })^{\prime }\Omega ^{-1}D(\lambda
_{F})+ \\ 
\qquad \left( 
\begin{array}{c}
1 \\ 
-\lambda _{F}%
\end{array}%
\right) ^{\prime }\left( 
\begin{array}{cc}
1 & 0 \\ 
0 & Q_{\bar{F}\bar{F}}^{-1}%
\end{array}%
\right) \left( 
\begin{array}{cc}
\lambda _{F}^{\ast } & I_{m}%
\end{array}%
\right) _{\perp }^{\prime }\delta ^{\prime }D(\lambda _{F}^{\ast })_{\perp
}^{\prime }D(\lambda _{F}),%
\end{array}
\label{der facpop}
\end{equation}%
which equals zero when $\lambda _{F}$ is the pseudo-true value but also at
the other stationary points. When there is no misspecification, $\delta =0$
and $D(\lambda _{F}^{\ast })=-\ddot{\beta}$ so 
\begin{equation}
\begin{array}{rl}
\sqrt{T}\mu _{f}(\lambda _{F})^{\prime }\Omega ^{-1}D\left( \lambda _{F}\right) = & 
\left( 
\begin{array}{c}
1 \\ 
-\lambda _{F}%
\end{array}%
\right) ^{\prime }\left( 
\begin{array}{cc}
\lambda _{F}^{\ast } & I_{m}%
\end{array}%
\right) ^{\prime }\ddot{\beta}^{\prime }\Omega ^{-1}D(\lambda _{F}) \\ 
= & \left( \lambda _{F}^{\ast }-\lambda _{F}\right) ^{\prime }\ddot{\beta}%
^{\prime }\Omega ^{-1}D(\lambda _{F}),%
\end{array}
\label{no miss}
\end{equation}%
and $\ddot{\beta}$ is the only nuisance parameter.

Andrews et al. (2006)\nocite{andms05} construct the two-sided power envelope
for testing the single structural parameter in a linear IV regression model with independent normal errors and a known value
of the reduced form covariance matrix. This power envelope directly extends
to the linear one factor asset pricing model with independent normal errors
and no misspecification. It is then of interest to determine if such a power
envelope can be constructed in case of misspecification. Andrews et al.
(2006) construct the power envelope using the maximal invariant, which is stated in Theorem 9 alongside its distribution for the one factor linear
asset pricing model with independent normal errors and known covariance
matrices of the errors and factors.

\paragraph{Theorem 9:}

The maximal invariant, $S=\left( 
\begin{array}{cc}
S_{\perp\perp} & S_{\lambda_{F}^{1}\perp}^{\prime} \\ 
S_{\lambda_{F}^{1}\perp} & S_{\lambda_{F}^{1}\lambda_{F}^{1}}%
\end{array}
\right) ,$ for testing H$_{0}:\lambda_{F}=\lambda_{F}^{1}$ in the one factor
linear asset pricing model with independent normal errors and known values
of the covariance matrices of the errors, $\Omega,$ and factors, $Q_{\bar{F}%
\bar{F}},$ is the quadratic form of:%

\begin{equation}
\begin{array}{l}
\sqrt{T}\Omega^{-\frac{1}{2}}\left( 
\begin{array}{cc}
\bar{R} & \hat{\beta}%
\end{array}
\right) \\ 
\left( 
\begin{array}{ccc}
\left( 
\begin{array}{c}
1 \\ 
-\lambda_{F}^{1}%
\end{array}
\right) (1+\lambda_{F}^{1\prime}Q_{\bar{F}\bar{F}}^{-1}\lambda_{F}^{1})^{-%
\frac{1}{2}} & \vdots & \left( 
\begin{array}{cc}
1 & 0 \\ 
0 & Q_{\bar{F}\bar{F}}%
\end{array}
\right) \left( 
\begin{array}{cc}
\lambda_{F}^{1} & I_{m}%
\end{array}
\right) ^{\prime}(Q_{\bar{F}\bar{F}}+\lambda_{F}^{1}\lambda_{F}^{1\prime
})^{-\frac{1}{2}}%
\end{array}
\right) .%
\end{array}
\label{max inv 1}
\end{equation}
When $m=1,$ it has a non-central Wishart distribution with $T$ degrees of
freedom, identity scale matrix and non-centrality parameter:%
\begin{equation}
\begin{array}{l}
\text{Correct specification:} \\ 
\left( 
\begin{array}{c}
(\lambda_{F}^{\ast}-\lambda_{F}^{1})(1+(\lambda_{F}^{1})^{2}Q_{\bar{F}\bar{F}%
}^{-1})^{-\frac{1}{2}} \\ 
(Q_{\bar{F}\bar{F}}+(\lambda_{F}^{1})^{2})^{-\frac{1}{2}}\left( Q_{\bar {F}%
\bar{F}}+\lambda_{F}^{\ast}\lambda_{F}^{1\prime}\right)%
\end{array}
\right) \ddot{\beta}^{\prime}\Omega^{-1}\ddot{\beta}\left( 
\begin{array}{c}
(\lambda_{F}^{\ast}-\lambda_{F}^{1})(1+(\lambda_{F}^{1})^{2}Q_{\bar{F}\bar{F}%
}^{-1})^{-\frac{1}{2}} \\ 
(Q_{\bar{F}\bar{F}}+(\lambda_{F}^{1})^{2})^{-\frac{1}{2}}\left( Q_{\bar {F}%
\bar{F}}+\lambda_{F}^{\ast}\lambda_{F}^{1\prime}\right)%
\end{array}
\right) ^{\prime} \\ 
\ \\
\text{Misspecification:} \\ 
\left( 
\begin{array}{c}
(\lambda_{F}^{\ast}-\lambda_{F}^{1})(1+(\lambda_{F}^{1})^{2}Q_{\bar{F}\bar{F}%
}^{-1})^{-\frac{1}{2}} \\ 
(Q_{\bar{F}\bar{F}}+(\lambda_{F}^{1})^{2})^{-\frac{1}{2}}\left( Q_{\bar {F}%
\bar{F}}+\lambda_{F}^{\ast}\lambda_{F}^{1\prime}\right)%
\end{array}
\right) D(\lambda_{F}^{\ast})^{\prime}\Omega^{-1}D(\lambda_{F}^{\ast})\left( 
\begin{array}{c}
(\lambda_{F}^{\ast}-\lambda_{F}^{1})(1+(\lambda_{F}^{1})^{2}Q_{\bar{F}\bar{F}%
}^{-1})^{-\frac{1}{2}} \\ 
(Q_{\bar{F}\bar{F}}+(\lambda_{F}^{1})^{2})^{-\frac{1}{2}}\left( Q_{\bar {F}%
\bar{F}}+\lambda_{F}^{\ast}\lambda_{F}^{1\prime}\right)%
\end{array}
\right) ^{\prime}+ \\ 
\left( 
\begin{array}{c}
(1+(\lambda_{F}^{1})^{2}Q_{\bar{F}\bar{F}}^{-1})^{-\frac{1}{2}}\left(
1+\lambda_{F}^{\ast}Q_{\bar{F}\bar{F}}^{-1}\lambda_{F}^{1}\right) \\ 
-(Q_{\bar{F}\bar{F}}+(\lambda_{F}^{1})^{2})^{-\frac{1}{2}}\left( \lambda
_{F}^{\ast}-\lambda_{F}^{1}\right)%
\end{array}
\right) (1+(\lambda_{F}^{\ast})^{2}Q_{\bar{F}\bar{F}}^{-1})^{-1}\delta^{%
\prime}\delta\left( 
\begin{array}{c}
(1+(\lambda_{F}^{1})^{2}Q_{\bar{F}\bar{F}}^{-1})^{-\frac{1}{2}}\left(
1+\lambda_{F}^{\ast}Q_{\bar{F}\bar{F}}^{-1}\lambda_{F}^{1}\right) \\ 
-(Q_{\bar{F}\bar{F}}+(\lambda_{F}^{1})^{2})^{-\frac{1}{2}}\left( \lambda
_{F}^{\ast}-\lambda_{F}^{1}\right)%
\end{array}
\right) ^{\prime},%
\end{array}
\label{non-central}
\end{equation}
where the specifications of $D(\lambda_{F}^{\ast})$ and $\delta$ are stated
in Theorem 8.\smallskip

\begin{proof}
See the Online Appendix. \smallskip 
\end{proof}

The elements of the maximal invariant in Theorem 9 are such that: 
\begin{equation}
\begin{array}{rl}
S_{\lambda_{F}^{1}\lambda_{F}^{1}}= & T\hat{D}(\lambda_{F}^{1})^{\prime}\hat{%
V}_{\theta\theta.f}(\lambda_{F}^{1})^{-1}\allowbreak\hat{D}(\lambda _{F}^{1})
\\ 
S_{\perp\perp}= & Tf_{T}(\lambda_{F}^{1},X)^{\prime}\hat{V}_{ff}(\lambda
_{F}^{1})^{-1}f_{T}(\lambda_{F},X) \\ 
S_{\lambda_{F}^{1}\perp}= & T\left( \hat{V}_{ff}(\lambda_{F}^{1})^{-\frac {1%
}{2}}f_{T}(\lambda_{F},X)\right) ^{\prime}\allowbreak\left( \hat {V}%
_{\theta\theta.f}(\lambda_{F}^{1})^{-\frac{1}{2}}\allowbreak\hat{D}%
(\lambda_{F}^{1})\right) .%
\end{array}
\label{max inv 2}
\end{equation}
Since $1+(\lambda_{F}^{1})^{2}Q_{\bar{F}\bar{F}}^{-1}$ is known, the
distribution of the maximal invariant in Theorem 9 is a function of three
unknown parameters: $D(\lambda_{F}^{\ast})^{\prime}\Omega^{-1}D(\lambda
_{F}^{\ast}),$ $\delta^{\prime}\delta$, and $\left(
\lambda_{F}^{\ast}-\lambda_{F}^{1}\right) .$ Under H$_{0}:\lambda_{F}=%
\lambda_{F}^{1}=\lambda_{F}^{\ast},$ $\lambda_{F}^{\ast}-\lambda_{F}^{1}=0$,
so one of these three parameters is pinned down.

\paragraph{Corollary 3.}

Under H$_{0}:\lambda_{F}=\lambda_{F}^{\ast},$ the non-centrality parameter
of the non-central Wishart distribution of the maximal invariant equals:%

\begin{equation}
\begin{array}{l}
\text{Correct specification: }\left( 
\begin{array}{c}
0 \\ 
1%
\end{array}
\right) (Q_{\bar{F}\bar{F}}+(\lambda_{F}^{\ast})^{2})\ddot{\beta}^{\prime
}\Omega^{-1}\ddot{\beta}\left( 
\begin{array}{c}
0 \\ 
1%
\end{array}
\right) ^{\prime} \\ 
\text{Misspecification: }\left( 
\begin{array}{c}
0 \\ 
1%
\end{array}
\right) (Q_{\bar{F}\bar{F}}+(\lambda_{F}^{\ast})^{2})D(\lambda_{F}^{\ast
})^{\prime}\Omega^{-1}D(\lambda_{F}^{\ast})\left( 
\begin{array}{c}
0 \\ 
1%
\end{array}
\right) ^{\prime}+\left( 
\begin{array}{c}
1 \\ 
0%
\end{array}
\right) \delta^{\prime}\delta\left( 
\begin{array}{c}
1 \\ 
0%
\end{array}
\right) ^{\prime}.%
\end{array}
\label{noncentral h0}
\end{equation}

Corollary 3 shows that under H$_{0}$ and correct specification, the three
different elements of the maximal invariant depend on only one unknown
parameter, $(Q_{\bar{F}\bar{F}}+(\lambda _{F}^{\ast })^{2})\ddot{\beta}%
^{\prime }\Omega ^{-1}\ddot{\beta}.$ Because the $S_{\lambda _{F}^{1}\lambda
_{F}^{1}}$-element of the maximal invariant is a sufficient statistic for it
and independently distributed of the other elements of the maximal
invariant, we can condition on $S_{\lambda _{F}^{1}\lambda _{F}^{1}}$ to
construct the power envelope and for optimally combining the two other
elements of the maximal invariant, $S_{\lambda _{F}^{1}\perp }$ and $%
S_{\perp \perp },$ to improve the power for testing H$_{0}$; see Andrews et
al. (2006).

Under misspecification, the three elements of the maximal invariant depend
on two parameters, $(Q_{\bar{F}\bar{F}}+(\lambda _{F}^{\ast })^{2})D(\lambda
_{F}^{\ast })^{\prime }\Omega ^{-1}D(\lambda _{F}^{\ast })$ and $\delta
^{\prime }\delta .$ These are estimated using $S_{\lambda _{F}^{1}\lambda
_{F}^{1}}$ and $S_{\perp \perp }$, so we can no longer use $S_{\perp \perp }$
to improve the power of tests of H$_{0}$ like in case of correct
specification. The $S_{\lambda _{F}^{1}\perp }$-element of the maximal
invariant, which represents the score, is then the only element which can be
used to test H$_{0}$ under misspecification. It is thus not obvious how to
improve the power of invariant tests of H$_{0}:\lambda _{F}=\lambda
_{F}^{\ast }$ compared to the score test in case of misspecification.

The non-centrality parameter of the score element of the distribution of the
maximal invariant, $S_{\lambda _{F}^{1}\perp },$ in (\ref{non-central}):%
\begin{equation}
\begin{array}{c}
\left( \lambda _{F}^{\ast }-\lambda _{F}^{1}\right) (Q_{\bar{F}\bar{F}%
}+(\lambda _{F}^{1})^{2})^{-\frac{1}{2}}(1+(\lambda _{F}^{1})^{2}Q_{\bar{F}%
\bar{F}}^{-1})^{-\frac{1}{2}} \\ 
\left[ \left( Q_{\bar{F}\bar{F}}+\lambda _{F}^{\ast }\lambda _{F}^{1\prime
}\right) D(\lambda _{F}^{\ast })^{\prime }\Omega ^{-1}D(\lambda _{F}^{\ast
})-(1+(\lambda _{F}^{\ast })^{2}Q_{\bar{F}\bar{F}}^{-1})^{-1}\delta ^{\prime
}\delta \left( 1+\lambda _{F}^{\ast }Q_{\bar{F}\bar{F}}^{-1}\lambda
_{F}^{1}\right) \right] 
\end{array}
\label{score non-central}
\end{equation}%
shows that the power of the DRLM test positively
depends on the strength of identification, $D(\lambda _{F}^{\ast })^{\prime
}\Omega ^{-1}D(\lambda _{F}^{\ast }),$ and negatively on the amount of
misspecification, $\delta ^{\prime }\delta .$ It further shows that under H$%
_{0}:\lambda _{F}=\lambda _{F}^{1}=\lambda _{F}^{\ast },$ the non-centrality
parameter is zero when $\delta ^{\prime }\delta =\left( Q_{\bar{F}\bar{F}%
}+\lambda _{F}^{\ast 2}\right) D(\lambda _{F}^{\ast })^{\prime }\Omega
^{-1}D(\lambda _{F}^{\ast })$, so $\lambda _{F}$ is not identified when the
identification strength equals the amount of misspecification; see also (\ref%
{cross theo 7}).

\subsection{Testing multiple and subsets of the structural parameter vector}

The expressions of the DRLM\ statistic apply as well to settings where the
structural parameter vector has multiple elements. The power enhancement
procedure directly extends as well. Hence, we can improve the power of
testing a hypothesis on the structural parameter vector  by also rejecting it when there are
significant values of the statistic on every line going from the
hypothesized parameter value to the CUE.

Many times, we are interested in constructing confidence sets on the
individual elements of the structural parameter vector. Subset DRLM tests of
hypotheses specified on a selection of the elements of the structural
parameter vector which result from substituting the CUE for the parameters
left unspecified under the hypothesis of interest, are not necessarily size
correct, see Guggenberger et al. (2012). Confidence sets with the correct
coverage therefore result by projecting the joint confidence set that
applies to all structural parameters on the different axes, see also Dufour
and Taamouti (2005).\nocite{duftaa03}

\section{Nonlinear GMM}

The DRLM\ test is applicable to general non-linear GMM\ settings with unrestricted
covariance matrices. In this section we present a small simulation study using the
non-linear moment equation resulting from a CRRA utility function, see e.g. Hansen and Singleton (1982),%
\nocite{hansin82} to illustrate the size and power properties of the DRLM\
test in a non-linear GMM setting.

\paragraph{Running example 3: Constant relative risk aversion (CRRA)}

The moment function resulting from the CRRA utility function (see e.g.
Hansen and Singleton (1982)) is: 
\begin{equation}
\begin{array}{c}
E\left[ \delta \left( \frac{C_{t+1}}{C_{t}}\right) ^{-\gamma }(\iota
_{N}+R_{t+1})-\iota _{N}\right] =\mu _{f}(\delta ,\gamma ),%
\end{array}
\label{poweru}
\end{equation}%
with $\delta $ the discount factor, which is kept fixed at the value used in
the simulation experiment, $\delta _{0}=0.95,$ $\gamma $ the relative rate
of risk aversion, $C_{t}$ consumption at time $t$, $R_{t+1}$ an $N$%
-dimensional vector of asset returns, and $\iota_N$ an $N$-dimensional vector of ones. The sample moment function and its
derivative therefore only depend on $\gamma :$ 
\begin{equation}
\begin{array}{rlcrl}
f_{T}(\gamma ,X)= & \frac{1}{T}\sum_{t=1}^{T}f_{t}(\gamma ), & \qquad  & 
f_{t}(\gamma )= & \delta _{0}\left( \frac{C_{t+1}}{C_{t}}\right) ^{-\gamma
}(\iota _{N}+R_{t+1})-\iota _{N}, \\ 
q_{T}(\gamma ,X)= & \frac{1}{T}\sum_{t=1}^{T}q_{t}(\gamma ), &  & 
q_{t}(\gamma )= & -\delta _{0}\ln \left( \frac{C_{t+1}}{C_{t}}\right) \left( 
\frac{C_{t+1}}{C_{t}}\right) ^{-\gamma }(\iota _{N}+R_{t+1}).%
\end{array}
\label{sample CRRA}
\end{equation}%
The covariance matrix estimators are the Eicker-White ones, see White (1980):%
\nocite{wh80}%
\begin{equation}
\begin{array}{rl}
\hat{V}_{ff}(\gamma )= & \frac{1}{T}\sum_{t=1}^{T}(f_{t}(\gamma
)-f_{T}(\gamma ,X))(f_{t}(\gamma )-f_{T}(\gamma ,X))^{\prime }, \\ 
\hat{V}_{\theta f}(\gamma )= & \frac{1}{T}\sum_{t=1}^{T}(q_{t}(\gamma
)-q_{T}(\gamma ,X))(f_{t}(\gamma )-f_{T}(\gamma ,X))^{\prime }, \\ 
\hat{V}_{\theta \theta }(\gamma )= & \frac{1}{T}\sum_{t=1}^{T}(q_{t}(\gamma
)-q_{T}(\gamma ,X))(q_{t}(\gamma )-q_{T}(\gamma ,X))^{\prime }, \\ 
\hat{V}_{\theta \theta .f}(\gamma )= & \hat{V}_{\theta \theta }(\gamma )-%
\hat{V}_{\theta f}(\gamma )\hat{V}_{ff}(\gamma )^{-1}\hat{V}_{\theta
f}(\gamma )^{\prime }.%
\end{array}
\label{crra moment}
\end{equation}%

We use a log-normal data generating process to jointly simulate consumption
growth and asset returns in accordance with the moment equation. Since the
discount factor is fixed at its true value, $\gamma $ is the single
structural parameter of interest; see, for example, Savov (2011)\nocite%
{sav11} and Kroencke (2017).\nocite{kro17} The population moment function
then reads:\footnote{See the Online Appendix for its construction and for further
details on the simulation setup.}
\begin{equation}
\begin{array}{rl}
\mu _{f}(\gamma )= & \left( 
\begin{array}{c}
\exp \left( \ln (\delta _{0})+\mu _{2,1,0}+\frac{1}{2}\left(
V_{rr,11,0}+\gamma ^{2}V_{cc,0}-2\gamma V_{rc,1,0}\right) \right)  \\ 
\vdots  \\ 
\exp \left( \ln (\delta _{0})+\mu _{2,N,0}+\frac{1}{2}\left(
V_{rr,NN,0}+\gamma ^{2}V_{cc,0}-2\gamma V_{rc,N,0}\right) \right) 
\end{array}%
\right) -\iota _{N},%
\end{array}
\label{pop mom crra}
\end{equation}%
with $\mu _{2,0}=(\mu _{2,1,0}\ldots \mu _{2,N,0})^{\prime }$ the mean of $%
r_{t+1}=\ln (1+R_{t+1}),$ $V_{cc,0}$ the (scalar) variance of $\triangle
c_{t+1}=\ln \left( \frac{C_{t+1}}{C_{t}}\right) ,$ $V_{rc,0}=V_{cr,0}^{%
\prime }=(V_{rc,1,0}\ldots V_{rc,N,0})^{\prime }$ the $N\times 1$
dimensional covariance between $r_{t+1}$ and $\triangle c_{t+1}$ and $%
V_{rr,0}=V_{rr,ij,0}:$ $i,j=1,\ldots ,N,$ the $N\times N$ dimensional
covariance matrix of $r_{t+1}.$ The Online Appendix provides the expression of
the population covariance matrix $V_{ff}(\gamma )$ needed to compute the
pseudo-true value $\gamma ^{\ast }$:%
\begin{equation}
\begin{array}{c}
\gamma ^{\ast }=arg\min_{\gamma }\mu _{f}(\gamma )^{\prime }V_{ff}(\gamma
)^{-1}\mu _{f}(\gamma ).%
\end{array}
\label{pop function crra}
\end{equation}%
Unlike for the linear  asset pricing model, we need to compute the
pseudo-true value $\gamma ^{\ast }$ numerically since no closed-form expression is available
when there is misspecification. This also explains why we use the log-normal
setting so we have an analytical expression of the population moment
function, and only use one structural parameter since numerical optimizing in
higher dimensions is both computationally demanding and can be imprecise. 

We
analyze GMM-AR and DRLM\ tests for correctly specified and misspecified settings.

\paragraph{Correct Specification and $N=5$}

Standard GMM\ operates under correct specification so (\ref{pop mom crra})
equals zero, which implies that: 
\begin{equation}
\begin{array}{c}
\mu _{2,0}=-\iota _{N}\ln (\delta _{0})-\frac{1}{2}\left[ \left( 
\begin{array}{c}
V_{rr,11,0} \\ 
\vdots  \\ 
V_{rr,NN,0}%
\end{array}%
\right) +\iota _{N}\gamma ^{2}V_{cc,0}-2\gamma V_{rc,0}\right] .%
\end{array}
\label{specification2}
\end{equation}%
We revisit the simulation study in Kleibergen and Zhan (2020), who examine
the GMM-AR test on $\gamma $. We augment their simulation study by the\ DRLM
test. Figure 12 shows the resulting power curves of  GMM-AR and DRLM
tests. It indicates that GMM-AR and DRLM are both size-correct with good
power in the correctly specified setting.

\begin{equation*}
\begin{array}{c}
\text{Figure 12: Simulated power curves of GMM-AR (solid blue) and DRLM
(dashed red) tests with } \\ 
\text{5\% significance under correct specification. The\ CRRA moment
condition is imposed in the } \\ 
\text{data generation process  with }\delta =0.95\text{ and }N=5.\text{ The null
hypothesis is H}_{0}:\gamma =15. \\ 
\raisebox{-0pt}{\includegraphics[
height=2.8826in,
width=4.0283in
]%
{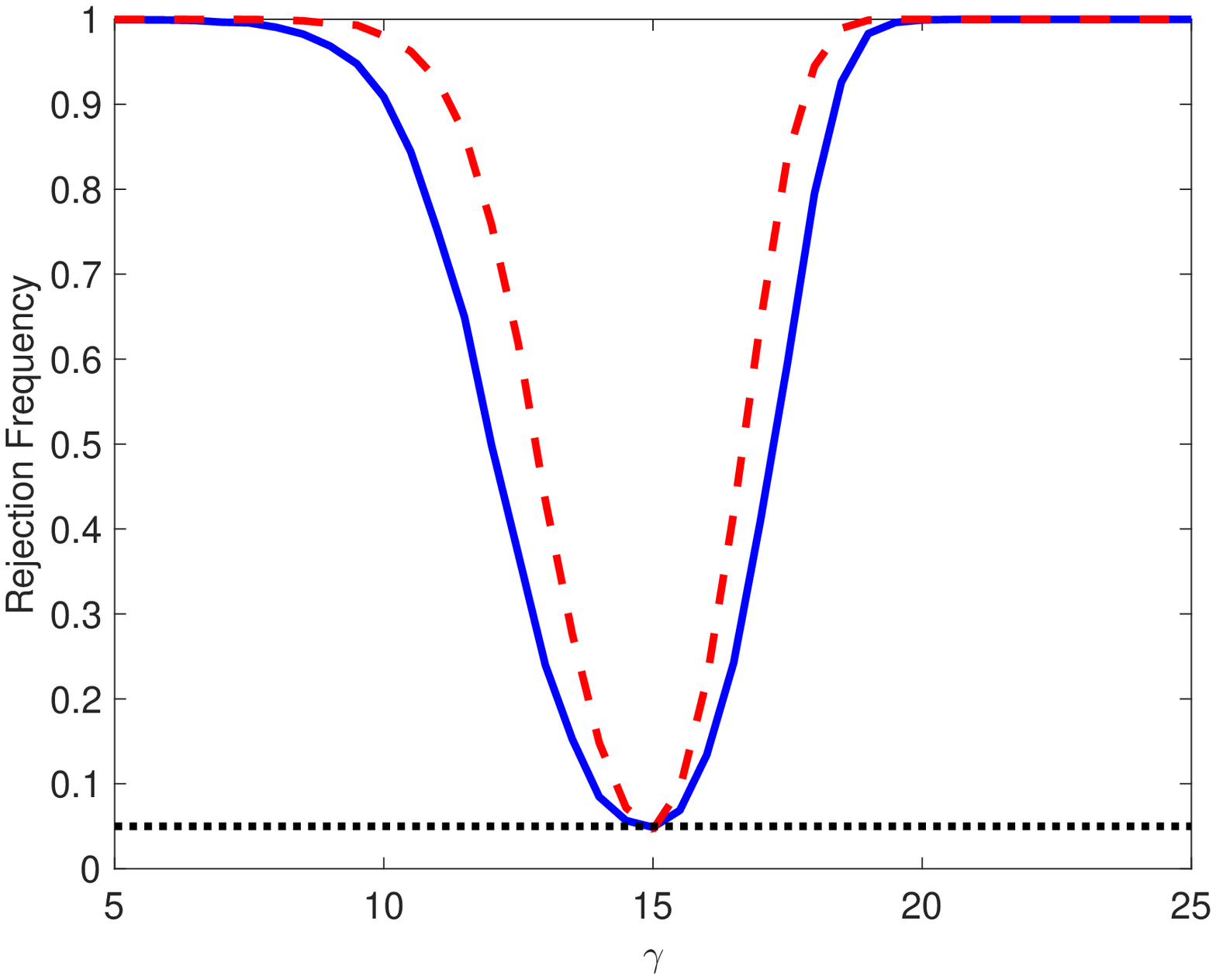}%
}\\
\end{array}%
\end{equation*}

In addition, since we consider $N=5$ in the data generation process (DGP), there is
over-identification, which helps explain the difference in power between the
GMM-AR and DRLM tests.

\paragraph{Misspecification and $N=5$}

For misspecification, we no longer impose (\ref{specification2}) in the DGP.
Instead, we just test for the pseudo-true value of $\gamma,$ denoted by $%
\gamma^{\ast}.$ Specifically, we start with an auxiliary $\tilde{\mu}_{2}$
that satisfies (\ref{specification2}), and then subtract a vector of
constants ($c$) to introduce misspecification in the DGP: 
\begin{equation}
\begin{array}{cl}
\tilde{\mu}_{2}= & -\iota_{N}\ln(\delta_{0})-\frac{1}{2}\left[ \left( 
\begin{array}{c}
V_{rr,11,0} \\ 
\vdots \\ 
V_{rr,NN,0}%
\end{array}
\right) +\iota_{N}\gamma^{2}V_{cc,0}-2\gamma V_{rc,0}\right], \\ 
\mu_{2,0}= & \tilde{\mu}_{2}-\iota_{N}\cdot c.%
\end{array}
\label{specification_mis}
\end{equation}

Figure 13 illustrates the simulation design. When $c=0$, $%
\gamma ^{\ast }=15$, and $\min \ \mu _{f}^{\prime }V_{ff}^{-1}\mu _{f}$ = $0$%
, as in the previous correct specification case. When $c$ deviates from
zero, the pseudo-true value $\gamma ^{\ast }$ starts to differ from 15 in Panel 13.1, and
the objective function $\mu _{f}^{\prime }V_{ff}^{-1}\mu _{f}$ in Panel
13.2 is no longer equal to zero at the pseudo-true value $\gamma ^{\ast }$.%

\begin{equation*}
\begin{array}{c}
\text{Figure 13: Pseudo-true value and population objective function as functions of the misspecification} \\ 
\begin{array}{cc}
\raisebox{-0pt}{\includegraphics[
height=2.026in,
width=2.3283in
]%
{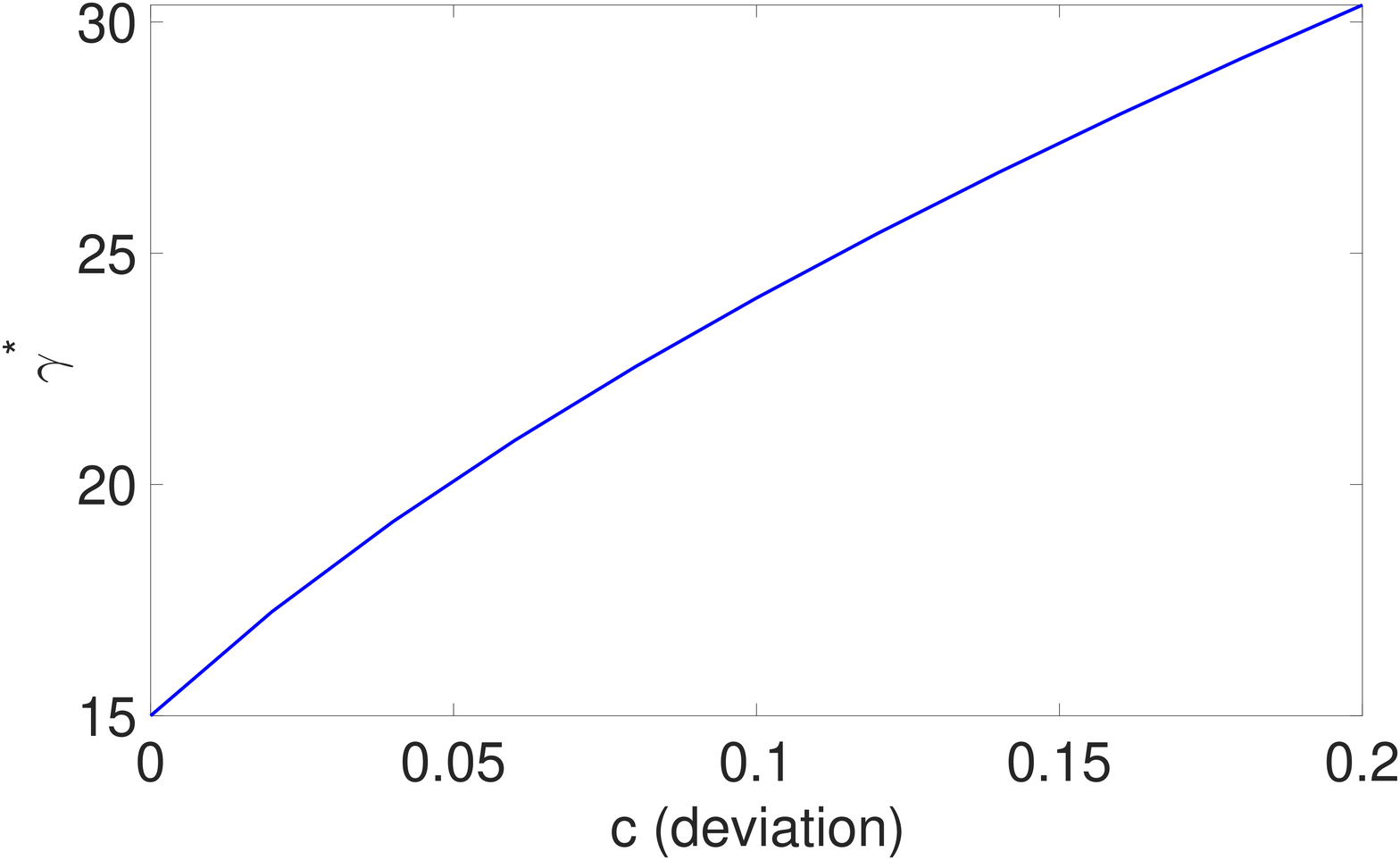}%
}
&
\raisebox{-0pt}{\includegraphics[
height=2.026in,
width=2.3283in
]%
{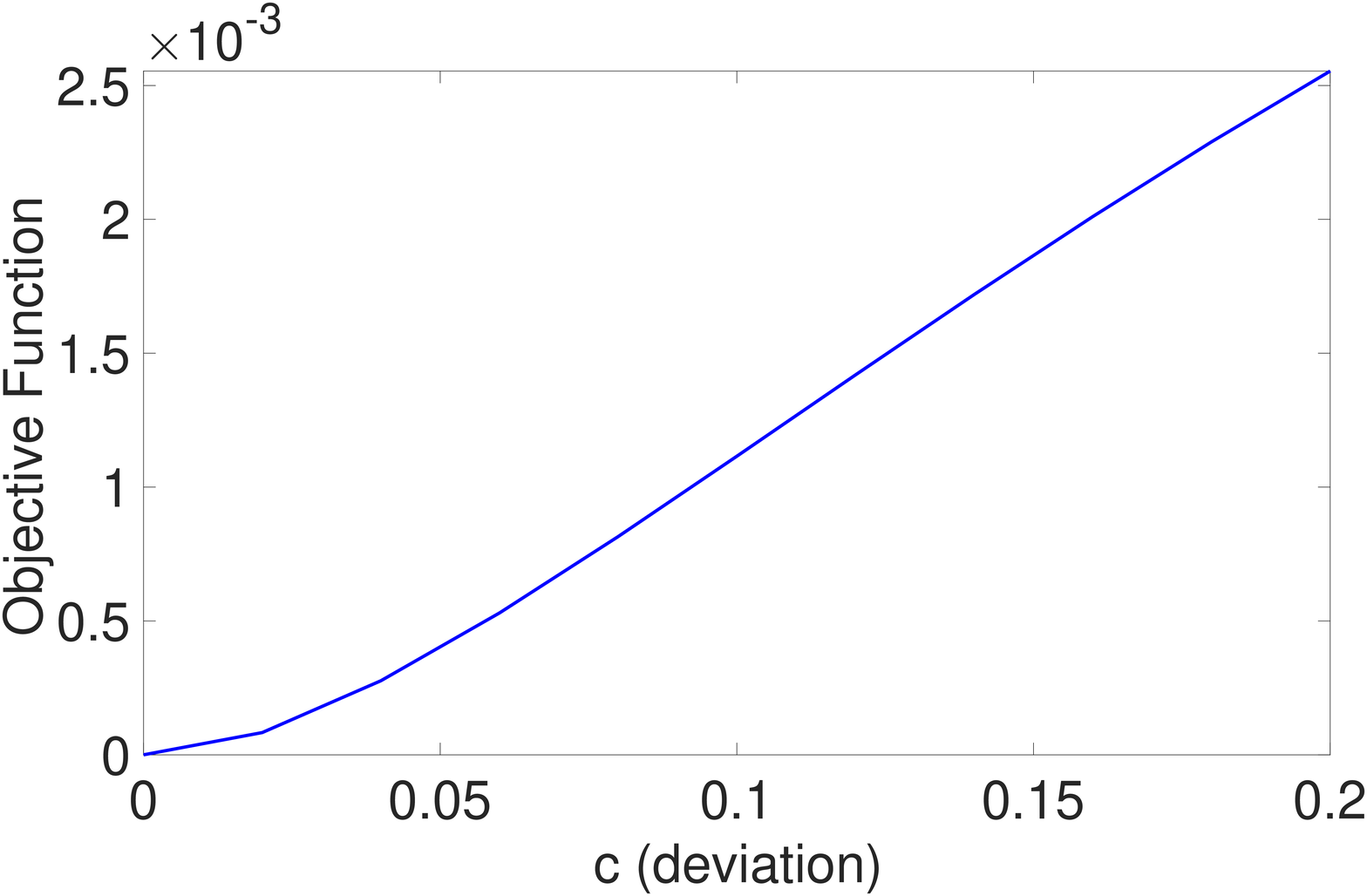}%
}\\ 
\text{Panel 13.1: Pseudo-true value function} & \text{Panel 13.2:
Population objective function at }\gamma ^{\ast }%
\end{array}%
\end{array}%
\end{equation*}

Figure 14 shows the rejection frequencies of GMM-AR and DRLM tests of H$%
_{0}:\gamma^{\ast}=24$ which corresponds, according to Panel 13.1, with a
degree of misspecification of 0.1. We consider a range of values of $c$ from
0 to 0.2 in the DGP while we test for H$_{0}:\gamma^{\ast}=24,$ or put
differently, H$_{0}:$ $c=0.1.$ Figure 14 shows that the GMM-AR test rejects
the null more often than the nominal significance level of 5\% to reflect
that the moment condition is misspecified. In contrast, since the DRLM test
allows for misspecification, it has the correct rejection frequency at the
hypothesized value.%
\begin{equation*}
\begin{array}{c}
\text{Figure 14: Simulated power curves of GMM-AR (solid blue) and DRLM
(dashed red) tests at the } \\ 
\text{5\% significance level under misspecification. The null hypothesis H}%
_{0}:\gamma=\gamma^{\ast}=24\text{ corresponds } \\ 
\text{with misspecification equal to }c=0.1\text{ where }c\text{ reflects
the deviation for misspecification.} \\ 
\raisebox{-0pt}{\includegraphics[
height=2.8826in,
width=4.0283in
]%
{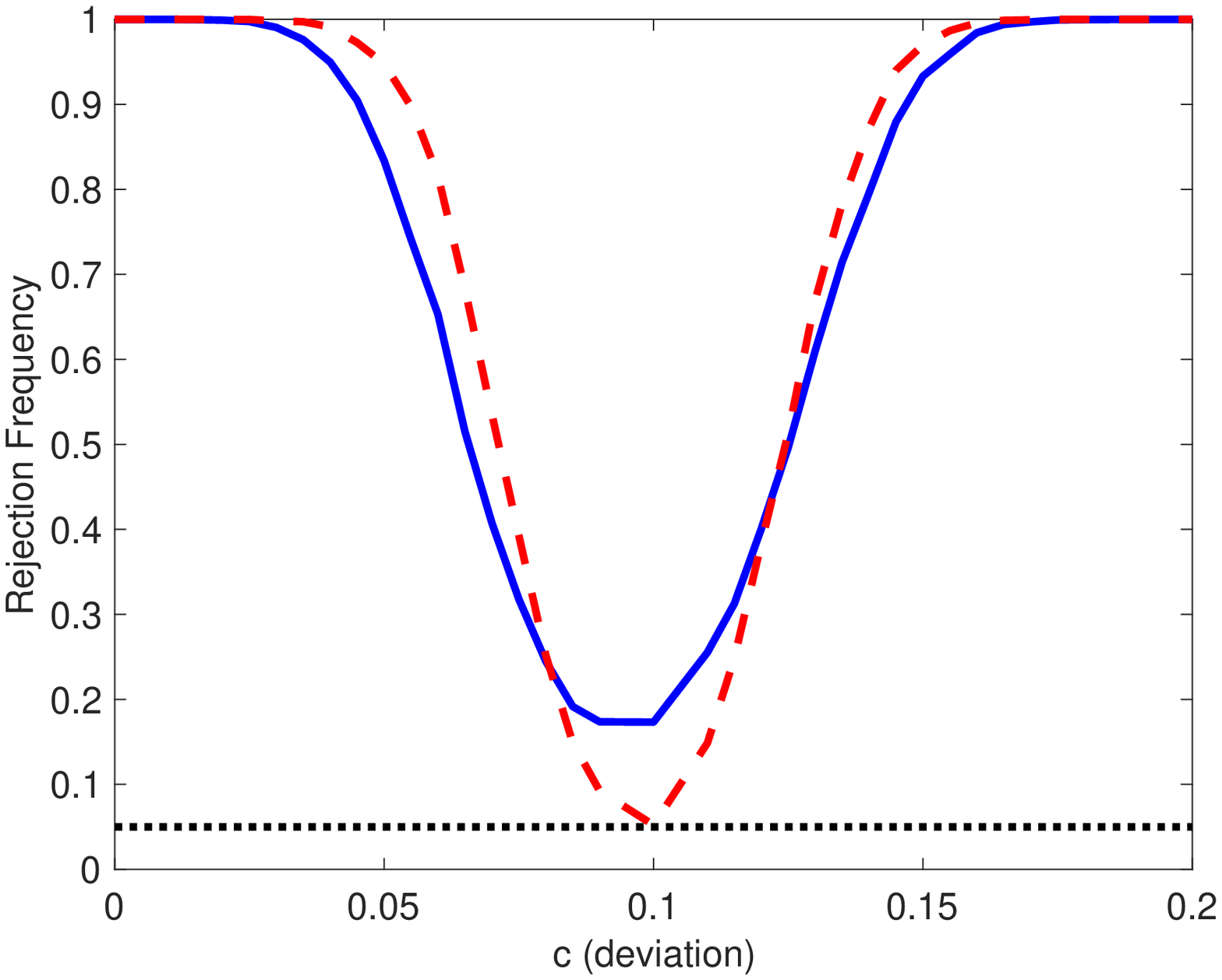}%
}
 \\ 
\end{array}%
\end{equation*}
\begin{equation*}
\begin{array}{c}
\text{Figure 15: Rejection frequencies of GMM-AR and DRLM tests of H}%
_{0}:\gamma=\gamma^{\ast}\text{ at the 5\% } \\ 
\text{significance level with }N=5\text{ as a function of the strengths of
identification, }\tilde{c},\text{ and misspecification }c. \\ 
\raisebox{-0pt}{\includegraphics[
height=2.5893in,
width=6.493in
]%
{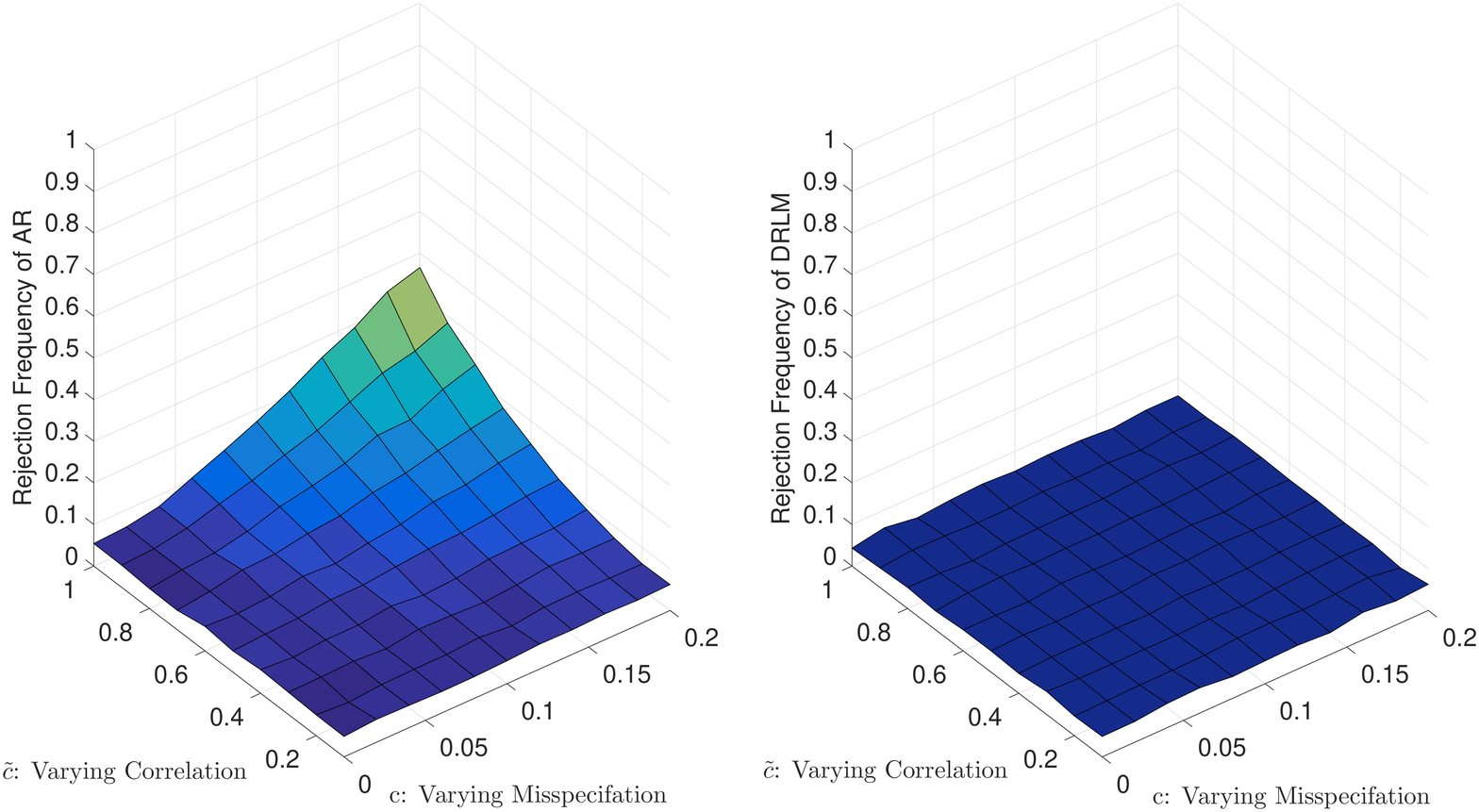}%
}\\ 
\end{array}%
\end{equation*}

\paragraph{Size of AR and DRLM tests with $N=5$}

Furthermore, Figure 15 shows the trade-off between the identification
strength and the amount of misspecification for the rejection frequencies of
GMM-AR and DRLM\ tests. The DGP\ is such that the correlation coefficient
between the log-consumption growth and the log asset returns, $\rho _{i}=%
\frac{V_{rc,i,0}}{\sqrt{V_{cc,0}V_{rr,ii,0}}},$ is scaled by a constant $%
\tilde{c}$ to vary identification. Figure 15 shows the rejection frequencies
of tests of H$_{0}:\gamma =\gamma ^{\ast }$ as a function of the
misspecification $c$ and strength of identification which is (partly)
reflected by $\tilde{c}$. We note that the pseudo-true value $\gamma ^{\ast }
$ is a function of $(c,$ $\tilde{c})$, so the reported rejection frequencies
in Figure 15 are for different hypothesized values of $\gamma ^{\ast }.$
Figure 15 shows that the GMM-AR test gets size distorted when the
misspecification increases. This is unlike the DRLM test, which remains size
correct for all values of the identification and misspecification strengths.

\section{Applications} 

We apply the DRLM\ test and the identification robust AR, KLM and LR tests
to data for two different models discussed previously: the linear asset
pricing model and the linear IV regression model.

\subsection{Running example 1: Linear asset pricing model}

We briefly revisit the linear factor models considered in Adrian et al.
(2014)\nocite{aem14} and He et al. (2017)\nocite{hkm17} using our DRLM test
and the identification robust factor AR, KLM and LR tests; see also Kleibergen (2009)
and Kleibergen and Zhan (2020).\nocite{kf09}

Adrian et al. (2014) propose a leverage risk factor (\textquotedblleft$%
LevFac $\textquotedblright) for asset pricing. The leverage level is the ratio of total assets over the difference between total assets and  liabilities, and the leverage risk factor equals its log change. The
empirical study of Adrian et al. (2014) uses quarterly data between 1968Q1
and 2009Q4. Following Lettau et al. (2019),\nocite{llm19} we extend the time
period to 1963Q3 - 2013Q4 and use $N=25$ size and book-to-market sorted portfolios
as test assets. Adrian et al. (2014) show that the leverage factor prices
the cross-section of many test portfolios, as reflected by the significant
Fama-MacBeth (FM)\nocite{fm73} (1973) and Kan-Robotti-Shanken (KRS) $t$%
-statistics on the risk premium reported in Table \ref{t_k2}. The KRS $t$%
-statistic is robust to misspecification but not to weak identification, see
Kan et al. (2013).\nocite{krs13}

He et al. (2017) propose the banking equity-capital ratio factor
(\textquotedblleft $EqFac$\textquotedblright ) for asset pricing. We
consider one of their specifications with \textquotedblleft $EqFac$%
\textquotedblright\ and the market return \textquotedblleft $R_{m}$%
\textquotedblright\ as the two factors. Table \ref{t_k2} shows significant FM and KRS $t$-statistics for the risk premium on
\textquotedblleft $EqFac$\textquotedblright\ .

For both Adrian et al. (2014) and He et al. (2017), the risk premia are, however, weakly identified, as indicated by the large $p$-values of both the $\chi^2$ and $F$ rank tests reported in  Table \ref{t_k2}.

\paragraph{DRLM: Adrian, Etula, and Muir (2014)}

Using the same data as for Table \ref{t_k2}, Figure 16 shows the $p$-values
for testing the risk premium on the leverage factor (horizontal line) using
the DRLM, AR, KLM, and LR tests. Most of the $p$-values in Figure 16 are
above the 5\% level, which implies that none of the DRLM, AR, KLM, and LR
tests leads to tight 95\% confidence intervals for the risk premium on the
leverage factor as shown in Table \ref{t_k2}. Given the smallish $p$-value
of the $J$-test, 0.20, and the weak identification of the risk premium on
the leverage factor reflected by the unbounded 95\% confidence sets, it is
likely that there is misspecification so it would be appropriate to use the
DRLM\ test.
\begin{equation*}
\begin{array}{c}
\text{Figure 16: Adrian, Etula and Muir (2014). }p\text{-value from the
DRLM\ (dashed red), AR\ (dashed blue), } \\ 
\text{KLM (solid black), LR (dash-dotted green) and the 5\% level (dotted
black). $J$-statistic} \\ 
\text{(=minimum AR) equals 28.42, with }p\text{-value of 0.20 resulting from 
}\chi ^{2}(N-2). \\ 
\raisebox{-0pt}{\includegraphics[
height=2.2826in,
width=4.0283in
]{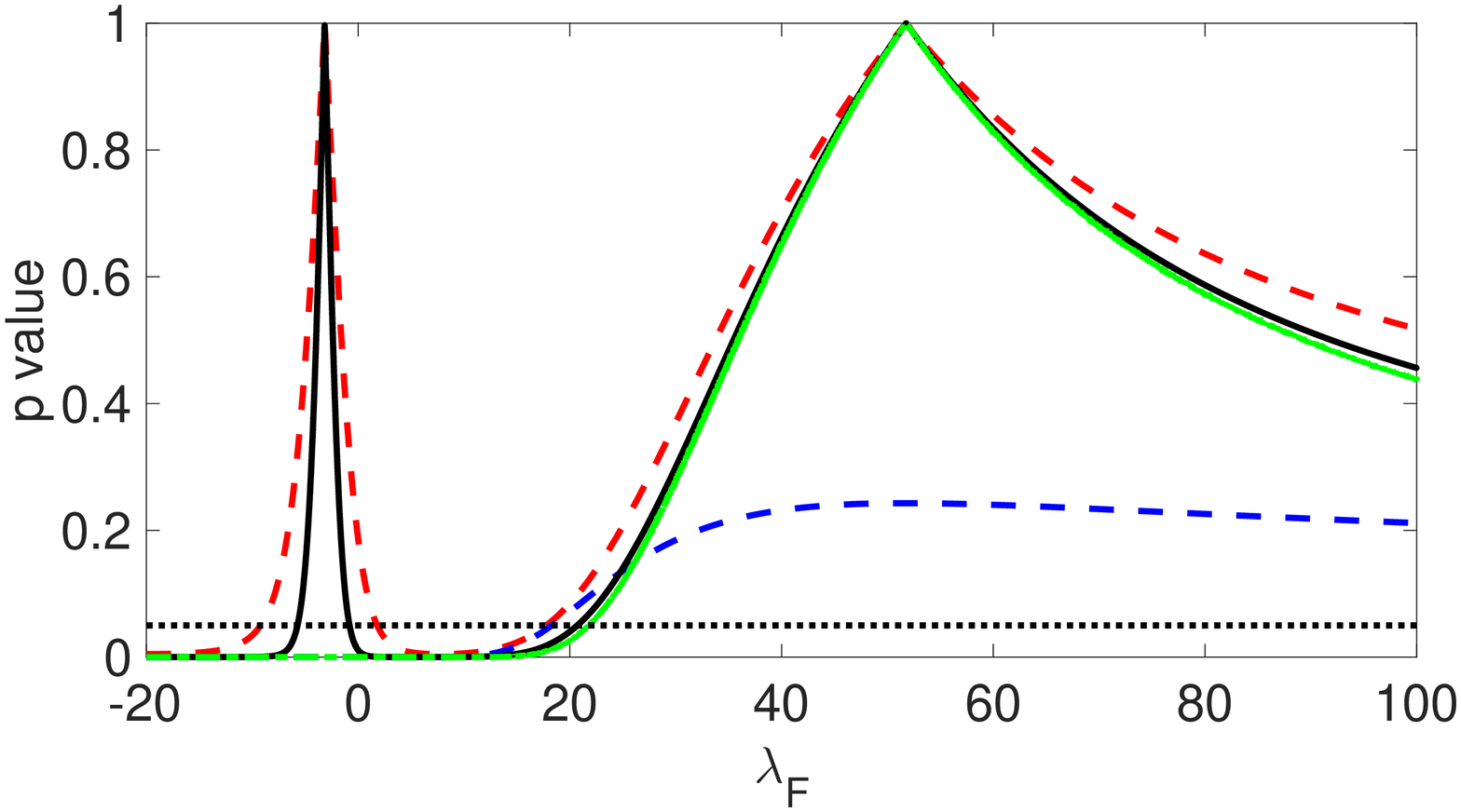}%
}
\end{array}%
\end{equation*}

\begin{landscape}
\begin{table}[htp]
\caption{\textbf{Inference on Risk Premia $\lambda_{F}$ in Adrian, Etula, and
Muir (2014) and He, Kelly, and Manela (2017)}}%

\bigskip

The test assets are the $N=25$ size and book-to-market portfolios from 1963Q3
to 2013Q4 taken from Lettau, Ludvigson, and Ma (2019).\nocite{llm19}
\textquotedblleft$LevFac$\textquotedblright\ is the leverage factor of Adrian,
Etula, and Muir (2014). \textquotedblleft$EqFac$\textquotedblright\ is the
banking equity-capital ratio factor of He, Kelly, and Manela (2017).
$``R_{m}\text{\textquotedblright}\ $is the market return. The estimate of
$\lambda_{F}$ and the FM $t$-statistic  result
from the Fama-MacBeth (1973) two-pass procedure. The KRS $t$-statistic is
based on the KRS $t$-test of Kan, Robotti, and Shanken (2013). The point estimates of $\lambda_{F}$ are identical to those reported in Lettau,
Ludvigson, and Ma (2019). The rank test is adopted from Kleibergen and Zhan (2020) for testing H$_0: rank(\beta)=m-1$, where $m$ is the number of risk factors. 
\par
\bigskip
\bigskip
\bigskip

\par
\centering{\centering}%
\begin{tabular}
[c]{lcccc}\hline
& Adrian, Etula, and Muir (2014) &  & \multicolumn{2}{c}{He, Kelly, and Manela
(2017)}\\\cline{2-2}\cline{4-5}
& $LevFac$ &  & $R_{m}$ & $EqFac$\\\hline
Estimate of $\lambda_{F}$ & \footnotesize{13.91} &  & \footnotesize{1.19} & \footnotesize{6.88}\\
FM $t$ & \footnotesize{3.58} &  & \footnotesize{0.81} & \footnotesize{2.14}\\
KRS $t$ & \footnotesize{2.55} &  & \footnotesize{0.77} & \footnotesize{2.10}\\\hline
CUE of $\lambda_{F}$ & \footnotesize{51.77} &  & \footnotesize{23.22} & \footnotesize{94.02}\\\hline
95\% confidence set &  &  &  & \\
FM $t$ & \footnotesize{(6.29, 21.54)} &  & \footnotesize{(-1.67, 4.05)} & \footnotesize{(0.57, 13.19)}\\
 & \\
KRS $t$ & \footnotesize{(3.22, 24.60)} &  & \footnotesize{(-1.84, 4.22)} & \footnotesize{(0.46, 13.30)}\\
 & \\
DRLM & \footnotesize{$(-\infty,-91.4)\cup(-9.2,1.7)\cup(17.8,+\infty)$} &  & \footnotesize{$(-\infty,+\infty)$} &\footnotesize{$(-\infty,+\infty)$}\\
 & \\
DRLM\ (power enh.) & \footnotesize{$(-\infty,-91.4)\cup(17.8,+\infty)$} &  & \footnotesize{$(-\infty,+\infty)$} & \footnotesize{$(-\infty,+\infty)$}\\
 & \\
AR & \footnotesize{$(-\infty,-101.4)\cup(18.3,+\infty)$} &  & \footnotesize{$(-\infty ,-64.6)\cup(8.1,+\infty)$} & \footnotesize{$(-\infty ,-244.1)\cup(37.7,+\infty)$}\\
 & \\
KLM & \footnotesize{$(-\infty,-185.7)\cup(-5.7,-0.9)\cup(20.8,+\infty)$} &  & \footnotesize{($-\infty,-7.2)\cup( -4.7, -0.3)\cup(1.0,+\infty)$}  &  \footnotesize{$(-\infty,-23.8)\cup(-8.1, 1.7)\cup(11.8,+\infty)$}\\
 & \\
LR & \footnotesize{$(-\infty,-274.2)\cup(21.8,+\infty)$} &  &  \footnotesize{$(-\infty ,-9.6)\cup(2.2,+\infty)$} & \footnotesize{$(-\infty ,-33.8)\cup(16.2,+\infty)$}\\ \hline
Rank test &\\
$\chi^2$-statistic ($p$-value) &  \footnotesize{31.97 (0.13)}  &&   \multicolumn{2}{c}{\footnotesize{ 35.88 (0.04) }}   \\
$F$-statistic ($p$-value) &  \footnotesize{1.17 (0.28)}  &&   \multicolumn{2}{c}{\footnotesize{ 1.33 (0.16) }}   \\
\hline
\end{tabular}
\label{t_k2}%
\end{table}
\end{landscape}

The $p$-values of the DRLM\ test in Figure 16 are equal to one at two
different points. The $p$-values of the AR test show that one of these two
points relates to the minimal value of the AR test and the other one to the
maximal value of the AR test. Using the power enhancement rule for the DRLM\
test, we can reject non-significant values that lie within the closed
interval indicated by the significant maximizers of the DRLM\ statistic that
does not contain the CUE, so the non-significant $p$-values of the DRLM\ test
which occur around the maximizer of the AR test can all be categorized as
significant ones according to the power enhancement rule. The resulting 95\%
confidence set for the DRLM\ test rejects a zero value of the risk premium
of the leverage factor and is reported in Table \ref{t_k2} alongside the one
which results from just applying the DRLM\ test. The FM\ and KRS $t$%
-statistics reported in Table 1 also reject a zero value of the risk premium,
but these tests are not reliable because of the weak identification of the
risk premium of the leverage factor and the likely misspecification
reflected by the smallish $p$-value of the $J$-test.%

The $\chi^2$ rank statistic reported in Table \ref{t_k2}, 31.97, corresponds with the sample analog of the $IS$ identification strength measure (\ref{rank 3}) and is always larger than or equal to the minimal value of the CUE objective function whose value corresponds with the $J$-statistic reported in Figure 16, 28.42. The just slightly larger value of the rank statistic implies that the CUE can very well result from a reduced rank value of the $\beta$'s which then further explains its huge value in Table \ref{t_k2}, 51.77, and the unbounded 95\% confidence sets. We thus have to be cautious with interpreting the CUE as reflecting the risk premium on the leverage factor. 

\paragraph{DRLM: He, Kelly, and Manela (2017)}

Figure 17 shows the joint 95\% confidence sets (shaded areas) of the risk
premia on the banking equity-capital ratio factor \textquotedblleft $EqFac$%
\textquotedblright\ and the market return \textquotedblleft $R_{m}$%
\textquotedblright , from using the DRLM, AR, KLM, and LR tests. The $p$%
-value of the $J$-test shows that misspecification is present, so it is
appropriate to use the DRLM\ test for the confidence set of the minimizer of
the population continuous updating objective function. The 95\% confidence
sets of the DRLM\ and KLM tests have two rather disjoint areas. The power
enhancement rule for the DRLM\ test shows that the smaller disjoint area can
be discarded for the joint 95\% confidence set that results from the DRLM
test. The resulting 95\%\ confidence set from the DRLM\ test includes a zero
value for the risk premium on \textquotedblleft $EqFac$\textquotedblright\ , 
which indicates that the pricing ability of \textquotedblleft $EqFac$%
\textquotedblright\ is under doubt.

The minimal value  of the CUE objective function reported in Figure 17 of 35.32, which equals the $J$-statistic, is just slightly below the $\chi^2$ rank statistic reported in Table \ref{t_k2}, 35.88. The sample analog of the $IS$ identification strength measure (\ref{rank 3}) is therefore just above the minimal value of the CUE objective function. This makes it likely that the CUE estimates result from a lower rank value of the  $\beta$-matrix of the risk factors which then further explains the very large values of the CUE estimates, (23.22, 94.02), and their unbounded 95\% confidence sets shown in Figure 17. It is thus difficult to interpret the CUE as reflecting the risk premium on the two risk factors. 

\begin{equation*}
\begin{array}{c}
\text{Figure 17: He, Kelly and Manela (2017). 95\% confidence sets from DRLM, 
} \text{AR, KLM\ and LR. } \\ 
J\text{-statistic (minimum of AR)} \text{ equals 35.32, with }p\text{-value
of 0.036 resulting from }\chi ^{2}(N-3). \\ 
\\ 
\begin{array}{ccc}
\raisebox{-0pt}{\includegraphics[
height=1.1092in,
width=2.629in
]%
{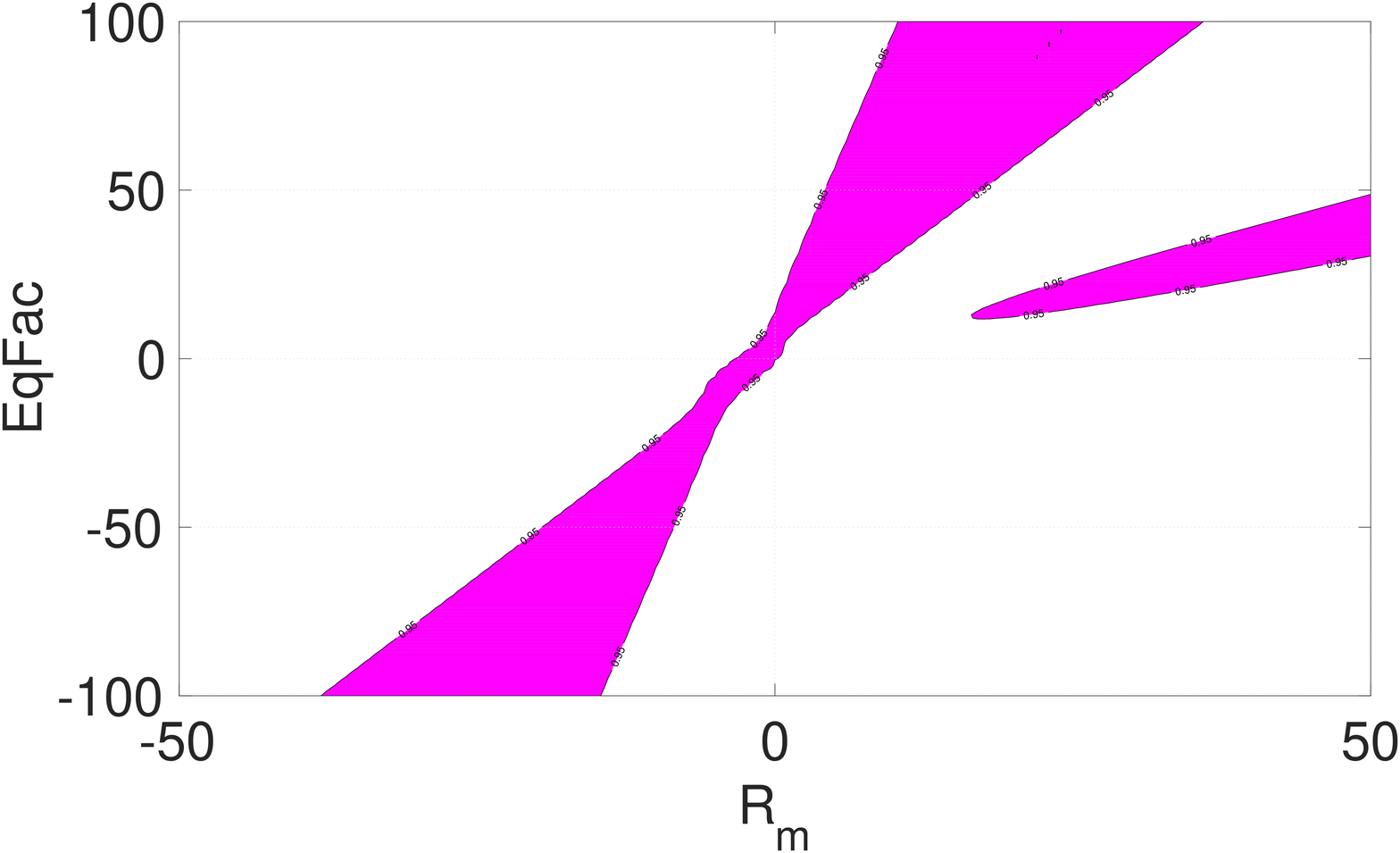}%
}
&  &
\raisebox{-0pt}{\includegraphics[
height=1.1092in,
width=2.629in
]%
{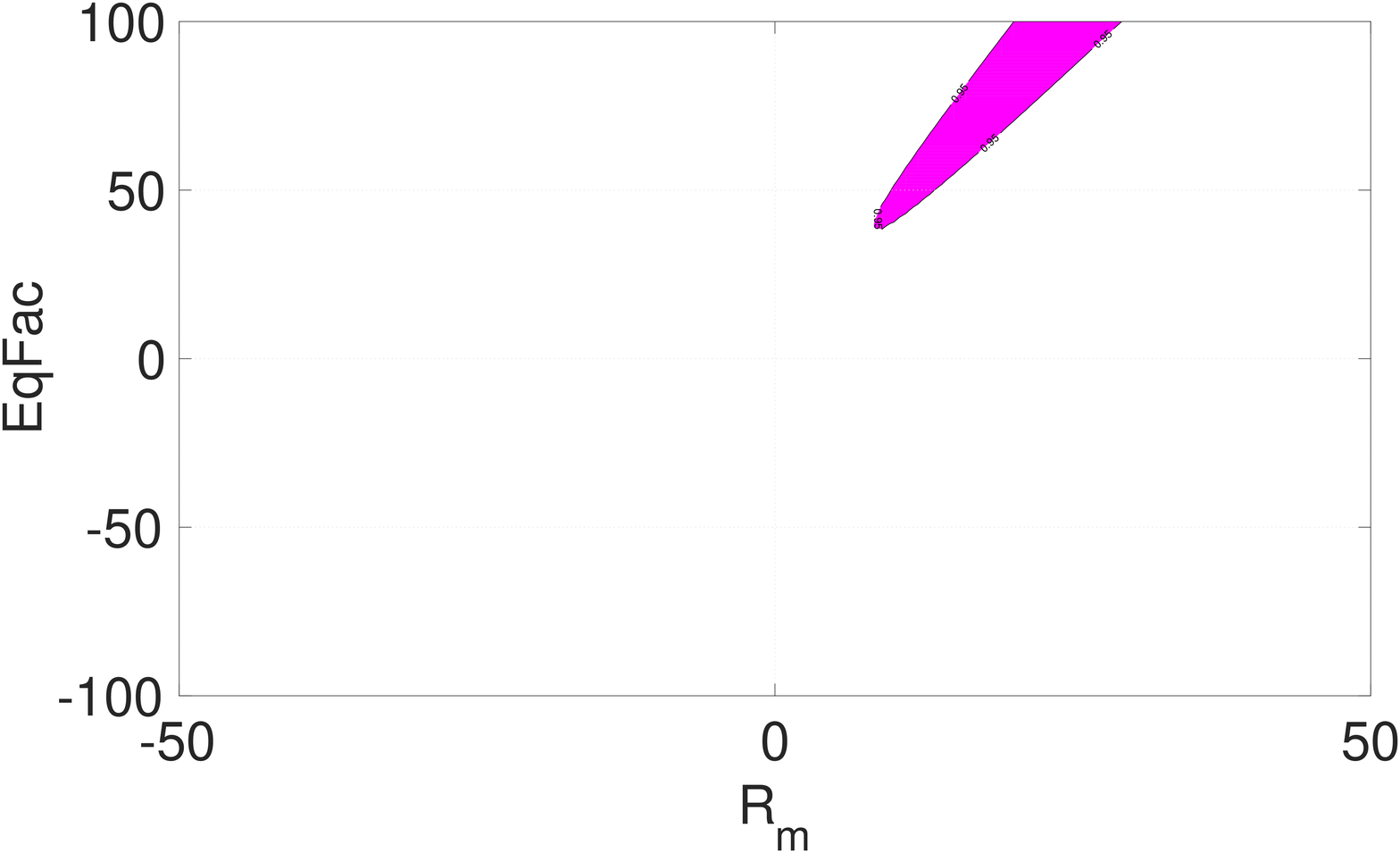}%
}
\\ 
\text{Panel 17.1: DRLM} &  & \text{Panel 17.2: AR} \\ 
&  &  \\ 
\raisebox{-0pt}{\includegraphics[
height=1.1092in,
width=2.629in
]%
{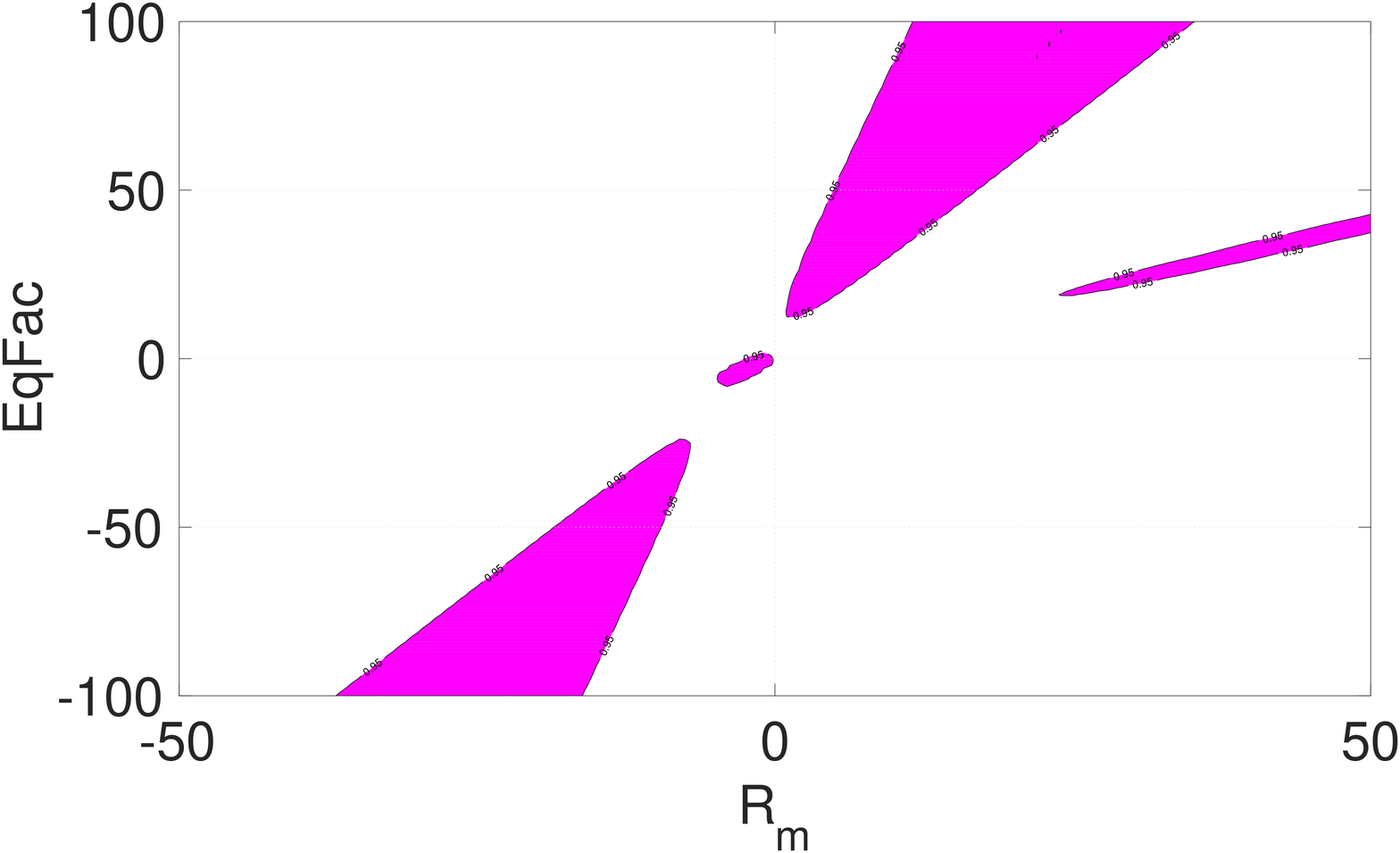}%
}
&  &
\raisebox{-0pt}{\includegraphics[
height=1.1092in,
width=2.629in
]%
{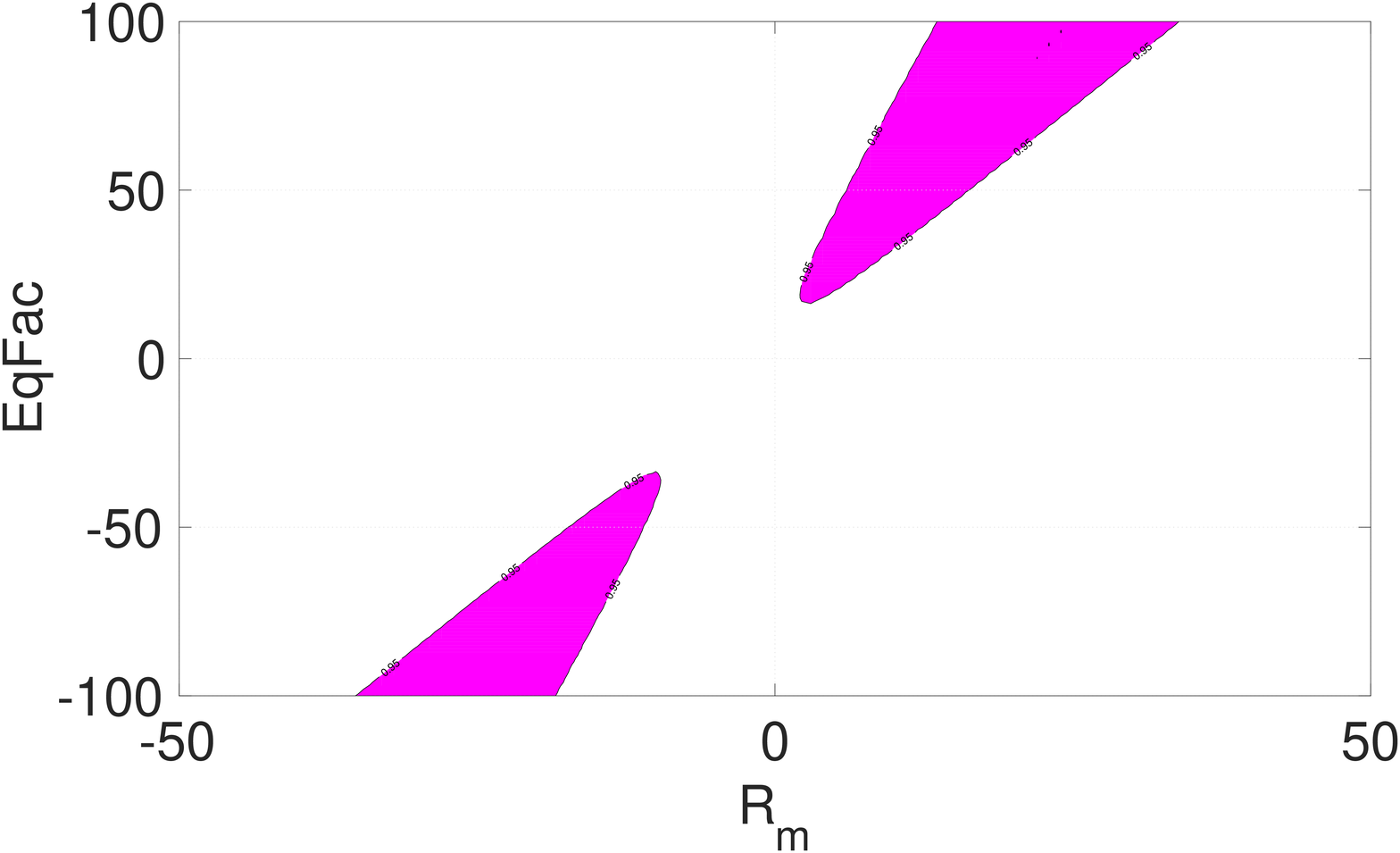}%
}
\\ 
\text{Panel 17.3: KLM} &  & \text{Panel 17.4: LR}%
\end{array}%
\end{array}%
\end{equation*}
\begin{equation*}
\begin{array}{c}
\text{Figure 18: }R_{m}\text{ and SMB. 95\% confidence sets from DRLM, } 
\text{AR, KLM\ and LR. } \\ 
J\text{-statistic (minimum of AR)} \text{ equals 59.34, with }p\text{-value
of 0.00 resulting from }\chi ^{2}(N-3). \\ 
\\ 
\begin{array}{ccc}
\raisebox{-0pt}{\includegraphics[
height=1.1092in,
width=2.629in
]%
{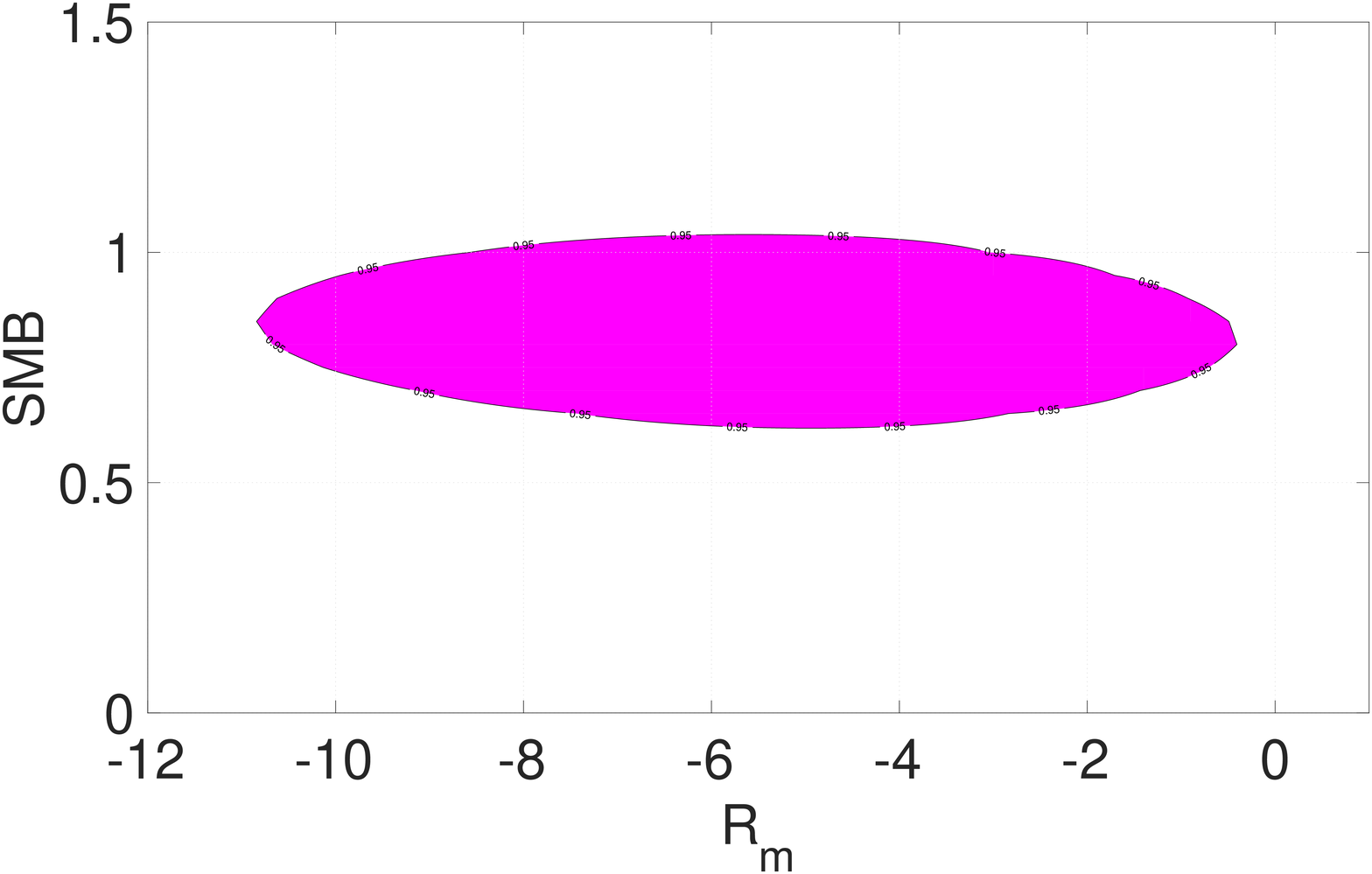}%
}
&  &
\raisebox{-0pt}{\includegraphics[
height=1.1092in,
width=2.629in
]%
{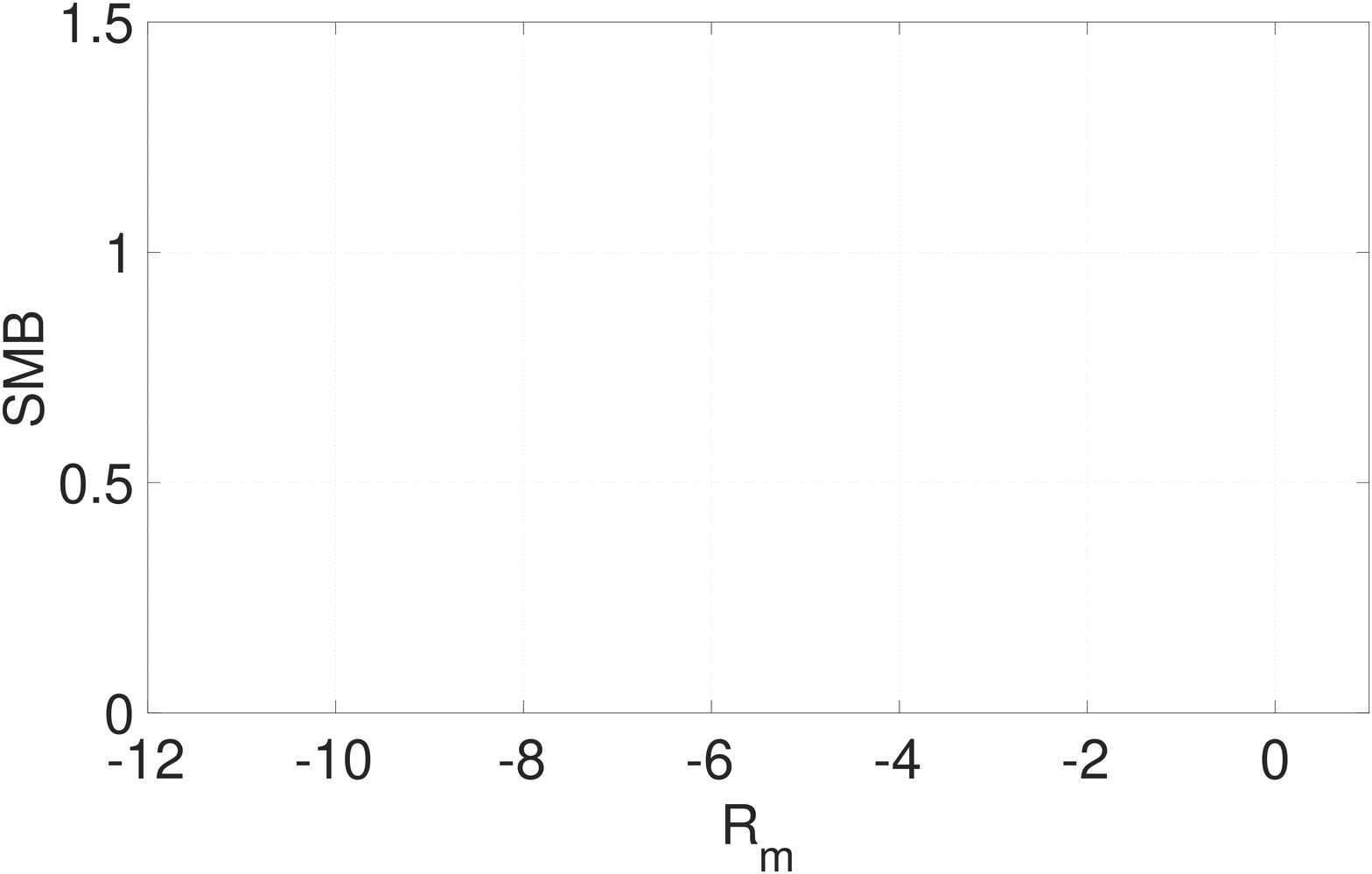}%
}

\\ 
\text{Panel 18.1: DRLM} &  & \text{Panel 18.2: AR} \\ 
&  &  \\ 
\raisebox{-0pt}{\includegraphics[
height=1.1092in,
width=2.629in
]%
{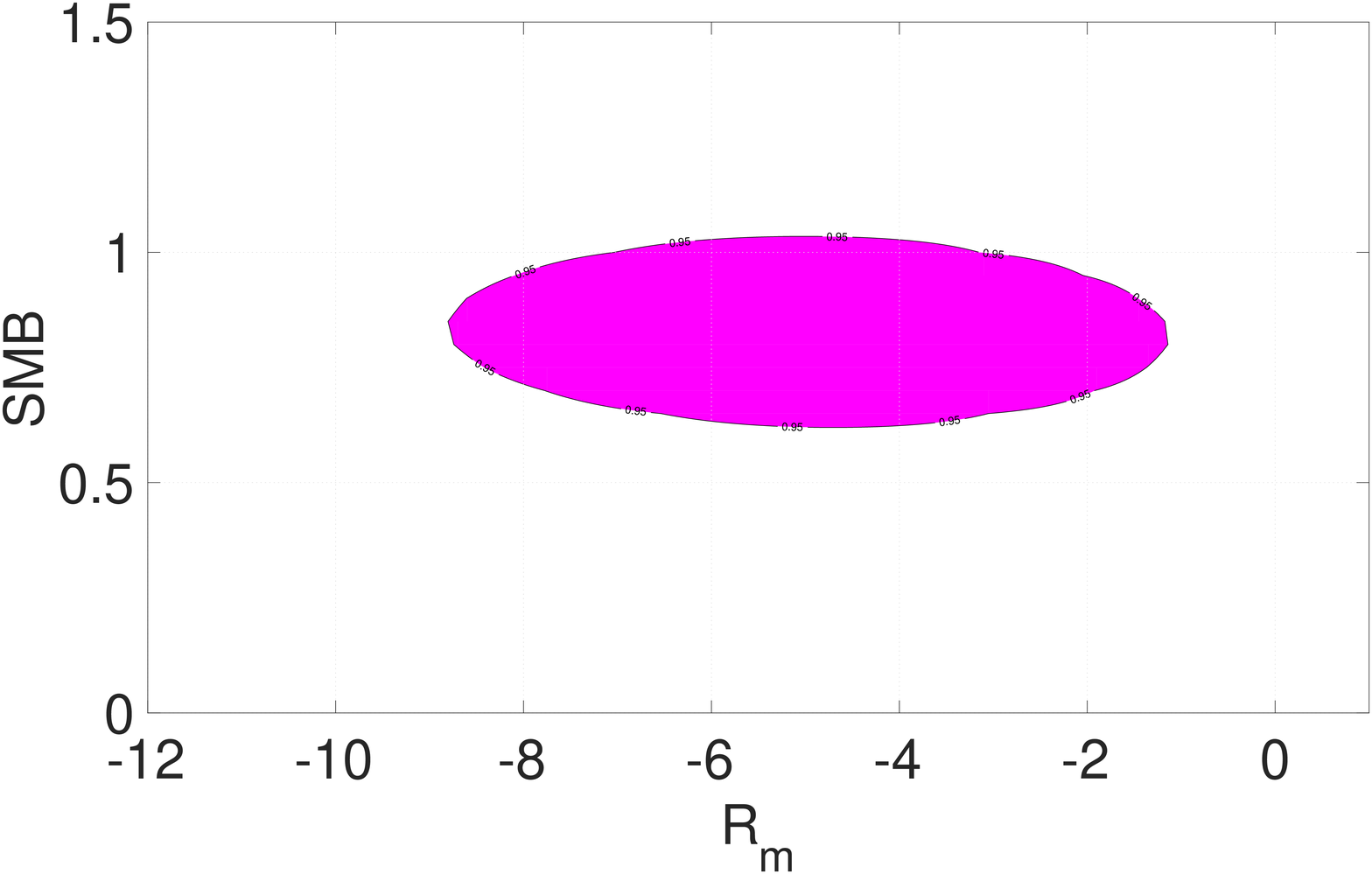}%
}
&  &
\raisebox{-0pt}{\includegraphics[
height=1.1092in,
width=2.629in
]%
{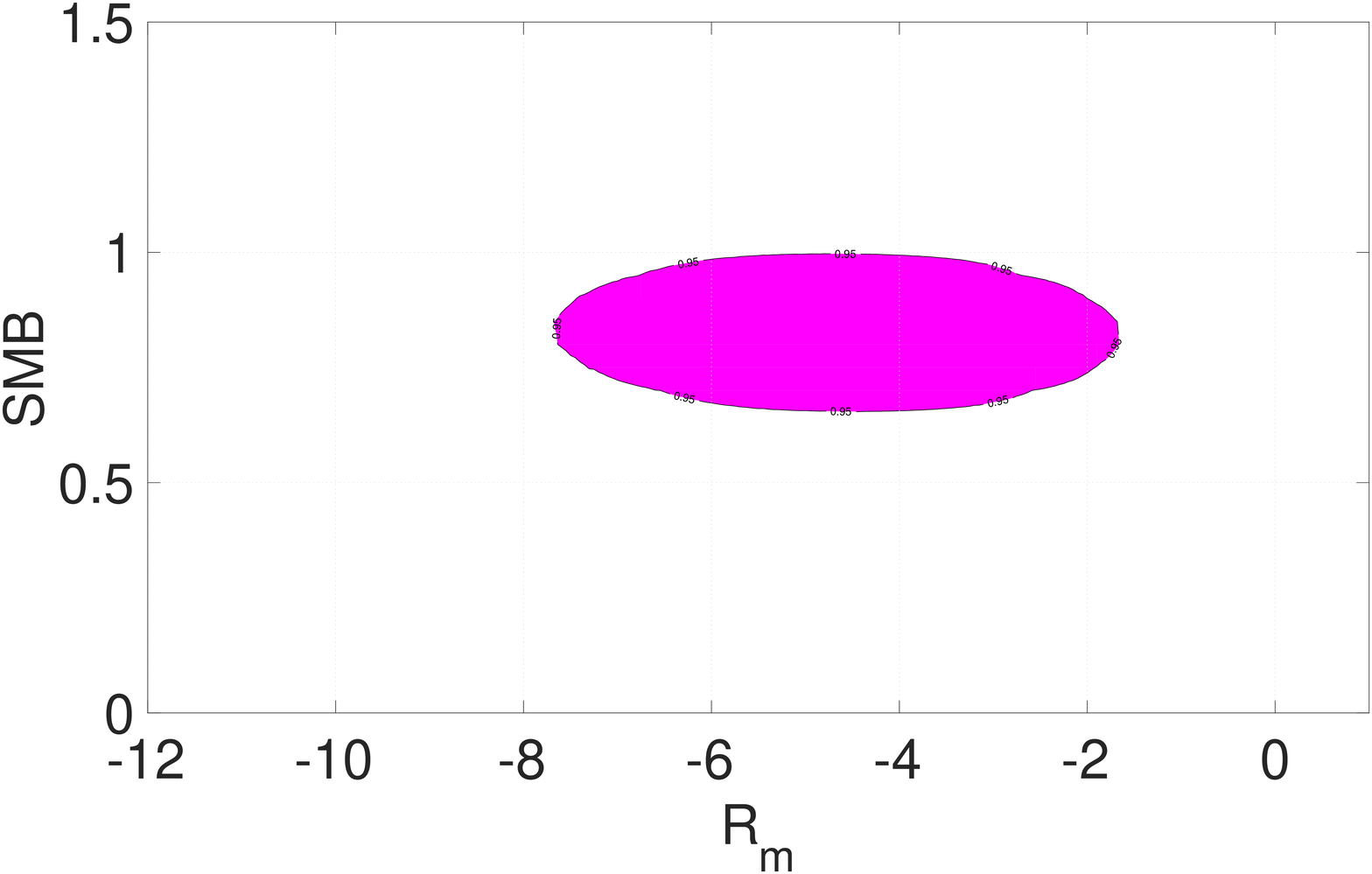}%
}
\\ 
\text{Panel 18.3: KLM} &  & \text{Panel 18.4: LR}%
\end{array}
\\ 
\end{array}%
\end{equation*}

To compare with Figure 17, we further replace the \textquotedblleft$EqFac$%
\textquotedblright\ risk factor with the \textquotedblleft
SMB\textquotedblright\ (small minus big) factor from Fama and French (1993)
and similarly construct Figure 18.  The risk premia on \textquotedblleft $R_{m}$%
\textquotedblright  \ and \textquotedblleft
SMB\textquotedblright\ are well-known to be strongly identified, and  the rank test of Kleibergen and Zhan (2020) yields a $p$-value close to zero for testing their $\beta$'s (with the rank test $\chi^2$-statistic equal to 128.35, $F$-statistic equal to 4.89).  The AR test now signals model
misspecification, since it rejects every hypothesized risk premia as shown
in Panel 18.2, so the 95\% confidence set that results from the AR test is
empty. Our DRLM test, which allows for misspecification, yields a tight
confidence set in Panel 18.1. This tight confidence set, in contrast with
the wide one in Panel 17.1, indicates that the pricing ability of
\textquotedblleft$EqFac$\textquotedblright\ differs substantially from
\textquotedblleft SMB\textquotedblright. Because of the misspecification,
the 95\% confidence sets resulting from the KLM and LR tests are not
representative for the minimizer of the population objective function.

The minimal value of the sample objective function of 59.34 reported in Figure 18, which equals the $J$-statistic, is now well below the sample measure of the identification strength of 128.35. The sample measure of the identification strength corresponds with the rank test $\chi^2$ statistic. Despite the misspecification, the CUE estimates can therefore be straightforwardly interpreted as representing the risk premia on \textquotedblleft $R_{m}$%
\textquotedblright  \ and \textquotedblleft
SMB\textquotedblright\ . This is further reflected by their convex 95\% confidence sets shown in Figure 18.

\subsection{Running example 2: Linear IV regression for
the return on education using Card (1995) data}

To further show the ease of implementing the DRLM test for applied work, we
use the return on education data from\ Card (1995).\nocite{car95} Card
(1995) uses proximity to college as the instrument in an IV regression of
(the log) wage on (length of) education. For more details on the data, we
refer to Card (1995). The instruments used in our specification are three
binary indicator variables which show the proximity to a two-year college, a
four-year college and a four-year public college, respectively. The included
exogenous variables are a constant term, age, age$^{2},$ and racial,
metropolitan, family and regional indicator variables. All three binary
instruments have their own local average treatment effects, which in case of
heterogeneous treatment effects leads to misspecification of the linear IV
regression model since it considers them to be identical, see Imbens and
Angrist (1994). 

Figure 19 presents the values of the\ AR, LR, KLM and DRLM\ statistics around
the CUE. It also shows their  critical value functions at the 5\% level. The other area of
small values of the DRLM\ statistic is left out since it would be discarded
by the power enhancement rule. The $J$-statistic, which equals the minimal
value of the AR statistic, is 2.99 with a $p$-value of 0.22. The first stage
$F$-statistic is 7.01 so the return on education is weakly identified, see
Stock and Yogo (2005),\nocite{sy01} which then also implies that the $J$-test does not
have much power. Its quite low $p$-value can thus as well indicate
misspecification, which would result from distinct local average treatment
effects for the different instruments. Lee (2018) constructs misspecification-robust standard errors for the two stage least
squares estimator when the local average treatment effects differ, but the
resulting $t$-test is not valid here because of the weak identification of
the return on education indicated by the small first stage $F$-statistic. This makes
the DRLM test more appealing, since it is robust to both misspecification and
weak identification. Kitagawa (2015)\nocite{kit15} further shows that the
validity of the instruments for the Card (1995) data depends on the specification
of the model. Figure 19 then shows that allowing for misspecification
further enlarges the identification-robust confidence set for the return on
education.
\begin{equation*}
\begin{array}{c}
\text{Figure 19: Tests of the return on education using Card (1995) data
with the DRLM (solid black),} \\ 
\text{KLM (dashed black), LR (solid red) and AR (solid blue)
statistics and their 95\% (conditional) } \\ 
\text{critical value lines (dotted in the color of the
test they refer to).} \\ 
\raisebox{-0pt}{\includegraphics[
height=2.6498in,
width=3.5198in
]%
{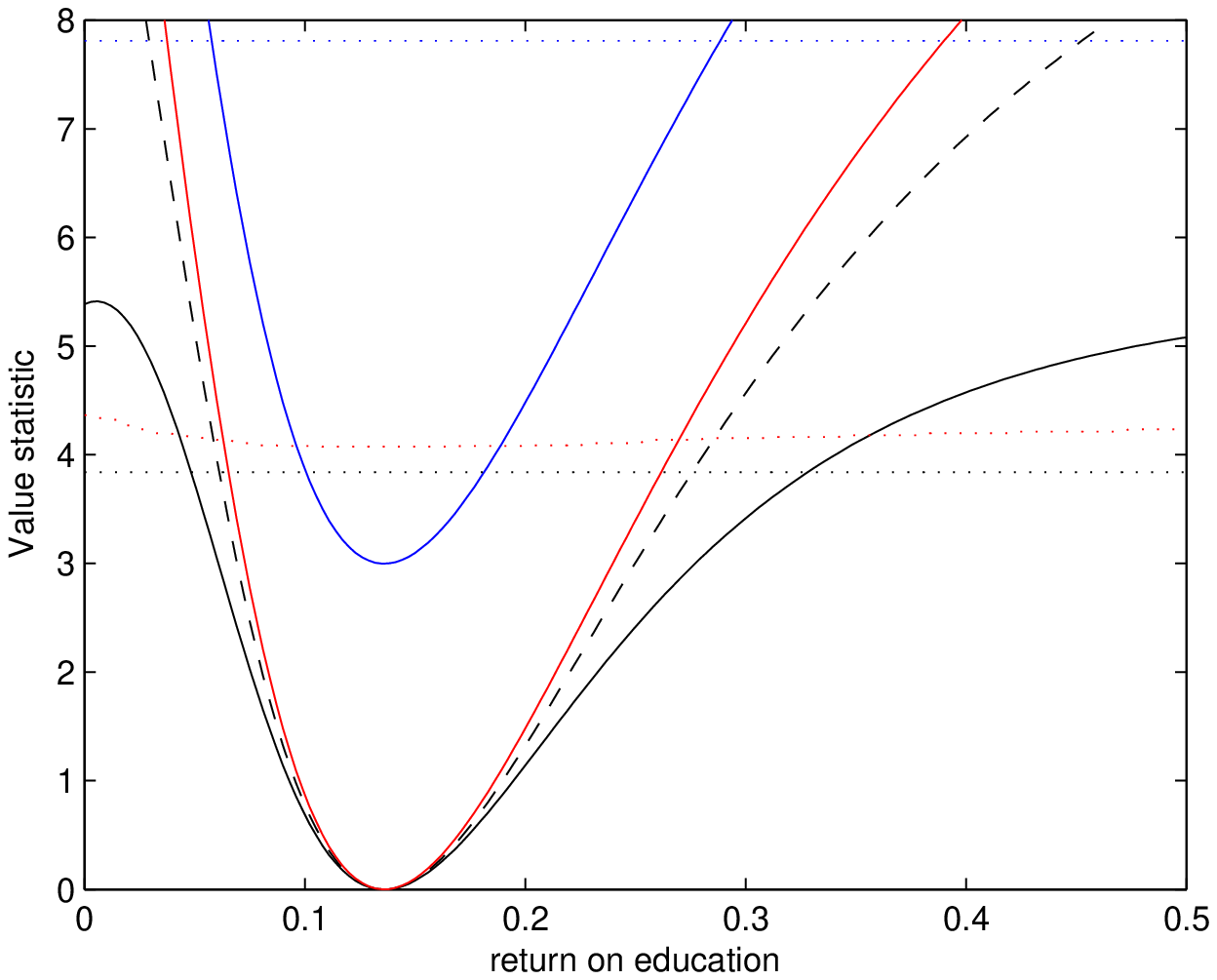}%
}
\\\text{ \ }%
\end{array}%
\end{equation*}

Since the number of instruments is equal to three, the sample analog of the $IS$ identification strength measure (\ref{rank 3}) of 21.03 (=three times the first stage $F$-statistic) is well above the minimal value of the CUE objective function, 2.99. A structural interpretation could therefore be rendered to the CUE, which is also reflected by the bounded 95\% confidence sets shown in Figure 19.

\section{Conclusions}

We show that it is generally feasible to conduct reliable inference on the
pseudo-true value of the structural parameters resulting from the population
continuous updating GMM\ objective function using the DRLM test. For linear moment equations, we also propose a measure of the identification strength that can be compared with the minimal value of the CUE objective function to gauge whether the pseudo-true value can be interpreted in a structural manner. While settings of weak identification paired with misspecification are empirically
relevant, it was so far not possible to conduct reliable inference in these
settings. This holds since weak identification robust tests are size
distorted when the model is misspecified, while the misspecification tests
which are typically used to detect misspecification, are virtually powerless
under weak identification. Hence, the DRLM test removes an important obstacle for conducting reliable inference in these empirically relevant settings.  We propose some straightforward power improvements for the DRLM\
test which make it work well, and hope to conduct further power improvements in future work. We also use the DRLM test to analyze data from three
studies which are plagued by both weak identification and misspecification
issues: Card (1995), Adrian et al. (2014), and He et al. (2017). It shows
that other inference procedures can seriously underestimate the uncertainty
concerning the structural parameters when both misspecification and weak
identification matter.\bigskip

\newpage \setstretch{1.70} 
\bibliographystyle{plain}
\bibliography{eqrand}

\newpage

\begin{center}

{\Large Online Appendix  for \\  \textquotedblleft Double robust inference for
continuous updating GMM\textquotedblright }
\end{center}

\bigskip

\

\setcounter{page}{1}

\doublespace\renewcommand{\thefootnote}{\arabic{footnote}} %
\setcounter{footnote}{0}\setcounter{section}{0}

\section{ \ Structural interpretation for misspecified settings}

In case of misspecification, the structural specification resulting from the
pseudo-true value depends on the involved population objective function. For
the linear asset pricing and instrumental variables regression models, it is
thus instructive to see how the population continuous updating objective
function comes to a structural specification at the pseudo-true value. We
therefore first lay out the unrestricted specification of the population
moments used by the population continuous updating objective function to
obtain its structural specification at the pseudo-true value for the linear
factor and instrumental variables regression models with i.i.d. errors:%
\begin{equation*}
\begin{array}{l}
\begin{array}{l}
\text{Factor model:} \\ 
\begin{array}{c}
\qquad \text{structural model\quad \qquad \qquad misspecification} \\ 
\begin{array}{rl}
\left( 
\begin{array}{cc}
\mu _{R} & \beta%
\end{array}%
\right) = & \overbrace{-D^{\ast }\left( 
\begin{array}{cc}
\lambda _{F}^{\ast } & I_{m}%
\end{array}%
\right) }+\overbrace{\Omega D_{\perp }^{\ast }\delta ^{\ast }\left( 
\begin{array}{cc}
\lambda _{F}^{\ast } & I_{m}%
\end{array}%
\right) _{\perp }\left( 
\begin{array}{cc}
1 & 0 \\ 
0 & Q_{\bar{F}\bar{F}}^{-1}%
\end{array}%
\right) } \\ 
= & -D^{\ast }\left( 
\begin{array}{cc}
\lambda _{F}^{\ast } & I_{m}%
\end{array}%
\right) +\left( 
\begin{array}{cc}
\gamma _{1} & \Gamma _{2}%
\end{array}%
\right)%
\end{array}%
\end{array}%
\end{array}
\\ 
\\ 
\begin{array}{l}
\text{Linear instrumental variables regression model:} \\ 
\begin{array}{c}
\quad \qquad \text{\quad \qquad \qquad structural model\quad \qquad
misspecification}\quad \quad \\ 
\begin{array}{rl}
\left( 
\begin{array}{cc}
\sigma _{Zy} & \Sigma _{ZX}%
\end{array}%
\right) = & \overbrace{-D^{\ast }\left( 
\begin{array}{cc}
\theta^{\ast } & I_{m}%
\end{array}%
\right) }+\overbrace{Q_{\bar{Z}\bar{Z}}D_{\perp }^{\ast }\delta ^{\ast
}\left( 
\begin{array}{cc}
\theta ^{\ast } & I_{m}%
\end{array}%
\right) _{\perp }\Omega} \\ 
= & -D^{\ast }\left( 
\begin{array}{cc}
\theta ^{\ast } & I_{m}%
\end{array}%
\right) +\left( 
\begin{array}{cc}
\gamma _{1} & \Gamma _{2}%
\end{array}%
\right)%
\end{array}%
\end{array}%
\end{array}%
\end{array}%
\end{equation*}%
where $D^{\ast }$ is an $N\times m$ dimensional matrix for the factor model
and a $k\times m$ dimensional matrix for the linear instrumental variables
regression model, $D_{\perp }^{\ast }$ is the orthogonal complement of $%
D^{\ast },$ so an $N\times (N-m)$ dimensional matrix for the factor model: $%
D^{\ast \prime }D_{\perp }^{\ast }\equiv 0,$ $D_{\perp }^{\ast \prime
}\Omega D_{\perp }^{\ast }\equiv I_{N-m};$ and a $k\times (k-m)$ dimensional
matrix for the linear instrumental variables regression model: $D^{\ast
\prime }D_{\perp }^{\ast }\equiv 0,$ $D_{\perp }^{\ast \prime }Q_{\bar{Z}%
\bar{Z}}D_{\perp }^{\ast }\equiv I_{k-m};$ in an identical manner: $\left( 
\begin{array}{cc}
\lambda _{F}^{\ast } & I_{m}%
\end{array}%
\right) _{\perp }=(1$ $-\lambda _{F}^{\ast \prime })(1+\lambda _{F}^{\ast
\prime }Q_{\bar{F}\bar{F}}^{-1}\lambda _{F}^{\ast })^{-\frac{1}{2}}$ and $%
\left( 
\begin{array}{cc}
\theta ^{\ast } & I_{m}%
\end{array}%
\right) _{\perp }=(1$ $-\theta^{\ast \prime })\left( \binom{1}{-\theta ^{\ast }%
}^{\prime }\Omega \binom{1}{-\theta ^{\ast }}\right) ^{-\frac{1}{2}}$ and $%
\delta ^{\ast }$ is an $(N-m)$ dimensional vector for the factor model and a 
$(k-m)$ dimensional vector for the linear instrumental variables regression
model reflecting the misspecification so in case of correct specification, $%
\delta ^{\ast }=0.$ The matrix $\left( 
\begin{array}{cc}
\gamma _{1} & \Gamma _{2}%
\end{array}%
\right) $ is $N\times (m+1)$ dimensional for the linear factor model and $%
k\times (m+1)$ dimensional for the instrumental variables regression model.
The above specification results from a singular value decomposition of the
normalized population moments, see e.g. Theorem 8 and Kleibergen and Paap
(2006). \nocite{kpaap02}

The unrestricted specifications show that the population continuous updating
objective function at the pseudo-true value equals: 
\begin{equation*}
\begin{array}{cll}
Q_{p}(\lambda _{F}^{\ast })= & \frac{1}{1+\lambda _{F}^{\prime }Q_{\bar{F}%
\bar{F}}^{-1}\lambda _{F}}(\mu _{R}-\beta \lambda _{F}^{\ast })^{\prime
}\Omega ^{-1}(\mu _{R}-\beta \lambda _{F}) & =\delta ^{\ast \prime }\delta
^{\ast } \\ 
Q_{p}(\theta ^{\ast })= & \frac{1}{\omega _{uu}-2\omega _{uV}\theta ^{\ast
}+\theta ^{\ast \prime }\Omega _{VV}\theta ^{\ast }}(\sigma _{Zy}-\Sigma
_{ZX}\theta ^{\ast })^{\prime }Q_{\bar{Z}\bar{Z}}^{-1}(\sigma _{Zy}-\Sigma
_{ZX}\theta ^{\ast }) & =\delta ^{\ast \prime }\delta ^{\ast },%
\end{array}%
\end{equation*}%
which further illustrates that $\delta ^{\ast \prime }\delta ^{\ast }$
equals the squared smallest singular value of either $\Omega ^{-\frac{1}{2}%
}\left( 
\begin{array}{cc}
\mu _{R} & \beta%
\end{array}%
\right) \allowbreak \left( 
\begin{array}{cc}
1 & 0 \\ 
0 & Q_{\bar{F}\bar{F}}^{\frac{1}{2}}%
\end{array}%
\right) ,$ factor model, or $Q_{\bar{Z}\bar{Z}}^{-\frac{1}{2}}\left( 
\begin{array}{cc}
\sigma _{Zy} & \Sigma _{ZX}%
\end{array}%
\right) \Omega ^{-\frac{1}{2}},$ linear instrumental variables regression
model.

For the unrestricted specification to have a structural interpretation, we
need that:%
\begin{equation*}
\begin{array}{l}
\begin{array}{l}
\text{Factor model:} \\ 
\gamma _{1}^{\prime }\Omega ^{-1}D^{\ast }=0,\text{ }\Gamma _{2}^{\prime
}\Omega ^{-1}D^{\ast }=0,\text{ }\left( 
\begin{array}{cc}
\gamma _{1} & \Gamma _{2}%
\end{array}%
\right) \left( 
\begin{array}{cc}
1 & 0 \\ 
0 & Q_{\bar{F}\bar{F}}^{-1}%
\end{array}%
\right) \left( 
\begin{array}{cc}
\lambda _{F}^{\ast } & I_{m}%
\end{array}%
\right) ^{\prime }=0.%
\end{array}
\\ 
\begin{array}{l}
\text{Linear instrumental variables regression model:} \\ 
\gamma _{1}^{\prime }Q_{\bar{Z}\bar{Z}}^{-1}D^{\ast }=0,\text{ }\Gamma
_{2}^{\prime }Q_{\bar{Z}\bar{Z}}^{-1}D^{\ast }=0,\text{ }\left( 
\begin{array}{cc}
\gamma _{1} & \Gamma _{2}%
\end{array}%
\right) \Omega \left( 
\begin{array}{cc}
\theta ^{\ast } & I_{m}%
\end{array}%
\right) ^{\prime }=0.%
\end{array}%
\end{array}%
\end{equation*}%
The restrictions for the linear instrumental variables regression model are
identical to those in Koles\'{a}r et al. (2015),\nocite{kcfgi15} except that
they also assume that $\Gamma _{2}=0,$\footnote{%
We note that Assumption 2 in Koles\'{a}r et al. (2015) is imposed on $\bar{Q}%
_{ZZ}^{-1}(\sigma _{Zy}$ $\Sigma _{ZX})$ so $\gamma _{1}^{\prime }Q_{\bar{Z}%
\bar{Z}}D^{\ast }=0$ in their specification.} who show that they allow for a
causal interpretation. Koles\'{a}r et al. (2015) motivate them by means of a
random coefficients assumption with potentially many instruments where
direct, channeled through $\gamma _{1},$ and indirect effects, channeled
through $D^{\ast },$ are independently distributed. In asset pricing, the
factors are often considered as proxies for true underlying risk factors.
The measurement error between the observed proxy risk factors and the true
underlying risk factors can then similarly be represented by a random
coefficient specification where the measurement error reflected by $\left( 
\begin{array}{cc}
\gamma _{1} & \Gamma _{2}%
\end{array}%
\right) $ is uncorrelated with the true risk factor $D^{\ast }$ after
correcting for the covariance matrix of the errors.

The unrestricted specifications crucially hinge on that the largest singular
values, identifying the structural specification, and the smallest one,
which represents the misspecification, differ considerably. The largest
singular values reflect the identification strength of the structural
parameters so when these are close to the singular value reflecting the
misspecification, the pseudo-true value is weakly identified. Furthermore,
when the singular value representing the misspecification exceeds (some of)
the singular values reflecting the identification strength, we can no longer
attribute a structural interpretation to the pseudo-true value.

\section{ \ Additional numerical results}

\subsection{Mild misspecification}

We increase the amount of misspecification to $\bar{\mu}^{\prime }\bar{%
\mu}=10,$ which is still quite small since there are twenty-five moment
equations. Figures A1, A2, and A3 show that the increased
misspecification exacerbates the size distortion of the AR, KLM\ and LR
tests compared to the setting of weak misspecification in the paper. 
\begin{equation*}
\begin{array}{c}
\text{Figure A1: Power of 5\% significance KLM and DRLM tests of} \\ 
\text{ H}_{0}:\lambda_{F}=0\text{ with misspecification, }\bar{\mu}^{\prime }%
\bar{\mu}=10,\text{ }N=25,\text{ }Q_{\bar{F}\bar{F}}=1 \\ 
\begin{array}{cc}
\raisebox{-0pt}{\includegraphics[
height=2.2139in,
width=2.9386in
]%
{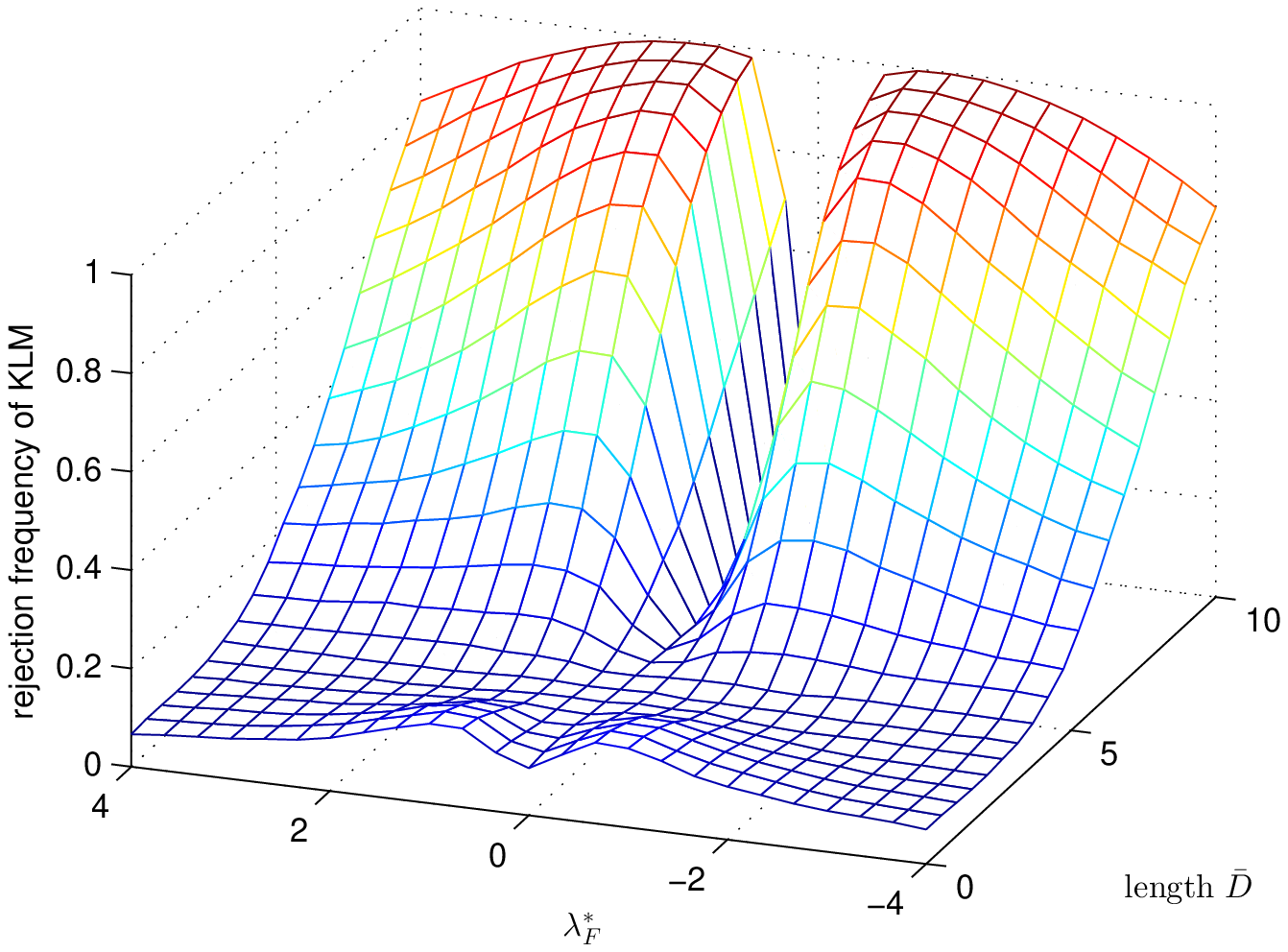}%
}
&
\raisebox{-0pt}{\includegraphics[
height=2.2139in,
width=2.9386in
]%
{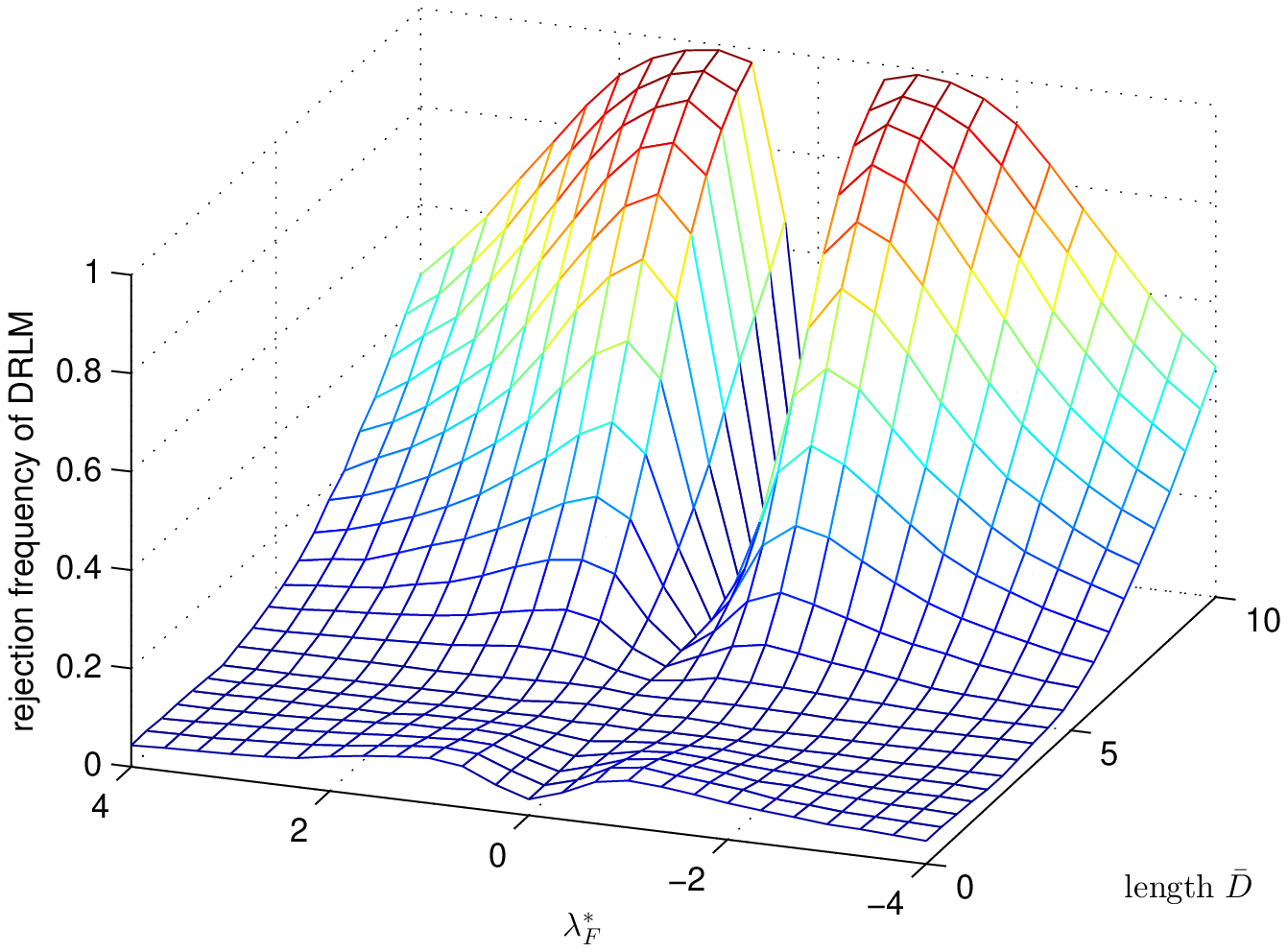}%
}
\\
\\ 
\text{Panel A1.1: KLM} & \text{Panel A1.2: DRLM}%
\end{array}%
\end{array}%
\end{equation*}

For the
conditional LR test, the rejection frequency at zero decreases from 30\% to
8\% when the identification strength increases. When the amount of
misspecification and the identification strength coincide, the rejection
frequency of the LR test is 27\% when $\lambda _{F}^{\ast }=0.$ For the KLM
test, the rejection frequency decreases from 10\%\ to 5\%. For the DRLM and
size and power improved DRLM test, we observe either no size distortion and
a rejection frequency of 8\% which decreases to 5\% when the identification
strength increases. The minor size distortion of the size and power improved
DRLM\ test only occurs when the amount of misspecification exceeds the
strength of identification, so the hypothesized value is not the minimizer
of the population objective function, and is not present when the
identification strength is larger than or equal to the amount of
misspecification. The rejection frequency of the AR test is equal to 36\%
for all identification strengths. When the amount of misspecification
exceeds the identification strength, the maximum of the population
continuous updating objective function is situated at $\lambda _{F}^{\ast
}=0,$ which explains why the rejection frequency of the AR and LR tests
decreases away from $\lambda _{F}^{\ast }=0$ for low values of the
identification strength. For values of the identification strength which
exceed the amount of misspecification, we see no decrease of the rejection
frequency when $\lambda _{F}^{\ast }$ moves away from zero.

\begin{equation*}
\begin{array}{c}
\text{Figure A2: Power of 5\% significance LR and size and power improved }
\\ 
\text{DRLM tests of H}_{0}:\lambda_{F}=0\text{ with misspecification, }\bar{%
\mu}^{\prime}\bar{\mu}=10,\text{ }N=25,\text{ }Q_{\bar{F}\bar{F}}=1 \\ 
\begin{array}{cc}
\raisebox{-0pt}{\includegraphics[
height=2.2139in,
width=2.9386in
]%
{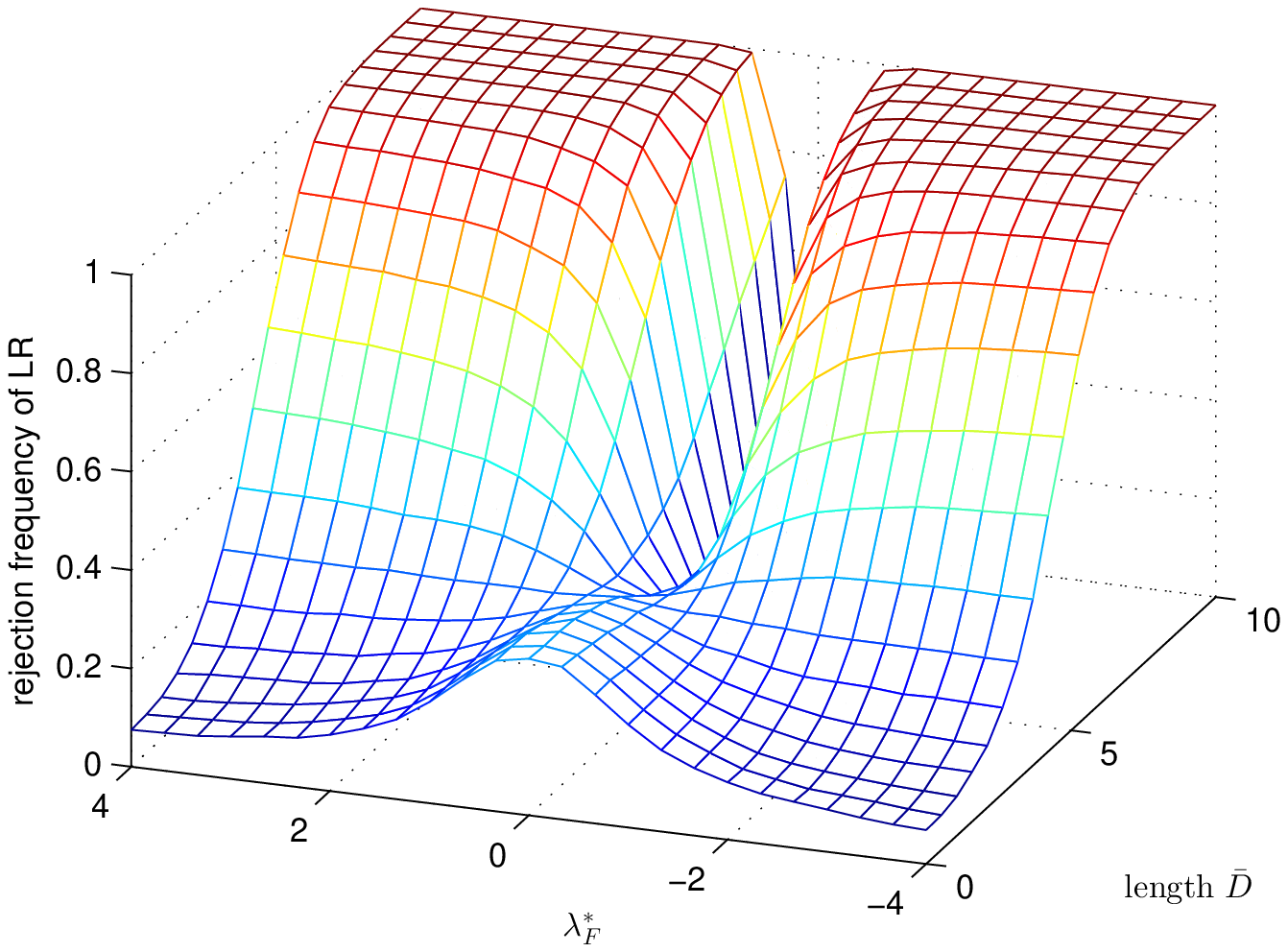}%
}
&
\raisebox{-0pt}{\includegraphics[
height=2.2139in,
width=2.9386in
]%
{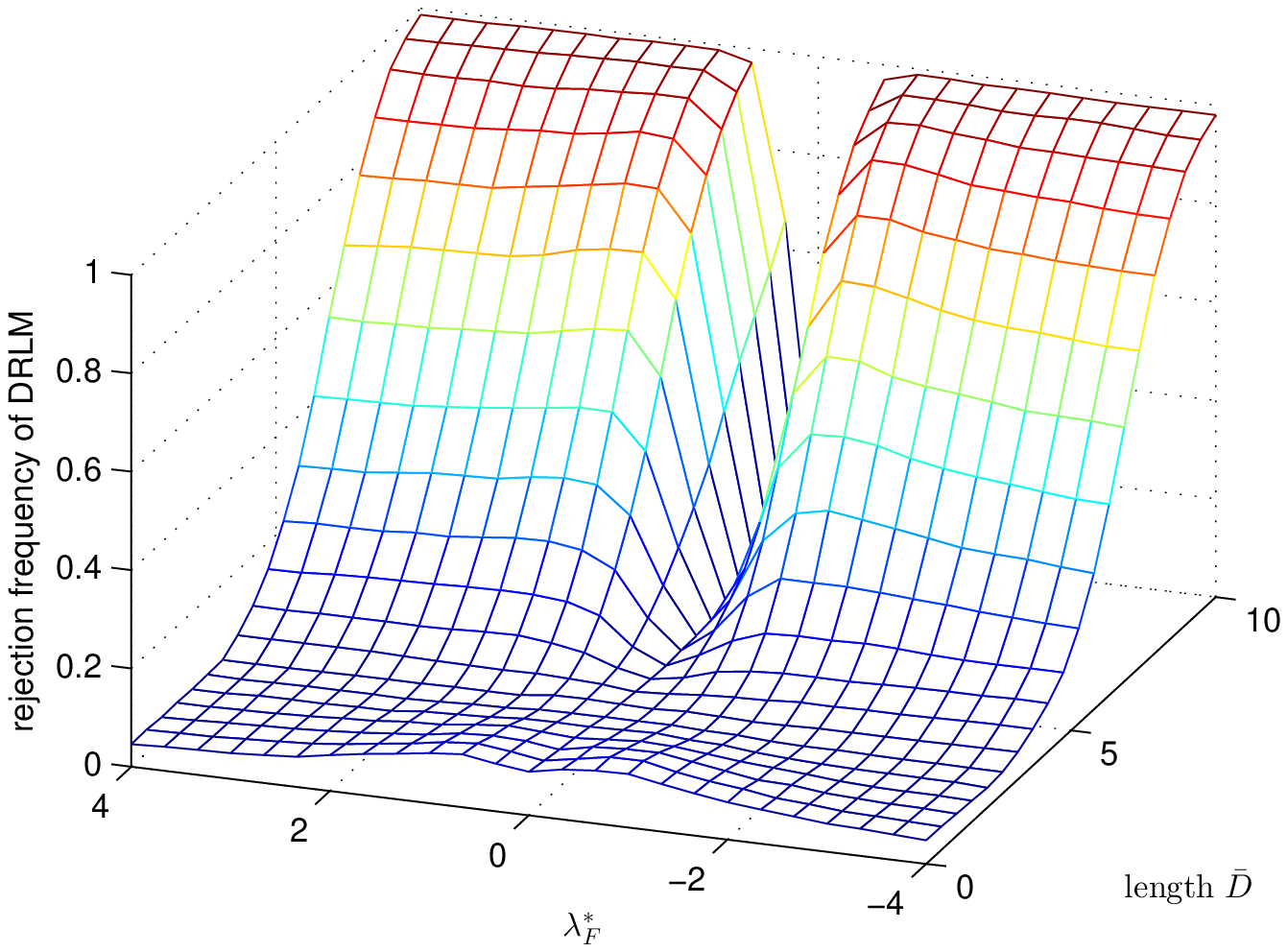}%
}
\\

\text{Panel A2.1: LR} & \text{Panel A2.2: DRLM with size and } \\ 
& \text{power improvements}%
\end{array}%
\end{array}%
\end{equation*}

\begin{equation*}
\begin{array}{c}
\text{Figure A3: Power of 5\% significance AR tests of H}_{0}:\lambda_{F}=0
\\ 
\text{with misspecification, }\bar{\mu}^{\prime}\bar{\mu}=10,\text{ }N=25,%
\text{ }Q_{\bar{F}\bar{F}}=1 \\ 
\begin{array}{c}
\raisebox{-0pt}{\includegraphics[
height=2.2139in,
width=2.9386in
]%
{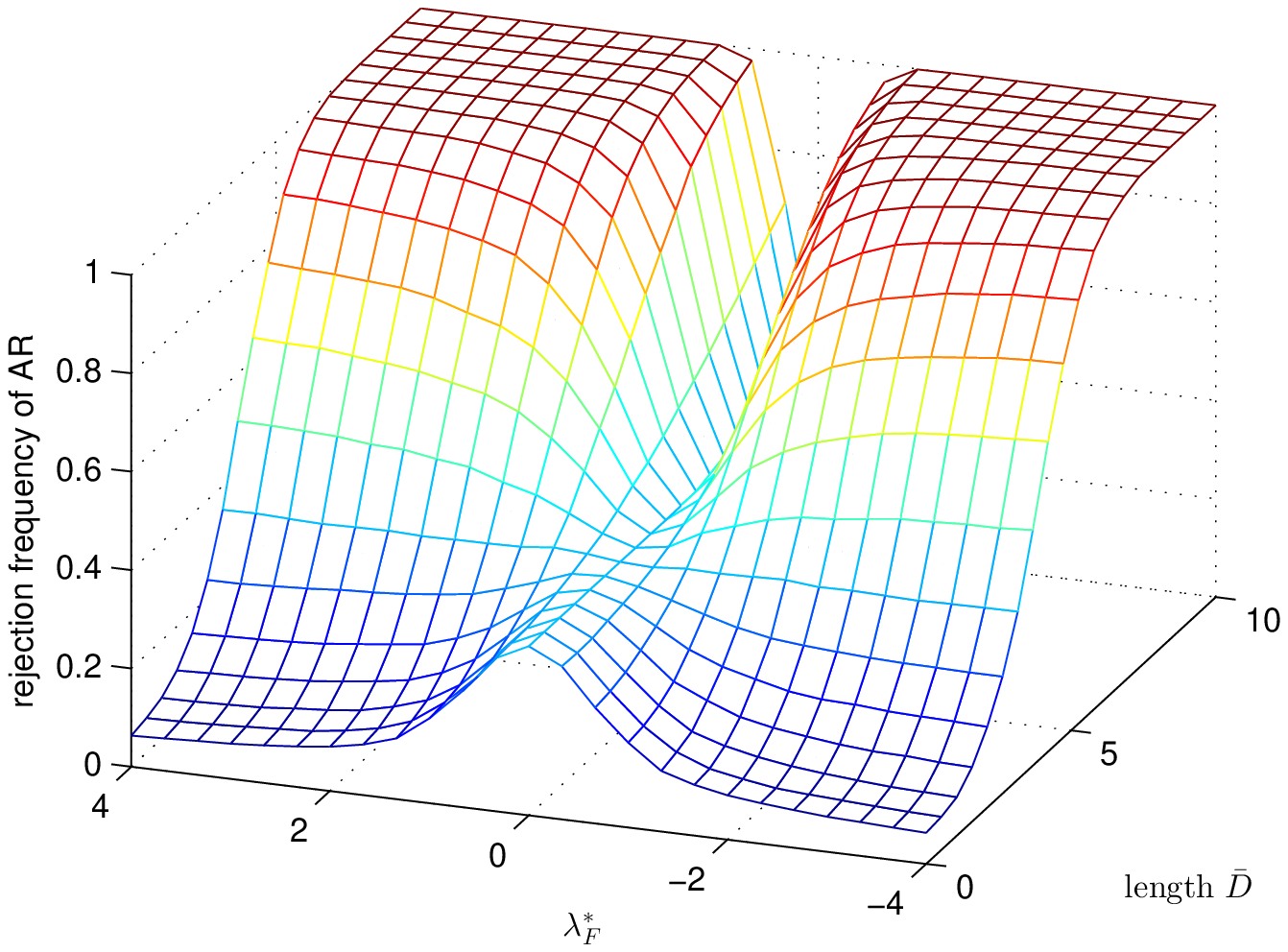}%
}
\end{array}%
\end{array}%
\end{equation*}

Figure A4 shows the distribution function of the misspecification $J$%
-statistic, which equals the minimal value of the AR statistic, when the
null hypothesis holds, so for values of $\lambda _{F}^{\ast }$ equal to
zero. It shows the distribution function for three different values of the
identification strength $\bar{D}^{\prime }\bar{D}:$ 0, 10 and 100.
Recognizing that the 95\% critical value of the $\chi ^{2}(24)$ distribution$%
,$ since $N-1=24,$ equals 36.42, Figure A4 shows that we never reject no
misspecification at the 5\% significance level when $\bar{D}^{\prime }\bar{D}
$ equals 0, 7\% of the times when $\bar{D}^{\prime }\bar{D}=10$ and 33\%
when $\bar{D}^{\prime }\bar{D}$ equals 100. This indicates the difficulty of
detecting the mild misspecification present in the simulated data. 
\begin{equation*}
\begin{array}{c}
\text{Figure A4: Distribution function of $J$-statistic for misspecification
when H}_{0}:\lambda _{F}=0\text{ holds, } \\ 
\text{solid line: }\bar{D}^{\prime }\bar{D}=0,\text{ dash-dot: }\bar{D}%
^{\prime }\bar{D}=10=\text{strength of misspecification, dashed: }\bar{D}%
^{\prime }\bar{D}=100. \\ 
\raisebox{-0pt}{\includegraphics[
height=2.0139in,
width=2.9386in
]%
{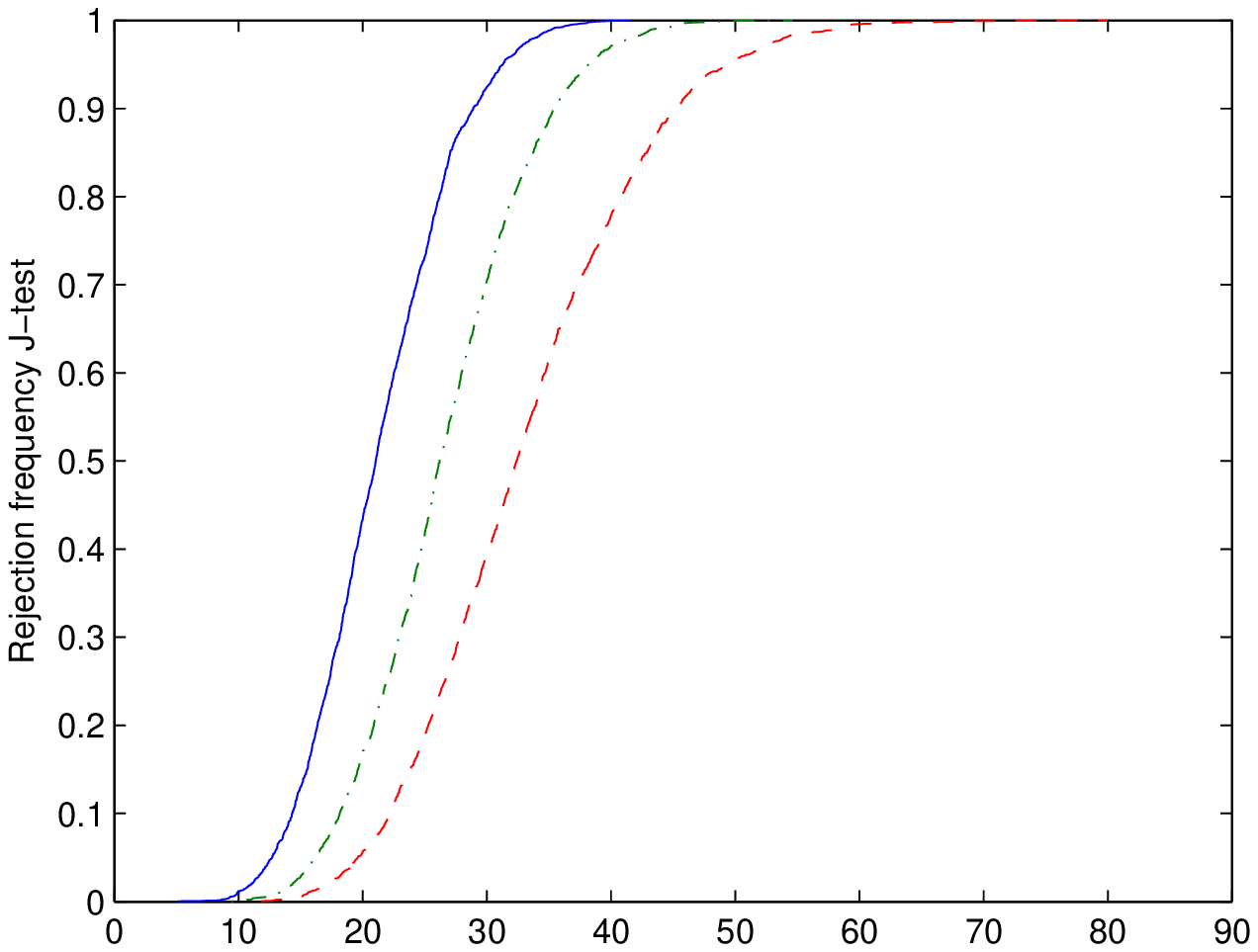}%
}

\end{array}%
\end{equation*}

To show
that the power issues discussed previously for both the identification
robust tests and the misspecification $J$-test do not result from the
somewhat large number of moment equations, 25, we next discuss a somewhat smaller
simulation experiment with fewer moment conditions. 

\subsection{Power of $J$-test and identification-robust tests with fewer
moment conditions}

To show that the low power of the $J$-test for misspecification is not just
resulting from the large number of moment equations, we repeat the
simulation exercise with fewer moment equations, $N=5,$ and weak
misspecification: $\bar{\mu}^{\prime }\bar{\mu}=2.5.$ Panels A5.1 and A5.2 in Figure 5 show the power curves for the conditional LR and size and power improved
DRLM\ tests. Panel A5.1 shows that the conditional LR test is size
distorted and its rejection frequency equals 17\% when the misspecification
and strength of identification are identical. The size and power improved
DRLM\ test shows no size distortion. Figure A6 shows the simulated
distribution function of the $J$-statistic. Since $N=5,$ the limiting
distribution of the $J$-statistic is a $\chi ^{2}(4)$ distribution whose 95\%
critical value equals 9.48. The simulated distribution function shows that
we never reject no misspecification when $\bar{D}^{\prime }\bar{D}=0,$ 2.5\%
of the times when $\bar{D}^{\prime }\bar{D}=2.5$ which equals the strength
of misspecification, and 20\% of the times when $\bar{D}^{\prime }\bar{D}%
=100.$ This reiterates the difficulty of detecting misspecification, which
leads to size distorted identification-robust tests, using the $J$-test when
the identification is weak; see also Gospodinov et al. (2017).\nocite{gkr17}

\begin{equation*}
\begin{array}{c}
\text{Figure A5: Power of 5\% significance LR and DRLM tests of} \\ 
\text{ H}_{0}:\lambda _{F}=0\text{ with misspecification, }\bar{\mu}^{\prime
}\bar{\mu}=2.5,\text{ }N=5,\text{ }Q_{\bar{F}\bar{F}}=1 \\ 
\begin{array}{cc}
\raisebox{-0pt}{\includegraphics[
height=2.0139in,
width=2.9386in
]%
{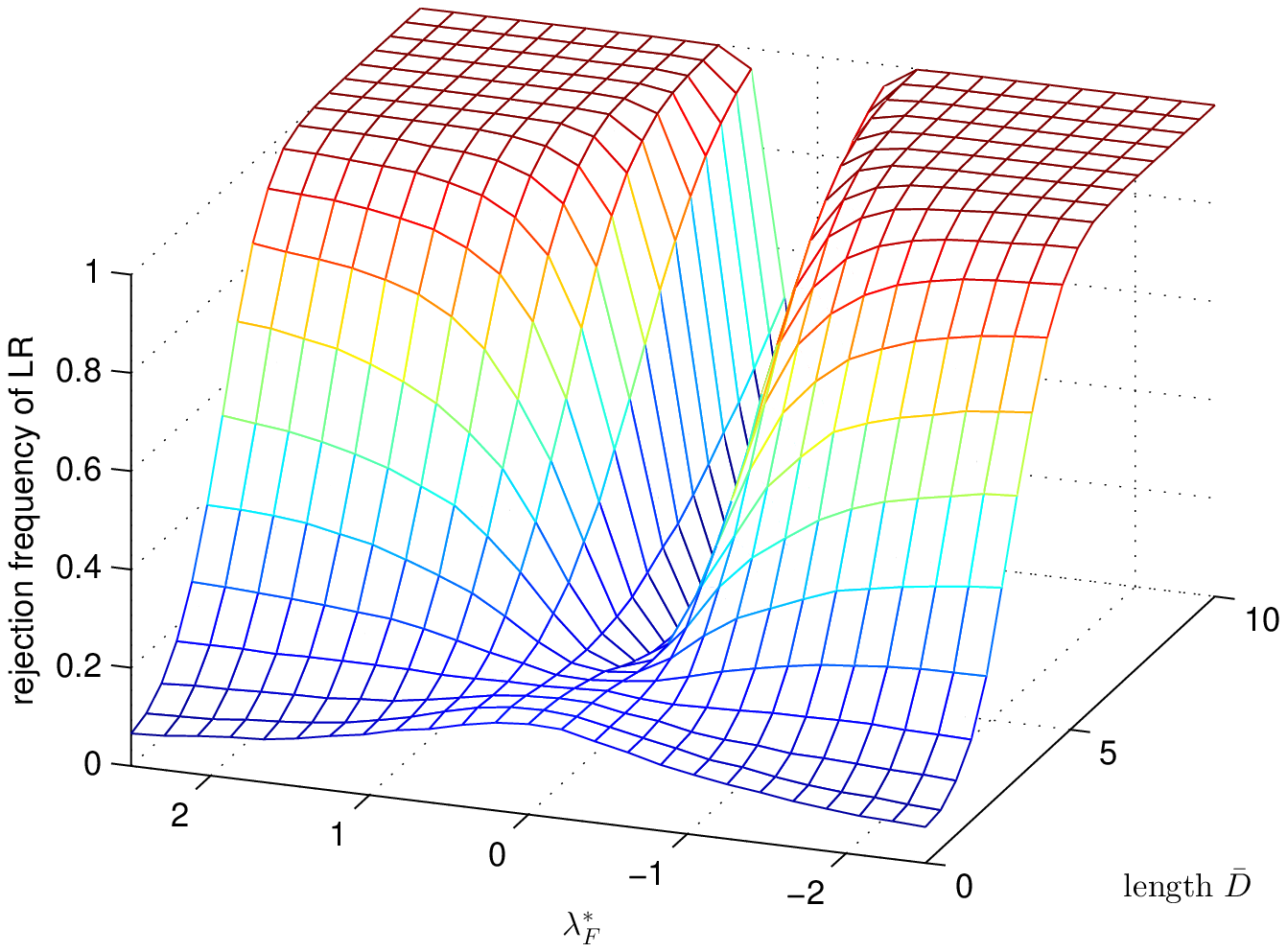}%
}
&
\raisebox{-0pt}{\includegraphics[
height=2.0139in,
width=2.9386in
]%
{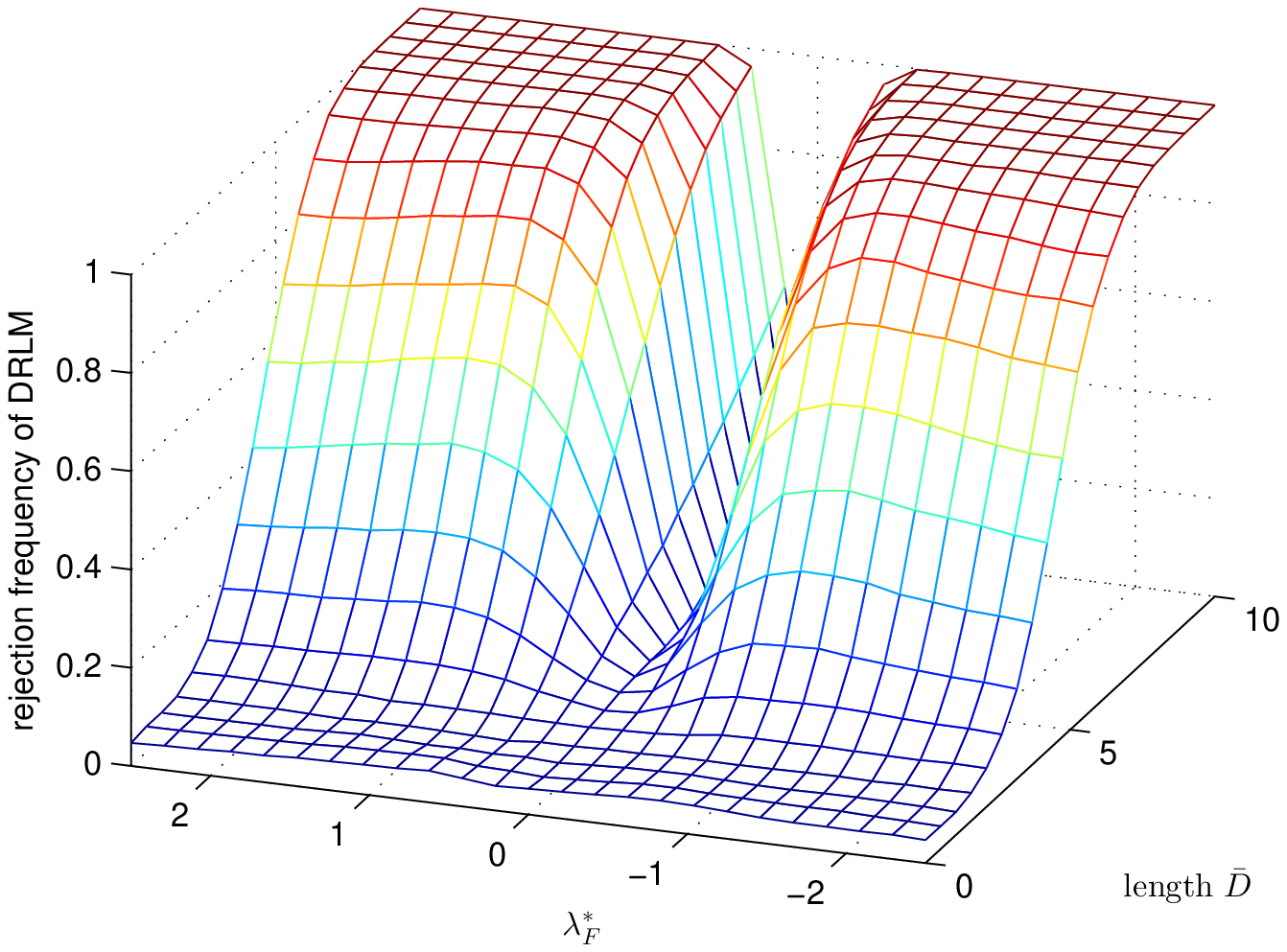}%
}
 \\ 
\text{Panel A5.1: LR} & \text{Panel A5.2: DRLM}%

\end{array}%
\end{array}%
\end{equation*}

\begin{equation*}
\begin{array}{c}
\text{Figure A6. Distribution function of $J$-statistic for misspecification when H}%
_{0}:\lambda _{F}=0\text{ holds, } \\ 
\text{solid line }\bar{D}^{\prime }\bar{D}=0,\text{ dash-dot: }\bar{D}%
^{\prime }\bar{D}=2.5=\text{strength of misspecification, dashed: }\bar{D}%
^{\prime }\bar{D}=100. \\ 
\raisebox{-0pt}{\includegraphics[
height=2.2139in,
width=2.9386in
]%
{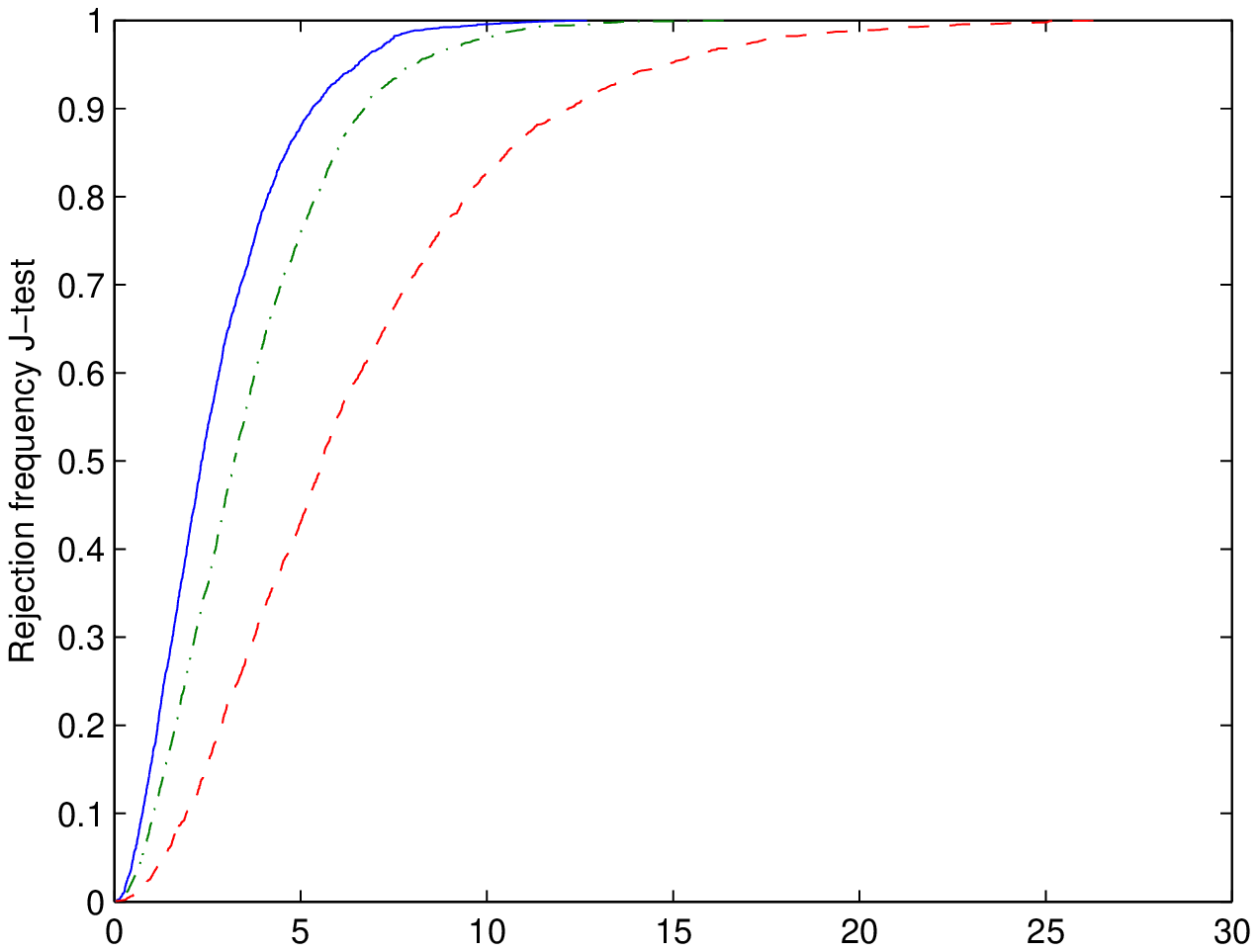}%
}\\
\end{array}%
\end{equation*}

\section{ \ Lemmas and Proofs}

\subsection{Lemma}

\paragraph{Lemma 1.}

The estimators $\bar{R}$ and $\hat{\beta}$ in the linear regression model:%
\begin{equation*}
R_{t}=c+\beta F_{t}+u_{t},
\end{equation*}
with $c$ an $N$-dimensional vector of constants, $F_{t}=G_{t}-\bar{G},$ with 
$G_{t}$ an $m$-dimensional vector of factors and $\bar{G}=\frac{1}{T}\sum
_{t=1}^{T}G_{t},$ so $\bar{F}=0,$ and $u_{t}$ an $N$-dimensional vector
which contains the errors which are i.i.d. distributed with mean zero and
covariance matrix $\Omega,$ are independently distributed in large samples.
\smallskip

\paragraph{Proof:}

Since $\bar{R}=\hat{c}+\hat{\beta}\bar{F}=\hat{c},$ and the joint limit
behavior of $\hat{c}$ and $\hat{\beta}$ accords with

\begin{equation*}
\begin{array}{cc}
\sqrt{T}\left[ \left( 
\begin{array}{c}
\hat{c} \\ 
\text{vec(}\hat{\beta})%
\end{array}
\right) -\left( 
\begin{array}{c}
c \\ 
\text{vec(}\beta)%
\end{array}
\right) \right] \underset{d}{\rightarrow} & \left( 
\begin{array}{c}
\psi_{c} \\ 
\psi_{\beta}%
\end{array}
\right) ,%
\end{array}%
\end{equation*}
with 
\begin{equation*}
\begin{array}{c}
\left( 
\begin{array}{c}
\psi_{c} \\ 
\psi_{\beta}%
\end{array}
\right) \sim N\left( 0,(Q^{-1}\otimes I_{N})\Sigma(Q^{-1}\otimes
I_{N})\right) ,%
\end{array}%
\end{equation*}
since $\frac{1}{T}\sum_{t=1}^{T}\left( 
\begin{array}{c}
1 \\ 
F_{t}%
\end{array}
\right) \left( 
\begin{array}{c}
1 \\ 
F_{t}%
\end{array}
\right) ^{^{\prime}}\underset{p}{\rightarrow}Q=\left( 
\begin{array}{cc}
1 & \mu_{F}^{\prime} \\ 
\mu_{F} & Q_{FF}%
\end{array}
\right) =\left( 
\begin{array}{cc}
1 & 0 \\ 
0 & Q_{FF}%
\end{array}
\right) ,$ $\mu_{F}=0,$ $Q_{FF}=E(F_{t}F_{t}^{\prime})=Q_{\bar{F}\bar{F}%
}+\mu_{F}\mu_{F}^{\prime},$ and $\frac{1}{T}\sum_{t=1}^{T}\left( \left( 
\begin{array}{c}
1 \\ 
F_{t}%
\end{array}
\right) \left( 
\begin{array}{c}
1 \\ 
F_{t}%
\end{array}
\right) ^{\prime}\otimes u_{t}u_{t}^{\prime}\right) \underset{p}{\rightarrow}%
\Sigma.$ When $u_{t}$ is i.i.d., $\Sigma=(Q\otimes\Omega),$ with $%
\Omega=var(u_{t}),$ so 
\begin{equation*}
\begin{array}{c}
\left( 
\begin{array}{c}
\psi_{c} \\ 
\psi_{\beta}%
\end{array}
\right) \sim N\left( 0,\left( 
\begin{array}{cc}
1 & 0 \\ 
0 & Q_{FF}^{-1}%
\end{array}
\right) \otimes\Omega\right) ,%
\end{array}%
\end{equation*}
so the limit behaviors of $\bar{R}=\hat{c}$ and $\hat{\beta}$ are
independent.

\paragraph{Lemma 2.}

\textbf{a. }When $\hat{V}_{ff}(\theta)^{-1}=\hat{V}_{ff}(\theta)^{-\frac{1}{2%
}\prime}\hat{V}_{ff}(\theta)^{-\frac{1}{2}},$ $\theta:1\times1,$ it holds
that 
\begin{equation*}
\begin{array}{c}
\frac{\partial}{\partial\theta}\hat{V}_{ff}(\theta)^{-\frac{1}{2}}=-\hat {V}%
_{ff}(\theta)^{-\frac{1}{2}}\hat{V}_{\theta f}(\theta)\hat{V}%
_{ff}(\theta)^{-1}.%
\end{array}%
\end{equation*}

\noindent\textbf{b. }%
\begin{equation*}
\begin{array}{c}
\frac{\partial}{\partial\theta}\hat{V}_{ff}(\theta)^{-\frac{1}{2}%
}f_{T}(\theta,X)=\hat{V}_{ff}(\theta)^{-\frac{1}{2}}\hat{D}(\theta).%
\end{array}%
\end{equation*}

\noindent\textbf{c.}%
\begin{equation*}
\begin{array}{rl}
\frac{\partial}{\partial\theta}\hat{V}_{ff}(\theta)^{-\frac{1}{2}%
}D_{T}(\theta,X)= & -2\hat{V}_{ff}(\theta)^{-\frac{1}{2}}\hat{V}_{\theta
f}(\theta)\hat{V}_{ff}(\theta)^{-1}D_{T}(\theta,X)-\hat{V}_{ff}(\theta )^{-%
\frac{1}{2}}\hat{V}_{\theta\theta.f}(\theta)\hat{V}_{ff}(\theta)^{-1}f_{T}(%
\theta,X).%
\end{array}%
\end{equation*}

\noindent\textbf{d.}%
\begin{equation*}
\begin{array}{l}
\begin{array}{rl}
\frac{\partial}{\partial\theta}f_{T}(\theta,X)^{\prime}\hat{V}%
_{ff}(\theta)^{-1}\hat{D}(\theta)= & \hat{D}(\theta)^{\prime}\hat{V}%
_{ff}(\theta)^{-1}\hat{D}(\theta)-2f_{T}(\theta,X)^{\prime}\hat{V}%
_{ff}(\theta)^{-1}\hat{V}_{\theta f}(\theta)\hat{V}_{ff}(\theta)^{-1}D_{T}(%
\theta,X)- \\ 
& f_{T}(\theta,X)^{\prime}\hat{V}_{ff}(\theta)^{-1}\hat{V}_{\theta\theta
.f}(\theta)\hat{V}_{ff}(\theta)^{-1}f_{T}(\theta,X).%
\end{array}%
\end{array}%
\end{equation*}

\noindent\textbf{e.}%
\begin{equation*}
\begin{array}{rl}
\frac{\partial}{\partial\theta}\left( D_{T}(\theta,X)^{\prime}\hat{V}%
_{ff}(\theta)^{-1}D_{T}(\theta,X)\right) = & -4D_{T}(\theta,X)^{\prime}\hat{V%
}_{ff}(\theta)^{-1}\hat{V}_{\theta f}(\theta)\hat{V}_{ff}(\theta
)^{-1}D_{T}(\theta,X)- \\ 
& 2D_{T}(\theta,X)^{\prime}\hat{V}_{ff}(\theta)^{-1}\hat{V}_{\theta\theta
.f}(\theta)\hat{V}_{ff}(\theta)^{-1}f_{T}(\theta,X).%
\end{array}%
\end{equation*}

\noindent\textbf{f.}%
\begin{equation*}
\begin{array}{rl}
\frac{\partial}{\partial\theta}V_{\theta\theta.f}(\theta)= & -\hat{V}%
_{\theta\theta.f}(\theta)\hat{V}_{ff}(\theta)^{-1}\hat{V}_{\theta
f}(\theta)^{\prime}-\hat{V}_{\theta f}(\theta)\hat{V}_{ff}(\theta)^{-1}\hat {%
V}_{\theta\theta.f}(\theta).%
\end{array}%
\end{equation*}

\noindent\textbf{g.}%
\begin{equation*}
\begin{array}{l}
\frac{\partial}{\partial\theta}\left( f_{T}(\theta,X)^{\prime}\hat{V}%
_{ff}(\theta)^{-1}\hat{V}_{\theta\theta.f}(\theta)\hat{V}_{ff}(\theta
)^{-1}f_{T}(\theta,X)\right) \\ 
\begin{array}{cl}
= & 2\hat{D}(\theta)^{\prime}\hat{V}_{ff}(\theta)^{-1}\hat{V}_{\theta\theta
.f}(\theta)\hat{V}_{ff}(\theta)^{-1}f_{T}(\theta,X)- \\ 
& 4f_{T}(\theta,X)^{\prime}\hat{V}_{ff}(\theta)^{-1}\hat{V}_{\theta
f}(\theta)\hat{V}_{ff}(\theta)^{-1}\hat{V}_{\theta\theta.f}(\theta)\hat{V}%
_{ff}(\theta)^{-1}f_{T}(\theta,X)%
\end{array}%
\end{array}%
\end{equation*}

\noindent\textbf{h.}%
\begin{equation*}
\begin{array}{l}
\frac{\partial}{\partial\theta}\left( D_{T}(\theta,X)^{\prime}\hat{V}%
_{ff}(\theta)^{-1}D_{T}(\theta,X)+f_{T}(\theta,X)^{\prime}\hat{V}%
_{ff}(\theta)^{-1}\hat{V}_{\theta\theta.f}(\theta)\hat{V}_{ff}(%
\theta)^{-1}f_{T}(\theta,X)\right) \\ 
\begin{array}{cl}
= & -4\left[ D_{T}(\theta,X)^{\prime}\hat{V}_{ff}(\theta)^{-1}\hat{V}%
_{\theta f}(\theta)\hat{V}_{ff}(\theta)^{-1}D_{T}(\theta,X)+\right. \\ 
& \left. f_{T}(\theta,X)^{\prime}\hat{V}_{ff}(\theta)^{-1}\hat{V}_{\theta
f}(\theta)\hat{V}_{ff}(\theta)^{-1}\hat{V}_{\theta\theta.f}(\theta)\hat {V}%
_{ff}(\theta)^{-1}f_{T}(\theta,X)\right] .%
\end{array}%
\end{array}%
\end{equation*}

\paragraph{Proof:}

\textbf{a. } Because $\hat{V}_{ff}(\theta)^{-1}=\hat{V}_{ff}(\theta)^{-%
\frac {1}{2}\prime}\hat{V}_{ff}(\theta)^{-\frac{1}{2}},$ $\hat{V}%
_{ff}(\theta)^{-\frac{1}{2}}\hat{V}_{ff}(\theta)\hat{V}_{ff}(\theta)^{-\frac{%
1}{2}\prime}=I_{k_{f}}$ and%
\begin{equation*}
\begin{array}{c}
\genfrac{(}{)}{}{}{\partial\hat{V}_{ff}(\theta)^{-\frac{1}{2}}}{\partial
\theta}%
\hat{V}_{ff}(\theta)\hat{V}_{ff}(\theta)^{-\frac{1}{2}\prime}+\hat{V}%
_{ff}(\theta)^{-\frac{1}{2}}%
\genfrac{(}{)}{}{}{\partial\hat{V}_{ff}(\theta)}{\partial\theta}%
\hat{V}_{ff}(\theta)^{-\frac{1}{2}\prime}+\hat{V}_{ff}(\theta)^{-\frac{1}{2}%
}\hat{V}_{ff}(\theta)%
\genfrac{(}{)}{}{}{\partial\hat{V}_{ff}(\theta)^{-\frac{1}{2}}}{\partial
\theta}%
^{\prime}=0,
\end{array}%
\end{equation*}
such that $\frac{\partial}{\partial\theta}\hat{V}_{ff}(\theta)^{-\frac{1}{2}%
}=-\hat{V}_{ff}(\theta)^{-\frac{1}{2}}\hat{V}_{\theta f}(\theta)\hat{V}%
_{ff}(\theta)^{-1}$ since $\frac{\partial\hat{V}_{ff}(\theta)}{%
\partial\theta }=\hat{V}_{\theta f}(\theta)+\hat{V}_{\theta
f}(\theta)^{\prime}$ which results from the definition of $q_{T}(\theta,X)=%
\frac{\partial}{\partial \theta}f_{T}(\theta,X).$

\noindent\textbf{b. }Using the product rule of differentation:%
\begin{equation*}
\begin{array}{rl}
\frac{\partial}{\partial\theta}\hat{V}_{ff}(\theta)^{-\frac{1}{2}%
}f_{T}(\theta,X)= & \left( \frac{\partial}{\partial\theta}\hat{V}%
_{ff}(\theta)^{-\frac{1}{2}}\right) f_{T}(\theta,X)+\hat{V}_{ff}(\theta )^{-%
\frac{1}{2}}\left( \frac{\partial}{\partial\theta}f_{T}(\theta,X)\right) \\ 
= & -\hat{V}_{ff}(\theta)^{-\frac{1}{2}}\hat{V}_{\theta f}(\theta)\hat{V}%
_{ff}(\theta)^{-1}f_{T}(\theta,X)+\hat{V}_{ff}(\theta)^{-\frac{1}{2}%
}q_{T}(\theta,X) \\ 
= & \hat{V}_{ff}(\theta)^{-\frac{1}{2}}\hat{D}(\theta).%
\end{array}%
\end{equation*}

\noindent\textbf{c. }The specification of $\hat{V}_{ff}(\theta)^{-\frac{1}{2}%
}\hat{D}(\theta)$ is $\hat{V}_{ff}(\theta)^{-\frac{1}{2}}\hat{D}(\theta)= 
\hat{V}_{ff}(\theta)^{-\frac{1}{2}}\left[ q_{T}(\theta,X)-\hat{V}_{\theta
f}(\theta )\hat{V}_{ff}(\theta)^{-1}f_{T}(\theta,X)\right] $, so:%
\begin{equation*}
\begin{array}{l}
\frac{\partial}{\partial\theta}\left( \hat{V}_{ff}(\theta)^{-\frac{1}{2}%
}D_{T}(\theta,X)\right) \\ 
\begin{array}{rl}
= & \left( \frac{\partial}{\partial\theta}\hat{V}_{ff}(\theta)^{-\frac{1}{2}%
}\right) D_{T}(\theta,X)+\hat{V}_{ff}(\theta)^{-\frac{1}{2}}\left( \frac{%
\partial}{\partial\theta}\left[ q_{T}(\theta,X)-\hat{V}_{\theta f}(\theta)%
\hat{V}_{ff}(\theta)^{-1}f_{T}(\theta,X)\right] \right) \\ 
= & -\hat{V}_{ff}(\theta)^{-\frac{1}{2}}\hat{V}_{\theta f}(\theta)\hat{V}%
_{ff}(\theta)^{-1}D_{T}(\theta,X)+\hat{V}_{ff}(\theta)^{-\frac{1}{2}}\left[ 
\frac{\partial}{\partial\theta}q_{T}(\theta,X)-\left( \frac{\partial }{%
\partial\theta}\hat{V}_{\theta f}(\theta)\right) \hat{V}_{ff}(\theta
)^{-1}f_{T}(\theta,X)-\right. \\ 
& \hat{V}_{\theta f}(\theta)\left( \frac{\partial}{\partial\theta}\hat {V}%
_{ff}(\theta)^{-1}\right) f_{T}(\theta,X)-\hat{V}_{\theta f}(\theta )\hat{V}%
_{ff}(\theta)^{-1}\left( \frac{\partial}{\partial\theta}f_{T}(\theta,X)) %
\right] \\ 
= & -\hat{V}_{ff}(\theta)^{-\frac{1}{2}}\hat{V}_{\theta f}(\theta)\hat{V}%
_{ff}(\theta)^{-1}D_{T}(\theta,X)-\hat{V}_{ff}(\theta)^{-\frac{1}{2}}\hat {V}%
_{\theta\theta}(\theta)\hat{V}_{ff}(\theta)^{-1}f_{T}(\theta,X)+ \\ 
& \hat{V}_{ff}(\theta)^{-\frac{1}{2}}\hat{V}_{\theta f}(\theta)\hat{V}%
_{ff}(\theta)^{-1}\hat{V}_{\theta f}(\theta)\hat{V}_{ff}(\theta)^{-1}f_{T}(%
\theta,X)+ \\ 
& \hat{V}_{ff}(\theta)^{-\frac{1}{2}}\hat{V}_{\theta f}(\theta)\hat{V}%
_{ff}(\theta)^{-1}\hat{V}_{\theta f}(\theta)^{\prime}\hat{V}_{ff}(\theta
)^{-1}f_{T}(\theta,X)-\hat{V}_{ff}(\theta)^{-\frac{1}{2}}\hat{V}_{\theta
f}(\theta)\hat{V}_{ff}(\theta)^{-1}q_{T}(\theta,X) \\ 
= & -2\hat{V}_{ff}(\theta)^{-\frac{1}{2}}\hat{V}_{\theta f}(\theta)\hat {V}%
_{ff}(\theta)^{-1}D_{T}(\theta,X)-\hat{V}_{ff}(\theta)^{-\frac{1}{2}}\hat{V}%
_{\theta\theta.f}(\theta)\hat{V}_{ff}(\theta)^{-1}f_{T}(\theta,X)%
\end{array}%
\end{array}%
\end{equation*}

\noindent\textbf{d. }%
\begin{equation*}
\begin{array}{l}
\frac{\partial}{\partial\theta}f_{T}(\theta,X)^{\prime}\hat{V}%
_{ff}(\theta)^{-1}\hat{D}(\theta) \\ 
\begin{array}{rl}
= & \left( \frac{\partial}{\partial\theta}\hat{V}_{ff}(\theta)^{-\frac{1}{2}%
}f_{T}(\theta,X)\right) ^{\prime}\hat{V}_{ff}(\theta)^{-\frac{1}{2}}\hat {D}%
(\theta)+f_{T}(\theta,X)^{\prime}\hat{V}_{ff}(\theta)^{-\frac{1}{2}}\left( 
\frac{\partial}{\partial\theta}\hat{V}_{ff}(\theta)^{-\frac{1}{2}}\hat {D}%
(\theta,X)\right) \\ 
= & \hat{D}(\theta)^{\prime}\hat{V}_{ff}(\theta)^{-1}\hat{D}(\theta
)-2f_{T}(\theta,X)^{\prime}\hat{V}_{ff}(\theta)^{-1}\hat{V}_{\theta
f}(\theta)\hat{V}_{ff}(\theta)^{-1}D_{T}(\theta,X)- \\ 
& f_{T}(\theta,X)^{\prime}\hat{V}_{ff}(\theta)^{-1}\hat{V}_{\theta\theta
.f}(\theta)\hat{V}_{ff}(\theta)^{-1}f_{T}(\theta,X).%
\end{array}%
\end{array}%
\end{equation*}

\noindent\textbf{e. }%
\begin{equation*}
\begin{array}{rl}
\frac{\partial}{\partial\theta}\left( \hat{D}(\theta)^{\prime}\hat{V}%
_{ff}(\theta)^{-1}\hat{D}(\theta)\right) = & 2\left( \hat{D}(\theta
)^{\prime}\hat{V}_{ff}(\theta)^{-\frac{1}{2}}\right) \left( \frac{\partial }{%
\partial\theta}\hat{V}_{ff}(\theta)^{-\frac{1}{2}}\hat{D}(\theta)\right) \\ 
= & -4\hat{D}(\theta)^{\prime}\hat{V}_{ff}(\theta)^{-1}\hat{V}_{\theta
f}(\theta)\hat{V}_{ff}(\theta)^{-1}\hat{D}(\theta)- \\ 
& 2\hat{D}(\theta)^{\prime}\hat{V}_{ff}(\theta)^{-1}\hat{V}_{\theta\theta
.f}(\theta)\hat{V}_{ff}(\theta)^{-1}f_{T}(\theta,X).%
\end{array}%
\end{equation*}

\noindent\textbf{f. }The specification of $V_{\theta\theta.f}(\theta
)=V_{\theta\theta}(\theta)-V_{\theta f}(\theta)V_{ff}(\theta)^{-1}V_{\theta
f}(\theta)^{\prime}$ is such that: 
\begin{equation*}
\begin{array}{l}
\frac{\partial}{\partial\theta}V_{\theta\theta.f}(\theta) \\ 
\begin{array}{rl}
= & \left( \frac{\partial}{\partial\theta}V_{\theta\theta}(\theta)\right)
-\left( \frac{\partial}{\partial\theta}\hat{V}_{\theta f}(\theta)\right) 
\hat{V}_{ff}(\theta)^{-1}\hat{V}_{\theta f}(\theta)^{\prime}-\hat{V}_{\theta
f}(\theta)\left( \frac{\partial}{\partial\theta}\hat{V}_{ff}(\theta
)^{-1}\right) \hat{V}_{\theta f}(\theta)^{\prime}- \\ 
& \hat{V}_{\theta f}(\theta)\hat{V}_{ff}(\theta)^{-1}\left( \frac{\partial }{%
\partial\theta}\hat{V}_{\theta f}(\theta)\right) ^{\prime} \\ 
= & -\hat{V}_{\theta\theta}(\theta)\hat{V}_{ff}(\theta)^{-1}\hat{V}_{\theta
f}(\theta)^{\prime}+\hat{V}_{\theta f}(\theta)\hat{V}_{ff}(\theta)^{-1}\hat {%
V}_{\theta f}(\theta)\hat{V}_{ff}(\theta)^{-1}\hat{V}_{\theta
f}(\theta)^{\prime}+ \\ 
& \hat{V}_{\theta f}(\theta)\hat{V}_{ff}(\theta)^{-1}\hat{V}_{\theta
f}(\theta)^{\prime}\hat{V}_{ff}(\theta)^{-1}\hat{V}_{\theta
f}(\theta)^{\prime }-\hat{V}_{\theta f}(\theta)\hat{V}_{ff}(\theta)^{-1}\hat{%
V}_{\theta\theta }(\theta) \\ 
= & -\hat{V}_{\theta\theta.f}(\theta)\hat{V}_{ff}(\theta)^{-1}\hat{V}%
_{\theta f}(\theta)^{\prime}-\hat{V}_{\theta f}(\theta)\hat{V}%
_{ff}(\theta)^{-1}\hat {V}_{\theta\theta.f}(\theta)%
\end{array}%
\end{array}%
\end{equation*}

\noindent\textbf{g. }The specification of $f_{T}(\theta,X)^{\prime}\hat {V}%
_{ff}(\theta)^{-1}\hat{V}_{\theta\theta.f}(\theta)\hat{V}_{ff}(\theta
)^{-1}f_{T}(\theta,X)$ is such that:%

\begin{equation*}
\begin{array}{l}
\frac{\partial}{\partial\theta}\left( f_{T}(\theta,X)^{\prime}\hat{V}%
_{ff}(\theta)^{-1}\hat{V}_{\theta\theta.f}(\theta)\hat{V}_{ff}(\theta
)^{-1}f_{T}(\theta,X)\right) \\ 
\begin{array}{rl}
= & 2\left( \frac{\partial}{\partial\theta}\hat{V}_{ff}(\theta)^{-\frac{1}{2}%
}f_{T}(\theta,X)\right) ^{\prime}\hat{V}_{ff}(\theta)^{-\frac{1}{2}}\hat{V}%
_{\theta\theta.f}(\theta)\hat{V}_{ff}(\theta)^{-1}f_{T}(\theta,X)+ \\ 
& 2f_{T}(\theta,X)^{\prime}\hat{V}_{ff}(\theta)^{-\frac{1}{2}}\left( \frac{%
\partial}{\partial\theta}\hat{V}_{ff}(\theta)^{-\frac{1}{2}}\right) \hat{V}%
_{\theta\theta.f}(\theta)\hat{V}_{ff}(\theta)^{-1}f_{T}(\theta,X)+ \\ 
& f_{T}(\theta,X)^{\prime}\hat{V}_{ff}(\theta)^{-1}\left( \frac{\partial }{%
\partial\theta}\hat{V}_{\theta\theta.f}(\theta)\right) \hat{V}%
_{ff}(\theta)^{-1}f_{T}(\theta,X) \\ 
= & 2\hat{D}(\theta)^{\prime}\hat{V}_{ff}(\theta)^{-1}\hat{V}_{\theta\theta
.f}(\theta)\hat{V}_{ff}(\theta)^{-1}f_{T}(\theta,X)- \\ 
& 2f_{T}(\theta,X)^{\prime}\hat{V}_{ff}(\theta)^{-1}\hat{V}_{\theta
f}(\theta)\hat{V}_{ff}(\theta)^{-1}\hat{V}_{\theta\theta.f}(\theta)\hat{V}%
_{ff}(\theta)^{-1}f_{T}(\theta,X)- \\ 
& f_{T}(\theta,X)^{\prime}\hat{V}_{ff}(\theta)^{-1}\hat{V}_{\theta\theta
.f}(\theta)\hat{V}_{ff}(\theta)^{-1}\hat{V}_{\theta f}(\theta)^{\prime}\hat {%
V}_{ff}(\theta)^{-1}f_{T}(\theta,X)- \\ 
& f_{T}(\theta,X)^{\prime}\hat{V}_{ff}(\theta)^{-1}\hat{V}_{\theta f}(\theta)%
\hat{V}_{ff}(\theta)^{-1}\hat{V}_{\theta\theta.f}(\theta)\hat{V}%
_{ff}(\theta)^{-1}f_{T}(\theta,X) \\ 
= & 2\hat{D}(\theta)^{\prime}\hat{V}_{ff}(\theta)^{-1}\hat{V}_{\theta\theta
.f}(\theta)\hat{V}_{ff}(\theta)^{-1}f_{T}(\theta,X)- \\ 
& 4f_{T}(\theta,X)^{\prime}\hat{V}_{ff}(\theta)^{-1}\hat{V}_{\theta
f}(\theta)\hat{V}_{ff}(\theta)^{-1}\hat{V}_{\theta\theta.f}(\theta)\hat{V}%
_{ff}(\theta)^{-1}f_{T}(\theta,X)%
\end{array}%
\end{array}%
\end{equation*}

\noindent\textbf{h. } It follows from \textbf{e} and \textbf{g} above.

\subsection{Proof of Theorem 1}

The derivative of $Q_{p}(\theta )$ with respect to $\theta $ consists of two
parts. The derivative of $\mu _{f}(\theta )$ with respect to $\theta :$ $%
J(\theta )=\frac{\partial }{\partial \theta ^{\prime }}\mu _{f}(\theta )$,
and the derivative of $V_{ff}(\theta )^{-1}$ with respect to $\theta .$ To
obtain the derivative of $V_{ff}(\theta )^{-1}$ with respect to $\theta ,$
we start out with the derivative of $V_{ff}(\theta )$ with respect to $%
\theta :$ 
\begin{align*}
& 
\begin{array}{c}
\text{vec(}V_{ff}(\theta ))=\lim_{T\rightarrow \infty }\text{vec(var}\left( 
\sqrt{T}f_{T}(\theta ,X)\right) )%
\end{array}
\\
& 
\begin{array}{rl}
= & \lim_{T\rightarrow \infty }\text{vec}\left( E\left( \frac{1}{T}%
\sum_{t=1}^{T}\sum_{j=1}^{T}\left( f_{t}(\theta )-\mu _{f}(\theta )\right)
\left( f_{j}(\theta )-\mu _{f}(\theta )\right) ^{\prime }\right) \right)  \\ 
= & \lim_{T\rightarrow \infty }E\left( \frac{1}{T}\sum_{t=1}^{T}%
\sum_{j=1}^{T}\left[ \left( f_{j}(\theta )-\mu _{f}(\theta )\right) \otimes
\left( f_{t}(\theta )-\mu _{f}(\theta )\right) \right] \right) 
\end{array}%
\end{align*}%
\begin{align*}
& 
\begin{array}{c}
\frac{\partial }{\partial \theta ^{\prime }}\text{vec(}V_{ff}(\theta
))=\lim_{T\rightarrow \infty }\frac{\partial }{\partial \theta ^{\prime }}%
E\left( \frac{1}{T}\sum_{t=1}^{T}\sum_{j=1}^{T}\left[ \left( f_{j}(\theta
)-\mu _{f}(\theta )\right) \otimes \left( f_{t}(\theta )-\mu _{f}(\theta
)\right) \right] \right) 
\end{array}
\\
& 
\begin{array}{rl}
= & \lim_{T\rightarrow \infty }E\left( \frac{1}{T}\sum_{t=1}^{T}%
\sum_{j=1}^{T}\left[ \left( \frac{\partial }{\partial \theta ^{\prime }}%
f_{j}(\theta )-\frac{\partial }{\partial \theta ^{\prime }}\mu _{f}(\theta
)\right) \otimes \left( f_{t}(\theta )-\mu _{f}(\theta )\right) \right]
\right) + \\ 
& \lim_{T\rightarrow \infty }E\left( \frac{1}{T}\sum_{t=1}^{T}\sum_{j=1}^{T}%
\left[ \left( f_{j}(\theta )-\mu _{f}(\theta )\right) \otimes \left( \frac{%
\partial }{\partial \theta ^{\prime }}f_{t}(\theta )-\frac{\partial }{%
\partial \theta ^{\prime }}\mu _{f}(\theta )\right) \right] \right)  \\ 
= & \lim_{T\rightarrow \infty }E\left( \frac{1}{T}\sum_{t=1}^{T}%
\sum_{j=1}^{T}\left[ \left( q_{j}(\theta )-J(\theta )\right) \otimes \left(
f_{t}(\theta )-\mu _{f}(\theta )\right) \right] \right) + \\ 
& \lim_{T\rightarrow \infty }E\left( \frac{1}{T}\sum_{t=1}^{T}\sum_{j=1}^{T}%
\left[ \left( f_{j}(\theta )-\mu _{f}(\theta )\right) \otimes \left(
q_{t}(\theta )-J(\theta )\right) \right] \right)  \\ 
= & \left( \text{vec}\left( V_{\theta _{1}f}(\theta )\right) \ldots \text{vec%
}\left( V_{\theta _{m}f}(\theta )\right) \right) +\left( \text{vec}\left(
V_{\theta _{1}f}(\theta )^{\prime }\right) \ldots \text{vec}\left( V_{\theta
_{m}f}(\theta )^{\prime }\right) \right) 
\end{array}%
\end{align*}%
with $q_{j}(\theta )=\frac{\partial }{\partial \theta ^{\prime }}%
f_{j}(\theta )=(q_{1,j}(\theta )\ldots q_{m,j}(\theta ))$ and%
\begin{equation*}
\begin{array}{rlll}
V_{\theta _{i}f}(\theta )= & \lim_{T\rightarrow \infty }E\left( T(\frac{%
\partial }{\partial \theta _{i}}(f_{T}(\theta ,X)-\mu _{f}(\theta )))\left(
f_{T}(\theta ,X)-\mu _{f}(\theta )\right) ^{\prime }\right) , &  & 
i=1,\ldots ,m.%
\end{array}%
\end{equation*}%
We can now specify the derivative of the objective function with respect to $%
\theta $:%
\begin{equation*}
\begin{array}{rl}
\frac{1}{2}\frac{\partial }{\partial \theta ^{\prime }}Q_{p}(\theta )= & \mu
_{f}(\theta )^{\prime }V_{ff}(\theta )^{-1}\frac{\partial \mu _{f}(\theta )}{%
\partial \theta ^{\prime }}-\frac{1}{2}((\mu _{f}(\theta )\otimes \mu
_{f}(\theta ))^{\prime }(V_{ff}(\theta )^{-1}\otimes V_{ff}(\theta )^{-1})%
\frac{\partial }{\partial \theta ^{\prime }}\text{vec(}V_{ff}(\theta )) \\ 
= & \mu _{f}(\theta )^{\prime }V_{ff}(\theta )^{-1}J(\theta )-\frac{1}{2}%
((\mu _{f}(\theta )\otimes \mu _{f}(\theta ))^{\prime }(V_{ff}(\theta
)^{-1}\otimes V_{ff}(\theta )^{-1}) \\ 
& \left( \text{vec}\left( V_{\theta _{1}f}(\theta )\right) \ldots \text{vec}%
\left( V_{\theta _{m}f}(\theta )\right) \right) +\left( \text{vec}\left(
V_{\theta _{1}f}(\theta )^{\prime }\right) \ldots \text{vec}\left( V_{\theta
_{m}f}(\theta )^{\prime }\right) \right)  \\ 
= & \mu _{f}(\theta )^{\prime }V_{ff}(\theta )^{-1}J(\theta )- \\ 
& \frac{1}{2}\left[ \left( \mu _{f}(\theta )^{\prime }V_{ff}(\theta
)^{-1}V_{\theta _{1}f}(\theta )V_{ff}(\theta )^{-1}\mu _{f}(\theta )\ldots
\mu _{f}(\theta )^{\prime }V_{ff}(\theta )^{-1}V_{\theta _{m}f}(\theta
)V_{ff}(\theta )^{-1}\mu _{f}(\theta )\right) +\right.  \\ 
& \left. \left( \mu _{f}(\theta )^{\prime }V_{ff}(\theta )^{-1}V_{\theta
_{1}f}(\theta )^{\prime }V_{ff}(\theta )^{-1}\mu _{f}(\theta )\ldots \mu
_{f}(\theta )^{\prime }V_{ff}(\theta )^{-1}V_{\theta _{m}f}(\theta )^{\prime
}V_{ff}(\theta )^{-1}\mu _{f}(\theta )\right) \right]  \\ 
= & \mu _{f}(\theta )^{\prime }V_{ff}(\theta )^{-1}\left[ J(\theta )-\left(
V_{\theta _{1}f}(\theta )V_{ff}(\theta )^{-1}\mu _{f}(\theta )\ldots
V_{\theta _{m}f}(\theta )V_{ff}(\theta )^{-1}\mu _{f}(\theta )\right) \right]
\\ 
= & \mu _{f}(\theta )^{\prime }V_{ff}(\theta )^{-1}D(\theta ),%
\end{array}%
\end{equation*}%
with $D(\theta )=J(\theta )-\left[ V_{\theta _{1}f}(\theta )V_{ff}(\theta
)^{-1}\mu _{f}(\theta )\ldots V_{\theta _{m}f}(\theta )V_{ff}(\theta
)^{-1}\mu _{f}(\theta )\right] $.

\subsection{Proof that the $IS$ identification measure is larger than or equal to the minimal value of the population continuous updating objective function}

The minimal value over $(\lambda_{F},$ $D)$ of%
\[%
\begin{array}
[c]{cl}%
Q_{p}(\lambda_{F},D)= & \left[  \text{vec}\left(  \left(  \mu_{R}\text{
}\vdots\text{ }\beta\right)  +D\left(  \lambda_{F}\text{ }\vdots\text{ }%
I_{m}\right)  \right)  \right]  ^{\prime}\left[  \text{Var}\left(  \sqrt
{T}\left(  \bar{R}^{\prime}\text{ }\vdots\text{ vec(}\hat{\beta})^{\prime
}\right)  ^{\prime}\right)  \right]  ^{-1}\\
& \left[  \text{vec}\left(  \left(  \mu_{R}\text{ }\vdots\text{ }\beta\right)
+D\left(  \lambda_{F}\text{ }\vdots\text{ }I_{m}\right)  \right)  \right]  
\end{array}
\]
equals the minimal value of
\[%
\begin{array}
[c]{cl}%
Q_{p}(\phi,A)= & \left[  \text{vec}\left(  \left(  \mu_{R}\text{ }\vdots\text{
}\beta\right)  +A(I_{m}\text{ }\vdots\text{ }\phi)\right)  \right]  ^{\prime
}\left[  \text{Var}\left(  \sqrt{T}\left(  \bar{R}^{\prime}\text{ }%
\vdots\text{ vec(}\hat{\beta})^{\prime}\right)  ^{\prime}\right)  \right]
^{-1}\\
& \left[  \text{vec}\left(  \left(  \mu_{R}\text{ }\vdots\text{ }\beta\right)
+A(I_{m}\text{ }\vdots\text{ }\phi)\right)  \right]  
\end{array}
\]
over $(A, \phi)$, with $A$ an $N\times m$ matrix and $\phi$ an $m$-dimensional vector. This
results since $D(  \lambda_{F}\text{ }\vdots\text{ }I_{m})  $ and
$A(I_{m}$ $\vdots$ $\phi)$ are equivalent representations of an $N\times(m+1)$
dimensional matrix of rank $m$ (except for a measure zero space). Restricting
the top element of $\phi=(\phi_{1}$ $\vdots$ $\phi_{2}^{\prime})^{\prime},$
with $\phi_{1}$ a scalar and $\phi_{2}$ an $(m-1)$-dimensional vector, to
zero, so $\phi_{1}=0,$ does not decrease the minimal value of the above
function. The resulting restricted specification reads%
\[%
\begin{array}
[c]{rl}%
Q_{p}(\phi_{2},A)= & \left[  \text{vec}\left(  \left(  \mu_{R}\text{ }%
\vdots\text{ }\beta\right)  +A\left(  I_{m}\text{ }\vdots\text{ }\binom
{0}{\phi_{2}}\right)  \right)  \right]  ^{\prime}\\
& \left(
\begin{array}
[c]{cc}%
\Sigma_{\bar{R}\bar{R}.\hat{\beta}}^{-1} & -\Sigma_{\bar{R}\bar{R}.\hat{\beta
}}^{-1}\Sigma_{\bar{R}\hat{\beta}}\Sigma_{\hat{\beta}\hat{\beta}}^{-1}\\
-\Sigma_{\hat{\beta}\hat{\beta}}^{-1}\Sigma_{\hat{\beta}\bar{R}}\Sigma
_{\bar{R}\bar{R}.\hat{\beta}}^{-1} & \Sigma_{\hat{\beta}\hat{\beta}}%
^{-1}+\Sigma_{\hat{\beta}\hat{\beta}}^{-1}\Sigma_{\hat{\beta}\bar{R}}%
\Sigma_{\bar{R}\bar{R}.\hat{\beta}}^{-1}\Sigma_{\bar{R}\hat{\beta}}%
\Sigma_{\hat{\beta}\hat{\beta}}^{-1}%
\end{array}
\right)  \\
& \left[  \text{vec}\left(  \left(  \mu_{R}\text{ }\vdots\text{ }\beta\right)
+A\left(  I_{m}\text{ }\vdots\text{ }\binom{0}{\phi_{2}}\right)  \right)
\right]  \\
= & \left[  \mu_{R}+a_{1}-\Sigma_{\bar{R}\hat{\beta}}\Sigma_{\hat{\beta}%
\hat{\beta}}^{-1}\text{vec}\left(  \beta+A_{2}\left(  I_{m-1}\text{  }%
\vdots\text{ }\phi_{2}\right)  \right)  \right]  ^{\prime}\Sigma_{\bar{R}%
\bar{R}.\hat{\beta}}^{-1}\\
& \left[  \mu_{R}+a_{1}-\Sigma_{\bar{R}\hat{\beta}}\Sigma_{\hat{\beta}%
\hat{\beta}}^{-1}\text{vec}\left(  \beta+A_{2}\left(  I_{m-1}\text{ }%
\vdots\text{ }\phi_{2}\right)  \right)  \right]  +\\
& \left[  \text{vec}\left(  \beta+A_{2}\left(  I_{m-1}\text{ }\vdots\text{
}\phi_{2}\right)  \right)  \right]  ^{\prime}\Sigma_{\hat{\beta}\hat{\beta}%
}^{-1}\left[  \text{vec}\left(  \beta+A_{2}\left(  I_{m-1}\text{ }\vdots\text{
}\phi_{2}\right)  \right)  \right]  ,
\end{array}
\]
where we used that $A=(a_{1}$ $\vdots$ $A_{2}),$ $a_{1}$ an $N$-dimensional
vector, $A_{2}$ an $N\times(m-1)$ dimensional matrix and the partitioned
inverse of Var$\left(  \sqrt{T}\left(  \bar{R}^{\prime}\text{ }\vdots\text{
vec(}\hat{\beta})^{\prime}\right)  ^{\prime}\right)  :$%
\[%
\begin{array}
[c]{rl}%
\text{Var}\left(  \sqrt{T}\left(  \bar{R}^{\prime}\text{ }\vdots\text{
vec(}\hat{\beta})^{\prime}\right)  ^{\prime}\right)  = & \left(
\begin{array}
[c]{cc}%
\Sigma_{\bar{R}\bar{R}} & \Sigma_{\bar{R}\hat{\beta}}\\
\Sigma_{\hat{\beta}\bar{R}} & \Sigma_{\hat{\beta}\hat{\beta}}%
\end{array}
\right)  ,\\
\left[  \text{Var}\left(  \sqrt{T}\left(  \bar{R}^{\prime}\text{ }\vdots\text{
vec(}\hat{\beta})^{\prime}\right)  ^{\prime}\right)  \right]  ^{-1}= & \left(
\begin{array}
[c]{cc}%
\Sigma_{\bar{R}\bar{R}.\hat{\beta}}^{-1} & -\Sigma_{\bar{R}\bar{R}.\hat{\beta
}}^{-1}\Sigma_{\bar{R}\hat{\beta}}\Sigma_{\hat{\beta}\hat{\beta}}^{-1}\\
-\Sigma_{\hat{\beta}\hat{\beta}}^{-1}\Sigma_{\hat{\beta}\bar{R}}\Sigma
_{\bar{R}\bar{R}.\hat{\beta}}^{-1} & \Sigma_{\hat{\beta}\hat{\beta}}%
^{-1}+\Sigma_{\hat{\beta}\hat{\beta}}^{-1}\Sigma_{\hat{\beta}\bar{R}}%
\Sigma_{\bar{R}\bar{R}.\hat{\beta}}^{-1}\Sigma_{\bar{R}\hat{\beta}}%
\Sigma_{\hat{\beta}\hat{\beta}}^{-1}%
\end{array}
\right)
\end{array}
\]
with $\Sigma_{\bar{R}\bar{R}},$ $\Sigma_{\bar{R}\hat{\beta}}=\Sigma
_{\hat{\beta}\bar{R}}^{\prime}$ and $\Sigma_{\hat{\beta}\hat{\beta}}$ $N\times
N,$ $N\times Nm$ and $Nm\times Nm$ dimensional matrices respectively, and
$\Sigma_{\bar{R}\bar{R}.\hat{\beta}}=\Sigma_{\bar{R}\bar{R}}^{-1}-\Sigma
_{\bar{R}\hat{\beta}}\Sigma_{\hat{\beta}\hat{\beta}}^{-1}\Sigma_{\hat{\beta
}\bar{R}}.$

Stepwise minimization of $Q_{p}(\phi_{2},A)=Q_{p}(\phi_{2},a_{1},A_{2})$ now
results in%

\[%
\begin{array}
[c]{rl}%
\hat{a}_{1}(\phi_{2},A_{2})= & \arg\min_{a_{1}\in\mathbb{R}^{N}}Q_{p}(\phi
_{2},a_{1},A_{2})\\
= & -\left(  \mu_{R}-\Sigma_{\bar{R}\hat{\beta}}\Sigma_{\hat{\beta}\hat{\beta
}}^{-1}\text{vec}\left(  \beta+A_{2}\left(  I_{m-1}\text{ }\vdots\text{ }%
\phi_{2}\right)  \right)  \right)
\end{array}
\]
so
\[%
\begin{array}
[c]{rl}%
Q_{p}(\phi_{2},A_{2})= & \min_{g_{1}\in\mathbb{R}^{N}}Q_{p}(\phi_{2}%
,a_{1},A_{2})\\
= & \left[  \text{vec}\left(  \beta+A_{2}\left(  I_{m-1}\text{ }\vdots\text{
}\phi_{2}\right)  \right)  \right]  ^{\prime}\Sigma_{\hat{\beta}\hat{\beta}%
}^{-1}\left[  \text{vec}\left(  \beta+A_{2}\left(  I_{m-1}\text{ }\vdots\text{
}\phi_{2}\right)  \right)  \right]  ,
\end{array}
\]
whose minimal value over $(\phi_{2},A_{2})$ corresponds with the $IS$
identification measure so it is always larger than or equal to the minimal
value of the population continuous updating objective function.

\subsection{Proof of Proposition 1}

We pre and post-multiply the matrices in the characteristic polynomial: 
\begin{equation*}
\begin{array}{cc}
\left\vert \tau\left( 
\begin{array}{cc}
1 & 0 \\ 
0 & Q_{\bar{F}\bar{F}}^{-1}%
\end{array}
\right) -\left( \mu_{R}\text{ }\vdots\text{ }\beta\right)
^{\prime}\Omega^{-1}\left( \mu_{R}\text{ }\vdots\text{ }\beta\right)
\right\vert & =0,%
\end{array}%
\end{equation*}
by%
\begin{equation*}
\left( 
\begin{array}{cc}
1 & 0 \\ 
-\lambda_{F} & I_{k}%
\end{array}
\right) ,
\end{equation*}
which since the determinant of this matrix equals one does not alter the
roots:%
\begin{equation*}
\begin{array}{cc}
\left\vert \tau\left( 
\begin{array}{cc}
1+\lambda_{F}^{\prime}Q_{\bar{F}\bar{F}}^{-1}\lambda_{F} & -\lambda
_{F}^{\prime}Q_{\bar{F}\bar{F}}^{-1} \\ 
-Q_{\bar{F}\bar{F}}^{-1}\lambda_{F} & Q_{\bar{F}\bar{F}}^{-1}%
\end{array}
\right) -\left( \mu_{R}-\beta\lambda_{F}\text{ }\vdots\text{ }\beta\right)
^{\prime}\Omega^{-1}\left( \mu_{R}-\beta\lambda_{F}\text{ }\vdots\text{ }%
\beta\right) \right\vert & =0.%
\end{array}%
\end{equation*}
We next do so again using:%
\begin{equation*}
\left( 
\begin{array}{cc}
1 & \lambda_{F}^{\prime}Q_{\bar{F}\bar{F}}^{-1}(1+\lambda_{F}^{\prime}Q_{%
\bar{F}\bar{F}}^{-1}\lambda_{F})^{-1} \\ 
0 & I_{k}%
\end{array}
\right) ,
\end{equation*}
to obtain:%
\begin{equation*}
\begin{array}{rc}
\left\vert \tau\left( 
\begin{array}{cc}
1+\lambda_{F}^{\prime}Q_{\bar{F}\bar{F}}^{-1}\lambda_{F} & 0 \\ 
0 & Q_{\bar{F}\bar{F}}^{-1}-Q_{\bar{F}\bar{F}}^{-1}\lambda_{F}(1+\lambda
_{F}^{\prime}Q_{\bar{F}\bar{F}}^{-1}\lambda_{F})^{-1}\lambda_{F}^{\prime }Q_{%
\bar{F}\bar{F}}^{-1}%
\end{array}
\right) -\right. &  \\ 
\left. \left( \mu_{R}-\beta\lambda_{F}\text{ }\vdots\text{ }-D(\lambda
_{F})\right) ^{\prime}\Omega^{-1}\left( \mu_{R}-\beta\lambda_{F}\text{ }%
\vdots\text{ }-D(\lambda_{F})\right) \right\vert & =0.%
\end{array}%
\end{equation*}
with $D(\lambda_{F})=-\beta-\left( \mu_{R}-\beta\lambda_{F}\right)
\lambda_{F}^{\prime}Q_{\bar{F}\bar{F}}^{-1}(1+\lambda_{F}^{\prime}Q_{\bar {F}%
\bar{F}}^{-1}\lambda_{F})^{-1}.$ For a value of $\lambda_{F},$ $\lambda
_{F}^{s},$ which satisfies the FOC, so $\left(
\mu_{R}-\beta\lambda_{F}^{s}\right)
^{\prime}\Omega^{-1}D(\lambda_{F}^{s})=0, $ the characteristic polynomial
then becomes:

\begin{equation*}
\begin{array}{lc}
\left\vert \left( 
\begin{array}{c}
\tau(1+\lambda_{F}^{s\prime}Q_{\bar{F}\bar{F}}^{-1}\lambda_{F}^{s})-\left(
\mu_{R}-\beta\lambda_{F}^{s}\right) ^{\prime}\Omega^{-1}\left(
\mu_{R}-\beta\lambda_{F}^{s}\right) \\ 
0%
\end{array}
\right. \right. &  \\ 
\left. \left. 
\begin{array}{c}
0 \\ 
\tau(Q_{\bar{F}\bar{F}}^{-1}-Q_{\bar{F}\bar{F}}^{-1}\lambda_{F}^{s}(1+%
\lambda_{F}^{s\prime}Q_{\bar{F}\bar{F}}^{-1}\lambda_{F}^{s})^{-1}%
\lambda_{F}^{s\prime}Q_{\bar{F}\bar{F}}^{-1})-D(\lambda_{F}^{s})^{\prime
}\Omega^{-1}D\left( \lambda_{F}^{s}\right)%
\end{array}
\right) \right\vert & =0.%
\end{array}%
\end{equation*}
We can further use that $Q_{\bar{F}\bar{F}}^{-1}-Q_{\bar{F}\bar{F}%
}^{-1}\lambda_{F}^{s}(1+\lambda_{F}^{s\prime}Q_{\bar{F}\bar{F}}^{-1}\lambda
_{F}^{s})^{-1}\lambda_{F}^{s\prime}Q_{\bar{F}\bar{F}}^{-1}=(Q_{\bar{F}\bar{F}%
}+\lambda_{F}^{s}\lambda_{F}^{s\prime})^{-1}.$

\subsection{Proof of Theorem 3}

The joint limit behavior of $f_{T}(\theta,X)$ and $q_{T}(\theta,X)$ at the
pseudo-true value $\theta^{\ast}$ reads:%
\begin{equation*}
\begin{array}{cc}
\sqrt{T}\left( 
\begin{array}{c}
f_{T}(\theta^{\ast},X)-\mu_{f}(\theta^{\ast}) \\ 
vec(q_{T}(\theta^{\ast},X)-J(\theta^{\ast}))%
\end{array}
\right) \underset{d}{\rightarrow} & \left( 
\begin{array}{c}
\psi_{f}(\theta) \\ 
\psi_{\theta}(\theta)%
\end{array}
\right) .%
\end{array}%
\end{equation*}
We pre-multiply it by%
\begin{equation*}
\begin{array}{c}
\hat{R}(\theta^{\ast})=\left( 
\begin{array}{cc}
I_{k_{f}} & 0 \\ 
-\hat{V}_{\theta f}(\theta^{\ast}) \hat{V}_{ff}(\theta^{\ast})^{-1} & 
I_{k_{f}m}%
\end{array}
\right) \underset{p}{\rightarrow}\left( 
\begin{array}{cc}
I_{k_{f}} & 0 \\ 
-V_{\theta f}(\theta^{\ast}) V_{ff}(\theta^{\ast})^{-1} & I_{k_{f}m}%
\end{array}
\right) =R(\theta^{\ast}),%
\end{array}%
\end{equation*}
to obtain%
\begin{equation*}
\begin{array}{rlc}
\sqrt{T}\left[ \hat{R}(\theta^{\ast})\left( 
\begin{array}{c}
f_{T}(\theta^{\ast},X) \\ 
vec(q_{T}(\theta^{\ast},X))%
\end{array}
\right) -R(\theta^{\ast})\left( 
\begin{array}{c}
\mu_{f}(\theta^{\ast}) \\ 
vec(J(\theta^{\ast}))%
\end{array}
\right) \right] \underset{d}{\rightarrow} & R(\theta^{\ast})\left( 
\begin{array}{c}
\psi_{f}(\theta^{\ast}) \\ 
\psi_{\theta}(\theta^{\ast})%
\end{array}
\right) & \Leftrightarrow \\ 
\sqrt{T}\left( 
\begin{array}{c}
f_{T}(\theta^{\ast},X)-\mu_{f}(\theta^{\ast}) \\ 
vec(\hat{D}(\theta^{\ast})-D(\theta^{\ast}))%
\end{array}
\right) \underset{d}{\rightarrow} & \left( 
\begin{array}{c}
\psi_{f}(\theta^{\ast}) \\ 
\psi_{\theta.f}(\theta^{\ast})%
\end{array}
\right) , & 
\end{array}%
\end{equation*}
with $\psi_{\theta.f}(\theta^{\ast})=\psi_{\theta}(\theta^{\ast})-V_{\theta
f}(\theta^{\ast})V_{ff}(\theta^{\ast})^{-1}\psi_{f}(\theta^{\ast})$ which is
independent of $\psi_{f}(\theta^{\ast})$ since%
\begin{equation*}
R(\theta^{\ast})V(\theta^{\ast})R(\theta^{\ast})^{\prime}=\left( 
\begin{array}{cc}
V_{ff}(\theta^{\ast}) & 0 \\ 
0 & V_{\theta\theta.f}(\theta^{\ast})%
\end{array}
\right) ,
\end{equation*}
where $V_{\theta\theta.f}(\theta^{\ast})=V_{\theta\theta}(\theta^{\ast
})-V_{\theta f}(\theta^{\ast})V_{ff}(\theta^{\ast})^{-1}V_{\theta
f}(\theta^{\ast})^{\prime},$ so $\psi_{f}(\theta^{\ast})$ and $\psi_{\theta
.f}(\theta^{\ast})$ are uncorrelated and independent since they are normal
distributed random variables.

\subsection{Proof of Theorem 4}

The joint limit behaviors of $f_{T}(\theta ^{\ast },X),$ $\hat{D}(\theta
^{\ast })$ and $\hat{V}_{ff}(\theta ^{\ast })$ are such that:%
\begin{equation*}
\begin{array}{rl}
Ts(\theta ^{\ast })= & \left( \sqrt{T}f_{T}(\theta ^{\ast },X)\right)
^{\prime }\hat{V}_{ff}(\theta ^{\ast })^{-1}\left( \sqrt{T}\hat{D}(\theta
^{\ast })\right) \\ 
\underset{d}{\rightarrow } & \left[ \bar{\mu}_{f}(\theta ^{\ast })+\psi
_{f}(\theta ^{\ast })\right] ^{\prime }V_{ff}(\theta ^{\ast })^{-1}\left[ 
\bar{D}(\theta ^{\ast })+\Psi _{\theta .f}(\theta ^{\ast })\right] \\ 
= & \bar{\mu}_{f}(\theta ^{\ast })^{\prime }V_{ff}(\theta ^{\ast })^{-1}\Psi
_{\theta .f}(\theta ^{\ast })+\psi _{f}(\theta ^{\ast })^{\prime
}V_{ff}(\theta ^{\ast })^{-1}\left[ \bar{D}(\theta ^{\ast })+\Psi _{\theta
.f}(\theta ^{\ast })\right] \\ 
= & \left( \bar{\mu}_{f}(\theta ^{\ast })+\psi _{f}(\theta ^{\ast })\right)
^{\prime }V_{ff}(\theta ^{\ast })^{-1}\Psi _{\theta .f}(\theta ^{\ast
})+\psi _{f}(\theta ^{\ast })^{\prime }V_{ff}(\theta ^{\ast })^{-1}\bar{D}%
(\theta ^{\ast }),%
\end{array}%
\end{equation*}
where vec($\Psi _{\theta .f}(\theta ^{\ast }))=\psi _{\theta .f},$ since $%
\bar{\mu}_{f}(\theta ^{\ast })^{\prime }V_{ff}(\theta ^{\ast })^{-1}\bar{D}%
(\theta ^{\ast })=0.$ Since $\psi _{f}(\theta ^{\ast })$ and $\psi _{\theta
.f}(\theta ^{\ast })$ are independently distributed, this shows that the
expected value of the limit of the score of the CUE sample objective
function equals zero at the pseudo-true value $\theta ^{\ast }.$

\subsection{Proof of Theorem 5}

We can specify the limit behavior of $Ts(\theta ^{\ast })$ as:%
\begin{equation*}
\begin{array}{rl}
Ts(\theta ^{\ast })^{\prime }\underset{d}{\rightarrow } & a+b+c,%
\end{array}%
\end{equation*}%
with $a=\Psi _{\theta .f}(\theta ^{\ast })^{\prime }V_{ff}(\theta ^{\ast
})^{-1}\bar{\mu}_{f}(\theta ^{\ast }),$ $b=\bar{D}(\theta ^{\ast })^{\prime
}V_{ff}(\theta ^{\ast })^{-1}\psi _{f}(\theta ^{\ast })$ and $c=\Psi
_{\theta .f}(\theta ^{\ast })^{\prime }V_{ff}(\theta ^{\ast })^{-1}\psi
_{f}(\theta ^{\ast }).$ To obtain the bound on the limiting distribution of
the DRLM\ statistic, we next further characterize the limit behavior of the
above components. We first do so for $m=1$.

\noindent \textbf{m=1: }We specify $a,$ $b$ and $c$ as:%
\begin{equation*}
\begin{array}{rl}
a= & \Psi _{\theta .f}^{\ast \prime }G^{\prime }\mu ^{\ast }, \\ 
b= & D^{\ast \prime }G^{\prime }\psi _{f}^{\ast }, \\ 
c= & \Psi _{\theta .f}^{\ast \prime }G^{\prime }\psi _{f}^{\ast },%
\end{array}%
\end{equation*}%
which results from a singular value decomposition of $V_{ff}(\theta ^{\ast
})^{-\frac{1}{2}}V_{\theta \theta .f}(\theta ^{\ast })^{\frac{1}{2}}:$%
\begin{equation*}
V_{ff}(\theta ^{\ast })^{-\frac{1}{2}}V_{\theta \theta .f}(\theta ^{\ast })^{%
\frac{1}{2}}=LGK^{\prime },
\end{equation*}%
with $L$ and $K$ $k\times k$ dimensional orthonormal matrices and $G$ a
diagonal $k\times k$ dimensional matrix with the non-negative singular
values in decreasing order on the main diagonal and we used that $\mu ^{\ast
}=L^{\prime }V_{ff}(\theta ^{\ast })^{-\frac{1}{2}}\bar{\mu}_{f}(\theta
^{\ast }),$ $D^{\ast }=K^{\prime }V_{\theta \theta .f}(\theta ^{\ast })^{-%
\frac{1}{2}}\bar{D}(\theta ^{\ast }),$ $\psi _{f}^{\ast }=L^{\prime
}V_{ff}(\theta ^{\ast })^{-\frac{1}{2}}\psi _{f}(\theta ^{\ast })\sim
N(0,I_{k}),$ $\Psi _{\theta .f}^{\ast }=K^{\prime }V_{\theta \theta
.f}(\theta ^{\ast })^{-\frac{1}{2}}\Psi _{\theta .f}(\theta ^{\ast })\sim
N(0,I_{k})$ and independent of $\psi _{f}^{\ast }.$

Using the above, the limit behavior of the DRLM\ statistic can be specified
as:%
\begin{equation*}
\begin{array}{rl}
DRLM(\theta ^{\ast })\underset{d}{\rightarrow } & \left[ \Psi _{\theta
.f}^{\ast \prime }G^{\prime }\mu ^{\ast }+D^{\ast \prime }G^{\prime }\psi
_{f}^{\ast }+\Psi _{\theta .f}^{\ast \prime }G^{\prime }\psi _{f}^{\ast }%
\right] ^{\prime } \\ 
& \left[ \left( \mu ^{\ast }+\psi _{f}^{\ast }\right) ^{\prime }GG^{\prime
}\left( \mu ^{\ast }+\psi _{f}^{\ast }\right) +\left( D^{\ast }+\Psi
_{\theta .f}^{\ast }\right) ^{\prime }G^{\prime }G\left( D^{\ast }+\Psi
_{\theta .f}^{\ast }\right) \right] ^{-1} \\ 
& \left[ \Psi _{\theta .f}^{\ast \prime }G^{\prime }\mu ^{\ast }+D^{\ast
\prime }G^{\prime }\psi _{f}^{\ast }+\Psi _{\theta .f}^{\ast \prime
}G^{\prime }\psi _{f}^{\ast }\right]  \\ 
= & \frac{\left[ \Psi _{\theta .f}^{\ast \prime }G\mu ^{\ast }+D^{\ast
\prime }G\psi _{f}^{\ast }+\Psi _{\theta .f}^{\ast \prime }G\psi _{f}^{\ast }%
\right] ^{2}}{\left[ \left( \mu ^{\ast }+\psi _{f}^{\ast }\right) ^{\prime
}G^{2}\left( \mu ^{\ast }+\psi _{f}^{\ast }\right) +\left( D^{\ast }+\Psi
_{\theta .f}^{\ast }\right) ^{\prime }G^{2}\left( D^{\ast }+\Psi _{\theta
.f}^{\ast }\right) \right] }.%
\end{array}%
\end{equation*}%
The limiting distribution of the DRLM\ statistic only depends on the $3k$
parameters present in $k,$ $G,$ $\mu ^{\ast }$ and $D^{\ast }.$ The $3k$
results since the limiting distribution is invariant to multiplying $G$ by a
positive scalar so the largest element of $G$, $G_{11},$ can be set to one.
This implies that $G$ contains $k-1$ non-negative elements which are not
preset to 0 or 1. The number of elements in both $\mu ^{\ast }$ and $D^{\ast
}$ equals $k.$

When $\mu ^{\ast }$ and $D^{\ast }$ equal zero, the limit behavior of $%
DRLM(\theta ^{\ast })$ becomes:%
\begin{equation*}
\begin{array}{rl}
DRLM(\theta ^{\ast })|_{\mu ^{\ast }=D^{\ast }=0}\underset{d}{\rightarrow }
& \frac{\left[ \Psi _{\theta .f}^{\ast \prime }G^{\prime }\psi _{f}^{\ast }%
\right] ^{2}}{\left[ \psi _{f}^{\ast \prime }G^{2}\psi _{f}^{\ast }+\Psi
_{\theta .f}^{\ast \prime }G^{2}\Psi _{\theta .f}^{\ast }\right] }\preceq
\chi ^{2}(1),%
\end{array}%
\end{equation*}%
since both $\frac{\left[ \Psi _{\theta .f}^{\ast \prime }G^{\prime }\psi
_{f}^{\ast }\right] ^{2}}{\psi _{f}^{\ast \prime }G^{2}\psi _{f}^{\ast }}%
\sim \chi ^{2}(1)$ and $\frac{\left[ \Psi _{\theta .f}^{\ast \prime
}G^{\prime }\psi _{f}^{\ast }\right] ^{2}}{\Psi _{\theta .f}^{\ast \prime
}G^{2}\Psi _{\theta .f}^{\ast }}\sim \chi ^{2}(1)$ and where
\textquotedblleft $\preceq $\textquotedblright\ indicates stochastically
dominated so for a continuous non-negative scalar random variable $u\preceq
\chi ^{2}(m):$ $\Pr \left[ u>cv_{\chi ^{2}(m)}(\alpha )\right] \leq \alpha ,$
for $\alpha \in (0,1]$ and with $cv_{\chi ^{2}(m)}(\alpha )$ the $(1-\alpha
)\times 100\%$ critical value for the $\chi ^{2}(m)$ distribution.

Similarly, when the length of $\mu ^{\ast }$ and/or $D^{\ast }$ goes to
infinity:%
\begin{equation*}
\begin{array}{cc}
\left. 
\begin{array}{r}
\lim_{\mu ^{\ast \prime }\mu ^{\ast }\rightarrow \infty }DRLM(\theta ^{\ast
}) \\ 
\lim_{D^{\ast \prime }D^{\ast }\rightarrow \infty }DRLM(\theta ^{\ast }) \\ 
\lim_{\mu ^{\ast \prime }\mu ^{\ast }\rightarrow \infty ,D^{\ast \prime
}D^{\ast }\rightarrow \infty }DRLM(\theta ^{\ast })%
\end{array}%
\right\}  & \underset{d}{\rightarrow }\chi ^{2}(1).%
\end{array}%
\end{equation*}%
The limit behavior is identical with respect to the different elements of $%
\mu ^{\ast }$ and $D^{\ast }.$ Figure A7 shows for a pre-specified fixed
value of $G$ that the distribution function associated with the limit
behavior of $DRLM(\theta ^{\ast })$ is a non-increasing function of either
the length of $\mu ^{\ast }$ or $D^{\ast }$. Figure A7 also shows the
difference with the $\chi ^{2}(1)$ distribution function which makes it
clear that the $\chi ^{2}(1)$ distribution dominates the limiting
distribution of the DRLM\ statistic for this specific value of $G$. Since $G$
is a diagonal matrix with only non-negative elements, this behavior holds
also for all other values of $G$ so the limit behavior of $DRLM(\theta
^{\ast })$ is bounded by the $\chi ^{2}(1)$ distribution:%
\begin{equation*}
\begin{array}{c}
\lim_{T\rightarrow \infty }\Pr \left[ DRLM(\theta ^{\ast })>cv_{\chi
^{2}(1)}(\alpha )\right] \leq \alpha .%
\end{array}%
\end{equation*}

\noindent \textbf{m $>$ 1: }We specify $a,$ $b$ and $c$ as:%
\begin{equation*}
\begin{array}{rl}
a= & \Psi _{\theta .f}^{\ast \prime }\mu ^{\ast }, \\ 
b= & D^{\ast \prime }\psi _{f}^{\ast }, \\ 
c= & \Psi _{\theta .f}^{\ast \prime }\psi _{f}^{\ast },%
\end{array}%
\end{equation*}%
with $G=(I_{m}\otimes V_{ff}(\theta ^{\ast })^{-\frac{1}{2}})V_{\theta
\theta .f}(\theta ^{\ast })^{\frac{1}{2}},$ $\Psi _{\theta .f}^{\ast
}=V_{ff}(\theta ^{\ast })^{-\frac{1}{2}}\Psi _{\theta .f}(\theta ^{\ast }),$
vec($\Psi _{\theta .f}^{\ast })=G\psi _{\theta .f}^{\ast },$ $\psi _{\theta
.f}^{\ast }\sim N(0,I_{km}),$ $\psi _{f}^{\ast }=V_{ff}(\theta ^{\ast })^{-%
\frac{1}{2}}\psi _{f}(\theta ^{\ast })\sim N(0,I_{k})$ and independent of $%
\psi _{\theta .f}^{\ast },$ $\mu ^{\ast }=V_{ff}(\theta ^{\ast })^{-\frac{1}{%
2}}\bar{\mu}_{f}(\theta ^{\ast }),$ $D^{\ast }=V_{ff}(\theta ^{\ast })^{-%
\frac{1}{2}}\bar{D}(\theta ^{\ast }),$ vec($D^{\ast })=G$vec($\bar{D}^{\ast
})$ and $\bar{D}^{\ast }$ is a $k\times m$ dimensional matrix.

\begin{equation*}
\begin{array}{c}
\text{Figure A7. Distribution function of the DRLM statistic for a } \\ 
\text{fixed value of }G\text{ as a function of the length of either }\mu
^{\ast }\text{ or }D^{\ast }. \\ 
\raisebox{-0pt}{\includegraphics[
height=8.1549in,
width=5.2698in
]%
{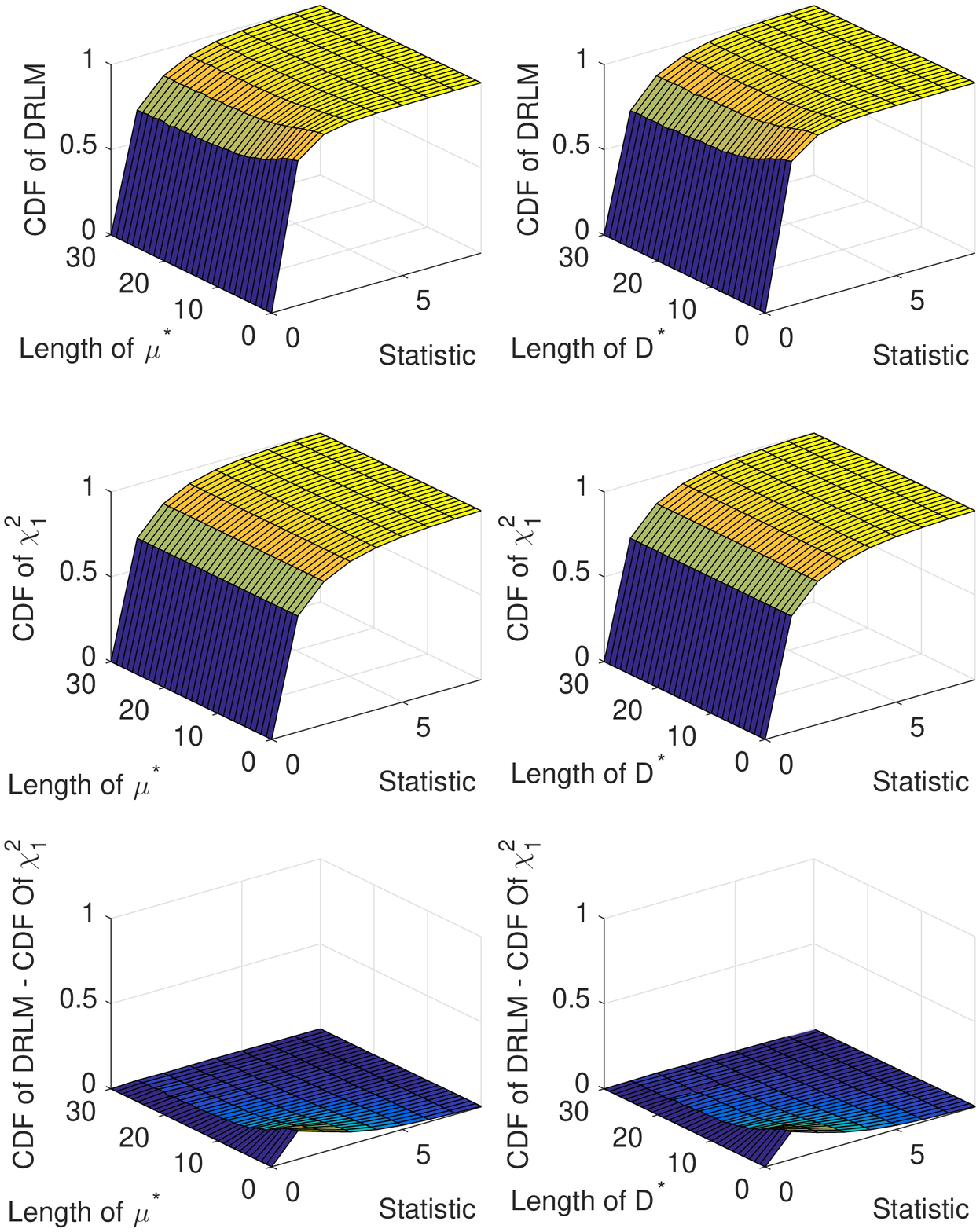}%
}
\end{array}%
\end{equation*}

Using the above, the limit behavior of the DRLM\ statistic can be specified
as:%
\begin{equation*}
\begin{array}{rl}
DRLM(\theta ^{\ast })\underset{d}{\rightarrow } & \left[ \Psi _{\theta
.f}^{\ast \prime }\mu ^{\ast }+D^{\ast \prime }\psi _{f}^{\ast }+\Psi
_{\theta .f}^{\ast \prime }\psi _{f}^{\ast }\right] ^{\prime } \\ 
& \left[ \left( I_{m}\otimes \left( \mu ^{\ast }+\psi _{f}^{\ast }\right)
\right) ^{\prime }GG^{\prime }\left( I_{m}\otimes \left( \mu ^{\ast }+\psi
_{f}^{\ast }\right) \right) +\left( D^{\ast }+\Psi _{\theta .f}^{\ast
}\right) ^{\prime }\left( D^{\ast }+\Psi _{\theta .f}^{\ast }\right) \right]
^{-1} \\ 
& \left[ \Psi _{\theta .f}^{\ast \prime }\mu ^{\ast }+D^{\ast \prime }\psi
_{f}^{\ast }+\Psi _{\theta .f}^{\ast \prime }\psi _{f}^{\ast }\right] .%
\end{array}%
\end{equation*}%
The limiting distribution of the DRLM\ statistic depends on the $%
k^{2}m^{2}+km+k+1$ parameters present in: $G,$ $D^{\ast },$ $\mu
^{\ast },$ $k$ and $m.$ Since the limiting distribution is invariant to
multiplying $G$ by a positive scalar, we normalize $G$ such that one
diagonal element of $G,$ say $G_{11},$ is equal to one. This explains the
number of parameters affecting the limiting distribution of the DRLM\
statistic.

When $\mu ^{\ast }$ and $D^{\ast }$ equal zero, the limit behavior of $%
DRLM(\theta ^{\ast })$ becomes:%
\begin{equation*}
\begin{array}{rl}
DRLM(\theta ^{\ast })|_{\mu ^{\ast }=D^{\ast }=0}\underset{d}{\rightarrow }
& \psi _{f}^{\ast \prime }\Psi _{\theta .f}^{\ast }\left[ \left(
I_{m}\otimes \psi _{f}^{\ast }\right) ^{\prime }GG^{\prime }\left(
I_{m}\otimes \psi _{f}^{\ast }\right) +\Psi _{\theta .f}^{\ast \prime }\Psi
_{\theta .f}^{\ast }\right] ^{-1}\Psi _{\theta .f}^{\ast \prime }\psi
_{f}^{\ast }\preceq \chi ^{2}(m),%
\end{array}%
\end{equation*}%
since $\psi _{f}^{\ast \prime }\Psi _{\theta .f}^{\ast }\left[ \Psi
_{\theta .f}^{\ast \prime }\Psi _{\theta .f}^{\ast }\right] ^{-1}\Psi
_{\theta .f}^{\ast \prime }\psi _{f}^{\ast }\sim \chi ^{2}(m)$ and $\psi
_{f}^{\ast \prime }\Psi _{\theta .f}^{\ast }\left[ \left( I_{m}\otimes \psi
_{f}^{\ast }\right) ^{\prime }GG^{\prime }\left( I_{m}\otimes \psi
_{f}^{\ast }\right) \right] ^{-1}\Psi _{\theta .f}^{\ast \prime }\psi
_{f}^{\ast }\sim \chi ^{2}(m).$ 

Similarly, when using a singular value
decomposition of $D^{\ast }:$%
\begin{equation*}
D^{\ast }=L_{D}G_{D}K_{D}^{\prime },
\end{equation*}%
with $L_{D}$ and $K_{D}$ $k\times k$ and $m\times m$ dimensional orthonormal
matrices and $G_{D}$ a diagonal $k\times m$ dimensional matrix with the
non-negative singular values in decreasing order on the main diagonal, we
can specify the limiting behavior of the DRLM statistic: 
\begin{equation*}
\begin{array}{rl}
DRLM(\theta ^{\ast })\underset{d}{\rightarrow } & \left[ K_{D}\bar{\Psi}%
_{\theta .f}^{\prime }L_{D}^{\prime }\mu ^{\ast }+K_{D}G_{D}^{\prime
}L_{D}^{\prime }\psi _{f}^{\ast }+K_{D}\bar{\Psi}_{\theta .f}^{\prime
}L_{D}^{\prime }\psi _{f}^{\ast }\right] ^{\prime } \\ 
& \left[ \left( I_{m}\otimes \left( \mu ^{\ast }+\psi _{f}^{\ast }\right)
\right) ^{\prime }GG^{\prime }\left( I_{m}\otimes \left( \mu ^{\ast }+\psi
_{f}^{\ast }\right) \right) +\right.  \\ 
& \left. \left( L_{D}G_{D}K_{D}^{\prime }+L_{D}\bar{\Psi}_{\theta
.f}K_{D}^{\prime }\right) ^{\prime }\left( L_{D}G_{D}K_{D}^{\prime }+L_{D}%
\bar{\Psi}_{\theta .f}K_{D}^{\prime }\right) \right] ^{-1} \\ 
& \left[ K_{D}\bar{\Psi}_{\theta .f}^{\prime }L_{D}^{\prime }\mu ^{\ast
}+K_{D}G_{D}^{\prime }L_{D}^{\prime }\psi _{f}^{\ast }+K_{D}\bar{\Psi}%
_{\theta .f}^{\prime }L_{D}^{\prime }\psi _{f}^{\ast }\right]  \\ 
= & \left[ \bar{\Psi}_{\theta .f}^{\prime }\bar{\mu}+G_{D}^{\prime }\bar{\psi%
}_{f}+\bar{\Psi}_{\theta .f}^{\prime }\bar{\psi}_{f}\right] ^{\prime }\left[
\left( I_{m}\otimes \left( \bar{\mu}+\bar{\psi}_{f}\right) \right) ^{\prime }%
\bar{G}\bar{G}^{\prime }\left( I_{m}\otimes \left( \bar{\mu}+\bar{\psi}%
_{f}\right) \right) +\right.  \\ 
& \left. \left( G_{D}+\bar{\Psi}_{\theta .f}\right) ^{\prime }\left( G_{D}+%
\bar{\Psi}_{\theta .f}\right) \right] ^{-1}\left[ \bar{\Psi}_{\theta
.f}^{\prime }\bar{\mu}+G_{D}^{\prime }\bar{\psi}_{f}+\bar{\Psi}_{\theta
.f}^{\prime }\bar{\psi}_{f}\right] ,%
\end{array}%
\end{equation*}%
where $\Psi _{\theta .f}^{\ast }=L_{D}\bar{\Psi}_{\theta .f}K_{D}^{\prime },$
$\bar{\psi}_{f}=L_{D}^{\prime }\psi _{f}^{\ast },$ $\bar{\mu}=L_{D}^{\prime
}\mu ^{\ast },$ $\bar{G}=(K_{m}\otimes L_{D}V_{ff}(\theta ^{\ast })^{-\frac{1%
}{2}})V_{\theta \theta .f}(\theta ^{\ast })^{\frac{1}{2}},$ vec($\bar{\Psi}%
_{\theta .f})=\bar{G}\psi _{\theta .f}^{\ast },$ $\psi _{\theta .f}^{\ast
}\sim N(0,I_{km}).$ The resulting limiting behavior is such that when the
length of $\mu ^{\ast }$ or the $m$ singular values in $G_{D}$ go to
infinity:%
\begin{equation*}
\begin{array}{cc}
\left. 
\begin{array}{r}
\lim_{\mu ^{\ast \prime }\mu ^{\ast }\rightarrow \infty }DRLM(\theta ^{\ast
}) \\ 
\lim_{G_{D,ii}^{\ast }\rightarrow \infty ,\text{ }i=1,\ldots ,m}DRLM(\theta
^{\ast }) \\ 
\lim_{\mu ^{\ast \prime }\mu ^{\ast }\rightarrow \infty ,G_{D,ii}^{\ast
}\rightarrow \infty ,\text{ }i=1,\ldots ,m}DRLM(\theta ^{\ast })%
\end{array}%
\right\}  & \underset{d}{\rightarrow }\chi ^{2}(m).%
\end{array}%
\end{equation*}%
Since $G$ is positive semi-definite, it can be verified numerically that for
any fixed $G,$ the distribution function associated with the limit behavior
of $DRLM(\theta ^{\ast })$ is non-increasing when any element of $\mu ^{\ast
}$ or $G_{D,ii}^{\ast }\rightarrow \infty ,$ $i=1,\ldots ,m$ increases. The
limit behavior of $DRLM(\theta ^{\ast })$ is therefore bounded by the $\chi
^{2}(m)$ distribution:%
\begin{equation*}
\begin{array}{c}
\lim_{T\rightarrow \infty }\Pr \left[ DRLM(\theta ^{\ast })>cv_{\chi
^{2}(m)}(\alpha )\right] \leq \alpha .%
\end{array}%
\end{equation*}

\noindent \textbf{m $>$ 1 and }$V_{ff}(\theta )=v_{ff}(\theta )\bar{%
V},$\textbf{\ }$V_{\theta \theta .f}(\theta )=\left( \Sigma _{\theta \theta
}(\theta )\otimes \bar{V}\right) ,$\textbf{\ }$\Sigma _{\theta \theta
}(\theta ):m\times m$\textbf{\ dimensional matrix: }We specify $a,$ $b$ and $%
c$ as:%
\begin{equation*}
\begin{array}{rl}
a= & \Sigma _{\theta \theta }(\theta )^{\frac{1}{2}}\Psi _{\theta .f}^{\ast
\prime }\mu ^{\ast }v_{ff}(\theta )^{\frac{1}{2}}, \\ 
b= & \Sigma _{\theta \theta }(\theta )^{\frac{1}{2}}D^{\ast }\psi _{f}^{\ast
}v_{ff}(\theta )^{\frac{1}{2}}, \\ 
c= & \Sigma _{\theta \theta }(\theta )^{\frac{1}{2}}\Psi _{\theta .f}^{\ast
\prime }\psi _{f}^{\ast }v_{ff}(\theta )^{\frac{1}{2}},%
\end{array}%
\end{equation*}%
with $\mu ^{\ast }=v_{ff}(\theta ^{\ast })^{-\frac{1}{2}}\bar{V}^{-\frac{1}{2%
}}\bar{\mu}_{f}(\theta ^{\ast }),$ $D^{\ast }=\bar{V}^{-\frac{1}{2}}\bar{D}%
(\theta ^{\ast })\Sigma _{\theta \theta }(\theta )^{-\frac{1}{2}},$ $\psi
_{f}^{\ast }=v_{ff}(\theta ^{\ast })^{-\frac{1}{2}}\bar{V}^{-\frac{1}{2}%
}\psi _{f}(\theta ^{\ast })\sim N(0,I_{k}),$ $\Psi _{\theta .f}^{\ast }=\bar{%
V}^{-\frac{1}{2}}\Psi _{\theta .f}(\theta ^{\ast })\Sigma _{\theta \theta
}(\theta )^{-\frac{1}{2}}\sim N(0,I_{km})$ and independent of $\psi
_{f}^{\ast }.$

Using the above, the limit behavior of the DRLM\ statistic can be specified
as:%
\begin{equation*}
\begin{array}{rl}
DRLM(\theta ^{\ast })\underset{d}{\rightarrow } & \left[ \Psi _{\theta
.f}^{\ast \prime }\mu ^{\ast }+D^{\ast \prime }\psi _{f}^{\ast }+\Psi
_{\theta .f}^{\ast \prime }\psi _{f}^{\ast }\right] ^{\prime } \\ 
& \left[ \left( \mu ^{\ast }+\psi _{f}^{\ast }\right) ^{\prime }\left( \mu
^{\ast }+\psi _{f}^{\ast }\right) I_{m}+\left( D^{\ast }+\Psi _{\theta
.f}^{\ast }\right) ^{\prime }\left( D^{\ast }+\Psi _{\theta .f}^{\ast
}\right) \right] ^{-1} \\ 
& \left[ \Psi _{\theta .f}^{\ast \prime }\mu ^{\ast }+D^{\ast \prime }\psi
_{f}^{\ast }+\Psi _{\theta .f}^{\ast \prime }\psi _{f}^{\ast }\right] .%
\end{array}%
\end{equation*}%
The limiting distribution of the DRLM\ statistic only depends on the $2+k+km$
parameters present in $k,$ $m,$ $\mu ^{\ast }$ and $D^{\ast }.$ To reduce
this further, we conduct a singular value decomposition of $D^{\ast }:$%
\begin{equation*}
D^{\ast }=L_{D^{\ast }}G_{D^{\ast }}K_{D^{\ast }}^{\prime },
\end{equation*}%
with $L_{D^{\ast }}$ and $K_{D^{\ast }}$ $k\times k$ and $m\times m$
dimensional orthonormal matrices and $G_{D^{\ast }}$ a diagonal $k\times m$
dimensional matrix with the non-negative singular values in decreasing order
on the main diagonal. Using next that $\bar{\Psi}_{\theta .f}=L_{D^{\ast
}}^{\prime }\Psi _{\theta .f}^{\ast }K_{D^{\ast }}\sim N(0,I_{km}),$ $\bar{%
\mu}=L_{D^{\ast }}^{\prime }\mu ^{\ast }$ and $\bar{\psi}_{f}=L_{D^{\ast
}}^{\prime }\psi _{f}^{\ast }\sim N(0,I_{k}),$ we can specify the limit
behavior as:%
\begin{equation*}
\begin{array}{rl}
DRLM(\theta ^{\ast })\underset{d}{\rightarrow } & \left[ \bar{\Psi}_{\theta
.f}^{\prime }\bar{\mu}+G_{D^{\ast }}^{\prime }\bar{\psi}_{f}+\bar{\Psi}%
_{\theta .f}^{\prime }\bar{\psi}_{f}\right] ^{\prime } \\ 
& \left[ \left( \bar{\mu}+\bar{\psi}_{f}\right) ^{\prime }\left( 
\bar{\mu}+\bar{\psi}_{f}\right) I_{m}+\left( G_{D^{\ast }}+\bar{\Psi}%
_{\theta .f}\right) ^{\prime }\left( G_{D^{\ast }}+\bar{\Psi}_{\theta
.f}\right) \right] ^{-1} \\ 
& \left[ \bar{\Psi}_{\theta .f}^{\prime }\bar{\mu}+G_{D^{\ast }}^{\prime
}\bar{\psi}_{f}+\bar{\Psi}_{\theta .f}^{\prime }\bar{\psi}_{f}\right] ,%
\end{array}%
\end{equation*}%
which only depends on the $m$ singular values in $G_{D^{\ast }}$ and the
length of $\bar{\mu}.$ The distribution function of the limit behavior is
again a non-decreasing function of the length of $\bar{\mu}$ and the $m$
singular values in $G_{D^{\ast }}$ so its limit behavior is bounded by the $%
\chi ^{2}(m)$ distribution. 

\paragraph{Definition of the parameter space}

In Andrews and\ Guggenberger (2017),\nocite{ag17} the asymptotic size of the
KLM\ test is proven to equal the nominal size and the accompanying parameter
space on the distributions of the observations is stated for both i.i.d. and
dependent data settings.

To start out with the i.i.d. setting, define for some $\kappa ,$ $\tau >0$
and $M<\infty ,$ the parameter space$:$%
\begin{equation*}
\begin{array}{cl}
\mathcal{F=} & \left\{ F:\left\{ X_{t}:t\geq 1\right\} \text{ are i.i.d.
under }F,\text{ }E(f_{t}(\theta ^{\ast }))=\mu _{f}(\theta ^{\ast }),\text{
for }\right. \\ 
& \theta ^{\ast }=\arg \min_{\theta \in 
\mathbb{R}
^{m}}\mu _{f}(\theta )^{\prime }V_{ff}(\theta )^{-1}\mu _{f}(\theta ),\text{ 
}V_{ff}(\theta )=E\left( \left( f_{t}(\theta )-\mu _{f}(\theta )\right)
\left( f_{t}(\theta )-\mu _{f}(\theta )\right) ^{\prime }\right) \\ 
& \left. E\left( \left\Vert \left( f_{t}(\theta ^{\ast })^{\prime }\text{ }%
\vdots \text{ }\left( \text{vec}\left( \frac{\partial }{\partial \theta
^{\prime }}f_{t}(\theta ^{\ast })\right) \right) ^{\prime }\right)
^{2+\kappa }\right\Vert \right)\leq M \text{ and }\lambda _{\min
}(V_{ff}(\theta ^{\ast }))\geq \tau \right\} ,%
\end{array}%
\end{equation*}
where $\lambda _{\min }(A)$ is the smallest characteristic root of the
matrix $A.$ The parameter space above is identical to the one in Andrews
and\ Guggenberger (2017) Equation (3.3) except that it is defined for the
pseudo-true value $\theta ^{\ast }$ defined as the minimizer of the
population continuous updating objective function for which $\mu _{f}(\theta
^{\ast })$ is not necessarily equal to zero.

Since we are after proving the size correctness of the DRLM\ test which
tests hypotheses specified on the pseudo-true value $\theta ^{\ast },$ we
define the recentered Jacobian:%
\begin{equation*}
\begin{array}{l}
D(\theta )=J(\theta )-\left[ V_{\theta _{1}f}(\theta )V_{ff}(\theta
)^{-1}\mu _{f}(\theta )\ldots V_{\theta _{m}f}(\theta )V_{ff}(\theta
)^{-1}\mu _{f}(\theta )\right] ,\text{ }J(\theta )=\frac{\partial }{\partial
\theta ^{\prime }}\mu _{f}(\theta ), \\ 
V_{\theta _{i}f}(\theta )=E\left[ (\frac{\partial }{\partial \theta _{i}}%
(f_{t}(\theta )-\mu _{f}(\theta )))\left( f_{t}(\theta )-\mu _{f}(\theta
)\right) ^{\prime }\right] ,\text{ }i=1,\ldots ,m, \\ 
V_{ff}(\theta )=E\left( \left( f_{t}(\theta )-\mu _{f}(\theta )\right)
\left( f_{t}(\theta )-\mu _{f}(\theta )\right) ^{\prime }\right) .%
\end{array}%
\end{equation*}%
The pseudo-true value is then such that%
\begin{equation*}
\begin{array}{c}
\mu _{f}(\theta ^{\ast })^{\prime }V_{ff}(\theta ^{\ast })^{-1}D(\theta
^{\ast })=0.%
\end{array}%
\end{equation*}%
To guarantee with probability one, a non-singular value of the limit value
of the sample analog of $V_{ff}(\theta ^{\ast })^{-1}D(\theta ^{\ast }),$ $%
\hat{V}_{ff}(\theta ^{\ast })^{-1}\hat{D}(\theta ^{\ast }),$ Andrews and
Guggenberger (2017) provide a number of additional conditions on the
parameter space $\mathcal{F}.$ Since we allow for misspecification, these
conditions have to hold when using the recentered Jacobian $D(\theta )$
instead of the Jacobian $J(\theta )$ as in Andrews and Guggenberger (2017).
Taken together these conditions imply that the singular values of $%
V_{ff}(\theta ^{\ast })^{-1}D(\theta ^{\ast })$ should be bounded away from
zero and the same applies for the quadratic form of the orthonormal vectors
resulting from the singular value decomposition of $V_{ff}(\theta ^{\ast
})^{-1}D(\theta ^{\ast })$ with respect to the covariance matrix of vec($%
\hat{D}(\theta ^{\ast })).$ We refer to Andrews and Guggenberger (2017) for
the definition of this reduced parameter space.

The parameter spaces in Andrews and Guggenberger (2017) imply Lemma 10.2 in\
their Supplementary Appendix which coincides with our Theorem 3 except that
Theorem 3 allows for a population mean function $\mu _{f}(\theta ^{\ast })$
different from zero. Jointly with some weak laws of large numbers, the
limiting distributions resulting from Lemma 10.2 in the Supplementary
Appendix of Andrews and Guggenberger (2017) provide the building blocks for
their Theorem 11.1 which states that the asymptotic size of the KLM\ test
equals the nominal size. Since the parameter spaces also imply our Theorem 3
whose resulting limiting distributions alongside some weak laws of large
numbers imply Theorems 4 and 5, which states that the limiting distribution
of the DRLM\ statistics is bounded by a $\chi ^{2}(m)$ distribution, the
parameter spaces thus also imply that the asymptotic size of the DRLM\ test
equals the nominal size.

For the dependent times-series setting, $\kappa ,$ $\tau >0,$ $d>(2+\kappa
)/\kappa $ and $M<\infty ,$ the space of distributions is defined by:%
\begin{equation*}
\begin{array}{cl}
\mathcal{F}_{ts}\mathcal{=} & \left\{ F:\left\{ X_{t}:t=0,1,\ldots \right\} 
\text{ are stationary and strong mixing under }F\text{ with strong }\right.
\\ 
& \text{mixing numbers }\{\alpha _{F}(p):p\geq 1\}\text{ that satisfy }%
\alpha _{F}(p)\leq Cp^{-d},\text{ }E(f_{t}(\theta ^{\ast }))=\mu _{f}(\theta
^{\ast }), \\ 
& \theta ^{\ast }=\arg \min_{\theta \in 
\mathbb{R}
^{m}}\mu _{f}(\theta )^{\prime }V_{ff}(\theta )^{-1}\mu _{f}(\theta ), \\ 
& V_{ff}(\theta )=E\left[ \lim_{T\rightarrow \infty }\frac{1}{T}%
\sum_{t=1}^{T}\sum_{j=1}^{T}\left( f_{t}(\theta )-\mu _{f}(\theta )\right)
\left( f_{j}(\theta )-\mu _{f}(\theta )\right) ^{\prime }\right] , \\ 
& \left. E\left( \left\Vert \left( f_{t}(\theta )^{\prime }\text{ }\vdots 
\text{ }\left( \text{vec}\left( \frac{\partial }{\partial \theta ^{\prime }}%
f_{t}(\theta )\right) \right) ^{\prime }\right) ^{2+\kappa
}\right\Vert\right) \leq M \text{ and }\lambda _{\min }(V_{ff}(\theta ^{\ast
}))\geq \tau \right\}%
\end{array}%
\end{equation*}%
which again, except for the usage of the pseudo-true value $\theta ^{\ast }$
and a possibly non-zero mean of $f_{t}(\theta ^{\ast }),$ is identical to
Equation (7.2) in\ Andrews and Guggenberger (2017). Identical to the i.i.d.
setting, Andrews and Guggenberger (2017) provide a number of additional
conditions on the parameter space $\mathcal{F}_{ts},$ to guarantee with
probability one, a non-singular value of the limit value of the sample
analog of $V_{ff}(\theta ^{\ast })^{-1}D(\theta ^{\ast }),$ $\hat{V}%
_{ff}(\theta ^{\ast })^{-1}\hat{D}(\theta ^{\ast }).$ Replacing the value of
the Jacobian, $J(\theta ),$ by the recentered Jacobian, $D(\theta ),$ in the
conditions from Andrews and Guggenberger (2017) then implies that also for
our setting the limit value of the $\hat{V}_{ff}(\theta ^{\ast })^{-1}\hat{D}%
(\theta ^{\ast })$ is non-singular with probability one. The resulting
parameter space then again implies our Theorem 3 from which Theorem 5
follows so the asymptotic size of the DRLM\ test coincides with the nominal
size.

\subsection{Proof of Theorem 6}

\textbf{a. }Starting out from a linear moment equation, like, for example,
the one for the linear asset pricing model, $f_{T}(\lambda_{F},X)=\bar{R}-%
\hat{\beta}\lambda_{F},$ which is WLOG:%
\begin{equation*}
\begin{array}{rl}
d= & \left( 
\begin{array}{c}
\bar{R} \\ 
\text{vec(}\hat{\beta})%
\end{array}
\right) ^{\prime}\widehat{\text{var}}\left( \sqrt{T}\left( 
\begin{array}{c}
\bar{R} \\ 
\text{vec(}\hat{\beta})%
\end{array}
\right) \right) ^{-1}\left( 
\begin{array}{c}
\bar{R} \\ 
\text{vec(}\hat{\beta})%
\end{array}
\right) \\ 
= & \left( 
\begin{array}{c}
\bar{R}-\hat{\beta}\lambda_{F} \\ 
\text{vec(}\hat{\beta})%
\end{array}
\right) ^{\prime}\left( \widehat{\text{var}}\left( \sqrt{T}\left( 
\begin{array}{c}
\bar{R}-\hat{\beta}\lambda_{F} \\ 
\text{vec(}\hat{\beta})%
\end{array}
\right) \right) \right) ^{-1}\left( 
\begin{array}{c}
\bar{R}-\hat{\beta}\lambda_{F} \\ 
\text{vec(}\hat{\beta})%
\end{array}
\right) \\ 
= & \left( \bar{R}-\hat{\beta}\lambda_{F}\right) ^{\prime}\left( \widehat{%
\text{var}}\left( \sqrt{T}\left( \bar{R}-\hat{\beta}\lambda _{F}\right)
\right) \right) ^{-1}\left( \bar{R}-\hat{\beta}\lambda _{F}\right) + \\ 
& \left( \text{vec(}\hat{D}(\lambda_{F}))\right) ^{\prime}\hat{V}%
_{\theta\theta.f}(\lambda_{F})^{-1}\left( \text{vec(}\hat{D}(\lambda
_{F}))\right) ,%
\end{array}%
\end{equation*}
which shows that, given a realized data set and since $d$ does not depend on 
$\lambda_{F},$ the sum of $f_{T}(\lambda_{F},X)^{\prime}\hat{V}%
_{ff}(\lambda_{F})^{-1}f_{T}(\lambda_{F},X)$ and $\left( \text{vec(}\hat{D}%
(\lambda_{F}))\right) ^{\prime}\hat{V}_{\theta\theta.f}(\lambda_{F})^{-1}%
\left( \text{vec(}\hat{D}(\lambda _{F}))\right) $ does not depend on $%
\lambda_{F}.$

\noindent\textbf{b. }Given the specifications of the derivatives in Lemma 2,
the derivative of DRLM($\theta)$ when $m=1$ and $f_{T}(\theta,X)$ is linear
in $\theta$ reads:%
\begin{equation*}
\begin{array}{l}
\frac{1}{2}\frac{\partial}{\partial\theta}DRLM(\theta) \\ 
\begin{array}{cl}
= & \frac{1}{2}T\frac{\partial}{\partial\theta}\left\{ f_{T}(\theta
,X)^{\prime}\hat{V}_{ff}(\theta)^{-1}\hat{D}(\theta)\left[
f_{T}(\theta,X)^{\prime}\hat{V}_{ff}(\theta)^{-1}\hat{V}_{\theta\theta.f}(%
\theta)\hat{V}_{ff}(\theta)^{-1}f_{T}(\theta,X)+\right. \right. \\ 
& \left. \left. \hat{D}(\theta)^{\prime}\hat{V}_{ff}(\theta)^{-1}\hat {D}%
(\theta)\right] ^{-1}\hat{D}(\theta)^{\prime}\hat{V}_{ff}(\theta
)^{-1}f_{T}(\theta,X)\right\} \\ 
= & T\left[ f_{T}(\theta,X)^{\prime}\hat{V}_{ff}(\theta)^{-1}\hat{V}%
_{\theta\theta.f}(\theta)\hat{V}_{ff}(\theta)^{-1}f_{T}(\theta,X)+\hat {D}%
(\theta)^{\prime}\hat{V}_{ff}(\theta)^{-1}\hat{D}(\theta)\right]
^{-1}f_{T}(\theta,X)^{\prime}\hat{V}_{ff}(\theta )^{-1}\hat{D}(\theta) \\ 
& \left( \frac{\partial}{\partial\theta}\hat{D}(\theta)^{\prime}\hat {V}%
_{ff}(\theta)^{-1}f_{T}(\theta,X)\right) - \frac{1}{2}T\left( \frac{%
f_{T}(\theta,X)^{\prime}\hat{V}_{ff}(\theta )^{-1}\hat{D}(\theta)}{\left[
f_{T}(\theta,X)^{\prime}\hat{V}_{ff}(\theta)^{-1}\hat{V}_{\theta\theta.f}(%
\theta)\hat{V}_{ff}(\theta)^{-1}f_{T}(\theta,X)+\hat{D}(\theta)^{\prime}\hat{%
V}_{ff}(\theta)^{-1}\hat {D}(\theta)\right] }\right) ^{2} \\ 
& \left( \frac{\partial}{\partial\theta}\left[ f_{T}(\theta,X)^{\prime}\hat{V%
}_{ff}(\theta)^{-1}\hat{V}_{\theta\theta.f}(\theta)\hat{V}%
_{ff}(\theta)^{-1}f_{T}(\theta,X)+\hat{D}(\theta)^{\prime}\hat{V}%
_{ff}(\theta )^{-1}\hat{D}(\theta)\right] \right) \\ 
= & T\left( \frac{f_{T}(\theta,X)^{\prime}\hat{V}_{ff}(\theta)^{-1}\hat {D}%
(\theta)}{\left[ f_{T}(\theta,X)^{\prime}\hat{V}_{ff}(\theta)^{-1}\hat {V}%
_{\theta\theta.f}(\theta)\hat{V}_{ff}(\theta)^{-1}f_{T}(\theta,X)+\hat {D}%
(\theta)^{\prime}\hat{V}_{ff}(\theta)^{-1}\hat{D}(\theta)\right] }\right)
\left\{ \hat{D}(\theta)^{\prime}\hat{V}_{ff}(\theta)^{-1}\hat{D}%
(\theta)-\right. \\ 
& 2f_{T}(\theta,X)^{\prime}\hat{V}_{ff}(\theta)^{-1}\hat{V}_{\theta
f}(\theta)\hat{V}_{ff}(\theta)^{-1}D_{T}(\theta,X)-f_{T}(\theta,X)^{\prime}%
\hat{V}_{ff}(\theta)^{-1}\hat{V}_{\theta\theta.f}(\theta)\hat{V}%
_{ff}(\theta)^{-1}f_{T}(\theta,X)+ \\ 
& \left. 2\frac{\left[ f_{T}(\theta,X)^{\prime}\hat{V}_{ff}(\theta)^{-1}\hat{%
V}_{\theta f}(\theta)\hat{V}_{ff}(\theta)^{-1}\hat{V}_{\theta\theta
.f}(\theta)\hat{V}_{ff}(\theta)^{-1}f_{T}(\theta,X)+\hat{D}(\theta)^{\prime }%
\hat{V}_{ff}(\theta)^{-1}\hat{V}_{\theta f}(\theta)\hat{V}_{ff}(\theta )^{-1}%
\hat{D}(\theta)\right] }{\left[ f_{T}(\theta,X)^{\prime}\hat{V}%
_{ff}(\theta)^{-1}\hat{V}_{\theta\theta.f}(\theta)\hat{V}_{ff}(\theta
)^{-1}f_{T}(\theta,X)+\hat{D}(\theta)^{\prime}\hat{V}_{ff}(\theta)^{-1}\hat {%
D}(\theta)\right] }f_{T}(\theta,X)^{\prime}\hat{V}_{ff}(\theta)^{-1}\hat {D}%
(\theta)\right\} .%
\end{array}%
\end{array}%
\end{equation*}

\noindent\textbf{c. }In case of i.i.d. data, $m=1,$ and $f_{T}(\theta,X)$
linear in $\theta,$ $\hat{V}(\theta)$ has a Kronecker product structure so $%
\hat {V}_{ff}(\theta)=\hat{v}_{ff}(\theta)\hat{V},$ $\hat{V}_{\theta
f}(\theta)=\hat{v}_{\theta f}(\theta)\hat{V}$ and $\hat{V}_{\theta\theta
.f}(\theta)=\hat{v}_{\theta\theta.f}(\theta)\hat{V},$ with $\hat{v}%
_{ff}(\theta),$ $\hat{v}_{\theta f}(\theta),$ $\hat{v}_{\theta\theta.f}(%
\theta)$ scalar and $\hat{V}$ a $k_{f}\times k_{f}$ matrix, the ratio in the
last line of the above expression simplifies to $\frac{\hat{v}_{\theta
f}(\theta)}{\hat{v}_{ff}(\theta)}$ so:%
\begin{equation*}
\begin{array}{l}
\frac{1}{2}\frac{\partial}{\partial\theta}DRLM(\theta) \\ 
\begin{array}{cl}
= & T \left( \frac{f_{T}(\theta,X)^{\prime}\hat{V}_{ff}(\theta)^{-1}\hat {D}%
(\theta)}{\left[ f_{T}(\theta,X)^{\prime}\hat{V}_{ff}(\theta)^{-1}\hat {V}%
_{\theta\theta.f}(\theta)\hat{V}_{ff}(\theta)^{-1}f_{T}(\theta,X)+\hat {D}%
(\theta)^{\prime}\hat{V}_{ff}(\theta)^{-1}\hat{D}(\theta)\right] }\right) \\ 
& \left( \hat{D}(\theta)^{\prime}\hat{V}_{ff}(\theta)^{-1}\hat{D}%
(\theta)-f_{T}(\theta,X)^{\prime}\hat{V}_{ff}(\theta)^{-1}\hat{V}%
_{\theta\theta.f}(\theta)\hat{V}_{ff}(\theta)^{-1}f_{T}(\theta,X)\right) \\ 
= & \left( \frac{\left( \hat{V}_{ff}(\theta)^{-\frac{1}{2}%
}f_{T}(\theta,X)\right) ^{\prime}\left( \hat{V}_{\theta\theta.f}(\theta )^{-%
\frac{1}{2}}\hat{D}(\theta)\right) }{\left[ f_{T}(\theta,X)^{\prime}\hat{V}%
_{ff}(\theta)^{-1}f_{T}(\theta,X)+\hat{D}(\theta)^{\prime}\hat {V}%
_{\theta\theta.f}(\theta)^{-1}\hat{D}(\theta)\right] }\right) \\ 
& \left( T\hat{D}(\theta)^{\prime}\hat{V}_{\theta\theta.f}(\theta)^{-1}\hat{D%
}(\theta)-Tf_{T}(\theta,X)^{\prime}\hat{V}_{ff}(\theta)^{-1}f_{T}(\theta,X)%
\right) \left( \frac{\hat{v}_{\theta\theta.f}(\theta)}{\hat {v}_{ff}(\theta)}%
\right) ^{\frac{1}{2}}.%
\end{array}%
\end{array}%
\end{equation*}

\subsection{Proof of Theorem 7}

We first construct the limit behavior of $\hat{D}(\lambda_{F}^{1})$ and $%
\hat{\mu}_{f}(\lambda_{F}^{1})$ when the (pseudo-) true value of $\lambda
_{F}$ equals $\lambda_{F}^{\ast}$ so we use that 
\begin{equation*}
R_{t}=\mu_{R}-\beta\lambda_{F}^{\ast}+\beta(\bar{F}_{t}+\lambda_{F}^{\ast
})+u_{t},
\end{equation*}
with $\frac{1}{T}\sum_{t=1}^{T}\bar{F}_{t}=0:$

\begin{equation*}
\begin{array}{rl}
-\hat{D}(\lambda_{F}^{1})= & \frac{1}{T}\sum_{t=1}^{T}R_{t}(\bar{F}%
_{t}+\lambda_{F}^{1})^{\prime}\left[ \frac{1}{T}\sum_{t=1}^{T}(\bar{F}%
_{t}+\lambda_{F}^{1})(\bar{F}_{t}+\lambda_{F}^{1})^{\prime}\right] ^{-1} \\ 
= & \frac{1}{T}\sum_{t=1}^{T}\left( \mu_{R}-\beta\lambda_{F}^{\ast}+\beta(%
\bar{F}_{t}+\lambda_{F}^{\ast})+u_{t}\right) (\bar{F}_{t}+\lambda
_{F}^{1})^{\prime} \\ 
& \left[ \frac{1}{T}\sum_{t=1}^{T}(\bar{F}_{t}+\lambda_{F}^{1})(\bar{F}%
_{t}+\lambda_{F}^{1})^{\prime}\right] ^{-1} \\ 
= & \frac{1}{T}\sum_{t=1}^{T}\left( \mu_{R}-\beta\lambda_{F}^{\ast}+\beta(%
\bar{F}_{t}+\lambda_{F}^{1}+\lambda_{F}^{\ast}-\lambda_{F}^{1})+u_{t}\right)
(\bar{F}_{t}+\lambda_{F}^{1})^{\prime} \\ 
& \left[ \frac{1}{T}\sum_{t=1}^{T}(\bar{F}_{t}+\lambda_{F}^{1})(\bar{F}%
_{t}+\lambda_{F}^{1})^{\prime}\right] ^{-1} \\ 
= & \beta+\frac{1}{T}\sum_{t=1}^{T}\left( \mu_{R}-\beta\lambda_{F}^{\ast
}+\beta(\lambda_{F}^{\ast}-\lambda_{F}^{1})+u_{t}\right) (\bar{F}%
_{t}+\lambda_{F}^{1})^{\prime} \\ 
& \left[ \frac{1}{T}\sum_{t=1}^{T}(\bar{F}_{t}+\lambda_{F}^{1})(\bar{F}%
_{t}+\lambda_{F}^{1})^{\prime}\right] ^{-1} \\ 
= & \beta+\frac{1}{T}\sum_{t=1}^{T}\left(
\mu_{R}-\beta\lambda_{F}^{1}+u_{t}\right) (\bar{F}_{t}+\lambda_{F}^{1})^{%
\prime}\left[ \frac{1}{T}\sum_{t=1}^{T}(\bar{F}_{t}+\lambda_{F}^{1})(\bar{F}%
_{t}+\lambda_{F}^{1})^{\prime}\right] ^{-1}%
\end{array}%
\end{equation*}
so 
\begin{equation*}
\begin{array}{rlll}
\sqrt{T}\left( \hat{D}(\lambda_{F}^{1})-D(\lambda_{F}^{1})\right) & \underset%
{d}{\rightarrow} & \psi_{\theta.f}(\lambda_{F}^{1}) & \Leftrightarrow \\ 
\sqrt{T}\left( \hat{D}(\lambda_{F}^{1})-D(\lambda_{F}^{\ast})-\left( \mu
_{R}-\beta\lambda_{F}^{\ast}\right) \lambda_{F}^{\ast\prime}(Q_{\bar{F}\bar{F%
}}+\lambda_{F}^{\ast}\lambda_{F}^{\ast\prime})^{-1}+\right. &  &  &  \\ 
\left. \left( \mu_{R}-\beta\lambda_{F}^{1}\right) \lambda_{F}^{1\prime }(Q_{%
\bar{F}\bar{F}}+\lambda_{F}^{1}\lambda_{F}^{1\prime})^{-1}\right) & \underset%
{d}{\rightarrow} & \psi_{\theta.f}(\lambda_{F}^{1}) & 
\end{array}%
\end{equation*}
with $-D(\lambda_{F})=\beta+\left( \mu_{R}-\beta\lambda_{F}\right)
\lambda_{F}^{\prime}(Q_{\bar{F}\bar{F}}+\lambda_{F}\lambda_{F}^{%
\prime})^{-1},$ $\psi_{\theta.f}(\lambda_{F}^{1})\sim N(0,(Q_{\bar {F}\bar{F}%
}+\lambda_{F}^{1}\lambda_{F}^{1\prime})^{-1}\otimes\Omega),$ and%
\begin{equation*}
\begin{array}{rl}
\bar{R}-\hat{\beta}\lambda_{F}^{1}= & \mu_{R}+\frac{1}{T}%
\sum_{t=1}^{T}u_{t}-\beta\lambda_{F}^{1}-\frac{1}{T}\sum_{t=1}^{T}u_{t}\bar{F%
}_{t}^{\prime }\left[ \frac{1}{T}\sum_{t=1}^{T}\bar{F}_{t}\bar{F}%
_{t}^{\prime}\right] ^{-1}\lambda_{F}^{1} \\ 
= & \mu_{R}-\beta\lambda_{F}^{\ast}+\beta(\lambda_{F}^{\ast}-%
\lambda_{F}^{1})+\frac{1}{T}\sum_{t=1}^{T}u_{t}-\frac{1}{T}%
\sum_{t=1}^{T}u_{t}\bar{F}_{t}^{\prime}\left[ \frac{1}{T}\sum_{t=1}^{T}\bar{F%
}_{t}\bar{F}_{t}^{\prime }\right] ^{-1}\lambda_{F}^{1} \\ 
= & \mu_{R}-\beta\lambda_{F}^{\ast}-D(\lambda_{F}^{\ast})(\lambda_{F}^{%
\ast}-\lambda_{F}^{1})-\left( \mu_{R}-\beta\lambda_{F}^{\ast}\right) \lambda
_{F}^{\ast\prime}(Q_{\bar{F}\bar{F}}+\lambda_{F}^{\ast}\lambda_{F}^{\ast
\prime})^{-1}(\lambda_{F}^{\ast}-\lambda_{F}^{1})+ \\ 
& \frac{1}{T}\sum_{t=1}^{T}u_{t}-\frac{1}{T}\sum_{t=1}^{T}u_{t}\bar{F}%
_{t}^{\prime}\left[ \frac{1}{T}\sum_{t=1}^{T}\bar{F}_{t}\bar{F}_{t}^{\prime }%
\right] ^{-1}\lambda_{F}^{1} \\ 
= & (\mu_{R}-\beta\lambda_{F}^{\ast})\left[ \left( 1-\lambda_{F}^{\ast
\prime}(Q_{\bar{F}\bar{F}}+\lambda_{F}^{\ast}\lambda_{F}^{\ast\prime})^{-1}%
\lambda_{F}^{\ast}\right) +\lambda_{F}^{\ast\prime}(Q_{\bar{F}\bar{F}%
}+\lambda_{F}^{\ast}\lambda_{F}^{\ast\prime})^{-1}\lambda_{F}^{1}\right] -
\\ 
& D(\lambda_{F}^{\ast})(\lambda_{F}^{\ast}-\lambda_{F}^{1})+\frac{1}{T}%
\sum_{t=1}^{T}u_{t}-\frac{1}{T}\sum_{t=1}^{T}u_{t}\bar{F}_{t}^{\prime}\left[ 
\frac {1}{T}\sum_{t=1}^{T}\bar{F}_{t}\bar{F}_{t}^{\prime}\right]
^{-1}\lambda _{F}^{1} \\ 
= & (\mu_{R}-\beta\lambda_{F}^{\ast})\left[ (1+\lambda_{F}^{\ast\prime }Q_{%
\bar{F}\bar{F}}^{-1}\lambda_{F}^{\ast})^{-1}+\lambda_{F}^{\ast\prime }(Q_{%
\bar{F}\bar{F}}+\lambda_{F}^{\ast}\lambda_{F}^{\ast\prime})^{-1}%
\lambda_{F}^{1}\right] -D(\lambda_{F}^{\ast})(\lambda_{F}^{\ast}-%
\lambda_{F}^{1})+ \\ 
& \frac{1}{T}\sum_{t=1}^{T}u_{t}-\frac{1}{T}\sum_{t=1}^{T}u_{t}\bar{F}%
_{t}^{\prime}\left[ \frac{1}{T}\sum_{t=1}^{T}\bar{F}_{t}\bar{F}_{t}^{\prime }%
\right] ^{-1}\lambda_{F}^{1},%
\end{array}%
\end{equation*}
\begin{equation*}
\begin{array}{rlll}
\sqrt{T}\left( (\bar{R}-\hat{\beta}\lambda_{F}^{1})-\left[
\mu_{f}(\lambda_{F}^{\ast})(1+\lambda_{F}^{\ast\prime}Q_{\bar{F}\bar{F}%
}^{-1}\lambda_{F}^{\ast})^{-1}-D(\lambda_{F}^{\ast})(\lambda_{F}^{\ast}-%
\lambda _{F}^{1})+\right. \right. &  &  &  \\ 
\left. \left. \mu_{f}(\lambda_{F}^{\ast})\lambda_{F}^{\ast\prime}(Q_{\bar {F}%
\bar{F}}+\lambda_{F}^{\ast}\lambda_{F}^{\ast\prime})^{-1}\lambda_{F}^{1}%
\right] \right) & \underset{d}{\rightarrow} & \psi_{f}(\lambda_{F}^{1}) & 
\end{array}%
\end{equation*}
with $\mu_{f}(\lambda_{F}^{\ast})=\mu_{R}-\beta\lambda_{F}^{\ast}$ and $%
\psi_{f}(\lambda_{F}^{1})\sim N(0,(1+\lambda_{F}^{1\prime}Q_{\bar{F}\bar{F}%
}^{-1}\lambda_{F}^{1})\Omega)$ and independent of $\psi_{\theta.f}(\lambda
_{F}^{1}).$

For testing H$_{0}:\lambda_{F}=0,$ so $\lambda_{F}^{1}=0,$ the above
expressions simplify to:%
\begin{equation*}
\begin{array}{rll}
\sqrt{T}\left( \hat{D}(\lambda_{F}^{1}=0)-\left[ D(\lambda_{F}^{\ast})+%
\mu_{f}(\lambda_{F}^{\ast})\lambda_{F}^{\ast\prime}(Q_{\bar{F}\bar{F}%
}+\lambda_{F}^{\ast}\lambda_{F}^{\ast\prime})^{-1}\right] \right) & \underset%
{d}{\rightarrow} & \psi_{\theta.f}(\lambda_{F}^{1}=0) \\ 
\sqrt{T}\left( \bar{R}-\left[ \mu_{f}(\lambda_{F}^{\ast})(1+\lambda
_{F}^{\ast\prime}Q_{\bar{F}\bar{F}}^{-1}\lambda_{F}^{\ast})^{-1}-D(\lambda
_{F}^{\ast})\lambda_{F}^{\ast}\right] \right) & \underset{d}{\rightarrow} & 
\psi_{f}(\lambda_{F}^{1}=0)%
\end{array}%
\end{equation*}
with $\psi_{\theta.f}(\lambda_{F}^{1}=0)\sim N(0, Q_{\bar{F}\bar{F}%
}^{-1}\otimes \Omega)$ and $\psi_{f}(\lambda_{F}^{1}=0)\sim N(0,\Omega).$ We
next use that $\mu^{\ast}=\lim_{T\rightarrow\infty}\sqrt{T}%
\mu_{f}(\lambda_{F}^{\ast }),$ $D^{\ast}=\lim_{T\rightarrow\infty}\sqrt{T}%
D(\lambda_{F}^{\ast}),$ $\bar{\mu}=\Omega^{-\frac{1}{2}}\mu^{\ast}(1+%
\lambda_{F}^{\ast\prime}Q_{\bar{F}\bar{F}}^{-1}\lambda_{F}^{\ast})^{-\frac{1%
}{2}},\mathbf{\ }\bar {D}=\Omega^{-\frac{1}{2}}D^{\ast}(Q_{\bar{F}\bar{F}%
}+\lambda_{F}^{\ast}\lambda_{F}^{\ast\prime})^{\frac{1}{2}}$ so for $m=1,$ $%
Q_{\bar{F}\bar{F}}=1: $%
\begin{equation*}
\begin{array}{rll}
\sqrt{T}\hat{\Omega}^{-\frac{1}{2}}\bar{R} & \underset{d}{\rightarrow} & 
\bar{\mu}(1+(\lambda_{F}^{\ast})^{2})^{-\frac{1}{2}}-\bar{D}(1+(\lambda
_{F}^{\ast})^{2})^{-\frac{1}{2}}\lambda_{F}^{\ast}+\psi_{f}^{\ast}(\lambda
_{F}^{1}=0) \\ 
\sqrt{T}\hat{\Omega}^{-\frac{1}{2}}\hat{D}(\lambda_{F}^{1}=0) & \underset{d}{%
\rightarrow} & \bar{D}(1+(\lambda_{F}^{\ast})^{2})^{-\frac{1}{2}}+\bar {\mu}%
(1+(\lambda_{F}^{\ast})^{2})^{-\frac{1}{2}}\lambda_{F}^{\ast}+\psi_{%
\theta.f}^{\ast}(\lambda_{F}^{1}=0),%
\end{array}%
\end{equation*}
with $\psi_{f}^{\ast}(\lambda_{F}^{1}=0)$ and $\psi_{\theta.f}^{\ast}(%
\lambda_{F}^{1}=0)$ independent standard normal $N$ dimensional random
vectors.

\subsection{Proof of Theorem 8}

We first specify: $\mathcal{U}_{1}\mathcal{S}_{1}\mathcal{V}_{1}^{\prime
}=-\Omega ^{-\frac{1}{2}}D(\lambda _{F}^{\ast })\left( 
\begin{array}{cc}
\lambda _{F}^{\ast } & I_{m}%
\end{array}%
\right) \left( 
\begin{array}{cc}
1 & 0 \\ 
0 & Q_{\bar{F}\bar{F}}^{\frac{1}{2}}%
\end{array}%
\right) ,\text{ }$ so $D(\lambda _{F}^{\ast })=-\Omega ^{\frac{1}{2}}%
\mathcal{U}_{1}S_{1}\mathcal{V}_{21}^{\prime }Q_{\bar{F}\bar{F}}^{-\frac{1}{2%
}}$,\ $\lambda _{F}^{\ast }$ = $Q_{\bar{F}\bar{F}}^{\frac{1}{2}}\mathcal{V}%
_{21}^{\prime -1}\mathcal{V}_{11}^{\prime }.$ We next specify: $\mathcal{U}%
_{2}\mathcal{S}_{2}\mathcal{V}_{2}^{\prime }=\Omega ^{\frac{1}{2}}D(\lambda
_{F}^{\ast })_{\perp }\delta \left( 
\begin{array}{cc}
\lambda _{F}^{\ast } & I_{m}%
\end{array}%
\right) _{\perp }\left( 
\begin{array}{cc}
1 & 0 \\ 
0 & Q_{\bar{F}\bar{F}}^{-\frac{1}{2}}%
\end{array}%
\right) ,\text{ }$ so for $D(\lambda _{F}^{\ast })=\Omega ^{\frac{1}{2}%
}(D(\lambda _{F}^{\ast })_{1}^{\prime }$ $D(\lambda _{F}^{\ast
})_{2}^{\prime })^{\prime },$ with $D(\lambda _{F}^{\ast })_{1}^{\prime}=-Q_{%
\bar{F}\bar{F}}^{-\frac{1}{2}\prime }\mathcal{V}_{21}\mathcal{S}_{1}\mathcal{%
U}_{11}^{\prime }:m\times m,$ $D(\lambda _{F}^{\ast })_{2}^{\prime}=-Q_{\bar{%
F}\bar{F}}^{-\frac{1}{2}\prime }\mathcal{V}_{21}\mathcal{S}_{1}\mathcal{U}%
_{21}^{\prime }: m \times (N-m):$%
\begin{equation*}
\begin{array}{rl}
D(\lambda _{F}^{\ast })_{\perp }= & \Omega ^{-\frac{1}{2}}\left( 
\begin{array}{l}
-\mathcal{U}_{11}^{\prime -1}\mathcal{S}_{1}^{-1}\mathcal{V}_{21}^{-1}Q_{%
\bar{F}\bar{F}}^{\frac{1}{2}\prime }Q_{\bar{F}\bar{F}}^{-\frac{1}{2}\prime }%
\mathcal{V}_{21}\mathcal{S}_{1}\mathcal{U}_{21}^{\prime } \\ 
\quad \quad \quad \quad \quad \quad I_{N-m}%
\end{array}%
\right) \\ 
& (I_{N-m}+\mathcal{U}_{21}\mathcal{S}_{1}^{-1}\mathcal{V}_{21}^{^{\prime }}%
\mathcal{V}_{21}^{-1\prime }\mathcal{S}_{1}^{-1}\mathcal{U}_{11}^{-1}%
\mathcal{U}_{11}^{\prime -1}\mathcal{S}_{1}^{-1}\mathcal{V}_{21}^{-1}%
\mathcal{V}_{21}\mathcal{S}_{1}\mathcal{U}_{21}^{\prime })^{-\frac{1}{2}} \\ 
= & \Omega ^{-\frac{1}{2}}\left( 
\begin{array}{l}
-\mathcal{U}_{11}^{\prime -1}\mathcal{U}_{21}^{\prime } \\ 
\quad I_{N-m}%
\end{array}%
\right) (I_{N-m}+\mathcal{U}_{21}\mathcal{U}_{11}^{-1}\mathcal{U}%
_{11}^{\prime -1}\mathcal{U}_{21}^{\prime })^{-\frac{1}{2}} \\ 
= & \Omega ^{-\frac{1}{2}}\left( 
\begin{array}{l}
\mathcal{U}_{12}\mathcal{U}_{22}^{-1} \\ 
\quad I_{N-m}%
\end{array}%
\right) (I_{N-m}+\mathcal{U}_{22}^{-1\prime }\mathcal{U}_{12}^{\prime }%
\mathcal{U}_{12}\mathcal{U}_{22}^{-1})^{-\frac{1}{2}} \\ 
= & \Omega ^{-\frac{1}{2}}\left( 
\begin{array}{l}
\mathcal{U}_{12} \\ 
\mathcal{U}_{22}%
\end{array}%
\right) \mathcal{U}_{22}^{-1}(\mathcal{U}_{22}^{-1\prime }(\mathcal{U}%
_{12}^{\prime }\mathcal{U}_{12}+\mathcal{U}_{22}^{\prime }\mathcal{U}_{22})%
\mathcal{U}_{22}^{-1})^{-\frac{1}{2}} \\ 
= & \Omega ^{-\frac{1}{2}}\left( 
\begin{array}{l}
\mathcal{U}_{12} \\ 
\mathcal{U}_{22}%
\end{array}%
\right) \mathcal{U}_{22}^{-1}(\mathcal{U}_{22}^{-1\prime }\mathcal{U}%
_{22}^{-1})^{-\frac{1}{2}} \\ 
= & \Omega ^{-\frac{1}{2}}\left( 
\begin{array}{l}
\mathcal{U}_{12} \\ 
\mathcal{U}_{22}%
\end{array}%
\right) \mathcal{U}_{22}^{-1}(\mathcal{U}_{22}\mathcal{U}_{22}^{\prime })^{%
\frac{1}{2}} \\ 
= & \Omega ^{-\frac{1}{2}}\mathcal{U}_{2}\mathcal{U}_{22}^{-1}(\mathcal{U}%
_{22}\mathcal{U}_{22}^{\prime })^{\frac{1}{2}}%
\end{array}%
\end{equation*}%
since $\mathcal{U}_{11}^{\prime }\mathcal{U}_{12}+\mathcal{U}_{21}^{\prime }%
\mathcal{U}_{22}=0$ (because of the orthogonality of $\mathcal{U}$), $%
\mathcal{U}_{12}\mathcal{U}_{22}^{-1}=-\mathcal{U}_{11}^{\prime -1}\mathcal{U%
}_{21}^{\prime },$ and $\mathcal{U}_{12}^{\prime }\mathcal{U}_{12}+\mathcal{U%
}_{22}^{\prime }\mathcal{U}_{22}=I_{N-m},$ and 
\begin{equation*}
\begin{array}{rl}
\left( 
\begin{array}{cc}
\lambda _{F}^{\ast } & I_{m}%
\end{array}%
\right) _{\perp }= & (1+\mathcal{V}_{11}\mathcal{V}_{21}^{-1}\mathcal{V}%
_{21}^{\prime -1}\mathcal{V}_{11}^{\prime })^{-\frac{1}{2}}\left( 
\begin{array}{ll}
1 & -\mathcal{V}_{11}\mathcal{V}_{21}^{-1}%
\end{array}%
\right) \left( 
\begin{array}{cc}
1 & 0 \\ 
0 & Q_{\bar{F}\bar{F}}^{\frac{1}{2}}%
\end{array}%
\right) \\ 
= & (1+\mathcal{V}_{12}^{-1\prime }\mathcal{V}_{22}^{\prime }\mathcal{V}_{22}%
\mathcal{V}_{12}^{-1})^{-\frac{1}{2}}\left( 
\begin{array}{ll}
1 & \mathcal{V}_{12}^{-1\prime }\mathcal{V}_{22}^{\prime }%
\end{array}%
\right) \left( 
\begin{array}{cc}
1 & 0 \\ 
0 & Q_{\bar{F}\bar{F}}^{\frac{1}{2}}%
\end{array}%
\right) \\ 
= & (\mathcal{V}_{12}^{-1\prime }(\mathcal{V}_{12}^{\prime }\mathcal{V}_{12}+%
\mathcal{V}_{22}^{\prime }\mathcal{V}_{22})\mathcal{V}_{12}^{-1})^{-\frac{1}{%
2}}\mathcal{V}_{12}^{-1\prime }\left( 
\begin{array}{ll}
\mathcal{V}_{12}^{\prime } & \mathcal{V}_{22}^{\prime }%
\end{array}%
\right) \left( 
\begin{array}{cc}
1 & 0 \\ 
0 & Q_{\bar{F}\bar{F}}^{\frac{1}{2}}%
\end{array}%
\right) \\ 
= & (\mathcal{V}_{12}^{-1\prime }\mathcal{V}_{12}^{-1})^{-\frac{1}{2}}%
\mathcal{V}_{12}^{-1\prime }\left( 
\begin{array}{ll}
\mathcal{V}_{12}^{\prime } & \mathcal{V}_{22}^{\prime }%
\end{array}%
\right) \left( 
\begin{array}{cc}
1 & 0 \\ 
0 & Q_{\bar{F}\bar{F}}^{\frac{1}{2}}%
\end{array}%
\right) \\ 
= & (\mathcal{V}_{12}\mathcal{V}_{12}^{\prime })^{\frac{1}{2}}\mathcal{V}%
_{12}^{-1\prime }\mathcal{V}_{2}^{\prime }\left( 
\begin{array}{cc}
1 & 0 \\ 
0 & Q_{\bar{F}\bar{F}}^{\frac{1}{2}}%
\end{array}%
\right)%
\end{array}%
\end{equation*}%
since $\mathcal{V}_{11}^{\prime }\mathcal{V}_{12}+\mathcal{V}_{21}^{\prime }%
\mathcal{V}_{22}=0,$ so $-\mathcal{V}_{21}^{\prime -1}\mathcal{V}%
_{11}^{\prime }=\mathcal{V}_{22}\mathcal{V}_{12}^{-1},$ and $\mathcal{V}%
_{12}^{\prime }\mathcal{V}_{12}+\mathcal{V}_{22}^{\prime }\mathcal{V}%
_{22}=1, $ from which it then results that 
\begin{equation*}
\delta =(\mathcal{U}_{22}\mathcal{U}_{22}^{\prime })^{-\frac{1}{2}}\mathcal{U%
}_{22}S_{2}\mathcal{V}_{12}^{\prime }(\mathcal{V}_{12}\mathcal{V}%
_{12}^{\prime })^{-\frac{1}{2}}.
\end{equation*}

\subsection{Proof of Theorem 9}

The proof that the quadratic form of 
\begin{equation*}
\begin{array}{l}
\Omega^{-\frac{1}{2}}\left( 
\begin{array}{cc}
\bar{R} & \hat{\beta}%
\end{array}
\right) \left( 
\begin{array}{cc}
\left( 
\begin{array}{c}
1 \\ 
-\lambda_{F}^{1}%
\end{array}
\right) (1+\lambda_{F}^{1\prime}Q_{FF}^{-1}\lambda_{F}^{1})^{-\frac{1}{2}} & 
\left( 
\begin{array}{cc}
1 & 0 \\ 
0 & Q_{\bar{F}\bar{F}}%
\end{array}
\right) \left( 
\begin{array}{cc}
\lambda_{F}^{1} & I_{m}%
\end{array}
\right) ^{\prime}(Q_{\bar{F}\bar{F}}+\lambda_{F}^{1}\lambda_{F}^{1\prime
})^{-\frac{1}{2}}%
\end{array}
\right) ,%
\end{array}%
\end{equation*}
is a maximal invariant follows along the lines of Andrews et al. (2006). It
uses that 
\begin{equation*}
\begin{array}{c}
\sqrt{T}\Omega^{-\frac{1}{2}}\left( 
\begin{array}{cc}
\bar{R} & \hat{\beta}%
\end{array}
\right) \left( 
\begin{array}{cc}
1 & 0 \\ 
0 & Q_{\bar{F}\bar{F}}^{\frac{1}{2}}%
\end{array}
\right) =\Omega^{-\frac{1}{2}}\left( 
\begin{array}{cc}
\ddot{\mu}_{R} & \ddot{\beta}%
\end{array}
\right) \left( 
\begin{array}{cc}
1 & 0 \\ 
0 & Q_{\bar{F}\bar{F}}^{\frac{1}{2}}%
\end{array}
\right) +\psi_{R\beta},%
\end{array}%
\end{equation*}

\noindent with vec($\psi_{R\beta})\sim N(0,I_{N(m+1)}),$ is post-multiplied
by the orthonormal matrices $\left( 
\begin{array}{cc}
1 & 0 \\ 
0 & Q_{\bar{F}\bar{F}}^{-\frac{1}{2}}%
\end{array}
\right) \left( 
\begin{array}{c}
1 \\ 
-\lambda_{F}^{1}%
\end{array}
\right) (1+\lambda_{F}^{1\prime}Q_{FF}^{-1}\lambda_{F}^{1})^{-\frac{1}{2}} $
and $\left( 
\begin{array}{cc}
1 & 0 \\ 
0 & Q_{\bar{F}\bar{F}}^{\frac{1}{2}}%
\end{array}
\right) \left( 
\begin{array}{cc}
\lambda_{F}^{1} & I_{m}%
\end{array}
\right) ^{\prime}(Q_{\bar{F}\bar{F}}+\lambda_{F}^{1}\lambda_{F}^{1})^{-\frac{%
1}{2}}.$ We next construct the distributions of the two elements in the
above expression for the cases of correct specification and
misspecification. For the latter we use the specification from Theorem 8.

\noindent\textbf{Correct specification. }Without misspecification, $\ddot{%
\mu }_{R}=\ddot{\beta}\lambda_{F}^{\ast}$ so $\Omega^{-\frac{1}{2}}\left( 
\begin{array}{cc}
\ddot{\mu}_{R} & \ddot{\beta}%
\end{array}
\right) =\Omega^{-\frac{1}{2}}\ddot{\beta}\left( 
\begin{array}{cc}
\lambda_{F}^{\ast} & I_{m}%
\end{array}
\right) $ and 
\begin{equation*}
\begin{array}{rl}
\Omega^{-\frac{1}{2}}\hat{\mu}(\lambda_{F})^{\ast}= & \Omega^{-\frac{1}{2}%
}\left( 
\begin{array}{cc}
\bar{R} & \hat{\beta}%
\end{array}
\right) \left( 
\begin{array}{cc}
1 & 0 \\ 
0 & Q_{\bar{F}\bar{F}}^{\frac{1}{2}}%
\end{array}
\right) \left( 
\begin{array}{cc}
1 & 0 \\ 
0 & Q_{\bar{F}\bar{F}}^{-\frac{1}{2}}%
\end{array}
\right) \left( 
\begin{array}{c}
1 \\ 
-\lambda_{F}^{1}%
\end{array}
\right) (1+\lambda_{F}^{1\prime}Q_{FF}^{-1}\lambda_{F}^{1})^{-\frac{1}{2}}
\\ 
= & \Omega^{-\frac{1}{2}}\left( \bar{R}-\hat{\beta}\lambda_{F}^{1}\right)
(1+\lambda_{F}^{1\prime}Q_{FF}^{-1}\lambda_{F}^{1})^{-\frac{1}{2}} \\ 
= & \Omega^{-\frac{1}{2}}\ddot{\beta}(\lambda_{F}^{\ast}-\lambda_{F}^{1})(1+%
\lambda_{F}^{\prime}Q_{FF}^{-1}\lambda_{F})^{-\frac{1}{2}}+\psi _{\perp},%
\end{array}%
\end{equation*}
with $\psi_{\perp}=\psi_{R\beta}\left( 
\begin{array}{cc}
1 & 0 \\ 
0 & Q_{\bar{F}\bar{F}}^{-\frac{1}{2}}%
\end{array}
\right) \left( 
\begin{array}{c}
1 \\ 
-\lambda_{F}^{1}%
\end{array}
\right) (1+\lambda_{F}^{1\prime}Q_{FF}^{-1}\lambda_{F}^{1})^{-\frac{1}{2}%
}\sim N(0,I_{N}),$ and 

\begin{equation*}
\begin{array}{rl}
\Omega^{-\frac{1}{2}}\hat{D}(\lambda_{F}^{1})^{\ast}= & \Omega^{-\frac{1}{2}%
}\left( 
\begin{array}{cc}
\bar{R} & \hat{\beta}%
\end{array}
\right) \left( 
\begin{array}{cc}
1 & 0 \\ 
0 & Q_{\bar{F}\bar{F}}%
\end{array}
\right) \left( 
\begin{array}{cc}
\lambda_{F}^{1} & I_{m}%
\end{array}
\right) ^{\prime}(Q_{\bar{F}\bar{F}}+\lambda_{F}^{1}\lambda_{F}^{1\prime
})^{-\frac{1}{2}} \\ 
= & \Omega^{-\frac{1}{2}}\ddot{\beta}\left( 
\begin{array}{cc}
\lambda_{F}^{\ast} & I_{m}%
\end{array}
\right) \left( 
\begin{array}{cc}
1 & 0 \\ 
0 & Q_{\bar{F}\bar{F}}%
\end{array}
\right) \left( 
\begin{array}{cc}
\lambda_{F}^{1} & I_{m}%
\end{array}
\right) ^{\prime}(Q_{\bar{F}\bar{F}}+\lambda_{F}^{1}\lambda_{F}^{1\prime
})^{-\frac{1}{2}}+\psi_{\lambda_{F}^{1}},%
\end{array}%
\end{equation*}
with $\psi_{\lambda_{F}^{1}}=\psi_{R\beta}\left( 
\begin{array}{cc}
1 & 0 \\ 
0 & Q_{\bar{F}\bar{F}}^{\frac{1}{2}}%
\end{array}
\right) \left( 
\begin{array}{cc}
\lambda_{F}^{1} & I_{m}%
\end{array}
\right) ^{\prime}(Q_{\bar{F}\bar{F}}+\lambda_{F}^{1}\lambda_{F}^{1})^{-\frac{%
1}{2}},$ vec($\psi_{\lambda_{F}^{1}})\sim N(0,I_{Nm}),$ and independent of $%
\psi_{\perp}.$ The maximal invariant is the quadratic form of the above two
components so it consists of the three elements:%
\begin{equation*}
\begin{array}{cl}
S_{\lambda_{F}^{1}\lambda_{F}^{1}}= & (Q_{\bar{F}\bar{F}}+\lambda_{F}^{1}%
\lambda_{F}^{1\prime})^{-\frac{1}{2}\prime}\left( Q_{\bar{F}\bar{F}%
}+\lambda_{F}^{\ast}\lambda_{F}^{1\prime}\right) ^{\prime}\ddot{\beta }%
^{\prime}\Omega^{-1}\ddot{\beta}\left( Q_{\bar{F}\bar{F}}+\lambda_{F}^{\ast
}\lambda_{F}^{1\prime}\right) (Q_{\bar{F}\bar{F}}+\lambda_{F}^{1}\lambda
_{F}^{1\prime})^{-\frac{1}{2}}+ \\ 
& (Q_{\bar{F}\bar{F}}+\lambda_{F}^{1}\lambda_{F}^{1\prime})^{-\frac{1}{2}%
\prime}\left( Q_{\bar{F}\bar{F}}+\lambda_{F}^{\ast}\lambda_{F}^{1\prime
}\right) ^{\prime}\ddot{\beta}^{\prime}\Omega^{-\frac{1}{2}%
\prime}\psi_{\lambda_{F}^{1}}+ \\ 
& \psi_{\lambda_{F}^{1}}^{\prime}\Omega^{-\frac{1}{2}}\ddot{\beta}\left( Q_{%
\bar{F}\bar{F}}+\lambda_{F}^{\ast}\lambda_{F}^{1\prime}\right) (Q_{\bar {F}%
\bar{F}}+\lambda_{F}^{1}\lambda_{F}^{1\prime})^{-\frac{1}{2}}+\psi
_{\lambda_{F}^{1}}^{\prime}\psi_{\lambda_{F}^{1}} \\
S_{\perp\perp}= & (1+\lambda_{F}^{\prime}Q_{FF}^{-1}\lambda_{F})^{-1}(%
\lambda_{F}^{\ast}-\lambda_{F}^{1})^{\prime}\ddot{\beta}^{\prime}\Omega ^{-1}%
\ddot{\beta}(\lambda_{F}^{\ast}-\lambda_{F}^{1})+ \\ 
& 2\psi_{\perp}^{\prime}\Omega^{-\frac{1}{2}}\ddot{\beta}(\lambda_{F}^{\ast
}-\lambda_{F}^{1})(1+\lambda_{F}^{1\prime}Q_{FF}^{-1}\lambda_{F}^{1})^{-%
\frac{1}{2}}+\psi_{\perp}^{\prime}\psi_{\perp}\\
S_{\lambda_{F}^{1}\perp}= & (Q_{\bar{F}\bar{F}}+\lambda_{F}^{1}\lambda
_{F}^{1\prime})^{-\frac{1}{2}\prime}\left( Q_{\bar{F}\bar{F}}+\lambda
_{F}^{\ast}\lambda_{F}^{1\prime}\right) ^{\prime}\ddot{\beta}%
^{\prime}\Omega^{-1}\ddot{\beta}(\lambda_{F}^{\ast}-\lambda_{F}^{1})(1+%
\lambda _{F}^{\prime}Q_{FF}^{-1}\lambda_{F})^{-\frac{1}{2}}+ \\ 
& (Q_{\bar{F}\bar{F}}+\lambda_{F}^{1}\lambda_{F}^{1\prime})^{-\frac{1}{2}%
\prime}\left( Q_{\bar{F}\bar{F}}+\lambda_{F}^{\ast}\lambda_{F}^{1\prime
}\right) ^{\prime}\ddot{\beta}^{\prime}\Omega^{-\frac{1}{2}%
\prime}\psi_{\perp}+ \\ 
& \psi_{\lambda_{F}^{1}}^{\prime}\Omega^{-\frac{1}{2}}\ddot{\beta}(\lambda
_{F}^{\ast}-\lambda_{F}^{1})(1+\lambda_{F}^{1\prime}Q_{FF}^{-1}%
\lambda_{F}^{1})^{-\frac{1}{2}}+\psi_{\lambda_{F}^{1}}^{\prime}\psi_{\perp}.%
\end{array}%
\end{equation*}

\noindent\textbf{Misspecification. }To specify the maximal invariant under
misspecification, we use the singular value decomposition from Theorem 8:%
\begin{equation*}
\begin{array}{rl}
\Omega^{-\frac{1}{2}}\hat{D}(\lambda_{F}^{1})^{\ast}= & \Omega^{-\frac{1}{2}%
}\left( 
\begin{array}{cc}
\bar{R} & \hat{\beta}%
\end{array}
\right) \left( 
\begin{array}{cc}
1 & 0 \\ 
0 & Q_{\bar{F}\bar{F}}^{\frac{1}{2}}%
\end{array}
\right) \left( 
\begin{array}{cc}
1 & 0 \\ 
0 & Q_{\bar{F}\bar{F}}^{\frac{1}{2}}%
\end{array}
\right) \left( 
\begin{array}{cc}
\lambda_{F}^{1} & I_{m}%
\end{array}
\right) ^{\prime}(Q_{\bar{F}\bar{F}}+\lambda_{F}^{1}\lambda_{F}^{1\prime
})^{-\frac{1}{2}} \\ 
= & \Omega^{-\frac{1}{2}}D(\lambda_{F}^{\ast})\left( 
\begin{array}{cc}
\lambda_{F}^{\ast} & I_{m}%
\end{array}
\right) \left( 
\begin{array}{cc}
1 & 0 \\ 
0 & Q_{\bar{F}\bar{F}}%
\end{array}
\right) \left( 
\begin{array}{cc}
\lambda_{F}^{1} & I_{m}%
\end{array}
\right) ^{\prime}(Q_{\bar{F}\bar{F}}+\lambda_{F}^{1}\lambda_{F}^{1\prime
})^{-\frac{1}{2}}+ \\ 
& \Omega^{\frac{1}{2}}D(\lambda_{F}^{\ast})_{\perp}\delta\left( 
\begin{array}{cc}
\lambda_{F}^{\ast} & I_{m}%
\end{array}
\right) _{\perp}\left( 
\begin{array}{cc}
\lambda_{F}^{1} & I_{m}%
\end{array}
\right) ^{\prime}(Q_{\bar{F}\bar{F}}+\lambda_{F}^{1}\lambda_{F}^{1\prime
})^{-\frac{1}{2}}+\psi_{\lambda_{F}^{1}} \\ 
= & \Omega^{-\frac{1}{2}}D(\lambda_{F}^{\ast})(Q_{\bar{F}\bar{F}}+\lambda
_{F}^{\ast}\lambda_{F}^{1\prime})(Q_{\bar{F}\bar{F}}+\lambda_{F}^{1}%
\lambda_{F}^{1\prime})^{-\frac{1}{2}}- \\ 
& \Omega^{\frac{1}{2}}D(\lambda_{F}^{\ast})_{\perp}\delta\left( \lambda
_{F}^{\ast}-\lambda_{F}^{1}\right) ^{\prime}(Q_{\bar{F}\bar{F}}+\lambda
_{F}^{1}\lambda_{F}^{1\prime})^{-\frac{1}{2}}(1+\lambda_{F}^{\ast%
\prime}Q_{FF}^{-1}\lambda_{F}^{\ast})^{-\frac{1}{2}}+\psi_{\lambda_{F}^{1}},%
\end{array}%
\end{equation*}
with vec($\psi_{\lambda_{F}^{1}})\sim N(0,I_{Nm}),$ and 
\begin{equation*}
\begin{array}{rl}
\Omega^{-\frac{1}{2}}\hat{\mu}(\lambda_{F})^{\ast}= & \Omega^{-\frac{1}{2}%
}\left( \bar{R}-\hat{\beta}\lambda_{F}^{1}\right) (1+\lambda_{F}^{1\prime
}Q_{FF}^{-1}\lambda_{F}^{1})^{-\frac{1}{2}} \\ 
= & \Omega^{-\frac{1}{2}}\left( 
\begin{array}{cc}
\bar{R} & \hat{\beta}%
\end{array}
\right) \left( 
\begin{array}{cc}
1 & 0 \\ 
0 & Q_{\bar{F}\bar{F}}^{\frac{1}{2}}%
\end{array}
\right) \left( 
\begin{array}{cc}
1 & 0 \\ 
0 & Q_{\bar{F}\bar{F}}^{-\frac{1}{2}}%
\end{array}
\right) \left( 
\begin{array}{c}
1 \\ 
-\lambda_{F}^{1}%
\end{array}
\right) (1+\lambda_{F}^{1\prime}\hat{Q}_{FF}^{-1}\lambda_{F}^{1})^{-\frac {1%
}{2}} \\ 
= & \Omega^{-\frac{1}{2}}D(\lambda_{F}^{\ast})\left( \lambda_{F}^{\ast
}-\lambda_{F}^{1}\right)
(1+\lambda_{F}^{1\prime}Q_{FF}^{-1}\lambda_{F}^{1})^{-\frac{1}{2}}+ \\ 
& \Omega^{\frac{1}{2}}D(\lambda_{F}^{\ast})_{\perp}\delta\left( 1+\lambda
_{F}^{\ast\prime}Q_{\bar{F}\bar{F}}^{-1}\lambda_{F}^{1}\right) (1+\lambda
_{F}^{1\prime}Q_{FF}^{-1}\lambda_{F}^{1})^{-\frac{1}{2}}(1+\lambda_{F}^{\ast%
\prime}Q_{FF}^{-1}\lambda_{F}^{\ast})^{-\frac{1}{2}}+\psi_{\perp},%
\end{array}%
\end{equation*}
with $\psi_{\perp}\sim N(0,I_{N})$ and independent of $\psi_{%
\lambda_{F}^{1}}.$ The maximal invariant is the quadratic form of the above
two components so it consists of the three elements:%
\begin{equation*}
\begin{array}{cl}
S_{\lambda_{F}^{1}\lambda_{F}^{1}}= & (Q_{\bar{F}\bar{F}}+\lambda_{F}^{1}%
\lambda_{F}^{1\prime})^{-\frac{1}{2}\prime}\left( Q_{\bar{F}\bar{F}%
}+\lambda_{F}^{\ast}\lambda_{F}^{1\prime}\right)
^{\prime}D(\lambda_{F}^{\ast})^{\prime}\Omega^{-1}D(\lambda_{F}^{\ast})%
\left( Q_{\bar{F}\bar{F}}+\lambda_{F}^{\ast}\lambda_{F}^{1\prime}\right) (Q_{%
\bar{F}\bar{F}}+\lambda_{F}^{1}\lambda_{F}^{1\prime})^{-\frac{1}{2}}+ \\ 
& (1+\lambda_{F}^{\ast\prime}Q_{FF}^{-1}\lambda_{F}^{\ast})^{-1}(Q_{\bar {F}%
\bar{F}}+\lambda_{F}^{1}\lambda_{F}^{1\prime})^{-\frac{1}{2}\prime}\left(
\lambda_{F}^{1}-\lambda_{F}^{\ast}\right) \delta^{\prime}D(\lambda_{F}^{\ast
})_{\perp}^{\prime}\Omega D(\lambda_{F}^{\ast})_{\perp} \\ 
& \delta\left( \lambda_{F}^{1}-\lambda_{F}^{\ast}\right) ^{\prime}(Q_{\bar{F}%
\bar{F}}+\lambda_{F}^{1}\lambda_{F}^{1\prime})^{-\frac{1}{2}%
}+\psi_{\lambda_{F}^{1}}^{\prime}\left[ \Omega^{-\frac{1}{2}}D(\lambda
_{F}^{\ast})(Q_{\bar{F}\bar{F}}+\lambda_{F}^{\ast}\lambda_{F}^{1\prime })(Q_{%
\bar{F}\bar{F}}+\lambda_{F}^{1}\lambda_{F}^{1\prime})^{-\frac{1}{2}}+\right.
\\ 
& \left. \Omega^{\frac{1}{2}}D(\lambda_{F}^{\ast})_{\perp}\delta\left(
\lambda_{F}^{1}-\lambda_{F}^{\ast}\right) ^{\prime}(Q_{\bar{F}\bar{F}%
}+\lambda_{F}^{1}\lambda_{F}^{1\prime})^{-\frac{1}{2}}(1+\lambda_{F}^{\ast%
\prime}Q_{FF}^{-1}\lambda_{F}^{\ast})^{-\frac{1}{2}}\right] + \\ 
& \left[ (Q_{\bar{F}\bar{F}}+\lambda_{F}^{1}\lambda_{F}^{1\prime})^{-\frac {1%
}{2}\prime}\left( Q_{\bar{F}\bar{F}}+\lambda_{F}^{\ast}\lambda_{F}^{1\prime}%
\right) ^{\prime}D(\lambda_{F}^{\ast})^{\prime}\Omega^{-\frac{1}{2}}+\right.
\\ 
& \left. (1+\lambda_{F}^{\ast\prime}Q_{\bar{F}\bar{F}}^{-1}\lambda_{F}^{\ast
})^{-\frac{1}{2}}(Q_{\bar{F}\bar{F}}+\lambda_{F}^{1}\lambda_{F}^{1\prime
})^{-\frac{1}{2}\prime}\left( \lambda_{F}^{1}-\lambda_{F}^{\ast}\right)
\delta^{\prime}D(\lambda_{F}^{\ast})_{\perp}^{\prime}\Omega^{\frac{1}{2}}%
\right] \psi_{\lambda_{F}^{1}}+\psi_{\lambda_{F}^{1}}^{\prime}\psi_{%
\lambda_{F}^{1}}\\
&\\
S_{\perp\perp}= & (1+\lambda_{F}^{1\prime}Q_{\bar{F}\bar{F}}^{-1}\lambda
_{F}^{1})^{-1}\left( \lambda_{F}^{\ast}-\lambda_{F}^{1}\right) ^{\prime
}D(\lambda_{F}^{\ast})^{\prime}\Omega^{-1}D(\lambda_{F}^{\ast})\left(
\lambda_{F}^{\ast}-\lambda_{F}^{1}\right) + \\ 
& \delta^{\prime}D(\lambda_{F}^{\ast})_{\perp}^{\prime}\Omega
D(\lambda_{F}^{\ast })_{\perp}\delta\left( 1+\lambda_{F}^{\ast\prime}Q_{\bar{%
F}\bar{F}}^{-1}\lambda_{F}^{1}\right) ^{2}(1+\lambda_{F}^{1\prime}Q_{\bar{F}%
\bar{F}}^{-1}\lambda_{F}^{1})^{-1}(1+\lambda_{F}^{\ast\prime}Q_{\bar{F}\bar{F%
}}^{-1}\lambda_{F}^{\ast})^{-1}+ \\ 
& 2\psi_{\perp}^{\prime}\left[ \Omega^{-\frac{1}{2}}D(\lambda_{F}^{\ast
})\left( \lambda_{F}^{\ast}-\lambda_{F}^{1}\right) (1+\lambda_{F}^{1\prime
}Q_{\bar{F}\bar{F}}^{-1}\lambda_{F}^{1})^{-\frac{1}{2}}\right. + \\ 
& \left. \Omega^{\frac{1}{2}}D(\lambda_{F}^{\ast})_{\perp}\delta\left(
1+\lambda_{F}^{\ast\prime}Q_{\bar{F}\bar{F}}^{-1}\lambda_{F}^{1}\right)
(1+\lambda_{F}^{1\prime}Q_{\bar{F}\bar{F}}^{-1}\lambda_{F}^{1})^{-\frac{1}{2}%
}(1+\lambda_{F}^{\ast\prime}Q_{\bar{F}\bar{F}}^{-1}\lambda_{F}^{\ast})^{-%
\frac{1}{2}}\right] +\psi_{\perp}^{\prime}\psi_{\perp} \\ 
& 
\end{array}%
\end{equation*}
\begin{equation*}
\begin{array}{cl}
S_{\lambda_{F}^{1}\perp}= & (Q_{\bar{F}\bar{F}}+\lambda_{F}^{1}\lambda
_{F}^{1\prime})^{-\frac{1}{2}\prime}\left( Q_{\bar{F}\bar{F}}+\lambda
_{F}^{\ast}\lambda_{F}^{1\prime}\right) ^{\prime}D(\lambda_{F}^{\ast
})^{\prime}\Omega^{-1}D(\lambda_{F}^{\ast})\left(
\lambda_{F}^{\ast}-\lambda_{F}^{1}\right) (1+\lambda_{F}^{1\prime}Q_{\bar{F}%
\bar{F}}^{-1}\lambda_{F}^{1})^{-\frac{1}{2}}- \\ 
& (1+\lambda_{F}^{\ast\prime}Q_{FF}^{-1}\lambda_{F}^{\ast})^{-1}(Q_{\bar {F}%
\bar{F}}+\lambda_{F}^{1}\lambda_{F}^{1\prime})^{-\frac{1}{2}\prime}\left(
\lambda_{F}^{\ast}-\lambda_{F}^{1}\right) \delta^{\prime}D(\lambda_{F}^{\ast
})_{\perp}^{\prime}\Omega D(\lambda_{F}^{\ast})_{\perp}\delta\left(
1+\lambda_{F}^{\ast\prime}Q_{\bar{F}\bar{F}}^{-1}\lambda_{F}^{1}\right) \\ 
& (1+\lambda_{F}^{1\prime}Q_{\bar{F}\bar{F}}^{-1}\lambda_{F}^{1})^{-\frac {1%
}{2}}+\left[ (Q_{\bar{F}\bar{F}}+\lambda_{F}^{1}\lambda_{F}^{1\prime })^{-%
\frac{1}{2}\prime}\left( Q_{\bar{F}\bar{F}}+\lambda_{F}^{\ast}\lambda_{F}^{1%
\prime}\right) ^{\prime}D(\lambda_{F}^{\ast})^{\prime}\Omega^{-\frac{1}{2}%
}+\right. \\ 
& \left. (1+\lambda_{F}^{\ast\prime}Q_{FF}^{-1}\lambda_{F}^{\ast})^{-\frac {1%
}{2}}(Q_{\bar{F}\bar{F}}+\lambda_{F}^{1}\lambda_{F}^{1\prime})^{-\frac{1}{2}%
\prime}\left( \lambda_{F}^{1}-\lambda_{F}^{\ast}\right) \delta^{\prime
}D(\lambda_{F}^{\ast})_{\perp}^{\prime}\Omega^{\frac{1}{2}}\right]
\psi_{\perp}+ \\ 
& \psi_{\lambda_{F}^{1}}^{\prime}\left[ \Omega^{-\frac{1}{2}}D(\lambda
_{F}^{\ast})\left( \lambda_{F}^{\ast}-\lambda_{F}^{1}\right) (1+\lambda
_{F}^{1\prime}Q_{\bar{F}\bar{F}}^{-1}\lambda_{F}^{1})^{-\frac{1}{2}}\right. +
\\ 
& \left. \Omega^{\frac{1}{2}}D(\lambda_{F}^{\ast})_{\perp}\delta\left(
1+\lambda_{F}^{\ast\prime}Q_{\bar{F}\bar{F}}^{-1}\lambda_{F}^{1}\right)
(1+\lambda_{F}^{1\prime}Q_{\bar{F}\bar{F}}^{-1}\lambda_{F}^{1})^{-\frac{1}{2}%
}(1+\lambda_{F}^{\ast\prime}Q_{\bar{F}\bar{F}}^{-1}\lambda_{F}^{\ast})^{-%
\frac{1}{2}}\right] +\psi_{\lambda_{F}^{1}}^{\prime}\psi_{\perp}.%
\end{array}%
\end{equation*}
Using further that $D(\lambda_{F}^{\ast})_{\perp}^{\prime}\Omega D(\lambda
_{F}^{\ast})_{\perp}=I_{N-m},$ $m=1$ so $(1+\lambda_{F}^{1\prime}Q_{\bar {F}%
\bar{F}}^{-1}\lambda_{F}^{1})=(1+(\lambda_{F}^{1})^{2}Q_{\bar{F}\bar{F}%
}^{-1})=Q_{\bar{F}\bar{F}}^{-1}(Q_{\bar{F}\bar{F}}+\lambda_{1}\lambda
_{1}^{\prime}),$ the above can be specified as:%
\begin{equation*}
\begin{array}{rl}
S_{\lambda_{F}^{1}\lambda_{F}^{1}}= & (Q_{\bar{F}\bar{F}}+(%
\lambda_{F}^{1})^{2})^{-1\prime}\left( Q_{\bar{F}\bar{F}}+\lambda_{F}^{\ast}%
\lambda _{F}^{1}\right)
^{2}D(\lambda_{F}^{\ast})^{\prime}\Omega^{-1}D(\lambda _{F}^{\ast})+ \\ 
& (1+\lambda_{F}^{\ast\prime}Q_{FF}^{-1}\lambda_{F}^{\ast})^{-1}(Q_{\bar {F}%
\bar{F}}+(\lambda_{F}^{1})^{2})^{-1}\left( \lambda_{F}^{1}-\lambda
_{F}^{\ast}\right) ^{2}\delta^{\prime}\delta+ \\ 
& 2(Q_{\bar{F}\bar{F}}+(\lambda_{F}^{1})^{2})^{-\frac{1}{2}}\psi_{\lambda
_{F}^{1}}^{\prime}\left[ \Omega^{-\frac{1}{2}}D(\lambda_{F}^{\ast})(Q_{\bar{F%
}\bar{F}}+\lambda_{F}^{\ast}\lambda_{F}^{1\prime})+\right. \\ 
& \left. \Omega^{\frac{1}{2}}D(\lambda_{F}^{\ast})_{\perp}\delta\left(
\lambda_{F}^{1}-\lambda_{F}^{\ast}\right)
(1+(\lambda_{F}^{\ast})^{2}Q_{FF}^{-1})^{-\frac{1}{2}}\right]
+\psi_{\lambda_{F}^{1}}^{\prime}\psi_{\lambda_{F}^{1}} \\ 
S_{\perp\perp}= & (1+(\lambda_{F}^{1})^{2}Q_{\bar{F}\bar{F}%
}^{-1})^{-1}\left( \lambda_{F}^{\ast}-\lambda_{F}^{1}\right)
^{\prime}D(\lambda_{F}^{\ast
})^{\prime}\Omega^{-1}D(\lambda_{F}^{\ast})\left(
\lambda_{F}^{\ast}-\lambda_{F}^{1}\right) + \\ 
& \delta^{\prime}\delta\left( 1+\lambda_{F}^{\ast}Q_{\bar{F}\bar{F}%
}^{-1}\lambda_{F}^{1}\right) ^{2}(1+(\lambda_{F}^{1})^{2}Q_{\bar{F}\bar{F}%
}^{-1})^{-1}(1+(\lambda_{F}^{\ast})^{2}Q_{\bar{F}\bar{F}}^{-1})^{-1}+ \\ 
& 2\psi_{\perp}^{\prime}\left[ \Omega^{-\frac{1}{2}}D(\lambda_{F}^{\ast
})\left( \lambda_{F}^{\ast}-\lambda_{F}^{1}\right)
(1+(\lambda_{F}^{1})^{2}Q_{\bar{F}\bar{F}}^{-1})^{-\frac{1}{2}}\right. + \\ 
& \left. \Omega^{\frac{1}{2}}D(\lambda_{F}^{\ast})_{\perp}\delta\left(
1+\lambda_{F}^{\ast\prime}Q_{\bar{F}\bar{F}}^{-1}\lambda_{F}^{1}\right)
(1+(\lambda_{F}^{1})^{2}Q_{\bar{F}\bar{F}}^{-1})^{-\frac{1}{2}}(1+(\lambda
_{F}^{\ast})^{2}Q_{\bar{F}\bar{F}}^{-1})^{-\frac{1}{2}}\right] +\psi_{\perp
}^{\prime}\psi_{\perp} \\ 
S_{\lambda_{F}^{1}\perp}= & \left( \lambda_{F}^{\ast}-\lambda_{F}^{1}\right)
(Q_{\bar{F}\bar{F}}+(\lambda_{F}^{1})^{2})^{-\frac{1}{2}}(1+(%
\lambda_{F}^{1})^{2}Q_{\bar{F}\bar{F}}^{-1})^{-\frac{1}{2}}\left[ \left( Q_{%
\bar{F}\bar{F}}+\lambda_{F}^{\ast}\lambda_{F}^{1\prime}\right)
D(\lambda_{F}^{\ast })^{\prime}\Omega^{-1}D(\lambda_{F}^{\ast})-\right. \\ 
& \left.
(1+(\lambda_{F}^{\ast})^{2}Q_{FF}^{-1})^{-1}\delta^{\prime}\delta\left(
1+\lambda_{F}^{\ast\prime}Q_{\bar{F}\bar{F}}^{-1}\lambda_{F}^{1}\right) %
\right] + \\ 
& \left[ (Q_{\bar{F}\bar{F}}+(\lambda_{F}^{1})^{2})^{-\frac{1}{2}\prime
}\left( Q_{\bar{F}\bar{F}}+\lambda_{F}^{\ast}\lambda_{F}^{1\prime}\right)
^{\prime}D(\lambda_{F}^{\ast})^{\prime}\Omega^{-\frac{1}{2}}+\right. \\ 
& \left. (1+(\lambda_{F}^{\ast})^{2}Q_{FF}^{-1})^{-\frac{1}{2}}(Q_{\bar {F}%
\bar{F}}+(\lambda_{F}^{1})^{2})^{-\frac{1}{2}\prime}\left(
\lambda_{F}^{1}-\lambda_{F}^{\ast}\right)
\delta^{\prime}D(\lambda_{F}^{\ast})_{\perp }^{\prime}\Omega^{\frac{1}{2}}%
\right]\psi_{\perp}+ \\ 
& \psi_{\lambda_{F}^{1}}^{\prime}\left[ \Omega^{-\frac{1}{2}}D(\lambda
_{F}^{\ast})\left( \lambda_{F}^{\ast}-\lambda_{F}^{1}\right) (1+(\lambda
_{F}^{1})^{2}Q_{\bar{F}\bar{F}}^{-1})^{-\frac{1}{2}}\right. + \\ 
& \left. \Omega^{\frac{1}{2}}D(\lambda_{F}^{\ast})_{\perp}\delta\left(
1+\lambda_{F}^{\ast\prime}Q_{\bar{F}\bar{F}}^{-1}\lambda_{F}^{1}\right)
(1+(\lambda_{F}^{1})^{2}Q_{\bar{F}\bar{F}}^{-1})^{-\frac{1}{2}}(1+(\lambda
_{F}^{\ast})^{2}Q_{\bar{F}\bar{F}}^{-1})^{-\frac{1}{2}}\right] +\psi
_{\lambda_{F}^{1}}^{\prime}\psi_{\perp}.%
\end{array}%
\end{equation*}
Since $\Omega^{-\frac{1}{2}}\hat{D}(\lambda_{F}^{1})^{\ast}$ and $\Omega^{-%
\frac{1}{2}}\hat{\mu}(\lambda_{F})^{\ast}$ are independently normal
distributed with identity covariance matrices, the quadratic form of $%
(\Omega^{-\frac{1}{2}}\hat{D}(\lambda_{F}^{1})^{\ast}$ $\vdots$ $\Omega^{-%
\frac{1}{2}}\hat{\mu}(\lambda_{F})^{\ast})$ with $T$ degrees of freedom,
identity scale matrices and a non-centrality parameter which is the
quadratic form of the mean of the distribution of $(\Omega^{-\frac{1}{2}}%
\hat{D}(\lambda_{F}^{1})^{\ast}$ $\vdots$ $\Omega^{-\frac{1}{2}}\hat{\mu }%
(\lambda_{F})^{\ast}),$ which read:

\text{Correct specification: }\newline
\begin{equation*}
\left( 
\begin{array}{c}
(\lambda_{F}^{\ast}-\lambda_{F}^{1})(1+(\lambda_{F}^{1})^{2}Q_{\bar{F}\bar{F}%
}^{-1})^{-\frac{1}{2}} \\ 
(Q_{\bar{F}\bar{F}}+(\lambda_{F}^{1})^{2})^{-\frac{1}{2}}\left( Q_{\bar {F}%
\bar{F}}+\lambda_{F}^{\ast}\lambda_{F}^{1\prime}\right)%
\end{array}
\right) \ddot{\beta}^{\prime}\Omega^{-1}\ddot{\beta}\left( 
\begin{array}{c}
(\lambda_{F}^{\ast}-\lambda_{F}^{1})(1+(\lambda_{F}^{1})^{2}Q_{\bar{F}\bar{F}%
}^{-1})^{-\frac{1}{2}} \\ 
(Q_{\bar{F}\bar{F}}+(\lambda_{F}^{1})^{2})^{-\frac{1}{2}}\left( Q_{\bar {F}%
\bar{F}}+\lambda_{F}^{\ast}\lambda_{F}^{1\prime}\right)%
\end{array}
\right) ^{\prime}\newline
\end{equation*}

\text{Misspecification:} \newline
\begin{equation*}
\begin{array}{l}
\left( 
\begin{array}{c}
(\lambda_{F}^{\ast}-\lambda_{F}^{1})(1+(\lambda_{F}^{1})^{2}Q_{\bar{F}\bar{F}%
}^{-1})^{-\frac{1}{2}} \\ 
(Q_{\bar{F}\bar{F}}+(\lambda_{F}^{1})^{2})^{-\frac{1}{2}}\left( Q_{\bar {F}%
\bar{F}}+\lambda_{F}^{\ast}\lambda_{F}^{1\prime}\right)%
\end{array}
\right) D(\lambda_{F}^{\ast})^{\prime}\Omega^{-1}D(\lambda_{F}^{\ast})\left( 
\begin{array}{c}
(\lambda_{F}^{\ast}-\lambda_{F}^{1})(1+(\lambda_{F}^{1})^{2}Q_{\bar{F}\bar{F}%
}^{-1})^{-\frac{1}{2}} \\ 
(Q_{\bar{F}\bar{F}}+(\lambda_{F}^{1})^{2})^{-\frac{1}{2}}\left( Q_{\bar {F}%
\bar{F}}+\lambda_{F}^{\ast}\lambda_{F}^{1\prime}\right)%
\end{array}
\right) ^{\prime}+ \\ 
\left( 
\begin{array}{c}
(1+(\lambda_{F}^{1})^{2}Q_{\bar{F}\bar{F}}^{-1})^{-\frac{1}{2}}\left(
1+\lambda_{F}^{\ast}Q_{\bar{F}\bar{F}}^{-1}\lambda_{F}^{1}\right) \\ 
-(Q_{\bar{F}\bar{F}}+(\lambda_{F}^{1})^{2})^{-\frac{1}{2}}\left( \lambda
_{F}^{\ast}-\lambda_{F}^{1}\right)%
\end{array}
\right) (1+(\lambda_{F}^{\ast})^{2}Q_{\bar{F}\bar{F}}^{-1})^{-1}\delta^{%
\prime}\delta\left( 
\begin{array}{c}
(1+(\lambda_{F}^{1})^{2}Q_{\bar{F}\bar{F}}^{-1})^{-\frac{1}{2}}\left(
1+\lambda_{F}^{\ast}Q_{\bar{F}\bar{F}}^{-1}\lambda_{F}^{1}\right) \\ 
-(Q_{\bar{F}\bar{F}}+(\lambda_{F}^{1})^{2})^{-\frac{1}{2}}\left( \lambda
_{F}^{\ast}-\lambda_{F}^{1}\right)%
\end{array}
\right) ^{\prime}.%
\end{array}%
\end{equation*}

\section{ \ Simulation setup for the CRRA moment function}

We use a log-normal data generating process to simulate consumption growth
and asset returns in accordance with the CRRA moment condition.

Let $%
\triangle c_{t+1}=\ln \left( \frac{C_{t+1}}{C_{t}}\right) $ and $r_{t+1}=\ln
(\iota _{N}+R_{t+1})$, which are i.i.d. normally distributed: 
\begin{eqnarray*}
\left[ 
\begin{array}{c}
\triangle c_{t+1} \\ 
r_{t+1}%
\end{array}%
\right] &\sim& NID(\mu ,V) \\
&\equiv& NID\left( \left[ 
\begin{array}{c}
0 \\ 
\mu _{2,0}%
\end{array}%
\right] ,\left[ 
\begin{array}{cc}
V_{cc,0} & V_{cr,0} \\ 
V_{rc,0} & V_{rr,0}%
\end{array}%
\right] \right) ,
\end{eqnarray*}%
with $\mu _{2,0}=(\mu _{2,1,0}\ldots \mu _{2,N,0})^{\prime }$ the mean of $%
r_{t+1},$ $V_{cc,0}$ the (scalar) variance of $\triangle c_{t+1},$ $%
V_{rc,0}=V_{cr,0}^{\prime }=(V_{rc,1,0}\ldots V_{rc,N,0})^{\prime }$ the $%
N\times 1$ dimensional covariance between $r_{t+1}$ and $\triangle c_{t+1}$
and $V_{rr,0}=(V_{rr,ij,0}):$ $i,j=1,\ldots ,N,$ the $N\times N$ dimensional
covariance matrix of $r_{t+1}.$

This DGP has also been used in Kleibergen and Zhan (2020), where the covariance matrix $%
V=[V_{cc,0},V_{cr,0};V_{rc,0},V_{rr,0}]$ is calibrated to data. We will change
the value of $\mu _{2,0}$ to vary the magnitude of the misspecification. We will
also alter the correlation coefficient of $\triangle c_{t+1}$ and $r_{t+1}$
to vary identification.\nocite%
{kz19}

Given pre-set values of $\delta _{0},$ $\mu
_{2,0},$ $V_{cc,0},$ $V_{rc,0}$ and $V_{rr,0},$ the CRRA moment equation is
such that: 
\begin{align*}
 \mu _{f}(\gamma )
& =E\left[ \delta _{0}\left( \frac{C_{t+1}}{C_{t}}\right)
^{-\gamma }(\iota _{N}+R_{t+1})-\iota _{N}\right]  \\
&=E\left[ \left( 
\begin{array}{c}
\exp \left( \ln (\delta _{0})-\gamma \triangle c_{t+1}+r_{t+1,1}\right) \\ 
\vdots \\ 
\exp \left( \ln (\delta _{0})-\gamma \triangle c_{t+1}+r_{t+1,N}\right)%
\end{array}%
\right) -\iota _{N}\right] \\
& =\left( 
\begin{array}{c}
\exp \left( \ln (\delta _{0})+\mu _{2,1,0}+\frac{1}{2}\left(
V_{rr,11,0}+\gamma ^{2}V_{cc,0}-2\gamma V_{rc,1,0}\right) \right) \\ 
\vdots \\ 
\exp \left( \ln (\delta _{0})+\mu _{2,N,0}+\frac{1}{2}\left(
V_{rr,NN,0}+\gamma ^{2}V_{cc,0}-2\gamma V_{rc,N,0}\right) \right)%
\end{array}%
\right) -\iota _{N}.
\end{align*}%

We also need the explicit expression of $V_{ff}(\gamma )$: 

\begin{align*}
V_{ff}(\gamma )& =E\left[ (f_{t}(\gamma )-\mu _{f}(\gamma ))(f_{t}(\gamma
)-\mu _{f}(\gamma ))^{\prime }\right]  =Var\left( e^{\ln (\delta )-\gamma \triangle c_{t+1}+r_{t+1}}\right) \\
& =\left( \left( 
\begin{array}{c}
\exp \left( \ln (\delta _{0})+\mu _{2,1,0}+\frac{1}{2}\left(
V_{rr,11,0}+\gamma ^{2}V_{cc,0}-2\gamma V_{rc,1,0}\right) \right) \\ 
\vdots \\ 
\exp \left( \ln (\delta _{0})+\mu _{2,N,0}+\frac{1}{2}\left(
V_{rr,NN,0}+\gamma ^{2}V_{cc,0}-2\gamma V_{rc,N,0}\right) \right)%
\end{array}%
\right) \right. \\
& \left. \left( 
\begin{array}{c}
\exp \left( \ln (\delta _{0})+\mu _{2,1,0}+\frac{1}{2}\left(
V_{rr,11,0}+\gamma ^{2}V_{cc,0}-2\gamma V_{rc,1,0}\right) \right) \\ 
\vdots \\ 
\exp \left( \ln (\delta _{0})+\mu _{2,N,0}+\frac{1}{2}\left(
V_{rr,NN,0}+\gamma ^{2}V_{cc,0}-2\gamma V_{rc,N,0}\right) \right)%
\end{array}%
\right) ^{\prime }\right) \odot \\
& \left( \exp \left( \left( -\gamma \iota _{N}\text{ }\vdots \text{ }%
I_{N}\right) \left[ 
\begin{array}{cc}
V_{cc,0} & V_{cr,0} \\ 
V_{rc,0} & V_{rr,0}%
\end{array}%
\right] \left( -\gamma \iota _{N}\text{ }\vdots \text{ }I_{N}\right)
^{\prime }\right) -\iota _{N}\iota _{N}^{\prime }\right) ,
\end{align*}%
where $\odot $ stands for element-by-element multiplication. \bigskip

\end{document}